\documentclass[twoside,fleqn]{ActaStyle1}
\usepackage{times,cite}
\usepackage{bm}

\usepackage{lscape}             
\usepackage{graphicx,epsfig}    
\usepackage[tbtags]{amsmath}    
\usepackage{amssymb}

\def\authorheading{S.~Scheel and S.Y.~Buhmann}
\def\shorttitle{Macroscopic QED --- concepts and applications}

\newcommand{\tens}[1]{\bm{#1}}
\newcommand{\vect}[1]{\mathbf{#1}}
\newcommand{\grad}{\bm{\nabla}}
\newcommand{\curl}{\bm{\nabla}\times}
\newcommand{\divv}{\bm{\nabla}\cdot}
\newcommand{\ueber}[2]{\genfrac{}{}{0pt}{}{#1}{#2}}
\newcommand{\dd}{\displaystyle}
\newcommand{\ten}[1]{\bm{#1}}
\renewcommand{\vec}[1]{\mathbf{#1}}

\newcommand{\veczero}{\mathbf{0}}
\newcommand{\sprod}{\cdot}
\newcommand{\tprod}{\otimes}
\newcommand{\vprod}{\times}
\newcommand{\trace}{{\operatorname{tr}}}
\newcommand{\trans}{{\operatorname{T}}}
\newcommand{\dif}{d}
\newcommand{\mi}{i} 
\newcommand{\me}{e}

\begin{document}

\pagerange{1}{133}
\title{Macroscopic quantum electrodynamics --- concepts and applications}

\author{Stefan~Scheel\email{s.scheel@imperial.ac.uk}
and Stefan~Yoshi~Buhmann\email{s.buhmann@imperial.ac.uk}}
{Quantum Optics and Laser Science, Blackett Laboratory,
Imperial College London, Prince Consort Road, London SW7 2AZ, United
Kingdom} 

\day{October 9, 2008}

\abstract{In this article, we review the principles of macroscopic
quantum electrodynamics and discuss a variety of applications of this
theory to medium-assisted atom-field coupling and dispersion
forces. The theory generalises the standard mode expansion of the
electromagnetic fields in free space to allow for the presence of
absorbing bodies. We show that macroscopic quantum electrodynamics
provides the link between isolated atomic systems and magnetoelectric
bodies, and serves as an important tool for the understanding of
surface-assisted atomic relaxation effects and the intimately connected
position-dependent energy shifts which give rise to Casimir--Polder
and van~der~Waals forces.}


\tableofcontents
\newpage


\section{Introduction}

In this article we review the basic principles, latest developments
and important applications of macroscopic quantum electrodynamics
(QED). This theory extends the well-established quantum optics in free
space (Sec.~\ref{sec:vacuumQED}) to include absorbing and dispersing
magnetoelectric bodies in its Hamiltonian description. In that way, a
connection is established between isolated atomic systems (atoms,
ions, molecules, Bose--Einstein condensates) and absorbing materials
(dielectrics, metals, superconductors). This is achieved by means of a
quantisation scheme for the medium-assisted electromagnetic fields
(Sec.~\ref{sec:causalQED}).

We set the scene by reviewing the basic elements of quantum optics in
free space (Sec.~\ref{sec:vacuumQED}). Beginning with the quantisation
of the electromagnetic field in free space in a Lagrangian formalism
(Sec.~\ref{sec:lagrangian}) and based on Maxwell's equations
(Sec.~\ref{sec:maxwell}), we briefly review two important
applications, the lossless beam splitter (Sec.~\ref{sec:beamsplitter})
and the mode summation approach to Casimir forces
(Sec.~\ref{sec:modecasimir}). We introduce minimal-coupling and
multipolar-coupling schemes (Sec.~\ref{sec:atomsinvacuum}) and discuss
important consequences of the quantised atom-field coupling,
spontaneous decay and the Lamb shift (Sec.~\ref{sec:vacuumse}). For
completeness, we briefly review optical Bloch equations
(Sec.~\ref{sec:bloch}) and the Jaynes--Cummings model
(Sec.~\ref{sec:jc}). 

The main part of this review deals with macroscopic quantum
electrodynamics (Sec.~\ref{sec:causalQED}) and its applications
(Secs.~\ref{sec:relaxation}--\ref{sec:cavityQED}). Macroscopic quantum
electrodynamics provides the foundations for investigations into
quantum-mechanical effects related to the presence of magnetoelectric
bodies or interfaces such as dispersion forces and medium-assisted
atomic relaxation and heating rates. The quantisation scheme is based
on an expansion of the electromagnetic field operators in terms of
dyadic Green functions, the fundamental solutions to the Helmholtz
equation (Sec.~\ref{sec:langevin}). We discuss the principles of
coupling atoms to the medium-assisted electromagnetic field by means
of the minimal-coupling and multipolar-coupling schemes
(Sec.~\ref{sec:interaction}), the latter of which will be used
extensively throughout the article. After deriving the basic
relations, we  first focus our attention on medium-assisted atomic
relaxation rates (Sec.~\ref{sec:relaxation}). We present examples of
modified spontaneous decay, near-field spin relaxation, heating and
local-field corrections. As we frequently refer to a number of
explicit formulas for multilayered media, we have collected some of
the most important relations in the Appendix (App.~\ref{sec:dgf}).

The Kramers--Kronig relations provide a close connection between the
relaxation rates (line widths) and the corresponding energy shifts
(Lamb shifts). The Lamb shift already exists in free space where the
bare atomic transition frequencies are modified due to the interaction
with the quantum vacuum. Because the quantum vacuum, i.e. the 
electromagnetic field fluctuations, are altered due to the presence of
magnetoelectric bodies, these energy shifts become position-dependent
and hence lead to dispersion forces. We develop the theory of Casimir,
Casimir--Polder (CP) and van~der~Waals (vdW) forces on the basis of
those field fluctuations (Sec.~\ref{sec:dispersion}). Amongst other 
examples, we discuss under which circumstances the results based on
mode summations and perfect boundary conditions (introduced in
Sec.~\ref{sec:modecasimir}) can be retrieved.

Finally, we apply the theory of macroscopic quantum electrodynamics to
strong atom-field coupling effects in microresonators
(Sec.~\ref{sec:cavityQED}). Here we discuss leaky optical cavities
from a field-theoretic point of view (Sec.~\ref{sec:extraction}) which
provides insight into input-output coupling at semi-transparent cavity
mirrors and generalises the Jaynes--Cummings model
(Sec.~\ref{sec:jc}). We further present an example for entanglement
generation between two atoms that utilises surface-guided modes on a
spherical microresonator (Sec.~\ref{sec:guidedmodes}). 


\section{Elements of vacuum quantum electrodynamics}
\label{sec:vacuumQED}

Let us begin our investigations of quantum electrodynamics by
revisiting some of the basics of this theory --- QED in free
space. There are essentially two ways of approaching the quantisation
of the electromagnetic field. In the quantum field theory literature
(see, e.g. \cite{Itzykson}), the formal route is taken in which a
Lagrangian is postulated that fulfils certain general requirements
such as relativistic covariance (Sec.~\ref{sec:lagrangian}). In order
to stress the intimate connection with classical optics
\cite{BornWolf}, we instead follow a second approach by sticking to
classical Maxwell theory as long as possible before quantising
(Sec.~\ref{sec:maxwell}). 

A simple application of the mode expansion that will be employed in
this section is the description of the lossless beam splitter
(Sec.~\ref{sec:beamsplitter}) which will be extended to lossy devices
in Sec.~\ref{sec:lossybeamsplitter}. The mode expansion approach will
also be used to discuss the Casimir force between two perfectly
conducting plates (Sec.~\ref{sec:modecasimir}). As we will later see
in Sec.~\ref{sec:lorentzcasimir}, this interpretation cannot be upheld
if the rather severe approximation of perfectly conducting plates is
being weakened. Instead, we will have to describe Casimir forces
(and related forces) in terms of fluctuating dipole forces.

Next, we consider the coupling of the quantised electromagnetic field
to charged particles (Sec.~\ref{sec:atomsinvacuum}). We will introduce
the notion of Kramers--Kronig relations and discuss simple atom-field
phenomena such as spontaneous decay, the Lamb shift, the optical 
Bloch equations and the Jaynes--Cummings model of cavity QED. These
examples will play a major role in our subsequent discussion of
macroscopic QED (Secs.~\ref{sec:relaxation} and \ref{sec:dispersion}).


\subsection{Quantisation of the electromagnetic field in free space}

In this section, we briefly describe the theory of quantum
electrodynamics in free space. We outline both the usual Lagrangian
formalism and the more {\em ad hoc} approach via Maxwell's equations
that highlights the connections with classical optics.


\subsubsection{Lagrangian formalism}
\label{sec:lagrangian}

Within the framework of U(1)-gauge theories, the coupling between the
fermionic matter fields and the electromagnetic field is described by
a gauge potential $A^\mu=(\phi/c,\vect{A})$ which is identified as
the four-vector of scalar and vector potentials. In order to determine
the dynamics of $A^\mu$ in the Lagrangian formalism, a Lorentz
covariant combination in terms of derivatives of $A^\mu$ has to be
sought, the simplest of which is the combination
\begin{equation}
\label{eq:lagrangiandensity}
L=\int d^4x {\cal L}= -\frac{1}{4\mu_0} \int d^4x F_{\mu\nu}F^{\mu\nu}
= \frac{1}{2} \int dt \int d^3r \left[
\varepsilon_0\vect{E}^2(\vect{r},t)
-\frac{1}{\mu_0}\vect{B}^2(\vect{r},t) \right]
\end{equation}
with the (covariant) anti-symmetric tensor
\begin{equation}
\label{eq:Fmunu}
F_{\mu\nu} = \partial_\mu A_\nu - \partial_\nu A_\mu 
= \left( \begin{array}{cccc}
0 & E_x/c & E_y/c & E_z/c \\
-E_x/c & 0 & -B_z & B_y \\
-E_y/c & B_z & 0 & -B_x \\
-E_z/c & -B_y & B_x & 0
\end{array} \right)\,.
\end{equation}
Recall that the contravariant components of a four-vector $x^\mu$ can
be derived from its covariant components $x_\mu$ by contraction with
the metric tensor $g^{\mu\nu}=\mathrm{diag}(1,-1,-1,-1)$,
$x^\mu=g^{\mu\nu}x_\nu$. 

The equations of motion that follow from the Lagrangian density
(\ref{eq:lagrangiandensity}),
\begin{equation}
\partial_\mu F^{\mu\nu} =0 \,,\qquad
\partial_\mu \epsilon^{\mu\nu\rho\sigma} F_{\rho\sigma} = 0
\end{equation}
[$\epsilon^{\mu\nu\rho\sigma}$: completely anti-symmetric symbol],
are equivalent to Maxwell's equations
\begin{eqnarray}
\label{eq:m1}
\divv\vect{B}(\vect{r},t) &=& 0\,,\\
\label{eq:m2}
\curl\vect{E}(\vect{r},t) &=& -\dot{\vect{B}}(\vect{r},t) \,,\\
\label{eq:m3}
\divv\vect{D}(\vect{r},t) &=& 0\,,\\
\label{eq:m4}
\curl\vect{H}(\vect{r},t) &=& \dot{\vect{D}}(\vect{r},t) \,.
\end{eqnarray}
They have to be supplemented with the free-space constitutive
relations 
\begin{eqnarray}
\label{eq:freeconstitutiveD}
\vect{D}(\vect{r},t) &=& \varepsilon_0 \vect{E}(\vect{r},t) \,,\\
\label{eq:freeconstitutiveH}
\vect{H}(\vect{r},t) &=& \frac{1}{\mu_0} \vect{B}(\vect{r},t) \,,
\end{eqnarray}
that connect the dielectric displacement 
field $\vect{D}(\vect{r},t)$ with the electric field
$\vect{E}(\vect{r},t)$ and the magnetic field $\vect{H}(\vect{r},t)$
with the induction field $\vect{B}(\vect{r},t)$.

Returning to the Lagrangian (\ref{eq:lagrangiandensity}), one
constructs the canonical momentum to the four-potential as
\begin{equation}
\Pi_\mu = \frac{\delta L}{\delta \dot{A}^\mu} \,.
\end{equation}
Its spatial components are proportional to the electric field,
$\bm{\Pi}=-\varepsilon_0\vect{E}$, whereas the component $\Pi_0$
vanishes due to the anti-symmetry of $F_{\mu\nu}$. This means that
there is no dynamical degree of freedom associated with the zero
component of the momentum field. Hence, the dynamics of the
electromagnetic field is constrained. Using the canonical momenta, one
introduces a Hamiltonian by means of a Legendre transform as
\begin{equation}
\label{eq:classicalH}
H = \frac{1}{2}\int dt \int d^3r \left[
\varepsilon_0\vect{E}^2(\vect{r},t)
+\frac{1}{\mu_0}\vect{B}^2(\vect{r},t) \right] \,.
\end{equation}

An additional complication arises due to the gauge freedom of
electrodynamics. From the definition of $F_{\mu\nu}$,
Eq.~(\ref{eq:Fmunu}), it is clear that adding the four-divergence of
an arbitrary scalar function $\Lambda$,
\begin{equation}
A_\mu \mapsto A_\mu +\partial_\mu \Lambda \,,
\end{equation}
does not alter the equations of motion, i.e. Maxwell's equations. One
is therefore free to choose a gauge function $\Lambda$ that is best
suited to simplify actual computations. Clearly, any physically
observable quantities derived from the electromagnetic fields are
independent of this choice of gauge. A particular choice that
preserves the relativistic covariance of Maxwell's equations is the
Lorentz gauge in which one imposes the constraint
$\partial_\mu A^\mu=0$. In quantum optics, where relativistic
covariance is not needed because the external sources envisaged there
rarely move with any appreciable speed, the Coulomb gauge is often
chosen. Here, one sets
\begin{equation}
\label{eq:coulomb}
\phi =0 \,,\qquad \divv \vect{A} =0\,,
\end{equation}
which obviously breaks relativistic covariance. Hence, in free space
there are only two independent components of the vector potential. The
scalar potential is identically zero; this is actually a consequence
of the requirement $\divv\vect{A}=0$ rather than a separate
constraint. In the Coulomb gauge, the Poisson bracket between the
dynamical variables and their respective canonical momenta reads
\begin{equation}
\left\{ \vect{A}(\vect{r},t), \bm{\Pi}(\vect{r}',t) \right\} =
\tens{\delta}^\perp(\vect{r}-\vect{r}') 
\end{equation}
where $\tens{\delta}^\perp(\vect{r})$ denotes the transverse $\delta$
function. With the relations $\vect{E}=-\bm{\Pi}/\varepsilon_0$ and
$\vect{B}=\curl\vect{A}$, the Poisson bracket for these fields simply
read
\begin{equation}
\label{eq:poisson}
\left\{ \vect{E}(\vect{r},t) , \vect{B}(\vect{r}',t)  \right\} =
-\frac{1}{\varepsilon_0} \curl \tens{\delta}^\perp(\vect{r}-\vect{r}')
\,, 
\end{equation}
which serves as the fundamental relation between the electromagnetic
fields. At this point, canonical field quantisation can be performed
in the usual way by means of the correspondence principle. Upon
quantisation, the Poisson bracket has to be replaced by
$(i\hbar)^{-1}$ times the commutator and Hamilton's equations of
motion have to replaced by Heisenberg's equations of motion.


\subsubsection{Maxwell's equations}
\label{sec:maxwell}

Instead of using the field-theoretic Lagrangian language, we adopt a
slightly simpler approach to quantisation that keeps aspects of the
classical theory for as long as possible. Maxwell's equations
(\ref{eq:m2}) and (\ref{eq:m4}) can equivalently be expressed in terms
of the vector potential [derivable from Eq.~(\ref{eq:Fmunu})]
\begin{equation}
\label{eq:classicalEB}
\vect{E}(\vect{r},t) = -\dot{\vect{A}}(\vect{r},t) \,,\qquad
\vect{B}(\vect{r},t) = \curl\vect{A}(\vect{r},t) \,.
\end{equation}
The vector potential $\vect{A}(\vect{r},t)$ in the Coulomb gauge
(\ref{eq:coulomb}) obeys the wave equation  
\begin{equation}
\label{eq:waveA}
\Delta\vect{A}(\vect{r},t) -\frac{1}{c^2}\ddot{\vect{A}}(\vect{r},t)
= \vect{0} \,.
\end{equation}
The solutions to Eq.~(\ref{eq:waveA}) can be found by separation of
variables, i.e. we make the {\em ansatz}
\begin{equation}
\vect{A}(\vect{r},t) = \sum\limits_\lambda \vect{A}_\lambda(\vect{r})
u_\lambda(t) 
\end{equation}
which amounts to a mode decomposition. The mode functions
$\vect{A}_\lambda(\vect{r})$ obey the Helmholtz equation
\begin{equation}
\label{eq:helmholtzA}
\Delta\vect{A}_\lambda(\vect{r}) +\frac{\omega_\lambda^2}{c^2}
\vect{A}_\lambda(\vect{r}) =\vect{0}
\end{equation}
where we defined the separation constant as $\omega_\lambda^2$ for
later convenience. One can read Eq.~(\ref{eq:helmholtzA}) as an
eigenvalue equation for the Hermitian operator $-\Delta$ having
eigenvalues $\omega_\lambda^2/c^2$ and eigenvectors
$\vect{A}_\lambda(\vect{r})$. Because of the Hermiticity of the
Laplace operator, the mode functions form a complete set of orthogonal
functions, albeit strictly only in a distributional sense. Hence,
\begin{equation}
\label{eq:csof}
\int d^3r\, \vect{A}^\ast_\lambda(\vect{r})\cdot
\vect{A}_{\lambda'}(\vect{r}) = \mathcal{N}_\lambda
\delta_{\lambda\lambda'}
\,,\qquad
\sum\limits_\lambda \frac{1}{\mathcal{N}_\lambda}
\vect{A}_\lambda(\vect{r}) \otimes \vect{A}^\ast_\lambda(\vect{r}')
= \tens{\delta}^\perp(\vect{r}-\vect{r}') 
\end{equation}
where $\mathcal{N}_\lambda$ denotes a normalisation factor.

The Helmholtz equation (\ref{eq:helmholtzA}) is easily solved in
cartesian coordinates. The solutions are plane waves
$\vect{A}_\lambda(\vect{r})=\vect{e}_\sigma(\vect{k})
e^{i\vect{k}\cdot\vect{r}}$ where 
the magnitude of the wavevector $\vect{k}$ obeys the dispersion
relation $k^2=\omega_\lambda^2/c^2$. For each wavevector $\vect{k}$
there are two orthogonal polarisations with unit vectors
$\vect{e}_\sigma(\vect{k})$ obeying 
$\vect{e}_\sigma(\vect{k})\cdot\vect{e}_{\sigma'}(\vect{k})$
$\!=\delta_{\sigma\sigma'}$ and
$\vect{e}_\sigma(\vect{k})\cdot\vect{k}=0$. Hence, the sum over
$\lambda$ has in fact the following meaning:
\begin{equation}
\sum\limits_\lambda \equiv \sum\limits_{\sigma=1}^2 \int
\frac{d^3k}{(2\pi)^{3/2}} \,.
\end{equation}
In cylindrical or spherical coordinates the solutions to the scalar
Helmholtz equation can be written in terms of cylindrical and
spherical Bessel functions, respectively (see
Appendix~\ref{sec:cylindricaldgf} and \ref{sec:sphericaldgf}).

The temporal part of the wave equation reduces to the differential
equation of a harmonic oscillator,
\begin{equation}
\ddot{u}_\lambda(t) +\omega_\lambda^2 u_\lambda(t) = 0\,,
\end{equation}
with solutions $u_\lambda(t)=e^{\pm i\omega_\lambda t}u_\lambda$.
Combining spatial and temporal parts, we obtain the mode decomposition
for the vector potential as ($\omega=kc$)
\begin{equation}
\label{eq:classicalA}
\vect{A}(\vect{r},t) = \sum\limits_{\sigma=1}^2 \int
\frac{d^3k}{(2\pi)^{3/2}} \, \vect{e}_\sigma(\vect{k}) \left[
u_{\vect{k}\sigma} e^{i(\vect{k}\cdot\vect{r}-\omega t)} +
u^\ast_{\vect{k}\sigma} e^{-i(\vect{k}\cdot\vect{r}-\omega t)} \right]
\,,
\end{equation}
where we have explicitly taken care of the reality of the vector
potential by imposing the condition
$u_{\vect{k}\sigma}(t)=u^\ast_{-\vect{k}\sigma}(t)$.

Writing the expressions (\ref{eq:classicalEB}) for the electric field
and the magnetic induction in terms of the vector potential
(\ref{eq:classicalA}),
\begin{eqnarray}
\vect{E}(\vect{r},t) &=& i \sum\limits_{\sigma=1}^2 \int
\frac{d^3k}{(2\pi)^{3/2}} \, \vect{e}_\sigma(\vect{k}) \omega \left[
u_{\vect{k}\sigma} e^{i(\vect{k}\cdot\vect{r}-\omega t)} -
u^\ast_{\vect{k}\sigma} e^{-i(\vect{k}\cdot\vect{r}-\omega t)} \right]
\,,\\
\vect{B}(\vect{r},t) &=& i \sum\limits_{\sigma=1}^2 \int
\frac{d^3k}{(2\pi)^{3/2}} \, \left[\vect{k}\times
\vect{e}_\sigma(\vect{k}) \right] \left[ u_{\vect{k}\sigma}
e^{i(\vect{k}\cdot\vect{r}-\omega t)} - u^\ast_{\vect{k}\sigma}
e^{-i(\vect{k}\cdot\vect{r}-\omega t)} \right] \,,
\end{eqnarray}
the Hamiltonian (\ref{eq:classicalH}) reads
\begin{eqnarray}
H &=& -\frac{1}{2} \sum\limits_{\sigma,\sigma'=1}^2 \iiint
\frac{d^3r\,d^3k\,d^3k'}{(2\pi)^3} \left[ \varepsilon_0
(\vect{e}_\sigma
\cdot \vect{e}_{\sigma'}) \omega\omega' +\frac{1}{\mu_0}
(\vect{k}\times\vect{e}_\sigma) \cdot
(\vect{k}'\times\vect{e}_{\sigma'}) \right]
\nonumber \\ && \hspace*{-3ex}
\times \left[ u_{\vect{k}\sigma} e^{i(\vect{k}\cdot\vect{r}-\omega t)}
-
u^\ast_{\vect{k}\sigma} e^{-i(\vect{k}\cdot\vect{r}-\omega t)} \right]
\left[ u_{\vect{k}'\sigma'} e^{i(\vect{k}'\cdot\vect{r}-\omega' t)} -
u^\ast_{\vect{k}'\sigma'} e^{-i(\vect{k}'\cdot\vect{r}-\omega' t)}
\right] \,.
\end{eqnarray}
Using the orthogonality of the polarisation vectors
$\vect{e}_\sigma\cdot\vect{e}_{\sigma'}=\delta_{\sigma\sigma'}$ as
well as the relation
$(\vect{k}\times\vect{e}_\sigma)\cdot(\vect{k}\times\vect{e}_{\sigma'}
)$
$\!=k^2(\vect{e}_\sigma\cdot\vect{e}_{\sigma'})$, and integrating over
$\vect{r}$ and $\vect{k}'$ leaves us with
\begin{equation}
H = 2\varepsilon_0 \sum\limits_{\sigma=1}^2 \int d^3k\, \omega^2
|u_{\vect{k}\sigma}|^2 \,.
\end{equation}
The complex-valued functions $u_{\vect{k}\sigma}$ can then be split
into their respective real and imaginary parts as
\begin{equation}
q_{\vect{k}\sigma} = \sqrt{\varepsilon_0} \left( u_{\vect{k}\sigma}
+u^\ast_{\vect{k}\sigma} \right) \,,\qquad
p_{\vect{k}\sigma} = -i\omega\sqrt{\varepsilon_0} \left(
u_{\vect{k}\sigma} -u^\ast_{\vect{k}\sigma} \right) \,,
\end{equation}
which finally yields the classical Hamiltonian in the form
\begin{equation}
\label{eq:harmonicoscillatorH}
H = \frac{1}{2} \sum\limits_{\sigma=1}^2 \int d^3k \left(
p^2_{\vect{k}\sigma}+\omega^2 q^2_{\vect{k}\sigma}  \right)  \,.
\end{equation}
In this way, we have converted the Hamiltonian (\ref{eq:classicalH})
of the classical electromagnetic field into an infinite sum of
uncoupled harmonic oscillators with frequencies $\omega=kc$.
The functions $q_{\vect{k}\sigma}$ and $p_{\vect{k}\sigma}$ are thus
analogous to the position and momentum of a classical particle of mass
$m$ attached to a spring with spring constant $k=m\omega^2$.

The conversion of a field Hamiltonian into a set of uncoupled harmonic
oscillators is the essence of every free-field quantisation scheme. In
cartesian coordinates it is equivalent to a decomposition into
uncoupled Fourier modes (or alternatively into Bessel--Fourier modes
if cylindrical or spherical coordinates are used). Note that the
introduction of mode functions with the completeness and orthogonality
relations (\ref{eq:csof}) circumvents the usual problem of having to
perform field quantisation in a space of finite extent, followed by
the limiting procedure to unbounded space at the end of the
calculation. The analogy with classical mechanics can be pushed even
further by noting that the functions $q_{\vect{k}\sigma}$ and
$p_{\vect{k}\sigma}$ obey the Poisson bracket relation
\begin{equation}
\label{eq:poissonbracket}
\left\{ q_{\vect{k}\sigma} , p_{\vect{k}'\sigma'} \right\}
= \delta(\vect{k}-\vect{k}') \delta_{\sigma\sigma'} \,.
\end{equation}

Quantisation is then performed by regarding the classical $c$-number
functions $q_{\vect{k}\sigma}$ and $p_{\vect{k}\sigma}$ as operators
in an abstract Hilbert space $\mathcal{H}$, and by replacing the
Poisson brackets (\ref{eq:poissonbracket}) by the respective
commutators times $(i\hbar)^{-1}$ \cite{VogelWelsch}:
\begin{equation}
q_{\vect{k}\sigma}\mapsto\hat{q}_{\vect{k}\sigma}
\,,\qquad
p_{\vect{k}\sigma}\mapsto\hat{p}_{\vect{k}\sigma}
\,,\qquad
\left[ \hat{q}_{\vect{k}\sigma} , \hat{p}_{\vect{k}'\sigma'} \right]
= i\hbar \delta(\vect{k}-\vect{k}') \delta_{\sigma\sigma'} \,.
\end{equation}
By returning to the complex amplitude functions, now with different
normalisation factors,
\begin{equation}
\label{eq:complexamplitude}
\hat{a}_\sigma(\vect{k}) = \sqrt{\frac{\omega}{2\hbar}}
\left( \hat{q}_{\vect{k}\sigma}
+\frac{i\hat{p}_{\vect{k}\sigma}}{\omega} \right) 
\,,\qquad
\hat{a}^\dagger_\sigma(\vect{k}) = \sqrt{\frac{\omega}{2\hbar}}
\left( \hat{q}_{\vect{k}\sigma}
-\frac{i\hat{p}_{\vect{k}\sigma}}{\omega} \right)\,,
\end{equation}
which obey the commutation rules
\begin{equation}
\left[ \hat{a}_\sigma(\vect{k}) , \hat{a}^\dagger_{\sigma'}(\vect{k}')
\right] = \delta(\vect{k}-\vect{k}') \delta_{\sigma\sigma'} \,,
\end{equation}
we can write the operator of the vector potential in the Schr\"odinger
picture as
\begin{equation}
\label{eq:cartesianA}
\hat{\vect{A}}(\vect{r}) = \sum\limits_{\sigma=1}^2 \int
\frac{d^3k}{(2\pi)^{3/2}} \sqrt{\frac{\hbar}{2\varepsilon_0\omega}}
\,\vect{e}_\sigma \left[ e^{i\vect{k}\cdot\vect{r}}
\hat{a}_\sigma(\vect{k}) +e^{-i\vect{k}\cdot\vect{r}}
\hat{a}^\dagger_\sigma(\vect{k})\right] \,.
\end{equation}
The plane-wave expansion (\ref{eq:cartesianA}) is a special case of
the more general mode expansion
\begin{equation}
\label{eq:vectorA}
\hat{\vect{A}}(\vect{r}) = \sum\limits_\lambda \left[
\vect{A}_\lambda(\vect{r})\hat{a}_\lambda +
\vect{A}^\ast_\lambda(\vect{r})\hat{a}^\dagger_\lambda \right] \,.
\end{equation}
The amplitude operators $\hat{a}_\lambda$ and
$\hat{a}^\dagger_\lambda$ then obey the commutation rules
\begin{equation}
\label{eq:commuatoraadagger}
\left[ \hat{a}_\lambda , \hat{a}^\dagger_\lambda \right] =
\delta_{\lambda\lambda'} \,.
\end{equation}
Finally, by introducing the amplitude operators via
Eq.~(\ref{eq:complexamplitude}), the Hamiltonian
(\ref{eq:harmonicoscillatorH}) is converted into diagonal form
\begin{equation}
\label{eq:freeH}
\hat{H} = \frac{1}{2}\sum\limits_\lambda \hbar\omega_\lambda \left(
\hat{a}^\dagger_\lambda \hat{a}_\lambda +\hat{a}_\lambda
\hat{a}^\dagger_\lambda \right)
= \sum\limits_\lambda \hbar\omega_\lambda \left(
\hat{a}^\dagger_\lambda \hat{a}_\lambda +\frac{1}{2}\right)
\end{equation}
where the second equality follows from application of the commutation
relation (\ref{eq:commuatoraadagger}). The last term in
(\ref{eq:freeH}) is an infinite, but additive, constant, the
quantum-mechanical ground-state energy.

With the expansion (\ref{eq:vectorA}) at hand, it is now
straightforward to write down the mode expansion of the operators of
the electric field and the magnetic induction as
\begin{eqnarray}
\label{eq:efeld}
\hat{\vect{E}}(\vect{r}) &=& i \sum\limits_\lambda \omega_\lambda
\left[ \vect{A}_\lambda(\vect{r})\hat{a}_\lambda -
\vect{A}^\ast_\lambda(\vect{r})\hat{a}^\dagger_\lambda \right] \,,\\
\hat{\vect{B}}(\vect{r}) &=& \sum\limits_\lambda \left[
\curl\vect{A}_\lambda(\vect{r})\hat{a}_\lambda +
\curl\vect{A}^\ast_\lambda(\vect{r})\hat{a}^\dagger_\lambda \right]
\,.
\end{eqnarray}
Using these expressions, we arrive at the (equal-time) commutation
relations for the electromagnetic field operators as
\begin{eqnarray}
\label{eq:commutatorEB}
\left[ \hat{\vect{E}}(\vect{r}) , \hat{\vect{B}}(\vect{r}') \right]
&=& i \sum\limits_\lambda \omega_\lambda
\left\{ \vect{A}_\lambda(\vect{r}) \cdot \left[
\curl\vect{A}^\ast_\lambda(\vect{r}') \right]
+\vect{A}^\ast_\lambda(\vect{r}) \cdot \left[
\curl\vect{A}_\lambda(\vect{r}') \right] \right\}
\nonumber \\ &=&
-\frac{i\hbar}{\varepsilon_0}
\curl\tens{\delta}^\perp(\vect{r}-\vect{r}')
\end{eqnarray}
where we have chosen a normalisation factor
$\mathcal{N}_\lambda=\hbar/(2\varepsilon_0\omega_\lambda)$ as in
Eq.~(\ref{eq:cartesianA}) and used the orthogonality relation
(\ref{eq:csof}). This commutator agrees with the canonical commutator
implied by the Poisson bracket (\ref{eq:poisson}) on imposing the
correspondence principle. The commutation rule (\ref{eq:commutatorEB})
tells us that the quantised electromagnetic field is a bosonic vector
field. Its elementary excitations, the photons, of polarisation
$\sigma$ and wavevector $\vect{k}$ are annihilated and created by the
amplitude operators $\hat{a}_{\vect{k}\sigma}$ and
$\hat{a}^\dagger_{\vect{k}\sigma}$.

Note that the operators of the electric field and the magnetic
induction describe the electromagnetic degrees of freedom alone. The
operators of the dielectric displacement $\hat{\vect{D}}(\vect{r})$
and the magnetic field $\hat{\vect{H}}(\vect{r})$, which in free space
are trivially connected to $\hat{\vect{E}}(\vect{r})$ and
$\hat{\vect{B}}(\vect{r})$ via Eqs.~(\ref{eq:freeconstitutiveD}) and
(\ref{eq:freeconstitutiveH}), in general contain both electromagnetic
as well as matter degrees of freedom. We will see later in the context
of macroscopic quantum electrodynamics that the same commutation rules
(\ref{eq:commutatorEB}) can be upheld even in the presence of 
magnetoelectric matter. The proof of the validity of this commutation
provides a cornerstone of macroscopic QED. 


\subsubsection{Lossless beam splitter}
\label{sec:beamsplitter}

Many propagation problems involving classical as well as quantised
light involves finding the eigenmodes of the geometric setup and
expanding the electromagnetic fields in terms of those modes. The
plane-wave expansion in empty space was the simplest case
imaginable. Manipulating light using passive optical elements such as
beam splitters, phase shifters or mirrors are classic examples of mode
matching problems at interfaces between dielectric or metallic bodies
and empty space that can be solved by mode expansion approaches.

\begin{figure}[!t!]
\begin{center}
\begin{picture}(100,100)(50,0)
\multiput(90,20)(0,2){41}{\line(1,0){20}}
\multiput(90,20)(2,0){11}{\line(0,1){80}}
\put(90,0){\line(0,1){20}}
\put(110,0){\line(0,1){20}}
\put(90,5){\vector(1,0){20}}
\put(110,5){\vector(-1,0){20}}
\put(30,75){\vector(1,0){60}}
\put(90,45){\vector(-1,0){60}}
\put(110,75){\vector(1,0){60}}
\put(170,45){\vector(-1,0){60}}
\put(97,10){$d$}
\put(50,80){$\hat{a}_1(\omega)$}
\put(50,50){$\hat{b}_2(\omega)$}
\put(130,80){$\hat{b}_1(\omega)$}
\put(130,50){$\hat{a}_2(\omega)$}
\put(130,5){\vector(1,0){50}}
\put(150,10){$x$}
\end{picture}
\end{center}
\caption{\label{fig:beamsplitter} Simple one-dimensional model of a
beam splitter of thickness $d$, consisting possibly of several layers
of different materials.}
\end{figure}
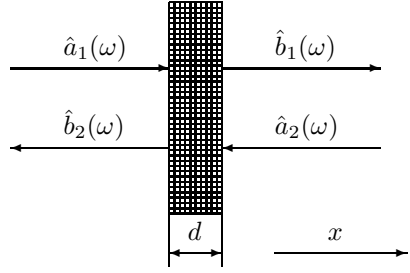

In most cases of interest, it is possible to restrict one's attention
to a one-dimensional propagation problem by choosing a particular
linear polarisation and considering one vector component of the
electromagnetic field only. For example, let us consider light
propagation along the $x$-direction in which case the electric-field
operator turns into a scalar operator
\begin{equation}
\hat{E}(x) = i \int dk\,c|k|A(k,x)\hat{a}(k) -i\int
dk\,c|k|A^\ast(k,x)\hat{a}^\dagger(k) \,.
\end{equation}
The mode functions $A(k,x)$ satisfy the one-dimensional Helmholtz
equation
\begin{equation}
\label{eq:1DHelmholtz}
\frac{d^2}{dx^2}A(k,x) +n^2(x) k^2 A(k,x) = 0
\end{equation}
with a spatially dependent (real) refractive index
$n(x)=\sqrt{\varepsilon(x)}$. The simplest model of a lossless beam
splitter involves assuming a refractive index profile with piecewise
constant refractive index (Fig.~\ref{fig:beamsplitter})
\begin{equation}
\label{eq:piecewisen}
n(x) = \left\{ \begin{array}{rl}
n, & |x|\le d/2, \\ 1, & |x|>d/2.
\end{array} \right.
\end{equation}

\begin{figure}[!t!]
\begin{picture}(200,85)(0,0)
\put(0,25){\line(1,0){60}}
\put(100,25){\line(1,0){60}}
\put(60,85){\line(1,0){40}}
\put(60,25){\line(0,1){60}}
\put(100,25){\line(0,1){60}}
\put(20,75){$e^{ikx}$}
\put(15,70){\vector(1,0){30}}
\put(15,45){$Re^{-ikx}$}
\put(45,40){\vector(-1,0){30}}
\put(118,60){$Te^{ikx}$}
\put(115,55){\vector(1,0){30}}
\put(60,15){\line(0,1){5}}
\put(100,15){\line(0,1){5}}
\put(45,0){$-d/2$}
\put(92,0){$d/2$}
\put(140,5){\vector(1,0){20}}
\put(132,2){$x$}
\end{picture}
\begin{picture}(200,85)(0,0)
\put(0,25){\line(1,0){60}}
\put(100,25){\line(1,0){60}}
\put(60,85){\line(1,0){40}}
\put(60,25){\line(0,1){60}}
\put(100,25){\line(0,1){60}}
\put(120,75){$e^{ikx}$}
\put(145,70){\vector(-1,0){30}}
\put(114,45){$R'e^{-ikx}$}
\put(115,40){\vector(1,0){30}}
\put(16,60){$T'e^{ikx}$}
\put(45,55){\vector(-1,0){30}}
\put(60,15){\line(0,1){5}}
\put(100,15){\line(0,1){5}}
\put(45,0){$-d/2$}
\put(92,0){$d/2$}
\put(140,5){\vector(1,0){20}}
\put(132,2){$x$}
\end{picture}
\caption{\label{fig:barrier} A plane wave $e^{ikx}$ with $k>0$ (left
figure) or $k<0$ (right figure) impinges onto a beam splitter from
the left or right, respectively, and splits into transmitted and
reflected parts.}
\end{figure}
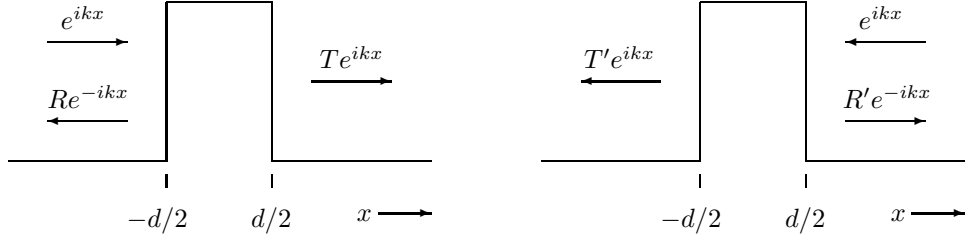

The solutions to the Helmholtz equation (\ref{eq:1DHelmholtz}) with
the refractive index profile (\ref{eq:piecewisen}) are once again
plane waves that can be constructed similar to the familiar
quantum-mechanical problem of wave scattering at a potential barrier
(Fig.~\ref{fig:barrier}). If an incoming plane wave $e^{ikx}$ from the
left ($k>0$) impinges onto the barrier, it will split into a reflected
wave $R(\omega)e^{-ikx}$ and a transmitted wave
$T(\omega)e^{ikx}$. Similarly, if a plane wave $e^{ikx}$ enters from
the right ($k<0$), it will split into a reflected wave
$R'(\omega)e^{-ikx}$ and a transmitted wave $T'(\omega)e^{ikx}$ with
as yet unspecified reflection and transmission coefficients
$R(\omega)$, $R'(\omega)$, $T(\omega)$ and $T'(\omega)$, respectively.
Hence, the mode functions $A(k,x)$ can be written as
\begin{eqnarray}
A(k,x) &=& \sqrt{\frac{\hbar}{4\pi\varepsilon_0\omega\mathcal{A}}}
\left\{ \begin{array}{ll}
e^{ikx} +R(\omega)e^{-ikx}, & x\le -\frac{d}{2} \\
T(\omega) e^{ikx}, &  x\ge \frac{d}{2}
\end{array} \right. \qquad k>0 \,,\\
A(k,x) &=& \sqrt{\frac{\hbar}{4\pi\varepsilon_0\omega\mathcal{A}}}
\left\{ \begin{array}{ll}
T'(\omega) e^{ikx}, & x\le -\frac{d}{2} \\
e^{ikx} +R'(\omega)e^{-ikx}, &  x\ge \frac{d}{2}
\end{array} \right. \qquad k<0 \,,
\end{eqnarray}
where $\mathcal{A}$ is a normalisation area.
%
%
The transmission and reflection coefficients can be obtained by
requiring continuity of the vector potential (the quantum-mechanical
wave function) and its first derivative at the beam-splitter
interfaces. This requirement is analogous to the well-known conditions
of continuity in classical electromagnetism. The result can be found
in textbooks (see, e.g., \cite{BornWolf}) as 
\begin{eqnarray}
\label{eq:airyT}
T(\omega) &=& \frac{1-r^2}{1-r^2 e^{2ind\omega/c}}
e^{i(n-1)d\omega/c} \,,\\ \label{eq:airyR}
R(\omega) &=&
-r e^{-id\omega/c}+r e^{ind\omega/c}T(\omega) \,,
\end{eqnarray}
where
$r=(n-1)/(n+1)=\frac{\sqrt{\varepsilon}-1}{\sqrt{\varepsilon}+1}$
is the Fresnel reflection coefficient for $p$-polarised waves (see
App.~\ref{sec:planardgf}). 

For a single dielectric plate, the transmission coefficients
$T(\omega)$ and $T'(\omega)$ and the reflection coefficients
$R(\omega)$ and $R'(\omega)$ are identical. For multilayered
dielectrics, this is not necessarily the case. There are, however, a
number of physical principles that restrict the form of these
coefficients. For example, Onsager reciprocity \cite{LL8} requires
the magnitudes of the transmission coefficients to be identical,
$|T(\omega)|=|T'(\omega)|$. In Sec.~\ref{sec:dgfproperties} we will
give details how this can be seen from general properties of the
dyadic Green tensor. Moreover, energy conservation (photon number
conservation, probability conservation) dictates that the squared
moduli of transmission and reflection coefficients must add up to
unity,
\begin{equation}
\label{eq:bsenergy}
|T(\omega)|^2 +|R(\omega)|^2  =1\,.
\end{equation}
The correctness of Eq.~(\ref{eq:bsenergy}) can be immediately
checked by applying Eqs.~(\ref{eq:airyT}) and (\ref{eq:airyR}).

We see from Fig.~\ref{fig:beamsplitter} that the electric field can be
decomposed into its incoming and outgoing parts associated with the
photonic amplitude operators $\hat{a}_i(\omega)$ and
$\hat{b}_i(\omega)$, respectively. The total electric field is thus
the sum of field components travelling to the right ($k>0$) and to the
left ($k<0$), whose amplitudes transform as
\begin{eqnarray}
\hat{b}_1(\omega) &=& T(\omega) \hat{a}_1(\omega) + R'(\omega)
\hat{a}_2(\omega) \,,\\
\hat{b}_2(\omega) &=& R(\omega) \hat{a}_1(\omega) + T'(\omega)
\hat{a}_2(\omega) \,.
\end{eqnarray}
For the transformed amplitude operators to represent photons, they
have to obey similar commutation rules as the untransformed operators,
hence we must have
\begin{equation}
\left[ \hat{b}_i(\omega) ,\hat{b}_j^\dagger(\omega') \right] =
\delta_{ij} \delta(\omega-\omega') \,.
\end{equation}
It follows that
$|T(\omega)|^2+|R'(\omega)|^2=|T'(\omega)|^2+|R(\omega)|^2=1$ and
$T(\omega)R^\ast(\omega)+R'(\omega)T^{'\ast}(\omega)=0$. These
requirements can be fulfilled if we set $T'(\omega)=T^\ast(\omega)$
and $R'(\omega)=-R^\ast(\omega)$. The transformation rules of the
photonic amplitude operators can thus be combined to a matrix equation
of the form
[$\hat{\vect{a}}(\omega)=[\hat{a}_1(\omega),\hat{a}_2(\omega)]^\trans$
,
$\hat{\vect{b}}(\omega)=[\hat{b}_1(\omega),\hat{b}_2(\omega)]^\trans$]
\begin{equation} 
\label{eq:ior}
\hat{\vect{b}}(\omega) = \tens{T}(\omega) \cdot \hat{\vect{a}}(\omega)
\end{equation}
where the transformation matrix
\begin{equation}
\tens{T}(\omega) = \left( \begin{array}{cc}
T(\omega) & R(\omega) \\ -R^\ast(\omega) & T^\ast(\omega)
\end{array} \right)
\end{equation}
is defined up to a global phase. Because of its structure,
$\tens{T}(\omega)$ is a unitary matrix, and in particular,
$\tens{T}(\omega)\in$~SU(2)
\cite{Yurke86,Prasad87,Ou87,Fearn87,Campos89,Leonhardt93}. The
unitarity of the transformation matrix reflects the energy
conservation requirement.

The input-output relations (\ref{eq:ior}) can be converted from a
matrix relation to an operator equation as
\begin{equation}
\label{eq:ior2}
\hat{\vect{b}}(\omega) = \hat{U}^\dagger \hat{\vect{a}}(\omega)
\hat{U}
\end{equation}
where the unitary operator $\hat{U}$ is given by
\begin{equation}
\hat{U} = \exp \left[ -i \int\limits_0^\infty d\omega\,
[\hat{\vect{a}}^\dagger(\omega)]^\trans\cdot\tens{\Phi}(\omega)\cdot
\hat{\vect{a}}(\omega) \right] \,,\qquad
\tens{T}(\omega) = \exp \left[ -i\tens{\Phi}(\omega) \right] \,.
\end{equation}
Using that operator, the quantum state of light impinging onto the
beam splitter transforms as
\begin{equation}
\label{eq:qst}
\hat{\varrho}_{\mathrm{out}} = \hat{U}
\hat{\varrho}_{\mathrm{in}} \hat{U}^\dagger \,.
\end{equation}
This can be easily verified by noting that the expectation value of
any operator that depends functionally on the amplitude operators
$\hat{\vect{a}}(\omega)$ and $\hat{\vect{a}}^\dagger(\omega)$ can be
computed either by transforming the amplitude operators using the
input-output relations (\ref{eq:ior2}), or by transforming the quantum
state using Eq.~(\ref{eq:qst}). The input-output relations would then
correspond to the Heisenberg picture, whereas the quantum-state
transformation could be regarded as its Schr\"odinger picture
equivalent.

Note that, despite the fact that the matrix $\tens{T}(\omega)$
describes an SU(2) transformation, the unitary operator $\hat{U}$ in
general does not. As the $n$-photon Fock space is the symmetric
subspace of the $n$-fold tensor product of single-photon Hilbert
spaces \cite{Bhatia}, the quantum-state transformation (\ref{eq:qst})
can be regarded as a transformation according to a subgroup of
SU(2$n$) where $n$ is the total number of photons impinging onto the
beam splitter. For example, in the basis
$\{|0,0\rangle,|1,0\rangle,|0,1\rangle,|2,0\rangle,|1,1\rangle,|0,
2\rangle\}$
the unitary operator $\hat{U}$ has the matrix representation
\begin{equation}
\label{eq:blockU}
\tens{U} = \left( \begin{array}{cccccc}
1&0&0&0&0&0\\
0&T&-R^\ast&0&0&0\\
0&R&T^\ast&0&0&0\\
0&0&0&T^2&\sqrt{2}T^\ast R^\ast&R^{\ast 2}\\
0&0&0&\sqrt{2}TR&(|T|^2-|R|^2)&-\sqrt{2}T^\ast R^\ast\\
0&0&0&R^2&-\sqrt{2}TR&T^{\ast 2}
\end{array}\right)
\end{equation}
which is block-diagonal with respect to the Fock layers of total
photon numbers $(0,1,2)$. This again expresses photon-number
conservation. The unitary matrix $\tens{U}$ thus has the structure of
a direct product,
\begin{equation}
\tens{U} = \bigoplus_{n=0}^\infty \tens{U}_n \,.
\end{equation}

The fact that the matrix transforming the quantum states acts on the
symmetric subspace of a tensor product Hilbert space means that it can
be constructed from permanents of the transmission matrix $\tens{T}$
\cite{Minc} that is responsible for the operator transformation
(\ref{eq:ior}). Let us define the set $G_{n,N}$ of all non-decreasing
integer sequences $\bm{\omega}$ as
\begin{equation}
G_{n,N} = \{ \bm{\omega} : 1\le \omega_1 \le \ldots \le \omega_n \le N
\}
\,. 
\end{equation}
Then the unitary transformation of an $N$-mode Fock state
$|n_1,\ldots,n_N\rangle$ with a total of $n$ photons can be written as
[$\bm{\Omega}=(1^{n_1},2^{n_2},\ldots,N^{n_N})$] \cite{Scheel04,QIV2}
\begin{equation}
\hat{U} |n_1,\ldots,n_N\rangle = \left( \prod\limits_i n_i!
\right)^{-1/2} \sum\limits_{\bm{\omega}\in G_{n,N}}
\frac{1}{\mu(\bm{\omega})} 
\textrm{per}\,\tens{T}[\bm{\omega}|\bm{\Omega}]
|m_1(\bm{\omega}),\ldots,m_N(\bm{\omega})\rangle 
\end{equation}
in which the $m_i(\bm{\omega})$ are the multiplicities of the
occurrence of the value $i$ in the non-decreasing integer sequence
$\bm{\omega}$ and $\mu_i(\bm{\omega})=\prod_im_i(\bm{\omega})!$. The
notation $\tens{T}[\bm{\omega}|\bm{\Omega}]$ thereby stands for the
matrix whose row and column indices are drawn from the non-decreasing
integer sequences $\bm{\omega}$ and $\bm{\Omega}$, respectively, and
whose elements are taken from the transformation matrix
$\tens{T}$. For example, the matrix $\tens{T}[(1,1)|(1,2)]$ contains
the elements
$\left(\begin{array}{cc}T_{11}&T_{12}\\T_{11}&T_{12}\end{array}
\right)$.
The symbol $\textrm{per}$ denotes the permanent of a matrix $\tens{T}$
which is defined as
\begin{equation}
\textrm{per}\,\tens{T}=\sum\limits_{\sigma\in S_n}
\prod\limits_{i=1}^n T_{i\sigma_i}
\end{equation}
where $S_n$ is the (symmetric) group of permutations.

For example, the probability amplitude of finding exactly one photon
in each beam splitter output when feeding two photons into its input
ports, is given by the permanent of the beam splitter transformation
matrix itself,
\begin{equation}
\langle 1,1|\hat{U}|1,1\rangle = \mathrm{per}\,\tens{T}
= T_{11} T_{22} +T_{12} T_{21} = |T|^2-|R|^2 \,. 
\end{equation}
For a symmetric beam splitter with vanishing permanent, this
probability is zero, and the Hong--Ou--Mandel quantum interference
effect is observed \cite{Hong87}. 
The appearance of a matrix function such as the permanent in the
context of single or networks of beam splitters is one particular
example of the links that exist between quantum optics and matrix
theory. 

For lossy beam splitters, neither the conservation law
(\ref{eq:bsenergy}) nor the direct product structure of the matrix
representation $\tens{U}$ of the unitary transformation operator
[Eq.~(\ref{eq:blockU})] hold. They will have to be replaced by a
generalised conservation law and a generalised unitary operator that
include material absorption (see Sec.~\ref{sec:lossybeamsplitter}). 


\subsubsection{Casimir force between two perfectly conducting parallel
plates}
\label{sec:modecasimir}

\begin{figure}[!t!]
\begin{center}
\begin{picture}(100,100)(0,0)
\put(20,20){\line(0,1){50}}
\put(40,40){\line(0,1){50}}
\put(20,20){\line(1,1){20}}
\put(20,70){\line(1,1){20}}
\put(50,20){\line(0,1){50}}
\put(70,40){\line(0,1){50}}
\put(50,20){\line(1,1){20}}
\put(50,70){\line(1,1){20}}
\put(20,15){\vector(1,0){30}}
\put(50,15){\vector(-1,0){30}}
\put(15,20){\vector(0,1){50}}
\put(15,70){\vector(0,-1){50}}
\put(15,75){\vector(1,1){20}}
\put(35,95){\vector(-1,-1){20}}
\put(32,5){$d$}
\put(5,42){$L$}
\put(15,87){$L$}
\end{picture}
\includegraphics[height=100pt]{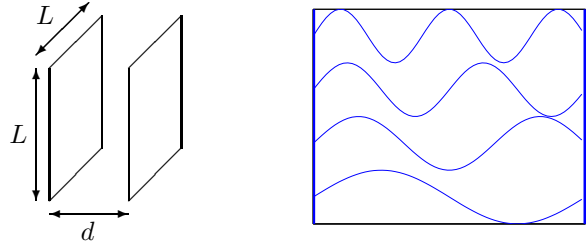}
\caption{\label{fig:casimir} Electromagnetic modes confined between
two
perfectly conducting parallel plates of separation $d$ lead to an
attractive force. The figure on the right shows some of the allowed
modes with the cavity walls.} 
\end{center}
\end{figure}

One of the most intriguing consequences of quantising the
electromagnetic field (or indeed any other field theory) is the
existence of an infinite ground-state energy
\begin{equation}
\label{eq:groundstateenergy}
E_0 = \frac{1}{2} \sum\limits_\lambda \hbar\omega_\lambda
\end{equation}
which is present even if no photon is. It might be argued that this
vacuum energy would not be physically relevant as all photon energies
can be referred to this base level. Because no absolute energy
measurements can be done, only relative with respect to this
ground-state energy, that energy would not be measurable. This
reasoning, however, is incorrect. To see why, it is instructive to
look at the quantities the ground-state energy depends on. Inspection
of Eq.~(\ref{eq:groundstateenergy}) reveals that is the mode structure
itself that determines its magnitude. In other words, the base level
from which photon energies are counted can be changed by altering the
number and the structure of the allowed electromagnetic modes. One way
to achieve that is to confine the electromagnetic field in a geometric
structure with appropriate boundary conditions that limits the number
of available modes, for example between two perfectly conducting
parallel plates (see Fig.~\ref{fig:casimir}).
%
%
Because of the boundary conditions for the electromagnetic field at
the plates surfaces, the modes perpendicular to the plates are
discrete with wave numbers $k_z=n\pi/d$, $n\in\mathbb{N}$. Hence, the
lowest wave number that is supported in the interspace between the
plates is $\propto 1/d$. After replacing the mode sum in
Eq.~(\ref{eq:groundstateenergy}) by an integral over wave vectors, we
find that the ground-state energy scales as $1/d^3$. Because the
ground-state energy clearly decreases with decreasing plate separation
$d$, there exists an attractive force between them. By dimensional
arguments, this force per unit area $L^2$ must be proportional to
\begin{equation}
\label{eq:casimirscaling}
\frac{F}{L^2} \propto -\frac{\hbar c}{d^4} \,.
\end{equation}

A more detailed calculation yields the correct numerical prefactor (we
will follow the derivation presented in Chap.~3.2.4 in
Ref.~\cite{Itzykson}). First note that the ground-state energy in a
box of volume $L^2d$ with $L\gg d$ is
\begin{equation}
E_0(d)= \frac{1}{2}\sum\limits_\lambda \hbar\omega_\lambda
= \frac{\hbar c}{2} \sum\limits_\lambda |\vect{k}_\lambda|
= \frac{\hbar c L^2}{2} \int \frac{d^2k_\|}{(2\pi)^2}
\left[ |\vect{k}_\|| +2\sum\limits_{n=1}^\infty \left( \vect{k}_\|^2
+\frac{n^2\pi^2}{d^2}  \right)^{1/2} \right]
\end{equation}
where we have used the fact that for $k_z>0$ there are two possible
polarisations $\sigma$ whereas there is only one independent
polarisation if $k_z=0$. This expression is, of course, infinite. To
render this expression finite, we subtract the contribution of free
space,
\begin{equation}
E_0(\infty) = \frac{\hbar c L^2}{2} \int \frac{d^2k_\|}{(2\pi)^2}
\int\limits_0^\infty dn \,2\sqrt{\vect{k}_\|^2+\frac{n^2\pi^2}{d^2}}
\end{equation}
where the double counting of the polarisation state at $n=0$ does not
influence the value of the integral. The ground-state energy per unit
area $L^2$ is thus, using polar coordinates,
\begin{equation}
\frac{E_0(d)-E_0(\infty)}{L^2}
= \frac{\hbar c}{2\pi} \int\limits_0^\infty dk\,k \left( \frac{k}{2}
+\sum\limits_{n=1}^\infty \sqrt{k^2+\frac{n^2\pi^2}{d^2}}
-\int\limits_0^\infty dn \sqrt{k^2+\frac{n^2\pi^2}{d^2}} \right) \,.
\end{equation}
This expression still seems to diverge for large wave numbers $k$ and
has to be regularised. For this purpose, we introduce a cut-off
function $f(k)$ such that
\begin{equation}
f(k) = \left\{ \begin{array}{ll}
1\,, &  k<k_\mathrm{max} \\ 0\,, & k\gg k_\mathrm{max}
\end{array}
\right.
\end{equation}
%
\begin{figure}[!t!]
\begin{center}
\begin{picture}(200,100)(0,0)
\put(10,10){\vector(1,0){190}}
\put(20,0){\vector(0,1){100}}
\put(20,80){\line(1,0){40}}
\qbezier(60,80)(100,80)(115,45)
\qbezier(115,45)(135,10)(160,10)
\put(195,0){$k$}
\put(-3,95){$f(k)$}
\put(10,76){$1$}
\put(65,6){\line(0,1){8}}
\put(58,0){$k_\mathrm{max}$}
\end{picture}
\end{center}
\caption{\label{eq:kmax} Smooth cut-off function $f(k)$ that falls to
zero for large enough wave numbers.}
\end{figure}
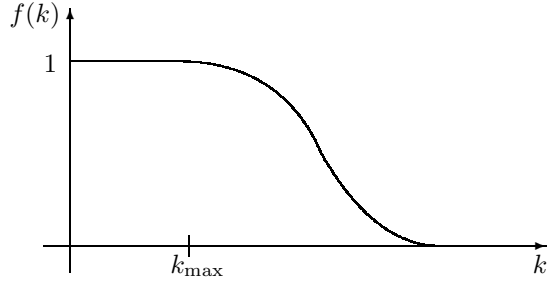
%
where the cut-off wave number $k_\mathrm{max}$ could be chosen to be
of the order of the inverse atomic size or, more appropriately, to
correspond to a frequency larger than the plasma frequency of the
material that the plates consist of (see Fig.~\ref{eq:kmax}). The
physical background to the latter requirement is that the permittivity
of a metal is for frequencies larger than the plasma frequency
$\omega_P$ well described by 
$\varepsilon(\omega)\approx 1-\omega_P^2/\omega^2$. Thus, for 
$\omega\gg\omega_P$ even metals become transparent and fail to provide
the required boundary conditions. 

After change of variables to $u=d^2k^2/\pi$, one obtains the
convergent expression
\begin{equation}
\label{eq:eulermclaurin1}
\frac{E_0(d)-E_0(\infty)}{L^2} = \frac{\hbar c\pi^2}{4d^3} \left[
\frac{1}{2}F(0)+\sum\limits_{n=1}^\infty F(n)
-\int\limits_0^\infty dn F(n) \right] 
\end{equation}
with
\begin{equation}
F(n) = \int\limits_0^\infty du \sqrt{u+n^2}
f\left(\pi\sqrt{u+n^2}/d\right)
= \int\limits_{n^2}^\infty du \sqrt{u}
f\left(\pi\sqrt{u}/d\right) \,.
\end{equation}
The expression in brackets in Eq.~(\ref{eq:eulermclaurin1}) can be
computed using the Euler-McLaurin resummation formula
\begin{equation}
\label{eq:eulermclaurin2}
\frac{1}{2}F(0)+\sum\limits_{n=1}^\infty F(n)
-\int\limits_0^\infty dn F(n) = -\sum\limits_{m=1}^\infty
\frac{B_{2m}}{(2m)!} F^{(2m-1)}(0) 
\end{equation}
where the $B_k$ are the Bernoulli numbers. Since we have constructed
the cut-off function such that $f(0)=1$ and $f^{(k)}(0)=0$, the only
non-zero contribution to Eq.~(\ref{eq:eulermclaurin2}) arises from
$F'''(0)=-4$, together with $B_4=-1/30$. Hence,
\begin{equation}
\frac{E_0(d)-E_0(\infty)}{L^2} = -\frac{\pi^2}{720} \frac{\hbar
c}{d^3}
\end{equation}
which leads to a force per unit area as
\begin{equation}
\label{eq:modecasimir}
\frac{F}{L^2} = -\frac{\pi^2}{240} \frac{\hbar c}{d^4}
\end{equation}
which is precisely Eq.~(\ref{eq:casimirscaling}) up to a numerical
factor of $\pi^2/240\approx 0.041$.

Similar calculations yield the Casimir forces for cylindrical and
spherical shells as shown in Tab.~\ref{tab:casimir}.
\begin{table}[!t!]
\begin{center}
\begin{tabular}{|c|c|c|}
\hline
geometry & Casimir force & Ref. \\ \hline\hline 
planar &
\parbox[h][1cm][c]{2cm}{$-\dd\frac{\pi^2 \hbar c L^2}{240d^4}$} &
\cite{Casimir48} \\ \hline 
cylinder &
\parbox[h][1cm][c]{2.2cm}{$-\dd\frac{0.02712\hbar c z}{R^3}$} &
\cite{deRaad81} \\ \hline
sphere & \parbox[h][1cm][c]{2cm}{$+\dd\frac{0.045\hbar c}{R^2}$} &
\cite{Boyer68}
\\ \hline 
\end{tabular}
\caption{\label{tab:casimir} Casimir forces for parallel plates and
cylindrical and spherical shells.}
\end{center}
\end{table}
Note here that the Casimir forces in both planar and cylindrical
geometries are attractive, whereas in case of a spherical shell it is
repulsive. The latter result seems to contradict our intuition that a
restriction of the number of modes always leads to an attractive
force. In order to resolve this conundrum, one needs to look closer at
the mode structure inside and outside a cylindrical or spherical
shell.

In the mode-summation approach that forms the basis of the
calculations referred to above, one has to regularise the wave number
integral at its upper limit by assuming a cut-off frequency above
which the plates have to become transparent (for cylindrical and
spherical shells one sometimes assumes two half-cylinders or
hemispheres whose separation provides the necessary regularisation).
This argument already suggests that the interpretation of the Casimir
force as a mode restriction between perfect conductors cannot be
upheld rigorously, and must be replaced by something that involves the
dielectric properties of the plates. 

Let us interrupt the flow of the argument at this point and mention a
classical analogue that can serve as an intuitive guidance to the
problem of Casimir energies: the problem of determining altitudes on
land. On literally all geophysical maps, the altitude of landmarks
such as mountains, lakes, and human dwellings is given in terms of its
altitude with respect to the average sea level. So one could say that
the average sea level represents the level of the infinite
ground-state energy. And in exactly the same way in which one is not
interested in the altitude of a mountain with respect to the sea
floor, we shall be content with measuring the photon energies from the
infinite ground-state level. On the other hand, one might ask the
question how the sea level can be properly defined given that there
are tides, wind and waves that distort that level. As we will see
later, it is exactly these fluctuations that are responsible for the
Casimir effect in the quantum-mechanical setting.


\subsection{Interaction of the quantised electromagnetic field with
atoms}
\label{sec:atomsinvacuum}

After we have determined how to quantise the electromagnetic field in
free space, we will now couple external sources to the field and focus
on the atomic degrees of freedom. For this purpose, let us begin again
with classical Maxwell's equations which, in the presence of external
sources, read
\begin{eqnarray}
\divv\vect{B}(\vect{r},t) &=& 0 \,,\\
\curl\vect{E}(\vect{r},t) &=& -\dot{\vect{B}}(\vect{r},t) \,,\\
\divv\vect{D}(\vect{r},t) &=& \rho(\vect{r},t) \,,\\
\curl\vect{H}(\vect{r},t) &=& \vect{j}(\vect{r},t)
+\dot{\vect{D}}(\vect{r},t) \,.
\end{eqnarray}
The charge density $\rho(\vect{r},t)$ and the current density
$\vect{j}(\vect{r},t)$ fulfil the equation of continuity
\begin{equation}
\label{eq:continuity}
\dot{\rho}(\vect{r},t) +\divv\vect{j}(\vect{r},t) =0
\end{equation}
which states that any change of the charge distribution within a
region of space is accompanied by a flow of current across the
boundary of that region.

The charge density for an ensemble of point charges $q_\alpha$ is
given by
\begin{equation}
\label{eq:chargedensity}
\rho(\vect{r},t) = \sum\limits_\alpha q_\alpha
\delta[\vect{r}-\vect{r}_\alpha(t)]
\end{equation}
where $\vect{r}_\alpha(t)$ denotes their classical trajectory. From
the continuity equation (\ref{eq:continuity}) it then follows that the
current density is
\begin{equation}
\label{eq:currentdensity}
\vect{j}(\vect{r},t) = \sum\limits_\alpha q_\alpha
\dot{\vect{r}}_\alpha(t) \delta[\vect{r}-\vect{r}_\alpha(t)]\,.
\end{equation}
In order to promote the charges to proper dynamical variables, we have
to supplement Maxwell's equations with Newton's equations of motion
for particles with mass $m_\alpha$,
\begin{equation}
m_\alpha \ddot{\vect{r}}_\alpha = q_\alpha \left[
\vect{E}(\vect{r}_\alpha,t) +\dot{\vect{r}}_\alpha \times
\vect{B}(\vect{r}_\alpha,t) \right] \,.
\end{equation}

Introducing scalar and vector potentials as in free space,
\begin{equation}
\vect{B}(\vect{r},t) = \curl\vect{A}(\vect{r},t) \,,\qquad
\vect{E}(\vect{r},t) =
-\dot{\vect{A}}(\vect{r},t)-\grad\phi(\vect{r},t) \,,
\end{equation}
we obtain their respective wave equations in the Coulomb gauge
$\divv\vect{A}(\vect{r},t)=0$ as
\begin{eqnarray}
\label{eq:scalarwave}
\Delta\phi(\vect{r},t) &=& -\frac{1}{\varepsilon_0} \rho(\vect{r},t)
\,,\\ \label{eq:vectorwave}
\Delta\vect{A}(\vect{r},t) -\frac{1}{c^2} \ddot{\vect{A}}(\vect{r},t)
&=& -\mu_0 \left[\vect{j}(\vect{r},t) -\varepsilon_0
\grad\dot\phi(\vect{r},t)  \right]\,.
\end{eqnarray}
Equation~(\ref{eq:scalarwave}) is Poisson's equation with the solution
\begin{equation}
\phi(\vect{r},t) = \frac{1}{4\pi\varepsilon_0} \int d^3r'
\frac{\rho(\vect{r},t)}{|\vect{r}-\vect{r}'|}
= \frac{1}{4\pi\varepsilon_0} \sum\limits_\alpha
\frac{q_\alpha}{|\vect{r}-\vect{r}_\alpha(t)|}
\end{equation}
where in the second equality we have used
Eq.~(\ref{eq:chargedensity}). The expression on the rhs of
Eq.~(\ref{eq:vectorwave}) is a transverse current density which can be
written in terms of the total current density as
\begin{equation}
\vect{j}^\perp(\vect{r},t) = \vect{j}(\vect{r},t)- \grad \int d^3r'
\frac{\divv\vect{j}(\vect{r}',t)}{4\pi|\vect{r}-\vect{r}'|} 
\end{equation}
using the continuity equation (\ref{eq:continuity}).

The above equations of motion for the electromagnetic field and the
charged particles can be derived from the classical Hamiltonian
function
\begin{equation}
\label{eq:minimalH}
H = \frac{1}{2} \int d^3r \left[ \varepsilon_0 \vect{E}^2(\vect{r},t)
+\frac{1}{\mu_0} \vect{B}^2(\vect{r},t) \right]
+\sum\limits_\alpha \frac{1}{2m_\alpha} \left[ \vect{p}_\alpha
-q_\alpha  \vect{A}(\vect{r}_\alpha) \right]^2
+\sum\limits_{\alpha\ne\alpha'} \frac{q_\alpha q_{\alpha'}}
{8\pi\varepsilon_0|\vect{r}_\alpha-\vect{r}_{\alpha'}|}
\end{equation}
in which the last term describes the Coulomb interaction between the
charged particles. Note that the particle momentum
$\vect{p}_\alpha$
$=m_\alpha\dot{\vect{r}}_\alpha+q_\alpha\vect{A}(\vect{r}_\alpha)$
is different from the mechanical momentum
$m_\alpha\dot{\vect{r}}_\alpha$ due to the interaction with the
electromagnetic field.

\begin{figure}[!t!]
\begin{center}
\begin{picture}(100,100)(0,0)
\put(80,80){\circle*{6}}
\put(70,70){\circle*{6}}
\put(60,75){\circle*{6}}
\put(85,65){\circle*{6}}
\put(70,55){\circle*{6}}
\put(60,65){\circle*{6}}
\put(0,0){\vector(1,1){68}}
\put(27,40){$\vect{r}_A$}
\end{picture}
\end{center}
\caption{\label{fig:multipolar} Individual charged particles are
combined to a coarse-grained system at position $\vect{r}_A$.}
\end{figure}
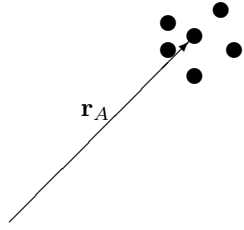

The Hamiltonian (\ref{eq:minimalH}) is referred to as the
minimal-coupling Hamiltonian because the electromagnetic field couples
to the microscopic degrees of freedom of the individual charged
particles such as position and momenta. This microscopic description
is often rather unwieldy, and an alternative approach in terms of
global quantities is sought. A particularly important situation arises
if the individual charged particles constitute an ensemble of bound
charges such as electrons and nuclei in an atom or a molecule. Let us
introduce a coarse-grained charge distribution $\bar{\rho}$ and
current density $\bar{\vect{j}}$ associated with that atomic system at
centre-of-mass position $\vect{r}_A=\sum_\alpha
(m_\alpha/m_A)\vect{r}_\alpha$ ($m_A=\sum_\alpha m_\alpha$:
total mass) (Fig.~\ref{fig:multipolar}),
\begin{equation}
\label{eq:coarsegrainedcharge}
\bar{\rho}(\vect{r},t) = \left( \sum\limits_\alpha q_\alpha\right)
\delta[\vect{r}-\vect{r}_A(t)] \,,\quad
\bar{\vect{j}} = \left( \sum\limits_\alpha q_\alpha\right)
\dot{\vect{r}}_A(t) \delta[\vect{r}-\vect{r}_A(t)] \,.
\end{equation}
%
%
Note that this charge density is zero for globally neutral
systems. In order to make up for the difference between the actual
charge distribution (\ref{eq:chargedensity}) and
Eq.~(\ref{eq:coarsegrainedcharge}), we define the microscopic
polarisation field via the implicit relation
\begin{equation}
\label{eq:microscopicP}
\divv\vect{P}_A(\vect{r},t) = -\rho(\vect{r},t)
+\bar{\rho}(\vect{r},t) \,.
\end{equation}
Adding this polarisation to the electric field, we define the modified
displacement field as
\begin{equation}
\vect{D}(\vect{r},t) = \varepsilon_0 \vect{E}(\vect{r},t)
+\vect{P}(\vect{r},t) 
\end{equation}
which obeys the modified Coulomb law
\begin{equation}
\divv\vect{D}(\vect{r},t) = \bar{\rho}(\vect{r},t) \,.
\end{equation}
Note that for globally neutral systems, the displacement field is a
transverse vector field. From the implicit relation
(\ref{eq:microscopicP}), one can show that the polarisation can be
written as \cite{Babiker74}
\begin{equation}
\label{eq:classicalpolarization}
\vect{P}_A(\vect{r}) = \sum\limits_\alpha q_\alpha
\bar{\vect{r}}_\alpha \int\limits_0^1 ds\,
\delta(\vect{r}-\vect{r}_A-s\bar{\vect{r}}_\alpha)
\end{equation}
($\bar{\vec{r}}_\alpha=\vec{r}_\alpha-\vec{r}_A$, relative particle
coordinates). Analogously, the magnetisation field is introduced via
the relation
\begin{equation}
\label{eq:classicalmagnetization}
\curl\vect{M}_A(\vect{r}) = \vect{j}(\vect{r}) -
\bar{\vect{j}}(\vect{r}) - \dot{\vect{P}}_A(\vect{r})
\end{equation}
which can be written as \cite{Babiker74}
\begin{equation}
\vect{M}_A(\vect{r}) = \sum\limits_\alpha q_\alpha \int\limits_0^1
ds\, s \bar{\vect{r}}_\alpha \times
\dot{\bar{\vect{r}}}_\alpha
\delta(\vect{r}-\vect{r}_A-s\bar{\vect{r}}_\alpha)
\,. 
\end{equation}

As before, field quantisation is performed by replacing the relevant
$c$-number quantities by Hilbert space operators and postulating their
canonical commutation rules. In contrast to relativistic quantum
electrodynamics, the charged particles are not treated within second
quantisation, i.e. in quantum optics electrons, atoms, molecules
etc. cannot be created or annihilated. Instead, their quantum
character is contained in their respective position and momenta, for
which we postulate the canonical commutation rules
\begin{equation}
\left[ \hat{\vec{r}}_{\alpha}, \hat{\vec{p}}_{j} \right] = i\hbar
\delta_{\alpha\alpha'} \ten{I} \,.
\end{equation}

At the moment, it seems as if we have not achieved anything other than
rewriting the charge density in terms of a new vector field that
itself, by construction, depends on the original microscopic
variables. To proceed, one either has to solve the microscopic
dynamics explicitly which is only possible for sufficiently small
systems, or one can invoke statistical arguments that relate the
polarisation field causally to the electric field by means of a (in
general nonlinear) response \textit{ansatz}. The latter approach leads
to the theory of macroscopic QED (Sec.~\ref{sec:causalQED}). 

A direct treatment of the microscopic dynamics can be considerably
simplified by casting the atom-field interactions appearing in the
minimal-coupling Hamiltonian~(\ref{eq:minimalH}) into its alternative
multipolar form. To that end, we transform the dynamical variables by
means of a Power--Zienau transformation
$\hat{O}'=\hat{U}\hat{O}\hat{U}^\dagger$, where the unitary operator
\cite{Power,Wolley,CraigThiru}
\begin{equation}
\label{eq:PowerZienauU1}
\hat{U} = \exp \left[ \frac{i}{\hbar} \int d^3r\,
\hat{\vect{P}}_A(\vect{r}) \cdot \hat{\vect{A}}(\vect{r}) \right]
\end{equation}
depends on the polarisation~(\ref{eq:classicalpolarization})
and the vector potential~(\ref{eq:vectorA}) of the electromagnetic
field. Expressing the Hamiltonian~(\ref{eq:minimalH}) in terms of the
transformed variables and applying a leading-order expansion in terms
of the relative particle coordinates, one obtains the electric-dipole
Hamiltonian for a neutral atomic system (cf. also
Sec.~\ref{sec:interaction}) 
\begin{equation}
\label{eq:multipolarHamiltonian2}
\hat{H}' = \sum\limits_\lambda \hbar\omega_\lambda \left(
\hat{a}^\dagger_\lambda \hat{a}_\lambda +\frac{1}{2} \right)
+\sum\limits_\alpha \frac{\hat{\vect{p}}_\alpha^{\prime 2}}{2m_\alpha}
 +\frac{1}{2\varepsilon_0}
\int d^3r \,\hat{\vect{P}}_A^2(\vect{r}) 
-\hat{\vect{d}}\cdot\hat{\vect{E}}'(\vect{r}_A).
\end{equation}
Here, 
\begin{equation}
\label{eq:dipolemoment}
\hat{\vect{d}} = \sum\limits_\alpha q_\alpha \hat{\vect{r}}_\alpha
= \sum\limits_\alpha q_\alpha \hat{\bar{\vect{r}}}_\alpha \,.
\end{equation}
is the electric dipole moment operator, and 
the transformed electric field
\begin{equation}
\label{Etransform}
\hat{\vect{E}}'(\vect{r}) = \hat{\vect{E}}(\vect{r})
+\frac{1}{\varepsilon_0} \hat{\vect{P}}_A^\perp(\vect{r})
\end{equation}
can be given in terms of the transformed bosonic operators
$\hat{a}'_\lambda$ via an expansion analogous to
Eq.~(\ref{eq:efeld}). The multipolar Hamiltonian is highly
advantageous in comparison to the minimal coupling
one~(\ref{eq:minimalH}) since the atom-field interaction 
$\hat{H}_\mathrm{int}=-\hat{\vect{d}}\cdot\hat{\vect{E}}(\vect{r}_A)$
consists of a single term. For this reason, we will exclusively employ
it throughout the remainder of this section and drop the primes
distinguishing the multipolar variables.


\subsubsection{Heisenberg equations of motion}
\label{sec:freespaceHeisenberg}

The electric-dipole interaction Hamiltonian
$\hat{H}_\mathrm{int}=-\hat{\vect{d}}\cdot\hat{\vect{E}}(\vect{r}_A)$,
can again be expanded in modes according to the description given
above. In particular, the operator of the electric field strength is
given by Eq.~(\ref{eq:efeld}). For the atomic system we choose to
expand its free Hamiltonian and the dipole moment in terms of its
energy eigenbasis $|n\rangle$. The atomic flip operators will be
denoted by $\hat{A}_{mn}\equiv|m\rangle\langle n|$, and they obey
the commutation rule
\begin{equation}
\label{eq:commutatorsigma}
\left[ \hat{A}_{mn} , \hat{A}_{kl} \right]
= \hat{A}_{ml}\delta_{nk} -\hat{A}_{kn}\delta_{lm} \,.
\end{equation}
With these preparations, the electric-dipole
Hamiltonian~(\ref{eq:multipolarHamiltonian2}) takes the form
\begin{eqnarray}
\hat{H} &=& \sum\limits_\lambda \hbar\omega_\lambda \left(
\hat{a}^\dagger_\lambda \hat{a}_\lambda +\frac{1}{2} \right)
 +\sum\limits_n \hbar\omega_n \hat{A}_{nn}
-i\sum\limits_{m,n} \omega_{mn}\hat{A}_{mn} \vect{d}_{mn}
\cdot \hat{\vect{A}}(\vect{r}_A)
\nonumber \\
&=& \sum\limits_\lambda \hbar\omega_\lambda \left(
\hat{a}^\dagger_\lambda \hat{a}_\lambda +\frac{1}{2} \right)
 +\sum\limits_n \hbar\omega_n \hat{A}_{nn} \nonumber \\ &&
-i\sum\limits_{m,n}\sum\limits_\lambda \omega_{mn}\hat{A}_{mn}
\vect{d}_{mn} \cdot \vect{A}_\lambda(\vect{r}_A) \hat{a}_\lambda
+\mbox{h.c.}
\end{eqnarray}

This Hamiltonian governs the dynamics of the atom-field system via
Heisenberg's equation of motion
\begin{equation}
\dot{\hat{O}} = \frac{1}{i\hbar} \left[ \hat{O},\hat{H} \right] .
\end{equation}
Applying the commutation rules (\ref{eq:commuatoraadagger}) and
(\ref{eq:commutatorsigma}), the equations of motions for the photonic
amplitude operators and the atomic flip operators read
\begin{eqnarray}
\dot{\hat{A}}_{mn} &=& i\omega_{mn} \hat{A}_{mn}
+\frac{1}{\hbar} \sum\limits_k \sum\limits_\lambda
(\omega_{nk}\vect{d}_{nk}\hat{A}_{mk}
-\omega_{km}\vect{d}_{km}\hat{A}_{kn})
\nonumber \\ && \times
\left( \vect{A}_\lambda(\vect{r}_A) \hat{a}_\lambda
-\vect{A}^\ast_\lambda(\vect{r}_A) \hat{a}_\lambda^\dagger \right), \\
\dot{\hat{a}}_\lambda &=& -i\omega_\lambda
\hat{a}_\lambda
+\frac{1}{\hbar}
\sum\limits_{m,n} \hat{A}_{mn} \vect{d}_{mn} \cdot
\vect{A}_\lambda^\ast(\vect{r}_A) .
\end{eqnarray}
In most cases of interest it is sufficient to concentrate on two out
of the potentially many atomic levels, a ground state $|g\rangle$ and
an excited state $|e\rangle$ separated by a transition frequency
$\omega_e-\omega_g=\omega_A$. The three relevant atomic flip
operators, the Pauli operators,
will be denoted by $\hat{\sigma}\equiv|g\rangle\langle e|$,
$\hat{\sigma}^\dagger\equiv|e\rangle\langle g|$,
$\hat{\sigma}_z\equiv|e\rangle\langle e|-|g\rangle\langle
g|$. Together with the identity operator
$\hat{I}\equiv|e\rangle\langle e|+|g\rangle\langle g|$ in that
two-dimensional Hilbert space, they generate the group SU(2).
Finally, Heisenberg's equations of motion reduce in the rotating-wave
approximation to
\begin{eqnarray}
\label{eq:heisenberg_sigma}
\dot{\hat{\sigma}} &=& -i\omega_A\hat{\sigma} -\frac{i}{\hbar}
\vect{d} \cdot \hat{\vect{E}}^{(+)}(\vect{r}_A) \hat{\sigma}_z,
\\ \label{eq:heisenberg_sigmaz}
\dot{\hat{\sigma}}_z &=& \frac{2i}{\hbar} \vect{d} \cdot
\hat{\vect{E}}^{(+)}(\vect{r}_A) \hat{\sigma}^\dagger  +\mbox{h.c.}, 
\\ \label{eq:heisenberg_alambda}
\dot{\hat{a}}_\lambda &=& -i\omega_\lambda \hat{a}_\lambda
+\frac{\omega_\lambda}{\hbar} \vect{d}^\ast \cdot
\vect{A}^\ast_\lambda(\vect{r}_A)  \hat{\sigma}\,,
\end{eqnarray}
where the positive-frequency part $\hat{\vect{E}}^{(+)}(\vect{r}_A)$
of the electric field is given by the first term in
Eq.~(\ref{eq:efeld}). We can attempt to solve
Eq.~(\ref{eq:heisenberg_alambda}) by formally integrating it,
\begin{equation}
\hat{a}_\lambda(t) = e^{-i\omega_\lambda t}\hat{a}_\lambda
+\frac{\omega_\lambda}{\hbar} \vect{d}^\ast \cdot
\vect{A}^\ast_\lambda(\vect{r}_A) \int\limits_0^t
dt'\,e^{-i\omega_\lambda(t-t')} \hat{\sigma}(t') \,.
\end{equation}
The first term in this equation is the free evolution of the photonic
amplitude operators whereas the second term is due to the interaction
with the two-level atom. The time integral contains the solution to
Eq.~(\ref{eq:heisenberg_sigma}) which itself is unknown. Thus, the
integral can only be computed approximately. For this purpose, we
split up the fast time evolution $e^{-i\omega_At}$ from the atomic
flip operator and define a slowly-varying quantity
$\hat{\tilde{\sigma}}(t)$ as
\begin{equation}
\hat{\tilde{\sigma}}(t) = \hat{\sigma}(t) e^{i\omega_At} \,.
\end{equation}
Then, the time integral can be approximated as
\begin{eqnarray} &&
\int\limits_0^t dt'\,e^{-i\omega_\lambda(t-t')}\hat{\sigma}(t')
= \int\limits_0^t dt'\,e^{-i\omega_\lambda(t-t')} e^{-i\omega_At'}
\hat{\tilde{\sigma}}(t')
\nonumber \\ && \approx \hat{\tilde{\sigma}}(t)
\int\limits_0^t dt'\,e^{-i\omega_\lambda(t-t')} e^{-i\omega_At'}
=  \hat{\sigma}(t)
\int\limits_0^t dt'\,e^{i(\omega_A-\omega_\lambda)(t-t')}
\end{eqnarray}
where the atomic flip operator has been taken out of the integral at
the upper time. This is only possible if the amplitude operator
$\hat{\tilde{\sigma}}(t)$ is almost constant over the time scale
$|\omega_A-\omega_\lambda|^{-1}$. This in fact also means that the
interaction between the electromagnetic field and the two-level atom
must not be too large. Because the atomic flip operator has been taken
out of the time integral, all memory effects of the atom-field
interaction have been neglected. This is known as the Markov
approximation.

\begin{figure}[!t!]
\begin{center}
\includegraphics[width=6cm]{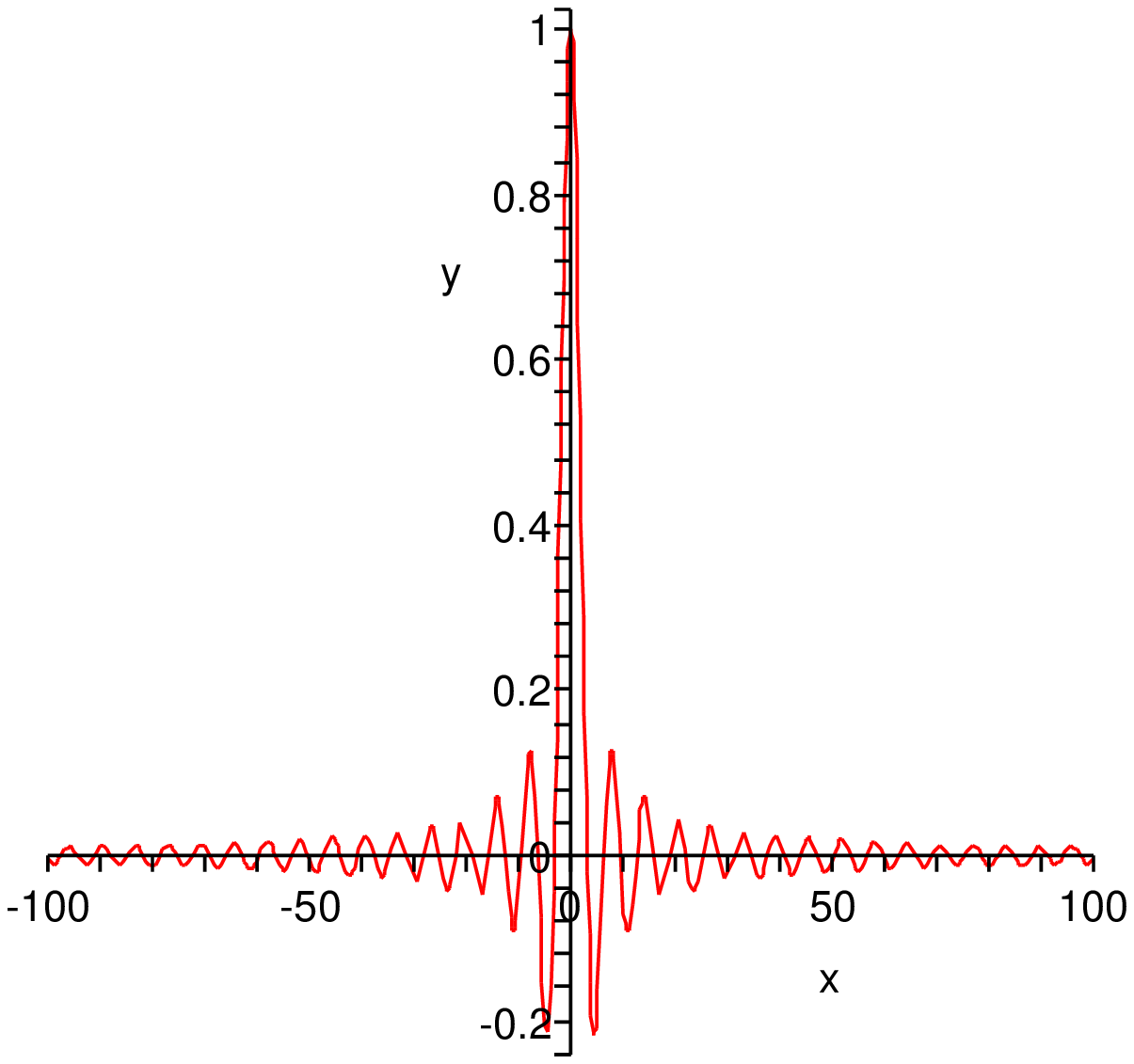}
\hfill
\includegraphics[width=6cm]{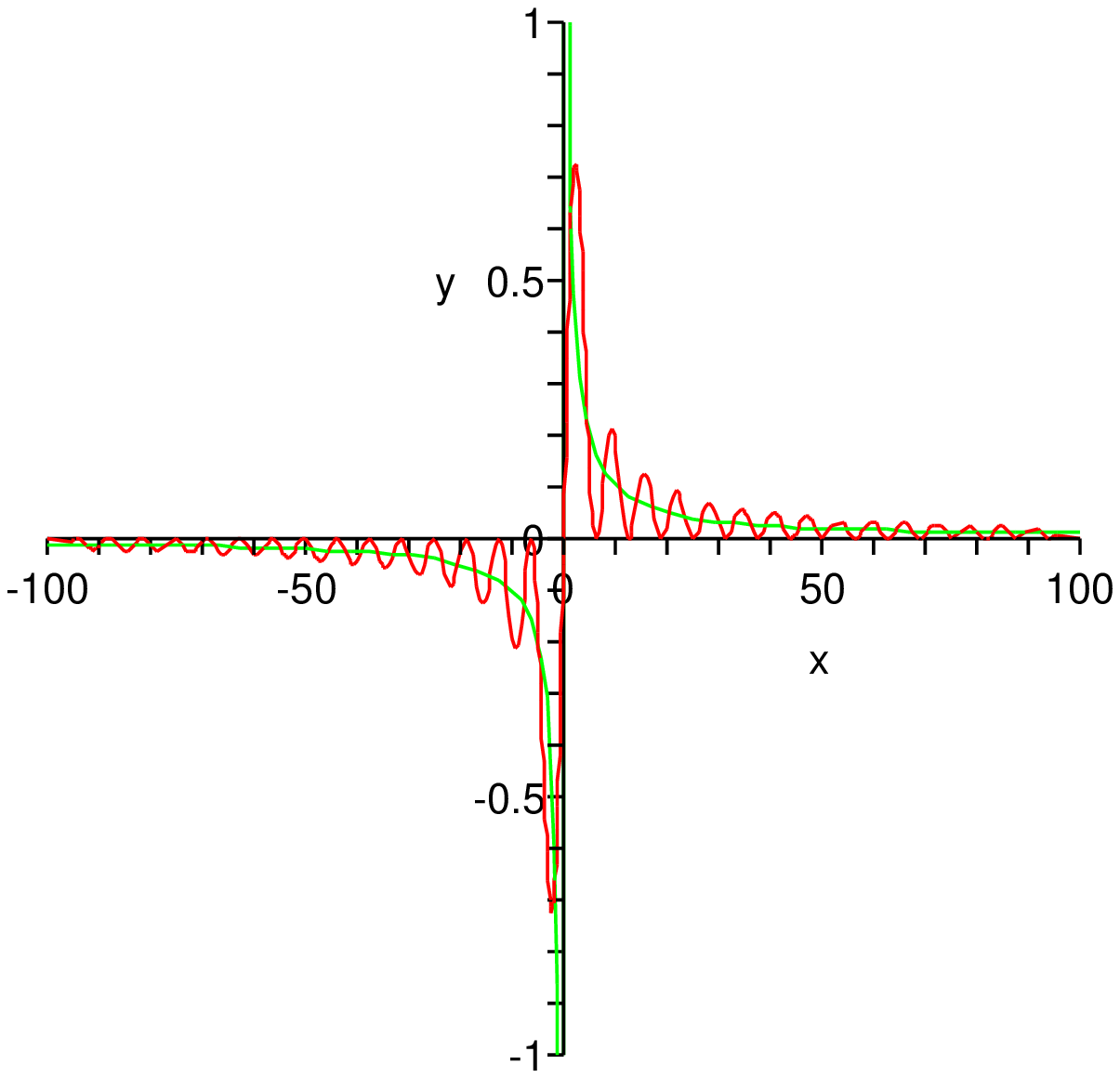}
\end{center}
\caption{\label{fig:sincos} Functions $s(x)$ and $c(x)$ showing how
they can be approximated by the functions $\pi\delta(x)$ and
$\mathcal{P}(1/x)$, respectively.}
\end{figure}

The remaining integral can be easily evaluated as
\begin{equation}
\label{eq:sincos}
\int\limits_0^t dt'\,e^{i(\omega_A-\omega_\lambda)(t-t')}
= \frac{\sin(\omega_A-\omega_\lambda)t}{\omega_A-\omega_\lambda}
+i\frac{[1-\cos(\omega_A-\omega_\lambda)t]}{\omega_A-\omega_\lambda}
\equiv s(\omega_A-\omega_\lambda) +ic(\omega_A-\omega_\lambda)
\end{equation}
which we have split into its real and imaginary parts. The function
$s(\omega_A-\omega_\lambda)$ is sharply peaked at
$\omega_\lambda=\omega_A$ (Fig.~\ref{fig:sincos}).
%
%
If all quantities that contain either of these functions are averaged
or cannot be resolved over sufficiently long times, then we can set
\begin{equation}
s(x) \mapsto \pi\delta(x) \,,\qquad c(x) \mapsto
\mathcal{P}\frac{1}{x} 
\end{equation}
[$\mathcal{P}$: principal value] thereby introducing little error.

If we re-insert the formal solution to
Eq.~(\ref{eq:heisenberg_alambda}) into
Eqs.~(\ref{eq:heisenberg_sigma}) and (\ref{eq:heisenberg_sigmaz}), we
obtain the equations of motion of the atomic flip operators in the
Markov approximation as
\begin{eqnarray}
\label{eq:markov_sigma}
\dot{\hat{\sigma}} &=&
-i\left(\omega_A+\delta\omega-i\frac{\Gamma}{2}\right) \hat{\sigma} 
-\frac{i}{\hbar} \hat{\sigma}_z
\hat{\vect{E}}^{(+)}_{\mathrm{free}}(\vect{r}_A)\cdot\vect{d} \,,\\
\label{eq:markov_sigmaz}
\dot{\hat{\sigma}}_z &=& -\Gamma \left( 1+\hat{\sigma}_z \right)
+\frac{2i}{\hbar} \hat{\sigma}^\dagger
\hat{\vect{E}}^{(+)}_{\mathrm{free}}(\vect{r}_A)\cdot\vect{d}
-\frac{2i}{\hbar} \hat{\sigma}
\hat{\vect{E}}^{(-)}_{\mathrm{free}}(\vect{r}_A)\cdot\vect{d}^\ast
\,,
\end{eqnarray}
where $\hat{\vect{E}}^{(\pm)}_{\mathrm{free}}(\vect{r}_A,t)$ denotes
the freely evolving electric field operator which, for example,
describes the action of an external (classical) driving field. The
symbols $\Gamma$ and $\delta\omega$ are abbreviations for the
following objects:
\begin{eqnarray}
\label{eq:vacuum_se}
\Gamma &=& \frac{2\pi}{\hbar^2} \sum\limits_\lambda \omega_\lambda^2
|\vect{A}_\lambda(\vect{r}_A)\cdot\vect{d}|^2
\delta(\omega_A-\omega_\lambda) \,,\\ \label{eq:vacuum_lamb}
\delta\omega &=& \frac{1}{\hbar^2} \sum\limits_\lambda \mathcal{P}
\left(\frac{\omega_\lambda^2}{\omega_A-\omega_\lambda}\right)
|\vect{A}_\lambda(\vect{r}_A)\cdot\vect{d}|^2 \,,
\end{eqnarray}
whose relevance will become clear in the next section.


\subsubsection{Spontaneous decay and Lamb shift}
\label{sec:vacuumse}

The equation of motion for the population inversion operator
$\hat{\sigma}_z$, Eq.~(\ref{eq:markov_sigmaz}), can be solved easily
if no external electric field is present. In this case the equation of
motion reduces to
\begin{equation}
\dot{\hat{\sigma}}_z = -\Gamma \left( 1+\hat{\sigma}_z \right) \,.
\end{equation}
Rewriting the inversion operator in terms of the projectors onto the
excited and ground states, $\hat{\sigma}_z$
$=|e\rangle\langle e|-|g\rangle\langle g|$
$\equiv\hat{\sigma}_{ee}-\hat{\sigma}_{gg}$, we find for the
excited-state projector the simple relation
\begin{equation}
\dot{\hat{\sigma}}_{ee} = -\Gamma \hat{\sigma}_{ee} \qquad \mapsto
\qquad \hat{\sigma}_{ee}(t) = e^{-\Gamma(t-t')} \hat{\sigma}_{ee}(t')
\,.
\end{equation}
Hence, the quantity $\Gamma$ determines the rate with which a
two-level atom decays spontaneously from its excited state to its
ground state. 

Using the expansion (\ref{eq:efeld}), the decay rate
Eq.~(\ref{eq:vacuum_se}) can be written as 
\begin{equation}
\label{eq:FGR}
\Gamma = \frac{2\pi}{\hbar^2} \sum\limits_\lambda
\vect{d} \cdot \langle 0| \hat{\vect{E}}^{(+)}(\vect{r}_A)
\otimes \hat{\vect{E}}^{(-)}(\vect{r}_A)  |0\rangle \cdot
\vect{d}^\ast \delta(\omega_A-\omega_\lambda)\,,
\end{equation}
which has to be understood in such a way that the $\delta$ function is
placed under the mode sum. Hence, the rate with which the atom loses
its excitation depends on the strength of the vacuum fluctuations of
the electric field at the frequency of the atomic transition. In a
certain sense, spontaneous decay can be viewed as stimulated emission
driven by the fluctuating electromagnetic field. Using the plane-wave
expansion (\ref{eq:cartesianA}), we obtain the well-known result for
the spontaneous decay rate in vacuum
\begin{equation}
\label{eq:vacuumse}
\Gamma_0 = \frac{2\pi}{\hbar^2} \sum\limits_\sigma
\int\limits_0^\infty
\frac{\omega^2 d\omega}{c^3(2\pi)^3} \int d\Omega
\frac{\hbar\omega}{2\varepsilon_0} |\vect{e}_\sigma \cdot\vect{d}|^2
\delta(\omega_A-\omega) =
\frac{\omega_A^3|\vect{d}|^2}{3\pi\hbar\varepsilon_0c^3} \,.
\end{equation}

\begin{figure}[!t!]
\begin{center}
\includegraphics[width=0.8\linewidth]{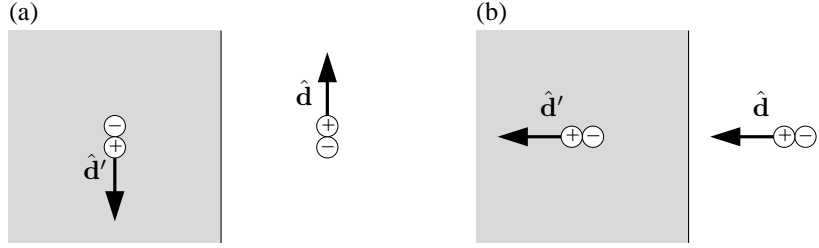}
\end{center}
\caption{
\label{fig:imagedecay}
An atomic dipole $\hat{\vec{d}}$ in front of a perfectly conducting
mirror and its image $\hat{\vec{d}}'$ for (a) parallel and (b)
perpendicular orientation of the dipole with respect to the mirror.
}
\end{figure}

From the above, it should have become clear that the rate of
spontaneous decay can be modified by altering the mode structure of
the electromagnetic field. We have seen previously in
Sec.~\ref{sec:modecasimir} that the presence of boundary conditions
for the electromagnetic field modifies the mode structure and hence
the strength of the vacuum field fluctuations at the atomic transition
frequency $\omega_A$. As a simple example, let us consider a radiating
dipole located close to a perfectly conducting mirror. Depending on
its orientation with respect to the mirror surface, its rate of
spontaneous decay is either completely suppressed or doubled with
respect to its free space rate (Fig.~\ref{fig:imagedecay}).
Suppression occurs when the dipole is parallel to the mirror and
hence the dipole and its image are antiparallel and cancel each other
[Fig.~\ref{fig:imagedecay}(a)]; doubling follows for perpendicular
orientation where the dipole and its image are parallel
[Fig.~\ref{fig:imagedecay}(b)].

The quantity $\delta\omega$ that arises in the context of weakly
interacting systems in the Markov approximation,
Eq.~(\ref{eq:vacuum_lamb}), induces a shift of the atomic transition
frequency, the Lamb shift. Rewriting the Lamb shift using the
plane-wave expansion (\ref{eq:cartesianA}), we find
\begin{equation}
\label{eq:vacuumLamb}
\delta\omega = \frac{1}{\hbar^2c^3} \sum\limits_\sigma
\mathcal{P}\int\limits_0^\infty \frac{d\omega}{(2\pi)^3}
\frac{\omega^2}{\omega_A-\omega} \int d\Omega
\frac{\hbar\omega}{2\varepsilon_0} |\vect{e}_\sigma \cdot\vect{d}|^2
= \frac{|\vect{d}|^2}{6\pi^2\hbar\varepsilon_0c^3}
\mathcal{P}\int\limits_0^\infty d\omega
\frac{\omega^3}{\omega_A-\omega} 
\end{equation}
which is clearly infinite. This is yet another artefact of quantum
theory in free space which can be remedied by mass
renormalisation (for details, see e.g. Ref.~\cite{Milonni}).
For our purposes it is sufficient
to argue that the `bare' atomic transition frequency $\omega_A$ is
unobservable because the interaction with the electromagnetic vacuum
field can never be switched off and hence the only observable quantity
is the renormalised frequency $\omega_A+\delta\omega$. However, in the
following section we will again show how the presence of boundary
conditions and, in particular, dielectric matter can modify the Lamb
shift by an additional finite (and thus measurable) amount.

At this point it is perhaps interesting to observe that spontaneous
decay and the Lamb shift are intimately connected by a causality
relation analogous to the Kramers--Kronig relations that we will
encounter in the next section. In fact, it is easy to see that
$\Gamma$ and $\delta\omega$, taken as functions of the atomic
transition frequency $\omega_A$, form a Hilbert transform
pair. Rewriting the integral (\ref{eq:sincos}) as
\begin{equation}
\lim_{t\to\infty} \int\limits_0^t dt'
\,e^{i(\omega_A-\omega_\lambda)(t-t')}
= \int\limits_0^\infty d\tau \,e^{i(\omega_A-\omega_\lambda)\tau}
= \int\limits_{-\infty}^\infty d\tau
\,e^{i(\omega_A-\omega_\lambda)\tau} \Theta(\tau) \,,
\end{equation}
one observes that this is nothing else than the Fourier transform of
the Heaviside step function $\Theta(t)$. This in turn can be
interpreted at the causal transform of the function $f(\omega)=1$, and
the functions $s(x)$ and $c(x)$ are its real and imaginary parts. Due
to the definition of a causal transform, and by Titchmarsh's theorem
\cite{Nussenzveig}, $s(x)$ and $c(x)$ and therefore $\Gamma$ and
$\delta\omega$ are Hilbert transform pairs and mutually connected via
Kramers--Kronig relations. Hence, knowledge of the spontaneous decay
rate at all frequencies gives access to the Lamb shift and vice versa.


\subsubsection{Optical Bloch equations}
\label{sec:bloch}

In this section we will return to Heisenberg's equations of motion for
the atomic quantities in Markov approximation,
Eqs.~(\ref{eq:markov_sigma}) and (\ref{eq:markov_sigmaz}), and solve
them under the assumption of an external driving field prepared in a
single-mode coherent state $|\alpha\rangle$ with frequency
$\omega_\lambda$. Hence, we set 
\begin{equation}
\langle\alpha| \hat{\vect{E}}^{(+)}_{\mathrm{free}}(\vect{r}_A,t)
|\alpha\rangle = i\omega_\lambda \vect{A}_\lambda(\vect{r}_A) \alpha
e^{-i\omega_\lambda t}
\,,\qquad
\langle\alpha| \hat{\vect{E}}^{(-)}_{\mathrm{free}}(\vect{r}_A,t)
|\alpha\rangle = -i\omega_\lambda \vect{A}_\lambda^\ast(\vect{r}_A)
\alpha^\ast e^{i\omega_\lambda t}\,.
\end{equation}
In a frame that co-rotates with the angular frequency
$\omega_\lambda$, Heisenberg's equations of motion reduce to
\begin{eqnarray}
\label{eq:blochsigma}
\dot{\hat{\sigma}} &=& i\delta \hat{\sigma} -\frac{\Gamma}{2}
\hat{\sigma} +\frac{\Omega}{2} \hat{\sigma}_z \,,\\
\label{eq:blochsigmaz}
\dot{\hat{\sigma}}_z &=& -\Gamma \left(1+\hat{\sigma}_z\right) -\Omega
\hat{\sigma}^\dagger - \Omega^\ast \hat{\sigma} \,,
\end{eqnarray}
where $\delta=\omega_\lambda-\omega_A-\delta\omega$ is the detuning
and
$\Omega=\frac{2\omega_\lambda\alpha}{\hbar}\vect{A}_\lambda(\vect{r}
_A)\cdot\vect{d}$
denotes the Rabi frequency.

The density operator of any two-level (or spin-$1/2$) system can be
written in terms of the Pauli operators as
\begin{equation}
\hat{\varrho} = \frac{1}{2} \left( \hat{I} +\vect{u}\cdot
\hat{\bm{\sigma}}  \right)
\end{equation}
where $\vect{u}=(u,v,w)^\trans$ is a real vector with norm
$|\vect{u}|\le 1$ and $\hat{\bm{\sigma}}$ is the vector of Pauli
matrices. Converting Eqs.~(\ref{eq:blochsigma}) and
(\ref{eq:blochsigmaz}) into equations of motion for the Bloch vector
$\vect{u}$ one finds the matrix equation
\begin{equation}
\left( \begin{array}{c}
\dot{u} \\ \dot{v} \\ \dot{w}
\end{array} \right)
=
\left( \begin{array}{ccc}
0 & -\delta & \Omega_R \\ \delta & 0 & \Omega_I \\ -\Omega_R &
-\Omega_I & 0 
\end{array} \right)
\left( \begin{array}{c}
u \\ v \\ w
\end{array} \right)
-\frac{\Gamma}{2} \left( \begin{array}{c}
u \\ v \\ 2(1+w)
\end{array}\right) \,.
\end{equation}

For negligible spontaneous decay $\Gamma$, the Bloch equations take on
the form of the gyroscopic equations
\begin{equation}
\dot{\vect{u}} = \bm{\beta} \times \vect{u}
\end{equation}
with $\bm{\beta}=(-\Omega_I,\Omega_R,\delta)^T$. Its solution is then
given by
\begin{eqnarray}
u(t) &=& -\frac{\Omega}{\sqrt{\delta^2+\Omega^2}}
\sin(\sqrt{\delta^2+\Omega^2}t) \,,\\
v(t) &=& -\frac{\delta\Omega}{\delta^2+\Omega^2}
\left[1- \cos(\sqrt{\delta^2+\Omega^2}t) \right] \,,\\
w(t) &=& -1+\frac{\Omega^2}{\delta^2+\Omega^2}
\left[1- \cos(\sqrt{\delta^2+\Omega^2}t) \right] \,.
\end{eqnarray}
Hence, the time evolution of the Bloch vector on time scales
$t\ll 1/\Gamma$ is described by precession of the Bloch vector with
frequency $\sqrt{\delta^2+\Omega^2}$ along the surface of a cone with
opening angle $\frac{1}{2}\arccos[\delta\Omega/(\delta^2+\Omega^2)]$.

In the long-time limit, $t\to\infty$, the solution of the optical
Bloch equations reaches its steady-state value. In the stationary
state, when $\dot{\vect{u}}=\vect{0}$, the solution to the full Bloch
equations is 
\begin{equation}
u_\mathrm{st} = -\frac{\delta\Omega}
{\frac{\Omega^2}{2}+\delta^2+(\frac{\Gamma}{2})^2}
\,,
v_\mathrm{st} = -\frac{\frac{\Gamma}{2}\Omega}
{\frac{\Omega^2}{2}+\delta^2+(\frac{\Gamma}{2})^2}
\,,
w_\mathrm{st} = -1+ \frac{\frac{\Omega^2}{2}}
{\frac{\Omega^2}{2}+\delta^2+(\frac{\Gamma}{2})^2}\,.
\end{equation}
In the weak driving limit, $\Omega\ll(\Gamma,\delta)$, the induced
atomic dipole moment $u_\mathrm{st}+iv_\mathrm{st}$ takes the form
\begin{equation}
\label{eq:induceddipole}
u_\mathrm{st}+iv_\mathrm{st} \approx -\frac{\Omega}{\delta-i\Gamma/2}
\end{equation}
whose real and imaginary parts fulfil the Kramers--Kronig
relations when integrated over the mode frequency
$\omega_\lambda$. Recall that both the spontaneous decay rate $\Gamma$
and the Lamb shift $\delta\omega$ (via the detuning $\delta$) are
contained in (\ref{eq:induceddipole}). In this weak-coupling regime,
the induced dipole is thus described by a linear susceptibility. Its
imaginary part, being proportional to the spontaneous decay rate
$\Gamma$, describes a loss channel for the incident field. The
concept of linear-response functions and their role in the
quantisation of the electromagnetic field in the presence of
magnetoelectric matter will be detailed in Sec.~\ref{sec:causalQED}. 


\subsubsection{Jaynes--Cummings model}
\label{sec:jc}

In a sense the opposite limit to the case described in
Sec.~\ref{sec:bloch} is obtained if one considers a situation in which
the mode structure of the electromagnetic field has been altered in
such a way that discrete field modes interact with the atomic
system. We have previously seen in connection with the Casimir effect
(Sec.~\ref{sec:modecasimir}) that this can be achieved in resonators
of Fabry-Perot type where the allowed modes in a cavity of length $d$
have discrete wave numbers $k_z=n\pi/d$. The half distance between two
neighboring modes, $c/(2d)$, is called the free spectral range.

If a two-level atom with transition frequency $\omega_A$ is almost
resonant with one of the cavity modes, one can treat the coupled
atom-field system approximately by a single-mode model described by
the Jaynes--Cummings Hamiltonian \cite{Jaynes63,0431,ShoreKnight}
\begin{equation}
\label{eq:jc}
\hat{H} = \hbar\omega\hat{a}^\dagger\hat{a} +\frac{1}{2}\hbar\omega_A
\hat{\sigma}_z -\hbar g \left(
\hat{a}\hat{\sigma}^\dagger+\hat{a}^\dagger\hat{\sigma} \right) \,.
\end{equation}
The coupling constant $g$, which we have given the dimension of a
frequency, can be read off from Heisenberg's equations of motion as
$g=\frac{i\omega_A}{\hbar}\vect{d}\cdot\vect{A}_\lambda(\vect{r}_A)$.
Because we have assumed near resonance between atomic transition and
the relevant cavity field mode, we have employed the rotating-wave
approximation and subsequently dropped counter-rotating terms in the
Hamiltonian (\ref{eq:jc}).

\begin{figure}[!t!]
\begin{picture}(200,100)(0,0)
\put(0,20){\line(1,0){50}}
\put(50,20){\line(0,1){80}}
\put(50,100){\line(1,0){20}}
\put(70,100){\line(0,-1){80}}
\put(70,20){\line(1,0){60}}
\put(130,20){\line(0,1){80}}
\put(130,100){\line(1,0){20}}
\put(150,100){\line(0,-1){80}}
\put(150,20){\line(1,0){50}}
\put(70,0){\vector(1,0){60}}
\put(130,0){\vector(-1,0){60}}
\put(100,5){$d$}
\end{picture}
\hfill
\includegraphics[height=100pt]{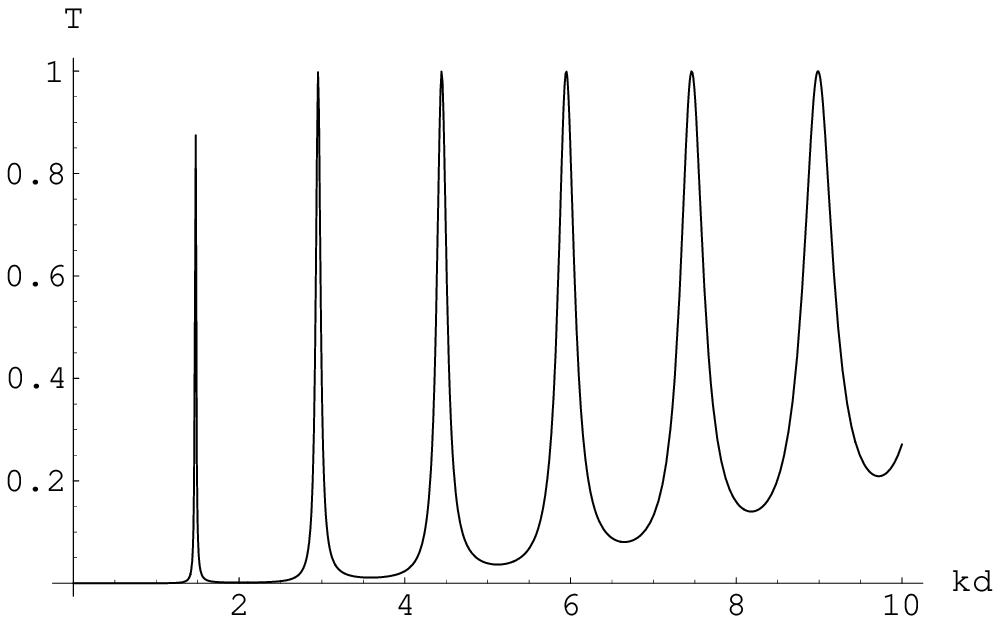}
\caption{\label{fig:doublebarrier} Equivalent potential model of a
cavity with highly reflecting walls (left panel). The transmission
coefficient is sharply peaked at the cavity resonances (right
panel). The line widths of these resonances decrease with increasing
reflectivity of the cavity walls.}
\end{figure}

The assumption that effectively only a single field mode interacts
with the two-level atom implies a sharply peaked, comb-like, density
of field modes. This requires a discretisation of the modes inside the
cavity that can only be achieved with (almost) perfectly reflecting
cavity walls. In reality, the material making up the cavity mirrors
shows some transmission, part of which is of course wanted in order to
be able to probe the cavity field from the outside. In effect, each of
the cavity mirrors can be treated as a beam splitter
(Sec.~\ref{sec:beamsplitter}). For one-dimensional light propagation
the equivalent potential encountered by the vector potential is a
double barrier (Fig.~\ref{fig:doublebarrier}).
%
%
Analytical solutions for the transmission coefficient have been
first obtained in Ref.~\cite{Yamamoto87}. However, the problem becomes
simpler by assuming that the barriers are $\delta$ functions with
strength $g$ \cite{Yanetka99}. Then the transmission coefficient near
one of the cavity resonances $\omega_C$ can be written as
\begin{equation}
T(\omega) \simeq
\frac{\frac{\Gamma}{2}}{\frac{\Gamma}{2}+i(\omega-\omega_C)} 
\end{equation}
where the line width $\Gamma$ is inversely proportional to the barrier
height $g$. Associating the barrier height with the squared index of
refraction of the mirror material, one finds that the better the
reflective properties of the mirrors [recall that
$r=(\sqrt{\varepsilon}-1)/(\sqrt{\varepsilon}+1)$] the 
narrower the resonances.

The Jaynes--Cummings model is one of the few exactly solvable models
of interacting quantum systems. The Hamiltonian (\ref{eq:jc}) is in
fact block diagonal in the basis $\{|n,e\rangle,|n+1,g\rangle\}$,
\begin{equation}
\hat{H}
\left( \begin{array}{c}
|n,e\rangle \\ |n+1,g\rangle
\end{array} \right)
=\hbar
\left( \begin{array}{cc}
n\omega+\frac{1}{2}\omega_A & -g\sqrt{n+1} \\
-g\sqrt{n+1} & (n+1)\omega-\frac{1}{2}\omega_A
\end{array} \right)
\left( \begin{array}{c}
|n,e\rangle \\ |n+1,g\rangle
\end{array} \right) \,.
\end{equation}
Its eigenfrequencies are
\begin{equation}
\omega_{n,\pm} = \left( n+\frac{1}{2}\right) \omega \pm \frac{1}{2}
\Delta_n
\end{equation}
where we have defined the Rabi splitting
$\Delta_n=\sqrt{\delta^2+\Omega_n^2}$ which depends on the detuning
$\delta=\omega_A-\omega$ and the $n$-photon Rabi frequency
$\Omega_n=2g\sqrt{n+1}$.
%
\begin{figure}[!t!]
\begin{center}
\begin{picture}(200,160)(0,0)
\put(0,0){\vector(0,1){160}}
\put(5,150){$E/\hbar$}
\put(60,10){\line(1,0){60}}
\put(60,50){\line(6,1){60}}
\put(60,50){\line(6,-1){60}}
\put(60,90){\line(5,1){60}}
\put(60,90){\line(5,-1){60}}
\put(60,130){\line(4,1){60}}
\put(60,130){\line(4,-1){60}}
\put(165,40){\vector(0,1){20}}
\put(165,60){\vector(0,-1){20}}
\put(165,78){\vector(0,1){24}}
\put(165,102){\vector(0,-1){24}}
\put(165,115){\vector(0,1){30}}
\put(165,145){\vector(0,-1){30}}
\linethickness{1pt}
\put(10,10){\line(1,0){50}}
\put(10,50){\line(1,0){50}}
\put(10,90){\line(1,0){50}}
\put(10,130){\line(1,0){50}}
\put(120,10){\line(1,0){50}}
\put(120,40){\line(1,0){50}}
\put(120,60){\line(1,0){50}}
\put(120,78){\line(1,0){50}}
\put(120,102){\line(1,0){50}}
\put(120,115){\line(1,0){50}}
\put(120,145){\line(1,0){50}}
\put(170,46){$\Omega_0=2g$}
\put(170,86){$\Omega_1=2\sqrt{2}g$}
\put(170,126){$\Omega_2=2\sqrt{3}g$}
\put(12,14){$\scriptstyle|0,g\rangle$}
\put(12,54){$\scriptstyle|0,e\rangle,|1,g\rangle$}
\put(12,94){$\scriptstyle|1,e\rangle,|2,g\rangle$}
\put(12,134){$\scriptstyle|2,e\rangle,|3,g\rangle$}
\put(122,14){$\scriptstyle|0,g\rangle$}
\put(122,44){$\scriptstyle|0,-\rangle$}
\put(122,64){$\scriptstyle|0,+\rangle$}
\put(122,82){$\scriptstyle|1,-\rangle$}
\put(122,106){$\scriptstyle|1,+\rangle$}
\put(122,119){$\scriptstyle|2,-\rangle$}
\put(122,149){$\scriptstyle|2,+\rangle$}
\end{picture}
\end{center}
\caption{\label{fig:jc} Energy level diagram of the resonant
Jaynes--Cummings model. Apart from the collective ground state
$|0,g\rangle$, the states $\{|n,e\rangle,|n+1,g\rangle\}$ are doubly
degenerate. This degeneracy is lifted by the interaction.}
\end{figure}
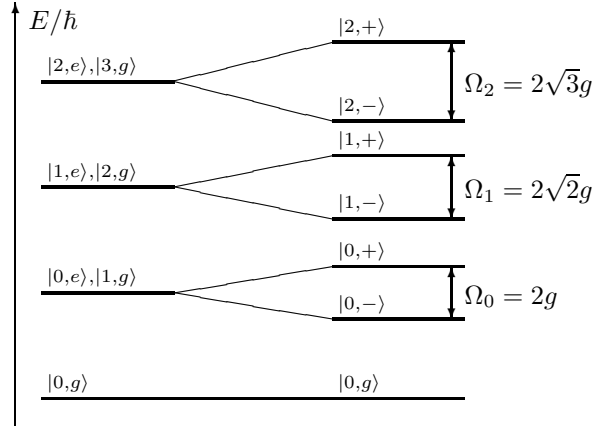
%
The eigenstates $|n,\pm\rangle$ of the Jaynes--Cummings Hamiltonian
are superpositions of the unperturbed eigenstates
$\{|n,e\rangle,|n+1,g\rangle\}$,
\begin{equation}
\left( \begin{array}{c}
|n,+\rangle \\ |n,-\rangle
\end{array}\right)
=\left( \begin{array}{cc}
\cos\Theta_n & -\sin\Theta_n \\ \sin\Theta_n & \cos\Theta_n
\end{array} \right)
\left( \begin{array}{c}
|n+1,g\rangle \\ |n,e\rangle
\end{array} \right) \,,
\end{equation}
where the rotation angles are
\begin{equation}
\sin\Theta_n = \frac{\Omega_n}{\sqrt{(\Delta_n-\delta)^2+\Omega_n^2}}
\,,\qquad
\cos\Theta_n =
\frac{\Delta_n-\delta}{\sqrt{(\Delta_n-\delta)^2+\Omega_n^2}} \,.
\end{equation}
For resonant interaction, $\delta=0$, the unperturbed eigenstates are
pairwise degenerate (Fig.~\ref{fig:jc}). This degeneracy is lifted by
the atom-field interaction. The level splitting $\Delta_n$ depends on
the number of photons. Note that even if there is initially no photon
present, the exact eigenstates will be split by an amount
$\Omega_0=2g$, the vacuum Rabi splitting. This splitting, or
`dressing', of the bare energy levels is equivalent to the Lamb shift
which we found in Sec.~\ref{sec:vacuumse}. However, it should be noted
that the vacuum Lamb shift was due to the interaction with
electromagnetic modes of all frequencies, whereas in the
Jaynes--Cummings model the level shift arises from the interaction
with a single discrete mode that has been selected by the resonator.

Because the Jaynes--Cummings Hamiltonian can be explicitly
diagonalised, the unitary evolution operator is also known explicitly
and reads
\begin{eqnarray}
\label{eq:jcU}
\hat{U}(t) &=& e^{-i\hat{H}t\hbar}
= e^{i\omega_At/2} |0,g\rangle\langle 0,g|
+\sum\limits_{\sigma=\pm}\sum\limits_{n=0}^\infty
e^{-i\omega_{n,\sigma}t} |n,\sigma\rangle\langle n,\sigma|
\nonumber \\ && \hspace*{-10ex} = 
e^{i\omega_At/2} |0,g\rangle\langle 0,g| +\sum\limits_{n=0}^\infty
e^{-i(n+1/2)\omega t} \left[ e^{-i\Delta_nt/2} |n,+\rangle\langle n,+|
+e^{i\Delta_nt/2} |n,-\rangle\langle n,-| \right]
\nonumber \\ && \hspace*{-10ex} =
e^{i\omega_At/2} |0,g\rangle\langle 0,g| +\sum\limits_{n=0}^\infty
e^{-i(n+1/2)\omega t} \left\{ \left[
\cos{\textstyle\frac{\Delta_nt}{2}}
+i{\textstyle\frac{\delta}{\Delta_n}}\sin{\textstyle\frac{\Delta_nt}{2
}}
\right] |n+1,g\rangle\langle n+1,g| \right.
\nonumber \\ && \hspace*{-10ex} \left.
+\left[ \cos{\textstyle\frac{\Delta_nt}{2}}
-i{\textstyle\frac{\delta}{\Delta_n}}
\sin{\textstyle\frac{\Delta_nt}{2}} \right]
|n,e\rangle\langle n,e| 
+i{\textstyle\frac{\Omega_n}{\Delta_n}}
\sin{\textstyle\frac{\Delta_nt}{2}} \left( 
|n+1,g\rangle\langle n,e|+|n,e\rangle\langle n+1,g| \right)  \right\}.
\nonumber \\
\end{eqnarray}
The last equality in Eq.~(\ref{eq:jcU}) expresses the unitary time
evolution in terms of the unperturbed eigenstates.

An important special case is the dispersive limit in which the
detuning $\delta$ is large compared to the relevant $n$-photon Rabi
frequency, $\delta\gg\Omega_n$. In this limit, the level splitting can
be approximated by $\Delta_n\simeq\delta+\Omega_n^2/(2\delta)$
$=\delta+2g^2(n+1)\delta$, and the exact eigenstates are approximately
the unperturbed eigenstates, $|n,+\rangle\simeq|n,e\rangle$ and
$|n,-\rangle\simeq|n+1,g\rangle$. Under this approximation, the
unitary evolution operator can be written as
\begin{eqnarray}
\hat{U} &\simeq& e^{i\omega_At/2} |0,g\rangle\langle 0,g|
+\sum\limits_{n=0}^\infty e^{-i(n+1/2)\omega t}
\nonumber \\ && \hspace*{-3ex} \times
\left[ e^{-i(\delta/2+g^2(n+1)/\delta)t} |n,e\rangle\langle n,e| +
e^{i(\delta/2+g^2(n+1)/\delta)t} |n+1,g\rangle\langle n+1,g| \right]
\end{eqnarray}
or, with the free Hamiltonian
$\hat{H}_0=\hbar\omega\hat{a}^\dagger\hat{a}
+\frac{1}{2}\hbar\omega_A\hat{\sigma}_z$,
\begin{equation}
\label{eq:dispersivelimit}
\hat{U} \simeq \exp\left(-\frac{i}{\hbar}\hat{H}_0t\right) \left[
\exp\left(-\frac{ig^2}{\delta}(\hat{n}+1)t\right)|e\rangle\langle e|
+\exp\left(\frac{ig^2}{\delta}\hat{n}t\right)|g\rangle\langle
g|\right] 
\end{equation}
It is instructive to note that the evolution operator the dispersive
limit, Eq.~(\ref{eq:dispersivelimit}), is quadratic in the interaction
strength $g$. This means that it results in an effective nonlinear
atom-field interaction. In order to investigate this claim in more
detail, let us rewrite Eq.~(\ref{eq:dispersivelimit}) in the following
form. The term in square brackets contains, apart from a linear Stark
shift of the excited state $|e\rangle$, a factor
$e^{-ig^2\hat{n}\hat{\sigma}_zt/\delta}$ which is the result of an
effective nonlinear Hamiltonian
\begin{equation}
\label{eq:dispersiveHeff}
\hat{H}_\mathrm{eff} = \hbar\frac{g^2}{\delta} \hat{a}^\dagger\hat{a}
\hat{\sigma}_z \,.
\end{equation}
Comparing Eq.~(\ref{eq:dispersiveHeff}) with the Jaynes--Cummings
Hamiltonian (\ref{eq:jc}) we see that the trilinear Hamiltonian
(\ref{eq:dispersiveHeff}) does not appear in the original
Jaynes--Cummings model. The appearance of such an effective
Hamiltonian is solely due the far off-resonant interaction. This is
just one example of a generic nonlinear interaction such as those
studied in Sec.~\ref{sec:nonlinear}.


\newpage
\section{Macroscopic quantum electrodynamics}
\label{sec:causalQED}

Having established the framework of quantum electrodynamics in free
space, we are now in a position to generalise the theory to
magnetoelectric background materials. Before we go ahead with our
program, let us first discuss the intrinsic difficulties associated
with magnetoelectric media. 

Let us assume we wanted to naively extend a plane-wave expansion of
the electromagnetic field to include dielectrics. We would then try to
replace the plane wave solutions $e^{i\vect{k}\cdot\vect{r}}$ to
the Helmholtz equation by $e^{in\vect{k}\cdot\vect{r}}$ where
$n\equiv n(\omega)$ is the index of refraction of the dielectric. In
order to conform with standard requirements from statistical physics,
the refractive index must be a complex function of frequency
$n(\omega)=\eta(\omega)+i\kappa(\omega)$ that satisfies the
Kramers--Kronig relations
\begin{equation}
\label{eq:KramersKronig}
\eta(\omega)-1 = \frac{1}{\pi}
\mathcal{P}\!\int\limits_{-\infty}^\infty 
d\omega' \frac{\kappa(\omega')}{\omega'-\omega} \,,\qquad
\kappa(\omega) = -\frac{1}{\pi}
\mathcal{P}\!\int\limits_{-\infty}^\infty
d\omega' \frac{\eta(\omega')-1}{\omega'-\omega} \,.
\end{equation} 
Due to the inevitable imaginary part of the refractive index, the
plane waves are generically damped. That in turn means that
they do not form a complete set of orthonormal functions needed to
perform a Fourier decomposition of the electromagnetic field. The
consequences of this failure are quite severe; either one insists on
bosonic commutation rules for the photonic amplitude operators
$\hat{a}_\lambda$ and  $\hat{a}_\lambda^\dagger$ which subsequently
lead to wrong commutation relations between the operators of the
electric field $\hat{\vect{E}}(\vect{r})$ and the magnetic
induction $\hat{\vect{B}}(\vect{r})$, or one postulates the
correctness of the latter and ends up with amplitude operators in
their Fourier decomposition that do not have the interpretation of
annihilation and creation operators of photonic modes. 

The reason for the failure of this naive quantisation scheme is easily
found. The introduction of the index of refraction $n(\omega)$ means
that there exists an underlying (microscopic) theory that couples the
free electromagnetic field to some dielectric matter, the effect of
which is taken into account only by means of the response function
$n(\omega)$. In doing so, the  matter-field coupling is hidden from
view but is nevertheless present. The damped plane waves
$e^{in\vect{k}\cdot\vect{r}}$ have therefore to be regarded as
eigensolutions of the combined field-matter system, and not of the
electromagnetic field alone. 


\subsection{Field quantisation in linear absorbing magnetoelectrics}

The above arguments necessarily lead one to consider the
electromagnetic field interacting with an atomic system coupled to a
reservoir that is responsible for absorption. An explicit matter-field
coupling theory that achieves field quantisation in dielectric matter
on the basis of a microscopic model has been developed by Huttner and
Barnett \cite{Huttner92a,Huttner92} (Sec.~\ref{sec:hb}). This
Hamiltonian model can be generalised to an effective Langevin noise
model (Sec.~\ref{sec:langevin}) which forms the basis of the remainder
of this article. 


\subsubsection{Huttner--Barnett model}
\label{sec:hb}

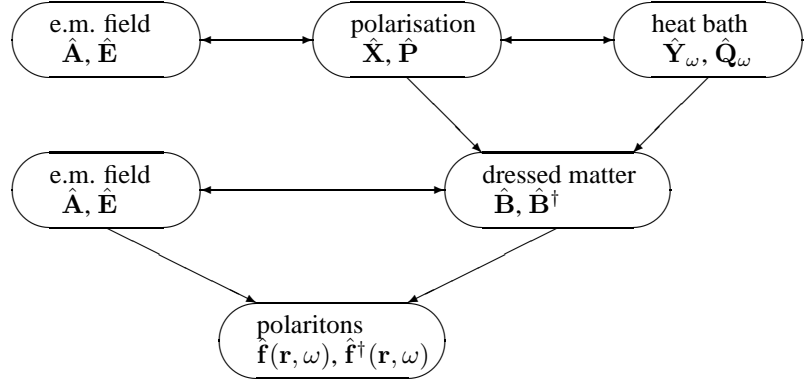
\begin{figure}[!t!]
\begin{center}
\setlength{\unitlength}{0.5cm}
\begin{picture}(21,12)(-1,0)
\put(2.5,11){\oval(5,2)}
\put(1,11.2){e.m. field}
\put(1.3,10.4){$\hat{\vect{A}}$, $\hat{\vect{E}}$}
\put(5,11){\vector(1,0){3}}
\put(8,11){\vector(-1,0){3}}
\put(10.5,11){\oval(5,2)}
\put(9,11.2){polarisation}
\put(9.3,10.4){$\hat{\vect{X}}$, $\hat{\vect{P}}$}
\put(13,11){\vector(1,0){3}}
\put(16,11){\vector(-1,0){3}}
\put(18.5,11){\oval(5,2)}
\put(17,11.2){heat bath}
\put(17.3,10.4){$\hat{\vect{Y}}_\omega$, $\hat{\vect{Q}}_\omega$}
\put(2.5,7){\oval(5,2)}
\put(1,7.2){e.m. field}
\put(1.3,6.4){$\hat{\vect{A}}$, $\hat{\vect{E}}$}
\put(5,7){\vector(1,0){6.5}}
\put(11.5,7){\vector(-1,0){6.5}}
\put(14.5,7){\oval(6,2)}
\put(12.5,7.2){dressed matter}
\put(12.8,6.4){$\hat{\vect{B}}$,
$\hat{\vect{B}}^\dagger$}
\put(10.5,10){\vector(1,-1){2}}
\put(18.5,10){\vector(-1,-1){2}}
\put(8.5,3){\oval(6,2)}
\put(6.5,3.2){polaritons}
\put(6.5,2.4){$\hat{\vect{f}}(\vect{r},\omega)$,
$\hat{\vect{f}}^\dagger(\vect{r},\omega)$}
\put(2.5,6){\vector(2,-1){4}}
\put(14.5,6){\vector(-2,-1){4}}
\end{picture}
\end{center}
\caption{\label{fig:hb} Two-step diagonalisation of the
Huttner--Barnett model. In the first step, the polarisation field and
the harmonic-oscillator heat bath form dressed-matter operators. These
are then combined in the second step with the free electromagnetic
field to form polariton operators.}
\end{figure}

Historically, the first successful attempt at quantising the
electromagnetic field in an absorbing dielectric material is due to
Huttner and Barnett \cite{Huttner92a,Huttner92}. They considered a
Hopfield model \cite{Hopfield58} of a homogeneous and isotropic bulk
dielectric in which a harmonic oscillator field representing the
medium polarisation is linearly coupled to a continuum of harmonic
oscillators standing for the reservoir (first line in
Fig.~\ref{fig:hb}). Such a model leads to an essentially
unidirectional energy flow --- from the medium polarisation to the
reservoir --- which means that the energy is absorbed. Strictly
speaking, because a single harmonic oscillator is coupled to a
continuum, the revival time is infinite, hence an excitation stored in
the continuum of harmonic oscillators will not return to the medium
polarisation in any finite time. The overall system of radiation,
matter polarisation, reservoir and their mutual couplings are regarded
as a Hamiltonian system whose Lagrangian reads
\begin{equation}
\label{eq:LHuttner}
L = \int d^3r \,\mathcal{L} = \int d^3r \,\left(
\mathcal{L}_\mathrm{em} +\mathcal{L}_\mathrm{mat}
+\mathcal{L}_\mathrm{int} \right)
\end{equation}
where 
\begin{eqnarray}
\label{eq:LemHuttner}
\mathcal{L}_\mathrm{em} &=& \frac{\varepsilon_0}{2} \left[ \left(
\dot{\vect{A}} +\grad \phi \right)^2 -c^2 \left( \curl\vect{A}
\right)^2 \right] \,,\\
\label{eq:LmatHuttner}
\mathcal{L}_\mathrm{mat} &=& \frac{\mu}{2} \left( \dot{\vect{X}}^2 -
\omega_0^2 \vect{X}^2\right)
+\frac{1}{2}\int\limits_0^\infty d\omega \,\mu \left(
\dot{\vect{Y}}_\omega^2 - \omega^2 \vect{Y}_\omega^2  \right)
\,,\\
\label{eq:LintHuttner}
\mathcal{L}_\mathrm{int} &=& -\alpha \left(
\vect{A}\cdot\dot{\vect{X}} +\phi \divv\vect{X} \right)
-\int\limits_0^\infty d\omega \,v(\omega) \vect{X}
\cdot\dot{\vect{Y}}_\omega \,.
\end{eqnarray}
Here $\mathcal{L}_\mathrm{em}$ and $\mathcal{L}_\mathrm{mat}$ are the
free Lagrangian densities of the radiation field and the matter,
respectively, where $\phi$ and $\vect{A}$ are the scalar and vector
potentials in the Coulomb gauge ($\divv\vect{A}=0$), and $\vect{X}$
and $\vect{Y}_\omega$ the medium and reservoir oscillator fields
with density $\mu$, respectively. In the interaction part,
$\mathcal{L}_\mathrm{int}$, $\alpha$ is the electric polarisability
and the medium-reservoir coupling constants $v(\omega)$ are assumed to
be square integrable.

Upon introducing the canonical momenta
\begin{eqnarray}
\label{eq:Pi}
\bm{\Pi} &=& \frac{\partial\mathcal{L}}{\partial\dot{\vect{A}}} =
\varepsilon_0 \dot{\vect{A}} \,,\\
\label{eq:P}
\vect{P} &=& \frac{\partial\mathcal{L}}{\partial\dot{\vect{X}}} =
\mu\dot{\vect{X}}-\alpha\vect{A} \,,\\
\label{eq:Q}
\vect{Q}_\omega &=&
\frac{\partial\mathcal{L}}{\partial\dot{\vect{Y}}_\omega} =
\mu\dot{\vect{Y}}_\omega -v(\omega) \vect{X} \,,
\end{eqnarray}
one can perform the Legendre transformation and construct a
Hamiltonian $H=H_\mathrm{em}+H_\mathrm{mat}+H_\mathrm{int}$. In
Fourier space,
\begin{eqnarray}
\label{eq:AFourier}
\vect{A}(\vect{r}) &\mapsto& \vect{A}(\vect{k}) =
\sum\limits_{\lambda=1}^2 A_\lambda(\vect{k})
\vect{e}_\lambda(\vect{k}) \,,\\
\label{XFourier}
\vect{X}(\vect{r}) &\mapsto& \vect{X}(\vect{k}) =
X_\|(\vect{k})\vect{e}_\vect{k} +\sum\limits_{\lambda=1}^2
X_\lambda(\vect{k}) \vect{e}_\lambda(\vect{k}) \,,
\end{eqnarray}
are the longitudinal and transverse components of the vector potential
and the matter polarisation, respectively. Similar decompositions are
made for all other fields.

As in free space, one introduces mode amplitudes according to
\begin{eqnarray}
\label{eq:alambda}
a_\lambda(\vect{k}) &=& \sqrt{\frac{\varepsilon_0}{2\hbar\tilde{k}c}}
\left[ \tilde{k}cA_\lambda(\vect{k}) +\frac{i}{\varepsilon_0}
\Pi_\lambda(\vect{k})\right] \,,\\
\label{eq:blambda}
b_\lambda(\vect{k}) &=& \sqrt{\frac{\mu}{2\hbar\tilde{\omega}}}
\left[ \tilde{\omega}X_\lambda(\vect{k})+\frac{i}{\mu}
Q_\lambda(\vect{k})\right] \,,\\
\label{eq:alambdaomega}
b_\lambda(\vect{k},\omega) &=& \sqrt{\frac{\mu}{2\hbar\omega}} \left[ 
-i\omega X_\lambda(\vect{k},\omega)+\frac{1}{\mu}
Q_\lambda(\vect{k},\omega)\right] 
\end{eqnarray}
where
\begin{equation}
\label{eq:shiftedfrequencies}
\tilde{k}^2 = k^2+\frac{\alpha^2}{\mu\varepsilon_0c^2} \,,\qquad
\tilde{\omega}^2 = \omega^2+\int\limits_0^\infty d\omega 
\frac{v^2(\omega)}{\mu^2} \,.
\end{equation}
The expressions (\ref{eq:shiftedfrequencies}) reflect the level shifts
due to the interaction between fields. Indeed, we have encountered
such shifts already in vacuum QED (Lamb shift, dressed energy levels
in the Jaynes--Cummings model etc.). Similar decompositions can be
made for the longitudinal fields which we will not consider here
\cite{Buch}.

The amplitude operators are then promoted to Hilbert space operators
with the usual bosonic commutation relations
\begin{eqnarray}
\label{eq:acommutator}
\left[ \hat{a}_\lambda(\vect{k}),
\hat{a}_{\lambda'}^\dagger(\vect{k}')
\right] &=& \delta_{\lambda\lambda'} \delta(\vect{k}-\vect{k}') \,,\\
\label{eq:bcommutator}
\left[ \hat{b}_\lambda(\vect{k}),
\hat{b}_{\lambda'}^\dagger(\vect{k}')
\right] &=& \delta_{\lambda\lambda'} \delta(\vect{k}-\vect{k}') \,,\\
\label{eq:bomegacommutator}
\left[ \hat{b}_\lambda(\vect{k},\omega), 
\hat{b}_{\lambda'}(\vect{k}',\omega') \right] &=&
\delta_{\lambda\lambda'}
\delta(\vect{k}-\vect{k}') \delta(\omega-\omega') \,.
\end{eqnarray}
The transverse Hamiltonian can be expressed in terms of the
annihilation and creation operators as
\begin{equation}
\label{eq:HHuttner}
\hat{H} = \hat{H}_\mathrm{em} +\hat{H}_\mathrm{mat}
+\hat{H}_\mathrm{int}
\end{equation}
with
\begin{eqnarray}
\label{eq:HemHuttner}
\hat{H}_\mathrm{em} &=& \sum\limits_{\lambda=1}^2 \int d^3k\, 
\hbar\tilde{k}c \,\hat{a}_\lambda^\dagger(\vect{k})
\hat{a}_\lambda(\vect{k}) \,,\\
\label{eq:HmatHuttner}
\hat{H}_\mathrm{mat} &=& \sum\limits_{\lambda=1}^2 \int d^3k\,
\bigg\{ \hbar\tilde{\omega}\,\hat{b}_\lambda^\dagger(\vect{k})
\hat{b}_\lambda(\vect{k})+\int\limits_0^\infty d\omega \,
\hbar\omega\, \hat{b}_\lambda^\dagger(\vect{k},\omega) 
\hat{b}_\lambda(\vect{k},\omega) 
\nonumber \\ &&
+\frac{\hbar}{2}\int\limits_0^\infty d\omega \,V(\omega) \left[
\hat{b}_\lambda^\dagger(\vect{k})+\hat{b}_\lambda(-\vect{k}) \right]
\left[ \hat{b}_\lambda^\dagger(-\vect{k},\omega) 
+\hat{b}_\lambda(\vect{k},\omega)\right] \bigg\} \,,\\
\label{eq:HintHuttner}
\hat{H}_\mathrm{int} &=& \frac{i\hbar}{2} \sum\limits_{\lambda=1}^2 
\int d^3k\,\Lambda(k) \left[ \hat{a}_\lambda^\dagger(-\vect{k}) 
+\hat{a}_\lambda(\vect{k}) \right] \left[
\hat{b}_\lambda^\dagger(\vect{k})+\hat{b}_\lambda(-\vect{k}) \right]
\end{eqnarray}
where $V(\omega)=[v(\omega)/\mu](\omega/\tilde{\omega})^{1/2}$ and
$\Lambda(k)=[\tilde{\omega}\alpha^2/(\mu
c\varepsilon_0\tilde{k})]^{1/2}$. The Hamiltonian is clearly bilinear
in all annihilation and creation operators, and can therefore be
diagonalised by a Bogoliubov (squeezing) transformation, that is, by a
linear transformation involving both annihilation and creation
operators. In the present context, the procedure is known as a
Fano-type diagonalisation \cite{Fano56}.
%
%
The diagonalisation is performed in two steps. In the first step, the
matter Hamiltonian $\hat{H}_\mathrm{mat}$ is diagonalised first 
(second line in Fig.~\ref{fig:hb}), leading to
\begin{equation}
\label{eq:HmatHuttnerdiag}
\hat{H}_\mathrm{mat} = \sum\limits_{\lambda=1}^2 \int d^3k 
\int\limits_0^\infty d\omega\, \hbar\omega\, 
\hat{B}_\lambda^\dagger(\vect{k},\omega) 
\hat{B}_\lambda(\vect{k},\omega) \,.
\end{equation}
In the second step, the dressed-matter operators
$\hat{B}_\lambda(\vect{k},\omega)$ and
$\hat{B}^\dagger_\lambda(\vect{k},\omega)$ are combined with the
photonic operators to the diagonal Hamiltonian 
\begin{equation}
\label{eq:HHuttnerdiag}
\hat{H} = \sum\limits_{\lambda=1}^2 \int d^3k \int\limits_0^\infty 
d\omega\, \hbar\omega\, \hat{f}_\lambda^\dagger(\vect{k},\omega) 
\hat{f}_\lambda(\vect{k},\omega) \,.
\end{equation}

Diagonalisation of the longitudinal field components can be achieved
analogously. Adding the resulting expression to
Eq.~(\ref{eq:HHuttnerdiag}) and  Fourier transforming gives
\begin{equation}
\label{eq:HHuttnerdiag2}
\hat{H} = \int d^3r \,\hbar\omega\,
\hat{\vect{f}}^\dagger(\vect{r},\omega) \cdot
\hat{\vect{f}}(\vect{r},\omega) 
\end{equation}
which is depicted in the last line in Fig.~\ref{fig:hb}. Due to the 
bosonic commutation relation of the amplitude operators, the
commutation rule for the new dynamical variables are
\begin{equation}
\label{eq:fHuttnercommutator}
\left[ \hat{\vect{f}}(\vect{r},\omega) , 
\hat{\vect{f}}^\dagger(\vect{r}',\omega') \right]
= \bm{\delta}(\vect{r}-\vect{r}') \delta(\omega-\omega') \,.
\end{equation}

Inverting the Bogoliubov transformation that has led to the
polariton-like operators $\hat{C}_\lambda(\vect{k},\omega)$ and
$\hat{C}_\lambda^\dagger(\vect{k},\omega)$ and subsequent Fourier
transformation leaves one with an expression for the vector potential
$\hat{\vect{A}}(\vect{r})$ and the polarisation field 
$\hat{\vect{P}}(\vect{r})$ in terms of the dynamical variables 
$\hat{\vect{f}}(\vect{r},\omega)$ and 
$\hat{\vect{f}}^\dagger(\vect{r},\omega)$. The expansion
coefficients turn out to be the dyadic Green tensor for a homogeneous
and isotropic bulk material with a dielectric permittivity that is
constructed from the microscopic coupling parameters $\alpha$,
$v(\omega)$ and $\mu$ \cite{Huttner92,Schmidt97}. Later, this theory
has been extended to inhomogeneous dielectrics where Laplace
transformation techniques have been used to solve the resulting
coupled differential equations \cite{Suttorp04}. However, neither of
these expressions for the resulting fields contains any hints towards
their underlying microscopic model, so it seems quite natural to start
from the source-quantity representation of the electromagnetic field
instead.


\subsubsection{Langevin-noise approach}
\label{sec:langevin}

From now on, we leave the microscopic models behind and concentrate on
the phenomenological Maxwell's equations, assuming that the relevant
response functions such as dielectric permittivity and magnetic
permeability are known from measurements. Maxwell's equations of
classical electromagnetism, in the presence of magnetoelectric
background media read, in the absence of other external sources or
currents,
\begin{eqnarray}
\label{eq:Maxwell1}
\divv\vect{B}(\vect{r}) &=& 0\,,\\
\label{eq:Maxwell2}
\curl\vect{E}(\vect{r}) &=& -\dot{\vect{B}}(\vect{r})\,,\\
\label{eq:Maxwell3}
\divv\vect{D}(\vect{r}) &=& 0\,,\\
\label{eq:Maxwell4}
\curl\vect{H}(\vect{r}) &=& \dot{\vect{D}}(\vect{r})\,.
\end{eqnarray}
They have to be supplemented by constitutive relations that connect
the electric and magnetic field components. Assuming for a moment that
the medium under consideration is not bianisotropic, we can write
\begin{equation}
\label{eq:constitutive}
\vect{D}(\vect{r}) = \varepsilon_0 \vect{E}(\vect{r}) 
+\vect{P}(\vect{r}) \,,\qquad
\vect{H}(\vect{r}) = \frac{1}{\mu_0} \vect{B}(\vect{r})
-\vect{M}(\vect{r})
\end{equation}
where $\vect{P}(\vect{r})$ and $\vect{M}(\vect{r})$ denote the
polarisation and magnetisation fields, respectively.

Polarisation and magnetisation are themselves complicated functions of
the electric field $\vect{E}(\vect{r})$ and the magnetic induction
$\vect{B}(\vect{r})$. Assuming that the medium responds linearly and 
locally to externally applied fields, the most general relations
between the fields that are consistent with causality and the linear
fluctuation-dissipation theorem can be cast into the form
\begin{eqnarray}
\label{eq:classicalP}
\vect{P}(\vect{r},t) = \varepsilon_0 \int\limits_0^\infty d\tau\,
\chi_\mathrm{e}(\vect{r},\tau) \vect{E}(\vect{r},t-\tau) 
+\vect{P}_\mathrm{N}(\vect{r},t)\,,\\
\label{eq:classicalM}
\vect{M}(\vect{r},t) = \frac{1}{\mu_0} \int\limits_0^\infty d\tau\,
\chi_\mathrm{m}(\vect{r},\tau) \vect{B}(\vect{r},t-\tau) 
-\vect{M}_\mathrm{N}(\vect{r},t)\,
\end{eqnarray}
where $\vect{P}_\mathrm{N}(\vect{r},t)$ and 
$\vect{M}_\mathrm{N}(\vect{r},t)$ are the noise polarisation and
magnetisation, respectively, that are associated with absorption in
the medium with electric and magnetic susceptibilities 
$\chi_\mathrm{e}(\vect{r},\tau)$ and $\chi_\mathrm{m}(\vect{r},\tau)$.

The Fourier transformed expressions (\ref{eq:classicalP}) and 
(\ref{eq:classicalM}) convert the constitutive relations 
(\ref{eq:constitutive}) into
\begin{equation}
\label{eq:constitutive2}
\vect{D}(\vect{r},\omega) = \varepsilon_0 \varepsilon(\vect{r},\omega)
\vect{E}(\vect{r},\omega) +\vect{P}_\mathrm{N}(\vect{r},\omega)\,,
\quad
\vect{H}(\vect{r},\omega) = \kappa_0 \kappa(\vect{r},\omega)
\vect{B}(\vect{r},\omega) -\vect{M}_\mathrm{N}(\vect{r},\omega)\,,
\end{equation}
[$\kappa(\vect{r},\omega)=\mu^{-1}(\vect{r},\omega)$] where
\begin{equation}
\label{eq:causaltransform}
\varepsilon(\vect{r},\omega) = 1+\int\limits_0^\infty d\tau\,
\chi_\mathrm{e}(\vect{r},\tau) e^{i\omega\tau} \,,\qquad
\kappa(\vect{r},\omega) = 1-\int\limits_0^\infty d\tau\,
\chi_\mathrm{m}(\vect{r},\tau) e^{i\omega\tau}
\end{equation}
are the relative dielectric permittivity and (inverse) magnetic
permability, respectively. An immediate consequence of the causal
relation (\ref{eq:causaltransform}) is the validity of Kramers--Kronig
(Hilbert transform) relations between the real and imaginary parts of
the susceptibilities in Fourier space,
\begin{equation}
\label{eq:kk}
\mathrm{Re}\,\chi(\vect{r},\omega) = \frac{1}{\pi} \mathcal{P}
\int\limits_{-\infty}^\infty d\omega' 
\frac{\mathrm{Im}\,\chi(\vect{r},\omega')}{\omega'-\omega} \,,\quad
\mathrm{Im}\,\chi(\vect{r},\omega) = -\frac{1}{\pi} \mathcal{P}
\int\limits_{-\infty}^\infty d\omega' 
\frac{\mathrm{Re}\,\chi(\vect{r},\omega')}{\omega'-\omega} \,.
\end{equation}

Using the expressions (\ref{eq:causaltransform}) for the dielectric
permittivity and the magnetic permeability, Maxwell's equations for
the Fourier components can thus be written as
\begin{eqnarray}
\label{eq:macroMax1}
\divv\vect{B}(\vect{r},\omega) &=& 0\,,\\
\label{eq:macroMax2}
\curl\vect{E}(\vect{r},\omega) &=& \vect{B}(\vect{r},\omega)\,,\\
\label{eq:macroMax3}
\varepsilon_0\divv\left[\varepsilon(\vect{r},\omega)
\vect{E}(\vect{r},\omega) \right] &=& 
\rho_\mathrm{N}(\vect{r},\omega)\,,\\
\label{eq:macroMax4}
\curl\left[\kappa(\vect{r},\omega)\vect{B}(\vect{r},\omega)\right] 
+i\frac{\omega}{c^2}\varepsilon(\vect{r},\omega)\vect{E}(\vect{r},
\omega)
&=& \mu_0 \vect{j}_\mathrm{N}(\vect{r},\omega)\,.
\end{eqnarray}
Here we have introduced the noise charge density
\begin{equation}
\label{eq:rhoN}
\rho_\mathrm{N}(\vect{r},\omega)  
= -\divv\vect{P}_\mathrm{N}(\vect{r},\omega)
\end{equation}
and noise current density
\begin{equation}
\label{eq:noisecurrent}
\vect{j}_\mathrm{N}(\vect{r},\omega) 
= -i\omega\vect{P}_\mathrm{N}(\vect{r},\omega)
+\curl\vect{M}_\mathrm{N}(\vect{r},\omega) \,,
\end{equation}
respectively, which by construction obey the continuity equation.

Equations~(\ref{eq:macroMax3}) and (\ref{eq:macroMax4}) now contain
source terms. Hence, the electromagnetic field in absorbing media is
driven by Langevin noise forces that are due to the presence of
absorption itself. Moreover, the particular combination in which the
noise polarisation and magnetisation enter Maxwell's equations, 
Eq.~(\ref{eq:noisecurrent}), suggests that dielectric and magnetic
properties cannot be uniquely distinguished. For example, one might
include the magnetisation in the transverse polarisation in which case
the constitutive relations (\ref{eq:constitutive}) would have to be 
altered. This simple observation implies that the constitutive
relations in the present form cannot be the fundamental relations.
Instead, the noise current density appears as the fundamental source
of the electromagnetic field. This becomes even more apparent if one
allows for spatial dispersion that makes the dielectric response
functions nonlocal in configuration space. It is therefore expedient
to rewrite Eqs.~(\ref{eq:macroMax3}) and (\ref{eq:macroMax4}) as
\begin{equation}
\curl\curl\vect{E}(\vect{r},\omega) -\frac{\omega^2}{c^2}
\vect{E}(\vect{r},\omega) = i\mu_0\omega \vect{j}(\vect{r},\omega) \,,
\end{equation}
and to consider the most general linear response relation between the
current density and the electric field in the form of a generalised
Ohm's law as 
\begin{equation}
\label{eq:ohm}
\vect{j}(\vect{r},\omega) = \int d^3r'\,
\tens{Q}(\vect{r},\vect{r}',\omega)\cdot \vect{E}(\vect{r}',\omega)
+\vect{j}_\mathrm{N}(\vect{r},\omega)
\end{equation}
where $\tens{Q}(\vect{r},\vect{r}',\omega)$ is the complex
conductivity tensor in the frequency domain \cite{MelroseMcPhedran}.

The Onsager--Lorentz reciprocity theorem demands the conductivity
tensor to be reciprocal, $\tens{Q}(\vect{r},\vect{r}',\omega)$
$=\tens{Q}^\trans(\vect{r}',\vect{r},\omega)$. The two spatial
arguments must be kept separate in general, except for translationally
invariant bulk media in which the conductivity only depends on the
difference $\vect{r}-\vect{r}'$, i.e. in this case
$\tens{Q}(\vect{r},\vect{r}',\omega)\equiv\tens{Q}(\vect{r}-\vect{r}',
\omega)$. We assume that, for chosen $\omega$, the conductivity tensor
is sufficiently well-behaved. By that we mean that it tends to zero
sufficiently rapidly as $|\vect{r}-\vect{r}'|\to\infty$ and has no
non-integrable singularities. However, $\delta$ functions and their
derivatives must be permitted to allow for the spatially
nondispersive limit. The real part of
$\tens{Q}(\vect{r},\vect{r}',\omega)$,
\begin{equation}
\label{eq:Resigma}
\tens{\sigma}(\vect{r},\vect{r}',\omega) =
\mathrm{Re}\,\tens{Q}(\vect{r},\vect{r}',\omega)
=\frac{1}{2}\left[ \tens{Q}(\vect{r},\vect{r}',\omega)
+\tens{Q}^+(\vect{r}',\vect{r},\omega) \right]\,,
\end{equation}
is connected with the dissipation of electromagnetic energy and for
absorbing media, as an integral kernel, associated with a positive
definite operator \cite{MelroseMcPhedran}. Under suitable assumptions
on the causality conditions satisfied by its temporal Fourier
transform $\tens{Q}(\vect{r},\vect{r}',t)$, the conductivity tensor is
analytic in the upper complex $\omega$ half-plane, satisfies
Kramers--Kronig (Hilbert transform) relations, and obeys the Schwarz
reflection principle
\begin{equation}
\label{eq:Schwarz}
\tens{Q}(\vect{r},\vect{r}',-\omega^\ast) =
\tens{Q}^\ast(\vect{r},\vect{r}',\omega) \,.
\end{equation}

We now identify the current density (\ref{eq:ohm}) as the one entering
macroscopic Maxwell's equations in the frequency domain. The
medium-assisted electric field thus satisfies an integro-differential
equation of the form
\begin{equation}
\label{eq:helmholtz}
\curl\curl\vect{E}(\vect{r},\omega) - \frac{\omega^2}{c^2}
\vect{E}(\vect{r},\omega) -i\mu_0\omega \int d^3r'
\,\tens{Q}(\vect{r},\vect{r}',\omega) \cdot \vect{E}(\vect{r}',\omega)
= i\mu_0 \omega \vect{j}_\mathrm{N}(\vect{r},\omega) \,.
\end{equation}
The unique solution to the Helmholtz equation (\ref{eq:helmholtz}) is
\begin{equation}
\label{eq:EGj}
\vect{E}(\vect{r},\omega) = i\mu_0\omega \int d^3r'\,
\tens{G}(\vect{r},\vect{r}',\omega) \cdot
\vect{j}_\mathrm{N}(\vect{r}',\omega) 
\end{equation}
where $\tens{G}(\vect{r},\vect{r}',\omega)$ is the classical Green
tensor that satisfies Eq.~(\ref{eq:helmholtz}) with a tensorial
$\delta$ function source,
\begin{equation}
\label{eq:helmholtzgreen}
\curl\curl\tens{G}(\vect{r},\vect{s},\omega) - \frac{\omega^2}{c^2}
\tens{G}(\vect{r},\vect{s},\omega) -i\mu_0\omega \int d^3r'
\,\tens{Q}(\vect{r},\vect{r}',\omega) \cdot
\tens{G}(\vect{r}',\vect{s},\omega) 
=\tens{\delta}(\vect{r}-\vect{s})
\end{equation}
together with the boundary conditions at infinity. It inherits all
properties such as analyticity in the upper complex $\omega$
half-plane, the validity of the Schwarz reflection principle, as well
as Onsager--Lorentz reciprocity from the conductivity tensor, viz.
\begin{eqnarray}
\label{eq:reciprocityG}
\tens{G}(\vect{r}',\vect{r},\omega) &=&
\tens{G}^\trans(\vect{r},\vect{r}',\omega) \,,\\
\label{eq:SchwarzG}
\tens{G}(\vect{r},\vect{r}',-\omega^\ast) &=&
\tens{G}^\ast(\vect{r},\vect{r}',\omega) \,.
\end{eqnarray}
In addition, the Green tensor satisfies an important integral
relation that can be derived as follows. The integro-differential
equation (\ref{eq:helmholtzgreen}) can be rewritten as
\begin{equation}
\label{eq:integralhelmholtz}
\int d^3s\,\tens{H}(\vect{r},\vect{s},\omega) \cdot 
\tens{G}(\vect{s},\vect{r}',\omega) =
\tens{\delta}(\vect{r}-\vect{r}')
\end{equation}
where the integral kernel $\tens{H}(\vect{r},\vect{r}',\omega)$ 
$\!=\curl\curl\tens{\delta}(\vect{r}-\vect{r}')$ 
$\!-\omega^2/c^2\tens{\delta}(\vect{r}-\vect{r}')$ 
$\!-i\mu_0\omega\tens{Q}(\vect{r},\vect{r}',\omega)$ is reciprocal,
from which Eq.~(\ref{eq:reciprocityG}) follows. With that, the complex
conjugate of Eq.~(\ref{eq:integralhelmholtz}) reads
\begin{equation}
\label{eq:integralhelmholtz2}
\int d^3s\, \tens{G}^+(\vect{r},\vect{s},\omega) \cdot 
\tens{H}^+(\vect{s},\vect{r}',\omega) =
\tens{\delta}(\vect{r}-\vect{r}') \,.
\end{equation}
If we now multiply Eq.~(\ref{eq:integralhelmholtz}) from the left with
$\tens{G}^+(\vect{s}',\vect{r},\omega)$ and integrate over $\vect{r}$,
then multiply Eq.~(\ref{eq:integralhelmholtz2}) from the right with 
$\tens{G}(\vect{r}',\vect{s}',\omega)$ and integrate over $\vect{r}'$,
and finally subtract the resulting two equations, we find that for
real $\omega$ the integral equation
\begin{equation}
\label{eq:generalmagicformula}
\mu_0\omega \int d^3s \int d^3s'\,\tens{G}(\vect{r},\vect{s},\omega)
\cdot \tens{\sigma}(\vect{s},\vect{s}',\omega) \cdot
\tens{G}^+(\vect{s}',\vect{r}',\omega) =
\mathrm{Im}\,\tens{G}(\vect{r},\vect{r}',\omega)
\end{equation}
holds \cite{Raabe07} (see also App.~\ref{sec:dgf}).

Up until this point, all our investigations regarded classical 
electrodynamics. In order to quantise the theory, we have to regard
the Langevin noise sources $\vect{j}_\mathrm{N}(\vect{r},\omega)$ as
operators with the commutation relation
\begin{equation}
\label{eq:commutatorj}
\left[ \hat{\vect{j}}_\mathrm{N}(\vect{r},\omega) , 
\hat{\vect{j}}_\mathrm{N}^\dagger(\vect{r}',\omega') \right] 
= \frac{\hbar\omega}{\pi} \delta(\omega-\omega') 
\tens{\sigma}(\vect{r},\vect{r}',\omega) \,,
\end{equation}
which follows from the fluctuation-dissipation theorem associated with
the linear response (\ref{eq:ohm}). In this way, the operator of the
electric field strength is given by the operator-valued version of
Eq.~(\ref{eq:EGj}) as
\begin{equation}
\label{eq:electricfield}
\hat{\vect{E}}(\vect{r}) = \int\limits_0^\infty d\omega\,
\hat{\vect{E}}(\vect{r},\omega) +\mbox{h.c.}\,,\quad
\hat{\vect{E}}(\vect{r},\omega) = i\mu_0\omega \int d^3r'\,
\tens{G}(\vect{r},\vect{r}',\omega) \cdot
\hat{\vect{j}}_\mathrm{N}(\vect{r}',\omega) \,.
\end{equation}

The consistency of this quantisation procedure can be proven by
checking the fundamental equal-time commutation relation between the
operators of the electric field and the magnetic induction. Using
Faraday's law, Eq.~(\ref{eq:macroMax2}), we find the frequency
components of the magnetic induction field as
\begin{equation}
\label{eq:magneticfield}
\hat{\vect{B}}(\vect{r}) = \int\limits_0^\infty d\omega\,
\hat{\vect{B}}(\vect{r},\omega) +\mbox{h.c.}\,,\quad
\hat{\vect{B}}(\vect{r},\omega) = \mu_0 \curl \int d^3r'\,
\tens{G}(\vect{r},\vect{r}',\omega) \cdot
\hat{\vect{j}}_\mathrm{N}(\vect{r}',\omega) \,.
\end{equation}
Hence, the equal-time commutator reads, on using the commutation
relation (\ref{eq:commutatorj}) and the integral formula
(\ref{eq:generalmagicformula}), as
\begin{equation}
\label{eq:EBcommutator}
\left[ \hat{\vect{E}}(\vect{r}), \hat{\vect{B}}(\vect{r}') \right]
= \frac{2i\hbar\mu_0}{\pi} \bm{\nabla}_{\vect{r}'}\times
\int\limits_0^\infty d\omega\,\omega\,
\mathrm{Im}\,\tens{G}(\vect{r},\vect{r}',\omega)
= \frac{\hbar}{\pi\varepsilon_0c^2} \bm{\nabla}_{\vect{r}'}\times
\int\limits_{-\infty}^\infty d\omega\,\omega\,
\tens{G}(\vect{r},\vect{r}',\omega) \,,
\end{equation}
where the second equality follows from the Schwarz reflection
principle. Using the analyticity properties of the Green tensor in the
upper complex $\omega$ half-plane, we then convert the integral along
the real $\omega$ axis into a large semi-circle in the upper
half-plane. From the integro-differential equation
(\ref{eq:helmholtzgreen}) and the properties of the conductivity
tensor $\tens{Q}(\vect{r},\vect{r}',\omega)$ we find the asymptotic
form of the Green tensor for large frequencies as
\begin{equation}
\label{eq:Ginfinity}
\tens{G}(\vect{r},\vect{r}',\omega)
\stackrel{|\omega|\to\infty}{\simeq} -\frac{c^2}{\omega^2}
\tens{\delta}(\vect{r}-\vect{r}') \,,
\end{equation}
so that the equal-time field commutator takes its final form of
\begin{equation}
\label{eq:commutatorEB2}
\left[ \hat{\vect{E}}(\vect{r}), \hat{\vect{B}}(\vect{r}') \right]
= -\frac{i\hbar}{\varepsilon_0} \curl
\tens{\delta}(\vect{r}-\vect{r}')
= -\frac{i\hbar}{\varepsilon_0} \curl
\tens{\delta}^\perp(\vect{r}-\vect{r}') \,.
\end{equation}

The striking feature is that the field commutator
(\ref{eq:commutatorEB2}) is exactly the same as in free-space quantum
electrodynamics [Eq.~(\ref{eq:commutatorEB})], despite the presence of
an absorbing dielectric background material. This fact reinforces the
view that the fields $\vect{E}$ and $\vect{B}$ represent the degrees
of freedom of the electromagnetic field alone and have little to do
with any material degrees of freedom. The apparent discrepancy with
the notion of the expansion (\ref{eq:electricfield}) as a
medium-assisted electric field is resolved by interpreting the Green
tensor as the integral kernel of a projection operator onto the
electromagnetic degrees of freedom. Finally, the Langevin noise
currents $\hat{\vect{j}}_\mathrm{N}(\vect{r},\omega)$ can be
renormalised to a bosonic vector field by taking the `square-root' of
the tensor $\tens{\sigma}(\vect{r},\vect{r}',\omega)$ (which exists
because of its positivity in case of absorbing media). Writing
\begin{equation}
\label{eq:sigma}
\tens{\sigma}(\vect{r},\vect{r}',\omega) = \int d^3s\,
\tens{K}(\vect{r},\vect{s},\omega) \cdot
\tens{K}^+(\vect{r}',\vect{s},\omega) \,,
\end{equation}
and defining
\begin{equation}
\label{eq:jN}
\hat{\vect{j}}_\mathrm{N}(\vect{r},\omega) = \left(
\frac{\hbar\omega}{\pi} \right)^{1/2} \int d^3r'\,
\tens{K}(\vect{r},\vect{r}',\omega) \cdot
\hat{\vect{f}}(\vect{r}',\omega)
\end{equation}
where
\begin{equation}
\label{eq:fcommutator}
\left[ \hat{\vect{f}}(\vect{r},\omega),
\hat{\vect{f}}^\dagger(\vect{r}',\omega') \right] =
\delta(\omega-\omega') \tens{\delta}(\vect{r}-\vect{r}') \,,
\end{equation}
the expansion (\ref{eq:electricfield}) of the frequency components of
the operator of the electric field strength finally becomes
\begin{equation}
\label{eq:generalelectricfield}
\hat{\vect{E}}(\vect{r},\omega) =
i\mu_0\omega\sqrt{\frac{\hbar\omega}{\pi}} \int d^3r' \int d^3s\,
\tens{G}(\vect{r},\vect{r}',\omega) \cdot
\tens{K}(\vect{r}',\vect{s},\omega) \cdot
\hat{\vect{f}}(\vect{s},\omega) \,.
\end{equation}


\paragraph{Hamiltonian:}
In order to complete the quantisation scheme, we need to introduce a
Hamiltonian as a function of the Langevin noise sources
$\hat{\vect{j}}_\mathrm{N}(\vect{r},\omega)$ and
$\hat{\vect{j}}_\mathrm{N}^\dagger(\vect{r},\omega)$ or, equivalently,
in terms of the bosonic dynamical variables
$\hat{\vect{f}}(\vect{r},\omega)$ and
$\hat{\vect{f}}^\dagger(\vect{r},\omega)$. Imposing the constraint
that the Hamiltonian should generate a time evolution according to
\begin{equation}
\label{eq:Hconstraint}
\left[ \hat{\vect{j}}_\mathrm{N}(\vect{r},\omega), \hat{H} \right] =
\hbar\omega \hat{\vect{j}}_\mathrm{N}(\vect{r},\omega) \,,
\end{equation}
the Hamiltonian must be of the form \cite{Raabe07}
\begin{equation}
\label{eq:HFgeneral}
\hat{H} = \pi \int\limits_0^\infty d\omega\, \int d^3r \int d^3r'\,
\hat{\vect{j}}_\mathrm{N}^\dagger(\vect{r},\omega) \cdot
\tens{\rho}(\vect{r},\vect{r}',\omega) \cdot
\hat{\vect{j}}_\mathrm{N}(\vect{r}',\omega) 
\end{equation}
where $\tens{\rho}(\vect{r},\vect{r}',\omega)$ is the inverse of the
integral operator associated with
$\tens{\sigma}(\vect{r},\vect{r}',\omega)$. In terms of the bosonic
dynamical variables, the Hamiltonian is diagonal,
\begin{equation}
\label{eq:freeHamiltonian}
\hat{H} = \int\limits_0^\infty d\omega \int d^3r\, \hbar\omega\,
\hat{\vect{f}}^\dagger(\vect{r},\omega) \cdot
\hat{\vect{f}}(\vect{r},\omega) 
\end{equation}
which is its most commonly used form \cite{Scheel98,Buch,Dung03}.
Perhaps surprisingly, it closely resembles its free-space counterpart,
Eq.~(\ref{eq:freeH}), in that it is bilinear in its dynamical
variables. The reason behind this behaviour is that any linear reponse
theory can be derived from an underlying microscopic model that is
bilinear in its constituent amplitude operators which, after a
suitable Bogoliubov-type transformation, leads to a Hamiltonian of the
form (\ref{eq:freeHamiltonian}). An example is provided by the
Huttner--Barnett model of a homogeneous, isotropic dielectric
(Sec.~\ref{sec:hb}). 


\paragraph{Spatially local, isotropic, inhomogeneous dielectric
media:}
We now apply the general theory to some special cases that are of
practical importance. Let us begin with the simplest, and historically
first, example of a spatially nondispersive, isotropic and
inhomogeneous dielectric material that shows no magnetic response. The
neglect of spatial dispersion makes the conductivity
tensor $\tens{Q}(\vect{r},\vect{r}',\omega)$ strictly local, so that
$\tens{\sigma}(\vect{r},\vect{r}',\omega)$
$\!=\tens{\sigma}(\vect{r},\omega)\delta(\vect{r}-\vect{r}')$.
Furthermore, isotropy means that
$\tens{\sigma}(\vect{r},\omega)=\sigma(\vect{r},\omega)\tens{I}$. If
we then make the identification
$\sigma(\vect{r},\omega)=\varepsilon_0\omega\mathrm{Im}\,\chi(\vect{r}
,\omega)$ where $\chi(\vect{r},\omega)$ is the dielectric
susceptibility, Eq.~(\ref{eq:generalelectricfield}) becomes
\begin{equation}
\label{eq:Edielectric}
\hat{\vect{E}}(\vect{r},\omega) =
i\sqrt{\frac{\hbar}{\pi\varepsilon_0}} \frac{\omega^2}{c^2} \int d^3r'
\,\sqrt{\mathrm{Im}\,\chi(\vect{r}',\omega)}\,
\tens{G}(\vect{r},\vect{r}',\omega) \cdot
\hat{\vect{f}}(\vect{r}',\omega)
\end{equation}
which yields the well-known quantisation scheme for a locally
responding dielectric material
\cite{Scheel98,Buch,Dung98,Barnett96a,Gruner96a,Matloob96,Matloob99,
DiStefano00}.


\paragraph{Spatially dispersive homogeneous bulk media:}
As a second example, we consider an infinitely extended homogeneous
material for which $\tens{Q}(\vect{r},\vect{r}',\omega)$ is
translationally invariant \cite{Raabe07,Suttorp07}. That is, it is
only a function of the difference $\vect{r}-\vect{r}'$. In this case,
we represent $\tens{\sigma}(\vect{r},\vect{r}',\omega)$ as the spatial
Fourier transform
\begin{equation}
\label{eq:sigmaFourier}
\tens{\sigma}(\vect{r},\vect{r}',\omega) = \int \frac{d^3k}{(2\pi)^3}
\,\tens{\sigma}(\vect{k},\omega)
e^{i\vect{k}\cdot(\vect{r}-\vect{r}')} \,.
\end{equation}
A similar decomposition can be made for the integral kernel
$\tens{K}(\vect{r},\vect{r}',\omega)$. For an isotropic medium without
optical activity, the Fourier components
$\tens{\sigma}(\vect{k},\omega)$ can be written as
\cite{MelroseMcPhedran}
\begin{equation}
\label{eq:dispersiveprojective}
\tens{\sigma}(\vect{k},\omega) = \sigma_\|(k,\omega)
\frac{\vect{k}\otimes\vect{k}}{k^2} +\sigma_\perp(k,\omega) \left(
\tens{I} - \frac{\vect{k}\otimes\vect{k}}{k^2} \right)
\end{equation}
and similarly for $\tens{K}(\vect{k},\omega)$, where the expansion
coefficients have to be replaced by their positive square-roots
$\sigma_\|^{1/2}(k,\omega)$ and $\sigma_\perp^{1/2}(k,\omega)$

Clearly, the decomposition (\ref{eq:dispersiveprojective}) is not
unique as $\tens{\sigma}(\vect{k},\omega)$ can be equivalently
decomposed into
\begin{equation}
\label{eq:sigmadecompose}
\tens{\sigma}(\vect{k},\omega) = \sigma_\|(k,\omega)\tens{I}
-\vect{k}\times\gamma(k,\omega)\tens{I}\times\vect{k}
\end{equation}
where
\begin{equation}
\label{eq:gamma}
\gamma(k,\omega) = \left[ \sigma_\perp(k,\omega)-\sigma_\|(k,\omega)
\right]/k^2 \,.
\end{equation}
Since both $\sigma_\|(k,\omega)$ and $\sigma_\perp(k,\omega)$ have to 
be real and positive to yield a positive definite integral kernel
$\tens{\sigma}(\vect{k},\omega)$, the new variable $\gamma(k,\omega)$
is real, too. However, it does not have a definite sign. If, on the
other hand, one forces $\gamma(k,\omega)$ to be positive, then the
tensor
\begin{equation}
\label{eq:K}
\tens{K}'(\vect{k},\omega) = \sigma_\|^{1/2}(k,\omega)\tens{I}
\pm\gamma^{1/2}(k,\omega)\tens{I}\times\vect{k}
\end{equation}
is the positive square-root of the integral kernel 
$\tens{\sigma}(\vect{k},\omega)$. The kernels
$\tens{K}'(\vect{k},\omega)$ and $\tens{K}(\vect{k},\omega)$, despite
being different, are related by a unitary transformation
\cite{Raabe07}. 


\paragraph{Spatially local magnetoelectric media:}
The local limit of the above theory has a rather interesting
structure. If one assumes that the functions $\sigma_\|(k,\omega)$ and
$\gamma(k,\omega)$ vary sufficiently slowly with $k$ and possess
well-defined long-wavelength limits
$\lim_{k\to 0}\sigma_\|(k,\omega)=\sigma_\|(\omega)>0$ and
$\lim_{k\to 0}\gamma(k,\omega)=\gamma(\omega)>0$, one finds the
approximation
\begin{equation}
\label{eq:sigmaprojective}
\tens{\sigma}(\vect{r},\vect{r}',\omega) =
\sigma_\|(\omega)\tens{\delta}(\vect{r}-\vect{r}') -\gamma(\omega)
\curl \tens{\delta}(\vect{r}-\vect{r}')
\times\overleftarrow{\bm{\nabla}}' \,.
\end{equation}
The full conductivity tensor associated with that real part (real and
imaginary parts are related by a Hilbert transform) is then
\begin{equation}
\label{eq:conductivity}
\tens{Q}(\vect{r},\vect{r}',\omega) = Q^{(1)}(\omega)
\tens{\delta}(\vect{r}-\vect{r}') -Q^{(2)}(\omega) \curl
\tens{\delta}(\vect{r}-\vect{r}') \times\overleftarrow{\bm{\nabla}}'
\end{equation}
with the identifications
\begin{equation}
\label{eq:Q12}
Q^{(1)}(\omega) = -i\varepsilon_0\omega\left[ \varepsilon(\omega)-1
\right] \,,\quad Q^{(2)}(\omega) = -i\kappa_0 \left[ 1-\kappa(\omega)
\right]/\omega \,,
\end{equation}
where $\varepsilon(\omega)$ is the dielectric permittivity and
$\mu(\omega)=1/\kappa(\omega)$ the (para-)magnetic permeability of the
medium. Note that the requirement $\gamma(\omega)>0$ implies that
$\mathrm{Im}\,\kappa(\omega)<0$ for $\omega>0$, from which it follows
that $\mu(\omega\to 0)>1$ \cite{LL8}. This in turn means that this
theory can only describe paramagnetic materials. Diamagnetic materials
are intrinsically nonlinear as their response functions themselves
depend on the magnetic field and thus are excluded from a
linear-response formalism.

The noise current density that is derived from the kernel
\begin{equation}
\label{eq:K'}
\tens{K}'(\vect{r},\vect{r}',\omega) = \sigma_\|^{1/2}(\omega)
\tens{\delta}(\vect{r}-\vect{r}') \mp \gamma^{1/2}(\omega) \curl
\tens{\delta}(\vect{r}-\vect{r}') 
\end{equation}
[which follows from Eq.~(\ref{eq:sigmaprojective})] can be decomposed 
into longitudinal and transverse parts according to
$\hat{\vect{j}}_\mathrm{N}(\vect{r},\omega)$
$=\hat{\vect{j}}_{\mathrm{N}\|}(\vect{r},\omega)$
$+\hat{\vect{j}}_{\mathrm{N}\perp}(\vect{r},\omega)$ with
\begin{eqnarray}
\label{eq:JNparallel}
\hat{\vect{j}}_{\mathrm{N}\|}(\vect{r},\omega) &\!=&\!
\sqrt{\frac{\hbar\varepsilon_0}{\pi}} \omega
\sqrt{\mathrm{Im}\,\varepsilon(\omega)}\,
\hat{\vect{f}}_\|(\vect{r},\omega) \,,\\
\label{eq:JNperpedicular}
\hat{\vect{j}}_{\mathrm{N}\perp}(\vect{r},\omega) &\!=&\!
\sqrt{\frac{\hbar\varepsilon_0}{\pi}} \omega
\sqrt{\mathrm{Im}\,\varepsilon(\omega)}
\hat{\vect{f}}_\perp(\vect{r},\omega)
\mp i\sqrt{\frac{\hbar\kappa_0}{\pi}} \curl \left[
\sqrt{\frac{\mathrm{Im}\,\mu(\omega)}{|\mu(\omega)|^2}} \,
\hat{\vect{f}}_\perp(\vect{r},\omega) \right] \,. \nonumber \\
\end{eqnarray}
The distinction between longitudinal and transverse components of the
noise current density (and subsequently the bosonic dynamical
variables) is essentially a projection formalism, and the
$\hat{\vect{f}}_{\|(\perp)}(\vect{r},\omega)$ are termed projective
variables \cite{Raabe07}.

Another, more frequently used decomposition is obtained by
redistributing the electric part of the transverse noise current
density. In this way, two new sets of independent bosonic variables,
$\hat{\vect{f}}_e(\vect{r},\omega)$ and
$\hat{\vect{f}}_m(\vect{r},\omega)$, are introduced that lead to an
equivalent decomposition of the noise current according to
\cite{Dung03,Buch}
\begin{eqnarray}
\label{eq:jNe}
\hat{\vect{j}}_{\mathrm{N}e}(\vect{r},\omega) &=&
\sqrt{\frac{\hbar\varepsilon_0}{\pi}} \omega 
\sqrt{\mathrm{Im}\,\varepsilon(\vect{r},\omega)}\,
\hat{\vect{f}}_e(\vect{r},\omega) \,,\\
\label{eq:jNm}
\hat{\vect{j}}_{\mathrm{N}m}(\vect{r},\omega) &=&
\mp i\sqrt{\frac{\hbar\kappa_0}{\pi}} \curl \left[
\sqrt{\frac{\mathrm{Im}\,\mu(\vect{r},\omega)}{|\mu(\vect{r},
\omega)|^2}}
\,\hat{\vect{f}}_m(\vect{r},\omega) \right] \,,
\end{eqnarray}
in which the possible spatial dependencies of the dielectric
permittivity and the paramagnetic permeability have been reinstated 
(see Ref.~\cite{Raabe07} for details). The Hamiltonian 
(\ref{eq:freeHamiltonian}) takes the form
\begin{equation}
\label{eq:meHamiltonian}
\hat{H} = \sum\limits_{\lambda=e,m} \int d^3r \int\limits_0^\infty
d\omega \, \hbar\omega \,
\hat{\vect{f}}^\dagger_\lambda(\vect{r},\omega)
\cdot \hat{\vect{f}}_\lambda(\vect{r},\omega) \,.
\end{equation}
Finally, the electric field (\ref{eq:generalelectricfield}) and the
magnetic induction can be written as 
\begin{eqnarray}
\label{eq:Eexpansion}
\hat{\vect{E}}(\vect{r},\omega) &=& \sum\limits_{\lambda=e,m} \int
d^3r'\,
\tens{G}_\lambda(\vect{r},\vect{r}',\omega) \cdot 
\hat{\vect{f}}_\lambda(\vect{r}',\omega) \,,\\
\label{eq:Bexpansion}
\hat{\vect{B}}(\vect{r},\omega) &=& \frac{1}{i\omega}
\sum\limits_{\lambda=e,m} \int d^3r'\,
\left[ \curl \tens{G}_\lambda(\vect{r},\vect{r}',\omega) \right] \cdot
\hat{\vect{f}}_\lambda(\vect{r}',\omega)
\end{eqnarray}
with the definitions
\begin{eqnarray}
\label{eq:Ge}
\tens{G}_e(\vect{r},\vect{r}',\omega) &=& i\frac{\omega^2}{c^2} 
\sqrt{\frac{\hbar}{\pi\varepsilon_0}
\mathrm{Im}\,\varepsilon(\vect{r},\omega)} \,
\tens{G}(\vect{r},\vect{r}',\omega) \,,\\
\label{eq:Gm}
\tens{G}_m(\vect{r},\vect{r}',\omega) &=& -i\frac{\omega}{c} 
\sqrt{\frac{\hbar}{\pi\varepsilon_0}\frac{\mathrm{Im}\,\mu(\vect{r},
\omega)}
{|\mu(\vect{r},\omega)|^2}} \,
\left[ \tens{G}(\vect{r},\vect{r}',\omega)
\times\overleftarrow{\bm{\nabla}}' \right] \,,
\end{eqnarray}
where $\tens{G}(\vect{r},\vect{r}',\omega)$ is the usual classical
Green tensor satisfying Eq.~(\ref{eq:helmholtzgreen}). The latter
Helmholtz equation condenses to (see also App.~\ref{sec:dgf})
\begin{equation}
\label{eq:Gintegral}
\curl\kappa(\vect{r},\omega)\curl\tens{G}(\vect{r},\vect{r}',\omega)
-\frac{\omega^2}{c^2}\varepsilon(\vect{r},\omega)
\tens{G}(\vect{r},\vect{r}',\omega) =
\tens{\delta}(\vect{r}-\vect{r}')\,.
\end{equation}
For completeness, we mention that the integral relation 
(\ref{eq:generalmagicformula}) can be cast into the form
\begin{equation}
\label{Glambdaintegral}
\sum\limits_{\lambda=e,m} \int d^3s\, 
\tens{G}_\lambda(\vect{r},\vect{s},\omega) \cdot
\tens{G}_\lambda^+(\vect{r}',\vect{s},\omega)
=\frac{\hbar}{\pi\varepsilon_0}\frac{\omega^2}{c^2} \,
\mathrm{Im}\,\tens{G}(\vect{r},\vect{r}',\omega) \,.
\end{equation}

Alternatively, the electric and magnetic noise current densities
(\ref{eq:jNe}) and (\ref{eq:jNm}) can be recast into the form of noise
polarisation and magnetisation fields
\begin{equation}
\hat{\vect{j}}_{\mathrm{Ne}}(\vect{r},\omega)
= -i\omega\hat{\vect{P}}_\mathrm{N}(\vect{r},\omega)  \,,\quad
\hat{\vect{j}}_{\mathrm{Nm}}(\vect{r},\omega) = \curl
\hat{\vect{M}}_\mathrm{N}(\vect{r},\omega) \,.
\end{equation}
Then, the electromagnetic fields read in terms of these noise fields
as
\begin{eqnarray}
\label{eq:meE}
\hat{\vect{E}}(\vect{r},\omega) &=& \frac{\omega^2}{\varepsilon_0c^2}
\int d^3r'\,\tens{G}(\vect{r},\vect{r}',\omega) \cdot
\hat{\vect{P}}_\mathrm{N}(\vect{r}',\omega) \nonumber \\ &&
-i\mu_0\omega \int d^3r'
\left[\tens{G}(\vect{r},\vect{r}',\omega)\times 
\overleftarrow{\grad}'\right] \cdot
\hat{\vect{M}}_\mathrm{N}(\vect{r}',\omega) \,,\\
\label{eq:meB}
\hat{\vect{B}}(\vect{r},\omega) &=& i\mu_0\omega \int d^3r'
\left[\curl\tens{G}(\vect{r},\vect{r}',\omega)\right] \cdot
\hat{\vect{P}}_\mathrm{N}(\vect{r}',\omega) \nonumber \\ &&
-\mu_0 \int d^3r' \left[\curl\tens{G}(\vect{r},\vect{r}',\omega)\times
\overleftarrow{\grad}'\right] \cdot
\hat{\vect{M}}_\mathrm{N}(\vect{r}',\omega) \,,\\
\label{eq:meD}
\hat{\vect{D}}(\vect{r},\omega) &=& -i\frac{\omega}{c^2}
\varepsilon(\vect{r},\omega) \int d^3r'
\left[\tens{G}(\vect{r},\vect{r}',\omega)\times\overleftarrow{\grad}'
\right] \cdot \hat{\vect{M}}_\mathrm{N}(\vect{r}',\omega) \nonumber \\
&&
+\frac{\omega^2}{c^2} \int d^3r' \left[ \varepsilon(\vect{r},\omega)
\tens{G}(\vect{r},\vect{r}',\omega) +\tens{\delta}(\vect{r}-\vect{r}')
\right] \cdot \hat{\vect{P}}_\mathrm{N}(\vect{r}',\omega) \,, \\
\label{eq:meH}
\hat{\vect{H}}(\vect{r},\omega) &=&
-i\frac{\omega}{\mu(\vect{r},\omega)} \int d^3r' \left[
\curl\tens{G}(\vect{r},\vect{r}',\omega) \right] \cdot
\hat{\vect{P}}_\mathrm{N}(\vect{r}',\omega) \nonumber \\ &&
- \int d^3r' \left[
\frac{\curl\tens{G}(\vect{r},\vect{r}',\omega)\times\overleftarrow{
\grad}'}
{\mu(\vect{r},\omega)} +\tens{\delta}(\vect{r}-\vect{r}')  \right]
\cdot \hat{\vect{M}}_\mathrm{N}(\vect{r}',\omega) \,.
\end{eqnarray}


\paragraph{Statistical properties:}
Thermal expectation values of the electromagnetic field can be
obtained from those of the dynamical variables. Assuming the
electromagnetic field in thermal equilibrium with temperature $T$, it
may be described by a (canonical) density operator
\begin{equation}
\hat{\varrho}_T =
\frac{e^{-\hat{H}_F/(k_BT)}}{\trace e^{-\hat{H}_F/(k_BT)}} 
\end{equation}
[$k_B$: Boltzmann constant]. Thermal averages
$\langle\ldots\rangle_T=\trace[\ldots\hat{\varrho}_T]$ of the
dynamical variables are thus given by
\begin{eqnarray}
\langle\hat{\vect{f}}_\lambda(\vect{r},\omega) \rangle_T &=& \vect{0}
=
\langle\hat{\vect{f}}_\lambda^\dagger(\vect{r},\omega) \rangle_T \,,\\
\langle\hat{\vect{f}}_\lambda(\vect{r},\omega)\otimes
\hat{\vect{f}}_{\lambda'}(\vect{r}',\omega')\rangle_T &=& \tens{0} =
\langle\hat{\vect{f}}_\lambda^\dagger(\vect{r},\omega)\otimes
\hat{\vect{f}}_{\lambda'}^\dagger(\vect{r}',\omega')\rangle_T \,,\\
\langle\hat{\vect{f}}_\lambda^\dagger(\vect{r},\omega)\otimes
\hat{\vect{f}}_{\lambda'}(\vect{r}',\omega')\rangle_T &=&
\bar{n}_\mathrm{th}(\omega) \delta_{\lambda\lambda'}
\delta(\vect{r}-\vect{r}') \delta(\omega-\omega') \,,\\
\langle\hat{\vect{f}}_\lambda(\vect{r},\omega)\otimes
\hat{\vect{f}}_{\lambda'}^\dagger(\vect{r}',\omega')\rangle_T &=&
\left[ \bar{n}_\mathrm{th}(\omega)+1\right] \delta_{\lambda\lambda'}
\delta(\vect{r}-\vect{r}') \delta(\omega-\omega') \,,
\end{eqnarray}
where
\begin{equation}
\bar{n}_\mathrm{th}(\omega) =
\frac{\sum_mme^{-m\hbar\omega/(k_BT)}}{\sum_me^{-m\hbar\omega/(k_BT)}}
=\frac{1}{e^{\hbar\omega/(k_BT)}-1}
\end{equation}
is the average thermal photon number. This translates into the
statistical properties of the electromagnetic fields as follows:
\begin{eqnarray}
\label{eq:expectE}
\langle\hat{\vect{E}}(\vect{r},\omega)\rangle_T &=& \vect{0}
= \langle\hat{\vect{E}}^\dagger(\vect{r},\omega)\rangle_T \,,\\
\langle\hat{\vect{E}}(\vect{r},\omega)\otimes
\hat{\vect{E}}(\vect{r}',\omega')\rangle_T &=& \tens{0}
= \langle\hat{\vect{E}}^\dagger(\vect{r},\omega)\otimes
\hat{\vect{E}}^\dagger(\vect{r}',\omega')\rangle_T \,,\\
\langle\hat{\vect{E}}^\dagger(\vect{r},\omega)\otimes
\hat{\vect{E}}(\vect{r}',\omega')\rangle_T &=&
\frac{\hbar}{\pi\varepsilon_0} \bar{n}_\mathrm{th}(\omega)
\frac{\omega^2}{c^2} \mathrm{Im}\,\tens{G}(\vect{r},\vect{r}',\omega)
\delta(\omega-\omega') \,,\\
\label{eq:EEdagger}
\langle\hat{\vect{E}}(\vect{r},\omega)\otimes
\hat{\vect{E}}^\dagger(\vect{r}',\omega')\rangle_T &=&
\frac{\hbar}{\pi\varepsilon_0} \left[
\bar{n}_\mathrm{th}(\omega)+1\right]
\frac{\omega^2}{c^2} \mathrm{Im}\,\tens{G}(\vect{r},\vect{r}',\omega)
\delta(\omega-\omega') \,,\\
\langle\hat{\vect{B}}(\vect{r},\omega)\rangle_T &=& \vect{0}
= \langle\hat{\vect{B}}^\dagger(\vect{r},\omega)\rangle_T \,,\\
\langle\hat{\vect{B}}(\vect{r},\omega)\otimes
\hat{\vect{B}}(\vect{r}',\omega')\rangle_T &=& \tens{0}
= \langle\hat{\vect{B}}^\dagger(\vect{r},\omega)\otimes
\hat{\vect{B}}^\dagger(\vect{r}',\omega')\rangle_T \,,\\
\langle\hat{\vect{B}}^\dagger(\vect{r},\omega)\otimes
\hat{\vect{B}}(\vect{r}',\omega')\rangle_T &=&
\frac{\hbar\mu_0}{\pi} \bar{n}_\mathrm{th}(\omega)
\mathrm{Im}\left[\curl\tens{G}(\vect{r},\vect{r}',\omega)\times
\overleftarrow{\bm{\nabla}} \right]
\delta(\omega-\omega') \,,\\
\label{eq:BBdagger}
\langle\hat{\vect{B}}(\vect{r},\omega)\otimes
\hat{\vect{B}}^\dagger(\vect{r}',\omega')\rangle_T &=&
\frac{\hbar\mu_0}{\pi} \left[ \bar{n}_\mathrm{th}(\omega)+1\right]
\mathrm{Im}\left[\curl\tens{G}(\vect{r},\vect{r}',\omega)\times
\overleftarrow{\bm{\nabla}} \right] \delta(\omega-\omega')
\,.\nonumber \\
\end{eqnarray}
These expressions will be needed for the calculation of relaxation
rates (Sec.~\ref{sec:relaxation}) and dispersion forces
(Sec.~\ref{sec:dispersion}).


\subsubsection{Duality transformations}

An important symmetry of the Maxwell's equations in free space is
duality where interchanging electric and magnetic fields yields the
same differential equations. Here we will show that this type of
symmetry can be established even within the framework of macroscopic 
quantum electrodynamics. At first, we consider macroscopic QED without
external charges and currents. We group the fields into dual pairs
and rewrite Maxwell's equations as
\begin{equation}
\label{eq:dualMax}
\divv \begin{pmatrix} \sqrt{\mu_0}\hat{\vect{D}} \\
\sqrt{\varepsilon_0} \hat{\vect{B}} \end{pmatrix}
=\begin{pmatrix} 0 \\ 0 \end{pmatrix} \,,\quad
\curl \begin{pmatrix} \sqrt{\varepsilon_0}\hat{\vect{E}} \\
\sqrt{\mu_0} \hat{\vect{H}} \end{pmatrix}
+\frac{\partial}{\partial t} \begin{pmatrix} 0 & 1 \\ -1 & 0 
\end{pmatrix} 
\begin{pmatrix} \sqrt{\mu_0}\hat{\vect{D}} \\
\sqrt{\varepsilon_0} \hat{\vect{B}} \end{pmatrix}
= \begin{pmatrix} \vect{0} \\ \vect{0} \end{pmatrix} \,,
\end{equation}
where the constitutive relations are combined to
\begin{equation}
\label{eq:dualfree}
\begin{pmatrix} \sqrt{\mu_0}\hat{\vect{D}} \\
\sqrt{\varepsilon_0} \hat{\vect{B}} \end{pmatrix}
= \frac{1}{c} \begin{pmatrix} \sqrt{\varepsilon_0}
\hat{\vect{E}} \\ \sqrt{\mu_0} \hat{\vect{H}} \end{pmatrix}
+\begin{pmatrix} \sqrt{\mu_0}\hat{\vect{P}} \\
\sqrt{\varepsilon_0}\mu_0 \hat{\vect{M}} \end{pmatrix} \,.
\end{equation}
A general rotation $\mathcal{D}(\theta)$ in the space of dual pairs
can be written as
\begin{equation}
\label{eq:dualrot}
\begin{pmatrix}\vect{x}\\ \vect{y}\end{pmatrix}^\star
 =\mathcal{D}(\theta)\begin{pmatrix}\vect{x}\\ \vect{y}\end{pmatrix},
 \quad\mathcal{D}(\theta)
 =\begin{pmatrix} \cos\theta & \sin\theta\\ 
 -\sin\theta & \cos\theta \end{pmatrix} \,.
\end{equation}
It is easily checked that Maxwell's equations (\ref{eq:dualMax}) in
free space with constitutive relations (\ref{eq:dualfree}) is
invariant under rotations of the form (\ref{eq:dualrot}).

In the presence of spatially local magnetoelectric materials, the
constitutive relations in frequency space can be further specified to
\begin{equation}
\label{eq:dualmedia}
\begin{pmatrix} \sqrt{\mu_0}\hat{\vect{D}} \\
\sqrt{\varepsilon_0}\hat{\vect{B}} \end{pmatrix}
=\frac{1}{c} \begin{pmatrix} \varepsilon & 0 \\ 0 & \mu \end{pmatrix}
\begin{pmatrix} \sqrt{\varepsilon_0}\hat{\vect{E}} \\
\sqrt{\mu_0}\hat{\vect{H}} \end{pmatrix}
+\begin{pmatrix} 1 & 0 \\ 0 & \mu \end{pmatrix}
\begin{pmatrix} \sqrt{\mu_0}\hat{\vect{P}}_\mathrm{N} \\
\sqrt{\varepsilon_0}\mu_0\hat{\vect{M}}_\mathrm{N}
\end{pmatrix} \,.
\end{equation}
Invariance of the constitutive relations (\ref{eq:dualmedia}) under
the duality transformation (\ref{eq:dualrot}) requires that
\begin{equation}
\begin{pmatrix} \varepsilon^\star & 0 \\ 0 & \mu^\star \end{pmatrix}
=\mathcal{D}(\theta)
\begin{pmatrix} \varepsilon & 0 \\ 0 & \mu \end{pmatrix}
\mathcal{D}^{-1}(\theta)\\ 
=\begin{pmatrix} \varepsilon\cos^2\theta+\mu\sin^2\theta
& (\mu-\varepsilon)\sin\theta\cos\theta \\
(\mu-\varepsilon)\sin\theta\cos\theta
& \varepsilon\sin^2\theta+\mu\cos^2\theta \end{pmatrix}
\end{equation}
which can be fulfilled in two ways. The first is obtained if the
dielectric permittivity of the material equals its magnetic
permeability, $\varepsilon=\mu$. This is achieved in free space as
well as by certain metamaterials, for example by a perfect lens with
$\varepsilon=\mu=-1$ \cite{Pendry00}. In this case duality is a
continuous symmetry which holds for all angles $\theta$.

Generally, duality holds only for discrete values of the rotation
angle, $\theta=n\pi/2$ with $n\in\mathbb{Z}$. In this case, the
transformation results in
\begin{equation}
\label{eq:dualdiscrete}
\begin{pmatrix} \varepsilon \\ \mu \end{pmatrix}^\star
\!=\begin{pmatrix} \cos^2\theta & \sin^2\theta \\
\sin^2\theta & \cos^2\theta \end{pmatrix}
\begin{pmatrix} \varepsilon \\ \mu \end{pmatrix}
,\;\;
\begin{pmatrix} \sqrt{\mu_0}\hat{\vect{P}}_\mathrm{N} \\
\sqrt{\varepsilon_0}\mu_0\hat{\vect{M}}_\mathrm{N}
\end{pmatrix}^\star
\!=\begin{pmatrix} \cos\theta & \mu\sin\theta \\ 
-\varepsilon^{-1}\sin\theta & \cos\theta \end{pmatrix}
\begin{pmatrix} \sqrt{\mu_0}\hat{\vect{P}}_\mathrm{N} \\
\sqrt{\varepsilon_0}\mu_0\hat{\vect{M}}_\mathrm{N}
\end{pmatrix}.
\end{equation}
It should be remarked that, not only are Maxwell's equations invariant
under the discrete duality transformation, but also the Hamiltonian
that generates them. 
%
\begin{table}[ht]
\begin{center}
\begin{tabular}{cccc}
\hline
Partners&\multicolumn{3}{c}{Transformation}\\
\hline
$\hat{\vect{E}}$, $\hat{\vect{H}}$:
&$\hat{\vect{E}}^\star=c\mu_0\hat{\vect{H}}$,&\hspace{1ex}&
$\hat{\vect{H}}^\star=-\hat{\vect{E}}/(c\mu_0)$\\
$\hat{\vect{D}}$, $\hat{\vect{B}}$:
&$\hat{\vect{D}}^\star=c\varepsilon_0\hat{\vect{B}}$,&\hspace{1ex}&
$\hat{\vect{B}}^\star=-\hat{\vect{D}}/(c\varepsilon_0)$\\
$\hat{\vect{P}}$, $\hat{\vect{M}}$:
&$\hat{\vect{P}}^\star=\hat{\vect{M}}/c$,&\hspace{1ex}&
$\hat{\vect{M}}^\star=-c\hat{\vect{P}}$\\
$\hat{\vect{P}}_A$, $\hat{\vect{M}}_A$:
&$\hat{\vect{P}}_A^\star=\hat{\vect{M}}_A/c$,&\hspace{1ex}&
$\hat{\vect{M}}_A^\star=-c\hat{\vect{P}}_A$\\
$\hat{\vect{d}}$, $\hat{\vect{m}}$:
&$\hat{\vect{d}}^\star=\hat{\vect{m}}/c$,&\hspace{1ex}&
$\hat{\vect{m}}^\star=-c\hat{\vect{d}}$\\
$\hat{\vect{P}}_\mathrm{N}$, $\hat{\vect{M}}_\mathrm{N}$:
&$\hat{\vect{P}}_\mathrm{N}^\star
=\mu\hat{\vect{M}}_\mathrm{N}/c$,&\hspace{1ex}
&$\hat{\vect{M}}_\mathrm{N}^\star
=-c\hat{\vect{P}}_\mathrm{N}/\varepsilon$\\
$\hat{\vect{f}}_e$, $\hat{\vect{f}}_m$:
&$\hat{\vect{f}}_e^\star
=-\mi(\mu/|\mu|)\hat{\vect{f}}_m$,&\hspace{1ex}
&$\hat{\vect{f}}_m^\star
=-\mi(|\varepsilon|/\varepsilon)\hat{\vect{f}}_e$\\
$\varepsilon$, $\mu$:
&$\varepsilon^\star=\mu$,&\hspace{1ex}
&$\mu^\star=\varepsilon$\\
$\alpha$, $\beta$:
&$\alpha^\star=\beta/c^2$,&\hspace{1ex}
&$\beta^\star=c^2\alpha$
\end{tabular}
\caption{\label{tab:duality} Effect of the duality transformation.}
\vspace{-3ex}
\end{center}
\end{table}%
%
To see this, one can derive the transformation properties of the
dynamical variables from the relations (\ref{eq:dualdiscrete}) which
read for $\theta=n\pi/2$ as
\begin{equation}
\begin{pmatrix} \hat{\vect{f}}_e \\ \hat{\vect{f}}_m
\end{pmatrix}^\star
=\begin{pmatrix} \cos\theta & -\mi(\mu/|\mu|)\sin\theta \\
-\mi(|\varepsilon|/\varepsilon)\sin\theta & \cos\theta \end{pmatrix}
\begin{pmatrix}\hat{\vect{f}}_e \\ \hat{\vect{f}}_m \end{pmatrix} \,.
\end{equation}
These transformations obviously leave the Hamiltonian
(\ref{eq:meHamiltonian}) invariant. Combining all relevant
transformation relations, we can collect the duality relations for all
electromagnetic fields, the linear response functions, and dipole
moments, in Tab.~\ref{tab:duality}. These relations allow one to
establish novel results for magnetic (electric) materials and atoms in
terms of already known results from their corresponding dual electric
(magnetic) counterparts. Finally, duality does not only hold on the
operator level in macroscopic QED without external charges and
currents, but can also be established for derived atomic quantities.
It is shown in Ref.~\cite{BuhmannScheel08} that dispersion forces as
well as decay rates are all duality invariant, provided that the
bodies are stationary and located in free space and that local-field
corrections are applied when considering atoms embedded in a medium.
The duality invariance of dispersion forces is further discussed and
exploited in Sec.~\ref{sec:dispersion}.


\subsection{Light propagation through absorbing dielectric devices}
\label{sec:lossybeamsplitter}

In a previous section (Sec.~\ref{sec:beamsplitter}) we developed the
theory of quantum-state transformation at lossless beam splitters
which accounts for a unitary transformation between the photonic
amplitude operators associated with incoming and outgoing light.
Unitarity is directly related to conservation of photon number during
the beam splitter transformation. It is already intuitively clear
that in the presence of losses, i.e. absorption, photon-number
conservation and thus unitarity cannot be upheld, at least not on the
level of the photonic amplitude operators. Having said that, because
we have constructed a bilinear Hamiltonian of the electromagnetic
field even in the presence of absorbing dielectrics,
Eq.~(\ref{eq:meHamiltonian}), there will be a unitary evolution
associated with the medium-assisted electromagnetic field, but not
with the (free) electromagnetic field alone. 

In order to see how the restricted evolution emerges, we consider
again a one-dimensional model of a beam splitter that consists of a
(planarly) multilayered dielectric structure surrounded by free space 
(see Fig.~\ref{fig:beamsplitter}). As opposed to a mode decomposition
in the lossless case, we seek the Green function associated with the
light scattering at the multilayered stack. In this one-dimensional
model, the Green function reduces to a scalar function which can be
constructed by fitting bulk Green functions at the interfaces between
regions of piecewise constant permittivity \cite{Gruner96}. Knowledge
of the Green function amounts to knowledge of the transmission,
reflection and absorption coefficients associated with impinging light
of frequency $\omega$. Similar decompositions can be made for
three-dimensional structures with translational invariance
\cite{Khanbekyan07}.

It turns out that the input-output relations (\ref{eq:ior}) have to
be amended by a term associated with absorption in the beam splitter,
\begin{equation}
\label{eq:lossyior}
\hat{\vect{b}}(\omega) = \tens{T}(\omega) \cdot
\hat{\vect{a}}(\omega) 
+\tens{A}(\omega) \cdot \hat{\vect{g}}(\omega) \,,
\end{equation}
where $\hat{g}_i(\omega)$ denote (bosonic) variables associated with 
excitations in the dielectric material (device operators) with complex
refractive index $n(\omega)=\eta(\omega)+i\kappa(\omega)$, and 
$\tens{A}(\omega)$ the absorption matrix. This expression is a direct
consequence of the expansion of the electromagnetic field operators in
terms of the dynamical variables. It is shown in Ref.~\cite{Gruner96}
that the device operators are integrated dynamical variables over the
beam splitter and read
\begin{equation}
\hat{g}_{1,2}(\omega) = i\sqrt{\frac{\omega}{2c\lambda_\pm(d,\omega)}}
e^{in(\omega)\omega d/(2c)} \int\limits_{-d/2}^{d/2} dx \left[
e^{in(\omega)\omega x/c} \pm e^{-in(\omega)\omega x/c}
\right] \hat{f}(x,\omega)
\end{equation}
where
\begin{equation}
\lambda_\pm(d,\omega) = e^{-\kappa(\omega)\omega d/c} \left\{
\frac{\sinh[\kappa(\omega)\omega d/c]}{\kappa(\omega)} \pm
\frac{\sin[\eta(\omega)\omega d/c]}{\eta(\omega)} 
\right\} \,.
\end{equation}

The transmission and absorption matrices obey the relation
\begin{equation}
\label{eq:lossybsenergy}
\tens{T}(\omega) \cdot \tens{T}^+(\omega) 
+ \tens{A}(\omega) \cdot \tens{A}^+(\omega) = \tens{I}
\end{equation}
which serves as the generalisation of the above-mentioned energy
conservation relation (\ref{eq:bsenergy}).
Equation~(\ref{eq:lossybsenergy}) says that the probabilities of a
photon being transmitted, reflected or absorbed add up to one. Hence,
photon numbers and thus energy is conserved only if one includes
absorption. For a single plate of thickness $d$ surrounded by vacuum,
the matrix elements read \cite{Gruner96,Buch}
\begin{eqnarray}
T_{11}(\omega) = T_{22}(\omega) &=& -e^{-i\omega d/c} r(\omega) \left[
1-t_1(\omega)e^{2in(\omega)\omega d/c} D(\omega) t_2(\omega) \right]
\,,\\
T_{12}(\omega) = T_{21}(\omega) &=& e^{-i\omega d/c} t_1(\omega)
e^{in(\omega)\omega d/c} D(\omega) t_2(\omega) \,,\\
A_{11}(\omega) = A_{21}(\omega) &=& \sqrt{\eta(\omega)\kappa(\omega)}
e^{-i\omega d/(2c)} t_1(\omega) D(\omega) \sqrt{\lambda_+(d,\omega)}
\nonumber \\ && \times
\left[ 1+e^{in(\omega)\omega d/c} r(\omega) \right] \,,\\
A_{12}(\omega) = -A_{22}(\omega) &=& \sqrt{\eta(\omega)\kappa(\omega)}
e^{-i\omega d/(2c)} t_1(\omega) D(\omega) \sqrt{\lambda_-(d,\omega)}
\nonumber \\ && \times
\left[ 1-e^{in(\omega)\omega d/c} r(\omega) \right] \,.
\end{eqnarray}
The interface reflection and transmission coefficients are functions
of the index of refraction and are defined as
\begin{equation}
r(\omega) = \frac{n(\omega)-1}{n(\omega)+1} \,,\quad
t_1(\omega) = \frac{2}{1+n(\omega)} \,,\quad
t_2(\omega) = \frac{2n(\omega)}{1+n(\omega)} \,,
\end{equation}
and the factor
$D(\omega)=[1-r^2(\omega)e^{2in(\omega)\omega d/c}]^{-1}$ accounts for
multiple reflections inside the plate. These coefficients are special
cases of the generalised Fresnel coefficients for $p$-polarisation and
normal incidence (see App.~\ref{sec:layeredmedia}).

The input-output relations (\ref{eq:lossyior}) translate into a 
generalised quantum-state transformation formula. For this purpose, we
need to look into an enlarged Hilbert space for the electromagnetic
field and the dielectric object. With the four-dimensional vectors
$\hat{\bm{\alpha}}(\omega)$
$=[\hat{\vect{a}}(\omega),\hat{\vect{g}}(\omega)]^\trans$
and $\hat{\bm{\beta}}(\omega)$
$=[\hat{\vect{b}}(\omega),\hat{\vect{h}}(\omega)]^\trans$,
the input-output relations can be extended to a unitary matrix 
transformation of the form
\begin{equation}
\label{eq:unitaryior}
\hat{\bm{\beta}}(\omega) = \tens{\Lambda}(\omega) \cdot 
\hat{\bm{\alpha}}(\omega)
\end{equation}
where the unitary $4\times 4$-matrix $\tens{\Lambda}(\omega)$ is an
element of the group SU(4) and can be expressed in terms of the
transmission and absorption matrices as \cite{Knoll99}
\begin{equation}
\tens{\Lambda}(\omega) = \left( 
\begin{array}{cc}
\tens{T}(\omega) & \tens{A}(\omega) \\
-\tens{S}(\omega)\cdot\tens{C}^{-1}(\omega)\cdot\tens{T}(\omega) &
\tens{C}(\omega)\cdot\tens{S}^{-1}(\omega)\cdot\tens{A}(\omega)
\end{array} \right)
\end{equation}
where
$\tens{C}(\omega)=\sqrt{\tens{T}(\omega)\cdot\tens{T}^+(\omega)}$
and $\tens{S}(\omega)=\sqrt{\tens{A}(\omega)\cdot\tens{A}^+(\omega)}$.
Given a density operator $\hat{\varrho}$ as a functional of the input
operators $\hat{\bm{\alpha}}$, $\hat{\varrho}_\mathrm{in}$
$=\hat{\varrho}_\mathrm{in}[\hat{\bm{\alpha}}(\omega),
\hat{\bm{\alpha}}^\dagger(\omega)]$, the transformed density operator
of the photonic degrees of freedom alone is then \cite{Knoll99}
\begin{equation}
\label{eq:lossyqst}
\hat{\varrho}_\mathrm{out}^{(F)} = \trace^{(D)}\left\{
\hat{\varrho}_\mathrm{in}\left[
\tens{\Lambda}^+(\omega)\cdot\hat{\bm{\alpha}}(\omega),
\tens{\Lambda}^\trans(\omega)\cdot\hat{\bm{\alpha}}^\dagger(\omega)
\right] \right\} \,,
\end{equation}
where $\trace^{(D)}$ denotes the trace over the device variables.
As a first illustrative example, we consider the transformation of
coherent states at an absorbing beam splitter. If we assume that both
the incoming electromagnetic field as well as the beam splitter are
prepared in two-mode coherent states $|\bm{\alpha}\rangle$ and
$|\bm{\beta}\rangle$ with respective amplitudes $\bm{\alpha}$ 
and $\bm{\beta}$, application of the input-output relations
(\ref{eq:lossyior}) [or equivalently, the quantum-state transformation
(\ref{eq:lossyqst})] reveals that the outgoing fields are prepared in
a two-mode coherent state
\begin{equation}
|\bm{\alpha}'\rangle = |\tens{T}\cdot\bm{\alpha}
+\tens{A}\cdot\bm{\beta} \rangle \,.
\end{equation}
Hence, the transformed amplitudes are determined not only by the
transmission matrix $\tens{T}$ but also by the absorption matrix
$\tens{A}$ \cite{Buch}. 

This theory has wide-ranging applications that include nonclassicality
studies of light propagation through optical elements and entanglement
degradation in optical fibres \cite{Scheelloss1,Scheelloss2}. We
mention here two important results relating to propagation of two-mode
quantum states of light through optical fibres. Consider a two-mode
squeezed vacuum state with squeezing parameter $\xi$ being sent
through identical optical fibres of length $l$ that are held at a
temperature $T$. We regard the optical fibres as essentially
one-dimensional objects whose effect on the quantum states of light
propagating through them can be described by the input-output theory
presented above. The optical fibres are characterised by their
absorption length $l_\mathrm{abs}$ which implies that we approximate
their transmission coefficient associated with them by
$|T|^2=e^{-l/l_\mathrm{abs}}$. Furthermore, a nonzero temperature $T$
gives rise to a mean thermal photon number
$\bar{n}_\mathrm{th}=[e^{\hbar\omega/(k_BT)}-1]^{-1}$.

As a two-mode squeezed vacuum is a Gaussian state, it is fully 
characterised by its first and second moments, the mean and
covariances of its Wigner function or characteristic function,
respectively. For the quantum state under consideration, the mean is
zero and its covariance matrix is
\begin{equation}
\tens{\Gamma} = \left( \begin{array}{cccc} 
c&0&s&0\\0&c&0&-s\\s&0&c&0\\0&-s&0&c
\end{array} \right) \,,\quad
\begin{array}{rcl}
c &=& \cosh 2\xi \,, \\ s &=& \sinh 2\xi \,.
\end{array}
\end{equation}
The covariance matrix $\tens{\Gamma}$ transforms under an arbitrary
completely positive map as
\begin{equation}
\label{eq:cpmap}
\tens{\Gamma} \mapsto \tens{\Gamma}' = \tens{A} \cdot \tens{\Gamma}
\cdot
\tens{A}^\trans + \tens{G}
\end{equation}
where $\tens{G}$ is a positive symmetric matrix and $\tens{A}$ an 
arbitrary matrix, provided that the resulting covariance matrix 
$\tens{\Gamma}'$ is a valid covariance matrix, i.e. obeys
Heisenberg's 
uncertainty relation $\tens{\Gamma}'+i\tens{\Sigma}\ge 0$
[$\tens{\Sigma}$: symplectic matrix]. The non-orthogonality of the
matrix $\tens{A}$ is a direct consequence of dissipation, and
$\tens{G}$ is the additional noise as required by the
fluctuation-dissipation theorem. The general structure
(\ref{eq:cpmap}) clearly also follows from the application of the 
quantum-state transformation formula (\ref{eq:lossyqst}).

Using the input-output relations (\ref{eq:lossyior}), we find that
the entries in the covariance matrix $\tens{\Gamma}$ change according
to \cite{QIV1}
\begin{equation}
c\mapsto c|T|^2+|R|^2+(2\bar{n}_\mathrm{th}+1)(1-|T|^2-|R|^2) \,,\quad
s\mapsto s|T|^2\,,
\end{equation}
which translates into a transformation of the covariance matrix as
\begin{equation}
\tens{\Gamma} \mapsto |T|^2\tens{\Gamma} +\left[ |R|^2 +
(2\bar{n}_\mathrm{th}+1)(1-|T|^2-|R|^2) \right] \tens{I}
\end{equation}
from which the matrices $\tens{A}$ and $\tens{G}$ can be read off as
$\tens{A}=|T|\tens{I}$ and 
$\tens{G}=[|R|^2+(2\bar{n}_\mathrm{th}+1)(1-|T|^2-|R|^2)]\tens{I}$,
respectively.

The entanglement content of a two-mode squeezed vacuum state,
expressed in terms of its negativity \cite{VidalWerner}, is
$E_N=2\xi$. At zero temperature, and neglecting coupling losses into
the fibres, the maximal amount of entanglement (in the limit of
infinite initial squeezing) that can be transmitted through the fibres
is $E_{N,\max}=-\ln\left(1-e^{-l/l_\mathrm{abs}}\right)$. For finite
initial squeezing, and at finite temperature $T$, the state becomes
separable, i.e. it loses all its entanglement, after the separability
length 
\begin{equation}
l_S = \frac{l_\mathrm{abs}}{2} \ln \left[
1+\frac{1}{2\bar{n}_\mathrm{th}} \left( 1-e^{-2\xi} \right) \right]
\,.
\end{equation}
Note that at zero temperature this separability length is infinite,
i.e. in this case it is always possible to transmit some entanglement
over arbitrary distances.


\subsection{Medium-assisted interaction of the quantised
electromagnetic field with atoms}
\label{sec:interaction}

Up until now, our emphasis has been on devising a quantisation scheme
for the electromagnetic field in the presence of magnetoelectric
background materials, but without external sources. The starting point
had been the definition of derived electromagnetic quantities such as
the displacement field $\vect{D}$ and the magnetic field $\vect{H}$ in
terms of the matter-related quantities $\vect{P}$ and $\vect{M}$, the
polarisation and magnetisation fields [Eq.~(\ref{eq:constitutive})].
However, the definition of the latter is not unique as parts of them
may be associated with magnetoelectric background media, whereas other
parts could be attributed to additional atomic or molecular sources
not included in the background matter. It is these additional sources
that give rise to a successful theory of microscopic-dielectric
interfaces.

In close analogy to the free-space case, the medium-assisted
electromagnetic field can be coupled to an atomic system consisting  
of non-relativistic spinless charged particles via the
minimal-coupling scheme. The dynamics of the combined atom-field
system is governed by the Hamiltonian 
\begin{eqnarray}
\label{eq:minimalHamiltonian}
\hat{H} &=& \sum\limits_{\lambda=e,m} \int d^3r \int\limits_0^\infty
d\omega \,\hbar\omega\,
\hat{\vect{f}}_\lambda^\dagger(\vect{r},\omega) \cdot 
\hat{\vect{f}}_\lambda(\vect{r},\omega) 
+\sum\limits_\alpha \frac{[\hat{\vect{p}}_\alpha
-q_\alpha\hat{\vect{A}}(\hat{\vect{r}}_\alpha)]^2}{2m_\alpha} 
\nonumber \\ &&
+\sum\limits_{\alpha\ne\alpha'} \frac{q_\alpha q_{\alpha'}}
{8\pi\varepsilon_0|\hat{\vect{r}}_\alpha-\hat{\vect{r}}_{\alpha'}|}
+\int d^3r\,\hat{\rho}_A(\vect{r}) \hat{\phi}(\vect{r})
\end{eqnarray}
where
\begin{equation}
\label{eq:rhoA}
\hat{\rho}_A(\vect{r}) = \sum\limits_\alpha q_\alpha 
\delta(\vect{r}-\hat{\vect{r}}_\alpha)
\end{equation}
is the charge density of the particles. The vector and scalar
potentials $\hat{\vect{A}}(\vect{r})$ and  $\hat{\phi}(\vect{r})$
of the medium-assisted electromagnetic field are expressed in terms of
the dynamical variables $\hat{\vect{f}}(\vect{r},\omega)$ and 
$\hat{\vect{f}}^\dagger(\vect{r},\omega)$ as
\begin{equation}
\label{eq:Aphi}
\hat{\vect{A}}(\vect{r}) = \int\limits_0^\infty d\omega
\frac{1}{i\omega}
\hat{\vect{E}}^\perp(\vect{r},\omega) +\mbox{h.c.} \,,\quad
-\grad\hat{\phi}(\vect{r}) = \int\limits_0^\infty d\omega\,
\hat{\vect{E}}^\|(\vect{r},\omega) +\mbox{h.c.} \,.
\end{equation}

The total electromagnetic fields are the sums of the medium-assisted
fields and the fields associated with the atomic system,
\begin{eqnarray}
\label{eq:Etot}
\hat{\mathcal{E}}(\vect{r}) &=& \hat{\vect{E}}(\vect{r}) 
-\grad\hat{\phi}_A(\vect{r}) \,,\\
\label{eq:Btot}
\hat{\mathcal{B}}(\vect{r}) &=& \hat{\vect{B}}(\vect{r}) \,,\\
\label{eq:Dtot}
\hat{\mathcal{D}}(\vect{r}) &=& \hat{\vect{D}}(\vect{r})
-\varepsilon_0\grad\hat{\phi}_A(\vect{r}) \,,\\
\label{eq:Htot}
\hat{\mathcal{H}}(\vect{r}) &=& \hat{\vect{H}}(\vect{r}) \,,
\end{eqnarray}
where
\begin{equation}
\label{eq:phiA}
\hat{\phi}_A(\vect{r}) = \int d^3r' 
\frac{\hat{\rho}_A(\vect{r}')}{4\pi\varepsilon_0|\vect{r}-\vect{r}'|}
\end{equation}
is the scalar potential of the charged particles.
In those cases in which the atomic system consists of sufficiently
localised particles such as in an atom or molecule, it is expedient to
introduce shifted particle coordinates 
$\hat{\bar{\vect{r}}}_\alpha=\hat{\vect{r}}_\alpha-\hat{\vect{r}}_A$ 
relative to the centre of mass $\hat{\vect{r}}_A=\sum_\alpha
(m_\alpha/m_A) \hat{\vect{r}}_\alpha$ 
[with the total mass $m_A=\sum_\alpha m_\alpha$]. Expanding the vector
and scalar potentials $\hat{\vect{A}}(\vect{r})$ and
$\hat{\phi}(\vect{r})$ around the centre of mass, the
Hamiltonian~(\ref{eq:minimalHamiltonian}) simplifies for globally
neutral atomic systems [$q_A=\sum_\alpha q_\alpha=0$] to the electric
dipole Hamiltonian
\begin{equation}
\label{eq:Hamiltonian}
\hat{H} = \hat{H}_F +\hat{H}_A +\hat{H}_{AF}
\end{equation}
where
\begin{eqnarray}
\label{eq:HFminimal}
\hat{H}_F &=& \sum\limits_{\lambda=e,m} \int d^3r \int\limits_0^\infty
d\omega 
\,\hbar\omega\, \hat{\vect{f}}_\lambda^\dagger(\vect{r},\omega) \cdot
\hat{\vect{f}}_\lambda(\vect{r},\omega) \,,\\
\label{eq:HAminimal}
\hat{H}_A &=& \sum\limits_\alpha
\frac{\hat{\vect{p}}_\alpha^2}{2m_\alpha}
+\sum\limits_{\alpha\ne\alpha'} \frac{q_\alpha q_{\alpha'}}
{8\pi\varepsilon_0|\hat{\vect{r}}_\alpha-\hat{\vect{r}}_{\alpha'}|}
\,,\\
\label{eq:HAFminimal}
\hat{H}_{AF} &=& \hat{\vect{d}}\cdot 
\left.\grad\hat{\phi}(\vect{r})\right|_{\vect{r}=\hat{\vect{r}}_A}
-\sum\limits_\alpha\frac{q_\alpha}{2m_\alpha} \hat{\vect{p}}_\alpha
\cdot
\hat{\vect{A}}(\hat{\vect{r}}_A) +\sum\limits_\alpha
\frac{q_\alpha^2}{2m_\alpha}
\hat{\vect{A}}^2(\hat{\vect{r}}_A) \,.
\end{eqnarray}
Recall that the electric dipole moment is given by
Eq.~(\ref{eq:dipolemoment}).

Equations~(\ref{eq:HFminimal})--(\ref{eq:HAFminimal}) could form the
starting point for investigations into interaction of the
medium-assisted electromagnetic field with atomic systems in the
long-wavelength approximation. However, as for free-space QED, the
treatment can be considerably simplified by transforming using the
alternative multipolar coupling scheme. To this end, we again apply a
Power--Zienau transformation \cite{Power,Wolley,CraigThiru}
\begin{equation}
\label{eq:PowerZienauU}
\hat{U} = \exp \left[ \frac{i}{\hbar} \int d^3r\,
\hat{\vect{P}}_A(\vect{r}) \cdot \hat{\vect{A}}(\vect{r}) \right]
\end{equation}
where the polarisation $\hat{\vect{P}}_A$ is still defined by
Eq.~(\ref{eq:classicalpolarization}), but the vector potential
$\hat{\vect{A}}$ is given by Eq.~(\ref{eq:Aphi}) in the presence of
magnetoelectrics, as opposed to the expansion~(\ref{eq:vectorA}) valid
in free space. The unitary operator transformation
$\hat{O}'=\hat{U}\hat{O}\hat{U}^\dagger$ with the
operator (\ref{eq:PowerZienauU}) leads to transformed dynamical
variables $\hat{\vect{f}}_\lambda'(\vect{r},\omega)$ as
\cite{Buhmann04}
\begin{equation}
\label{eq:Ftransform}
\hat{\vect{f}}_\lambda'(\vect{r},\omega) =
\hat{\vect{f}}_\lambda(\vect{r},\omega) +\frac{1}{\hbar\omega} \int
d^3r'\, \hat{\vect{P}}_A^\perp(\vect{r}') \cdot
\tens{G}_\lambda^+(\vect{r},\vect{r}',\omega) 
\end{equation}
and thus transformed electric fields~(\ref{Etransform}). The magnetic
induction field remains unchanged by this transformation,
$\hat{\vect{B}}'(\vect{r})=\hat{\vect{B}}(\vect{r})$, because it
clearly commutes with the operator of the vector potential. Other
quantities that remain unchanged are those that depend solely on the
atomic position operators
$\hat{\vect{r}}_\alpha'=\hat{\vect{r}}_\alpha$ such as the atomic
scalar potential,
$\hat{\phi}_A'(\vect{r})=\hat{\phi}_A(\vect{r})$, and the
polarisation and magnetisation fields. The atomic momentum operators,
however, transform as \cite{Buhmann04}
\begin{equation}
\label{eq:ptransform}
\hat{\vect{p}}_\alpha' = \hat{\vect{p}}_\alpha -q_\alpha
\hat{\vect{A}}(\hat{\vect{r}}_\alpha) -\int d^3r\,
\hat{\vect{\Xi}}_\alpha(\vect{r}) \times \hat{\vect{B}}(\vect{r})
\end{equation}
where we have defined the auxiliary vectors
\begin{eqnarray}
\label{Xi}
\hat{\vect{\Xi}}_\alpha(\vect{r}) &=& q_\alpha
\hat{\vect{\Theta}}_\alpha(\vect{r}) -\frac{m_\alpha}{m_A}
\sum\limits_\beta q_\beta \hat{\vect{\Theta}}_\beta(\vect{r})
+\frac{m_\alpha}{m_A} \hat{\vect{P}}_A(\vect{r}) \,,\\
\label{eq:Theta}
\hat{\vect{\Theta}}_\alpha(\vect{r}) &=& \hat{\bar{\vect{r}}}_\alpha
\int\limits_0^1 ds\,
s\delta(\vect{r}-\hat{\vect{r}}_A-s\hat{\bar{\vect{r}}}_\alpha) \,.
\end{eqnarray}

With these preparations, we can now write down the Hamiltonian
(\ref{eq:minimalHamiltonian}) in terms of the transformed variables as
\begin{eqnarray}
\label{eq:multipolarHamiltonian}
\hat{H}' &=& \sum\limits_{\lambda=e,m} \int d^3r \int\limits_0^\infty
d\omega \, \hbar\omega\,
\hat{\vect{f}}_\lambda'{}^{\!\dagger}(\vect{r},\omega) \cdot
\hat{\vect{f}}_\lambda'(\vect{r},\omega) +\frac{1}{2\varepsilon_0}
\int d^3r \,\hat{\vect{P}}_A^2(\vect{r}) \nonumber \\ &&
-\int d^3r\, \hat{\vect{P}}_A(\vect{r}) \cdot
\hat{\vect{E}}'(\vect{r})
+\sum\limits_\alpha \frac{1}{2m_\alpha} \left[ \hat{\vect{p}}_\alpha'
+\int d^3r\, \hat{\vect{\Xi}}_\alpha(\vect{r}) \times
\hat{\vect{B}}(\vect{r}) \right]^2 \,.
\end{eqnarray}
It should be mentioned that, due to the unitary nature of the
Power--Zienau transformation, the commutation relations between the
transformed atomic position and momentum operators
$\hat{\vect{r}}_\alpha'$ and $\hat{\vect{p}}_\alpha'$ as well as
between the transformed dynamical variables
$\hat{\vect{f}}_\lambda'(\vect{r},\omega)$ and
$\hat{\vect{f}}_\lambda'{}^{\!\dagger}(\vect{r},\omega)$ are
unchanged. The fields $\hat{\vect{E}}'(\vect{r})$ and
$\hat{\vect{B}}'(\vect{r})$ have to be thought of as being expanded in
terms of the transformed dynamical variables in the same way as the
untransformed fields are expanded in terms of the untransformed
dynamical variables.

In the long-wavelength approximation, one replaces the function
$\delta(\vect{r}-\hat{\vect{r}}_A-s\hat{\bar{\vect{r}}}_\alpha)$ by
its value at $s=0$, $\delta(\vect{r}-\hat{\vect{r}}_A)$, so that the
polarisation and the auxiliary fields reduce to
\begin{equation}
\label{eq:longwavelength}
\hat{\vect{P}}_A(\vect{r}) = \hat{\vect{d}}
\delta(\vect{r}-\hat{\vect{r}}_A) \,,\quad
\hat{\vect{\Theta}}_\alpha(\vect{r}) = \frac{1}{2}
\hat{\bar{\vect{r}}}_\alpha \delta(\vect{r}-\hat{\vect{r}}_A) \,,\quad
\hat{\vect{\Xi}}_\alpha(\vect{r}) =
q_\alpha\hat{\vect{\Theta}}_\alpha(\vect{r}) +\frac{m_\alpha}{2m_A}
\hat{\vect{P}}_A(\vect{r}) \,.
\end{equation}
Upon using these simplifications, we obtain the multipolar Hamiltonian
(\ref{eq:multipolarHamiltonian}) in long-wavelength approximation as
\begin{equation}
\label{eq:Hmultipolar}
\hat{H}' = \hat{H}_F' +\hat{H}_A' +\hat{H}_{AF}'
\end{equation}
where
\begin{eqnarray}
\label{eq:multipolarHF}
\hat{H}_F' &=& \sum\limits_{\lambda=e,m} \int d^3r
\int\limits_0^\infty
d\omega \, \hbar\omega\,
\hat{\vect{f}}_\lambda'{}^{\!\dagger}(\vect{r},\omega) \cdot 
\hat{\vect{f}}_\lambda'(\vect{r},\omega) \,,\\
\label{eq:multipolarHA}
\hat{H}_A' &=& \sum\limits_\alpha
\frac{\hat{\vect{p}}_\alpha'{}^{\!2}}{2m_\alpha}
+\frac{1}{2\varepsilon_0} \int d^3r \, \hat{\vect{P}}_A^2(\vect{r})
\,,\\
\label{eq:multipolarHAF}
\hat{H}_{AF}' &=&
-\hat{\vect{d}}\cdot\hat{\vect{E}}'(\hat{\vect{r}}_A)
-\hat{\vect{m}}'\cdot\hat{\vect{B}}'(\hat{\vect{r}}_A)
+\sum\limits_\alpha
\frac{q_\alpha^2}{8m_\alpha} \left[\hat{\bar{\vect{r}}}_\alpha \times
\hat{\vect{B}}'(\hat{\vect{r}}_A) \right]^2 \nonumber \\ &&
+\frac{3}{8m_A} \left[\hat{\vect{d}} \times
\hat{\vect{B}}'(\hat{\vect{r}}_A) \right]^2
+\frac{\hat{\vect{p}}_A'}{m_A}  \cdot \left[
\hat{\vect{d}} \times \hat{\vect{B}}'(\hat{\vect{r}}_A) \right] \,,
\end{eqnarray}
where
\begin{equation}
\hat{\vect{m}}' = \frac{1}{2} \sum\limits_\alpha
\frac{q_\alpha}{m_\alpha} \hat{\bar{\vect{r}}}_\alpha \times
\hat{\bar{\vect{p}}}_\alpha'
\end{equation}
is the magnetic dipole operator,
$\hat{\vect{p}}_A'=\sum_\alpha\hat{\vect{p}}_\alpha'$, and
$\hat{\bar{\vect{p}}}_\alpha'$ 
$=\hat{\vect{p}}_\alpha'-(m_\alpha/m_A)\hat{\vect{p}}_A'$.

At the moment, the magnetic dipole moment operator comprises only the
contribution from the angular momentum. Spin can be included via a
Pauli interaction term in Eq.~(\ref{eq:minimalHamiltonian}) leading to
\begin{equation}
\hat{\vect{m}}' = \sum\limits_\alpha \left[
\frac{q_\alpha}{2m_\alpha} \hat{\bar{\vect{r}}}_\alpha \times
\hat{\bar{\vect{p}}}_\alpha' +\gamma_\alpha \hat{\vect{s}}_\alpha
\right] 
\end{equation}
($\hat{\vect{s}}_\alpha$: particle spin, $\gamma_\alpha$: gyromagnetic
ratio) \cite{Hassan08}.

The scheme can be easily extended to include more than one atomic
subensemble. In the minimal-coupling scheme, this leads to interatomic
Coulomb interactions in Eq.~(\ref{eq:minimalHamiltonian}) which is why
the multipolar-coupling scheme is strongly preferable. In this case,
Eq.~(\ref{eq:Hmultipolar}) generalises to \cite{0009}
\begin{equation}
\hat{H}' = \hat{H}_F' +\sum\limits_A \left[ \hat{H}_A' +\hat{H}_{AF}'
\right] \,,
\end{equation}
where the dynamics of each atom is given by a Hamiltonian of the form
(\ref{eq:multipolarHA}) and each atom couples individually to the
electromagnetic field via coupling Hamiltonians of the form
(\ref{eq:multipolarHAF}).

As in the free-space case, we will henceforth treat all atom-field
couplings within the framework of the multipolar coupling scheme and
drop all primes denoting multipolar variables. 


\subsection{Nonlinear quantum electrodynamics}
\label{sec:nonlinear}

In all previous (as well as all subsequent) sections we have
concentrated on magnetoelectric background materials whose dielectric
(and magnetic) response to an external perturbation can be described
within the framework of linear-response theory. That is, polarisation
and magnetisation fields are linearly and causally related to the
primary electromagnetic fields. For purely dielectric media, this
means that [cf. Eq.~(\ref{eq:classicalP})]
\begin{equation}
\vect{P}(\vect{r},t) = \varepsilon_0 \int\limits_0^\infty d\tau\,
\chi(\vect{r},\tau) \vect{E}(\vect{r},t-\tau)
+\vect{P}_\mathrm{N}(\vect{r},t)
\end{equation}
where $\chi(\vect{r},\tau)$ is the dielectric susceptibility. However,
many materials show nonlinear behaviour, i.e. their dielectric
response has to be described by a polarisation with a component that
depends quadratically (cubically, quartically etc.) on the external
electric field. In free space where dispersion and absorption are
disregarded and a mode expansions of the electromagnetic field can be
used, the effect of these nonlinear polarisations is to add
interaction Hamiltonians that are cubic (quartic, quintic etc.) in the
photonic amplitude operators (see, e.g. \cite{SchubertWilhelmi}).
These effective Hamiltonians arise from off-resonant interactions with
atomic systems in rotating-wave approximation [see, e.g.
Eq.~(\ref{eq:dispersiveHeff})]. Our aim is to extend this theory to
nonlinear processes in absorbing matter where mode expansions
generally do not hold. 

For this purpose, we consider an interaction Hamiltonian in its most
general normal-ordered form corresponding to a $\chi^{(2)}$ medium.
Using the abbreviation $\vect{k}$ as a short-hand for the collection
of spatial and frequency variables
$\vect{k}\equiv(\vect{r}_k,\omega_k)$, this interaction Hamiltonian
can be written as \cite{Scheel06a}
\begin{equation}
\label{eq:HNL}
\hat{H}_\mathrm{NL} = \int d\vect{1}\,d\vect{2}\,d\vect{3}\,
\alpha_{ijk}(\vect{1},\vect{2},\vect{3}) \hat{f}_i^\dagger(\vect{1})
\hat{f}_j(\vect{2}) \hat{f}_k(\vect{3}) +\mbox{h.c.}
\end{equation}
with an as yet unknown coupling tensor 
$\alpha_{ijk}(\vect{1},\vect{2},\vect{3})$ that will eventually have
to be (linearly) related to the second-order nonlinear susceptibility 
$\chi_{ijk}^{(2)}$. The integration initially ranges over all
frequencies, even including those frequencies that would not guarantee
energy conservation. 

The Hamiltonian (\ref{eq:HNL}), together with the Hamiltonian
$\hat{H}_\mathrm{L}$ of the linear theory (we include the index
$\mathrm{L}$ here to distinguish it from the nonlinear interaction
Hamiltonian), is now used to construct the time-dependent Maxwell
equations such as Faraday's law 
$\curl\hat{\vect{E}}(\vect{r})=-\dot{\hat{\vect{B}}}(\vect{r})$ as
\begin{equation}
\curl\hat{\vect{E}}(\vect{r}) = -\frac{1}{i\hbar} \left[ 
\hat{\vect{B}}(\vect{r}), \hat{H}_\mathrm{L}+\hat{H}_\mathrm{NL}
\right] \,.
\end{equation}
The fields $\hat{\vect{E}}(\vect{r})$ and $\hat{\vect{B}}(\vect{r})$
have to be thought of as being expanded in terms of the dynamical
variables $\hat{\vect{f}}(\vect{r},\omega)$ and 
$\hat{\vect{f}}^\dagger(\vect{r},\omega)$. Since Faraday's law is
valid irrespective of the presence of matter, and by definition
$[\hat{\vect{B}}(\vect{r}), \hat{H}_\mathrm{L}]/(i\hbar)$
$\!=\dot{\hat{\vect{B}}}(\vect{r})$, we must have
\begin{equation}
\label{eq:constraintB}
\left[ \hat{\vect{B}}(\vect{r}), \hat{H}_\mathrm{NL} \right] 
=\veczero\,,
\end{equation}
which is a condition that is needed in the next step. We write 
Amp\`{e}re's law
$\curl\hat{\vect{H}}(\vect{r})=\dot{\hat{\vect{D}}}(\vect{r})$,
using Faraday's law, as
\begin{equation}
\curl\curl\hat{\vect{E}}(\vect{r}) = -\mu_0
\ddot{\hat{\vect{D}}}(\vect{r})
= -\mu_0 \ddot{\hat{\vect{D}}}_\mathrm{L}(\vect{r})
-\mu_0 \ddot{\hat{\vect{P}}}_\mathrm{NL}(\vect{r})
\end{equation}
where we split up the total displacement field
$\hat{\vect{D}}(\vect{r})$ into its linear part,
\begin{equation}
\label{eq:linearD}
\hat{\vect{D}}_\mathrm{L}(\vect{r}) = \varepsilon_0 
\varepsilon(\vect{r},\omega) \hat{\vect{E}}(\vect{r}) 
+\hat{\vect{P}}_\mathrm{L}^{(N)}(\vect{r}) \,,
\end{equation}
and some nonlinear polarisation
$\hat{\vect{P}}_\mathrm{NL}(\vect{r})$. Heisenberg's equations of
motion then imply that
\begin{eqnarray}
\label{eq:AmpereNL}
\curl\curl\hat{\vect{E}}(\vect{r}) &=& \frac{\mu_0}{\hbar^2} 
\left[\left[\hat{\vect{D}}_\mathrm{L}(\vect{r}),\hat{H}_\mathrm{L}
\right],
\hat{H}_\mathrm{L}\right] +\frac{\mu_0}{\hbar^2}
\left[\left[\hat{\vect{D}}_\mathrm{L}(\vect{r}),\hat{H}_\mathrm{L}
\right], 
\hat{H}_\mathrm{NL}\right] \nonumber \\ &&
+\frac{\mu_0}{\hbar^2}
\left[\left[\hat{\vect{D}}_\mathrm{L}(\vect{r}),\hat{H}_\mathrm{NL}
\right],
\hat{H}_\mathrm{L}\right] +\frac{\mu_0}{\hbar^2}
\left[\left[\hat{\vect{P}}_\mathrm{NL}(\vect{r}),\hat{H}_\mathrm{L}
\right], 
\hat{H}_\mathrm{L}\right]
\end{eqnarray}
where we kept only those terms that are at most linear in the coupling
tensor $\alpha_{ijk}$. The terms that have been left out have to be
included into higher order nonlinear processes. The first term on the
rhs of Eq.~(\ref{eq:AmpereNL}) is by definition equal to the lhs of
the same equation. The second term on its rhs vanishes because of the
constraint (\ref{eq:constraintB}). The remaining two terms have to
satisfy
\begin{equation}
\label{eq:particularsolution}
\left[\hat{\vect{D}}_\mathrm{L}(\vect{r}),\hat{H}_\mathrm{NL}\right]
=-\left[\hat{\vect{P}}_\mathrm{NL}(\vect{r}),\hat{H}_\mathrm{L}\right]
\end{equation}
which yields a solution for the nonlinear polarisation 
$\hat{\vect{P}}_\mathrm{NL}(\vect{r})$. Note that the double
commutator has been reduced to a single commutator as it turns out
that a general solution would have to include functionals that commute
with $\hat{H}_\mathrm{L}$ whose contributions can be shown to diverge 
\cite{Scheel06b} and thus have to be discarded.

The way to solve Eq.~(\ref{eq:particularsolution}) is to view its rhs
as being the Liouvillian generated by $\hat{H}_\mathrm{L}$, i.e. 
$\hat{\mathcal{L}}_\mathrm{L}\bullet$
$=(i\hbar)^{-1}[\bullet,\hat{H}_\mathrm{L}]$,
whose inverse can be formally written as
\begin{equation}
\hat{\vect{P}}_\mathrm{NL}(\vect{r}) = -\frac{1}{i\hbar} 
\hat{\mathcal{L}}_\mathrm{L}^{-1} 
\left[\hat{\vect{D}}_\mathrm{L}(\vect{r}),\hat{H}_\mathrm{NL}\right]
\,.
\end{equation}
The action of the inverse Liouvillian on an operator $\hat{O}$ is
given by
\begin{equation}
\hat{\mathcal{L}}_\mathrm{L}^{-1} \hat{O} = \lim_{s\to 0}
\int\limits_0^\infty
dt\,e^{-st} e^{-i\hat{H}_\mathrm{L}t/\hbar} \hat{O} 
e^{i\hat{H}_\mathrm{L}t/\hbar}
\end{equation}
which can be checked by direct calculation, with the result that
\begin{equation}
\hat{\vect{P}}_\mathrm{NL}(\vect{r}) = -\frac{1}{i\hbar} \lim_{s\to
0} 
\int\limits_0^\infty dt\,e^{-st} e^{-i\hat{H}_\mathrm{L}t/\hbar}
\left[\hat{\vect{D}}_\mathrm{L}(\vect{r}),\hat{H}_\mathrm{NL}\right]
e^{i\hat{H}_\mathrm{L}t/\hbar} \,.
\end{equation}
Before we continue solving this equation, we remark that due to the
decomposition of the linear displacement field (\ref{eq:linearD}) into
a reactive part and a Langevin noise contribution 
$\hat{\vect{P}}_\mathrm{L}(\vect{r})$, the nonlinear polarisation also
contains a contribution, $\hat{\vect{P}}_\mathrm{NL}^{(N)}(\vect{r})$
$\!=-(i\hbar)^{-1}\hat{\mathcal{L}}_\mathrm{L}^{-1}$
$\![\hat{\vect{P}}_\mathrm{L}^{(N)}(\vect{r}),\hat{H}_\mathrm{L}]$,
that disappears identically with vanishing absorption and can thus be
regarded as the nonlinear noise polarisation. The result of the
Liouvillian inversion can be cast into the form
\cite{Scheel06a,Scheel06b}
\begin{eqnarray}
\label{eq:nonlinpol}
\hat{P}_{\mathrm{NL},m}(\vect{r}) &\!=&\! \frac{1}{i\hbar}
\sqrt{\frac{\hbar\varepsilon_0}{\pi}} \int
d\vect{0}\,d\vect{2}\,d\vect{3}\,
\frac{\sqrt{\mathrm{Im}\,\varepsilon(\vect{0})}}{\omega_2+\omega_3}
\alpha_{njk}(\vect{0},\vect{2},\vect{3})
\frac{\omega^2}{c^2}\varepsilon(\vect{r},\omega)
G_{mn}(\vect{r},\vect{0}) \hat{f}_j(\vect{2}) \hat{f}_k(\vect{3})
\nonumber \\ && \hspace*{-3ex}
+\frac{1}{i\hbar}
\sqrt{\frac{\hbar\varepsilon_0}{\pi}} \int
d\vect{0}\,d\vect{1}\,d\vect{3}\,
\frac{\sqrt{\mathrm{Im}\,\varepsilon(\vect{0})}}{\omega_1-\omega_3}
\alpha_{imk}^\ast(\vect{1},\vect{0},\vect{3})
\frac{\omega^2}{c^2}\varepsilon(\vect{r},\omega)
G_{mn}(\vect{r},\vect{0}) \hat{f}_k^\dagger(\vect{3})
\hat{f}_i(\vect{1}) \nonumber \\ && \hspace*{-3ex}
+\mbox{h.c.} +\hat{P}_{\mathrm{NL},m}^{(N)}(\vect{r})
\end{eqnarray}
with
\begin{eqnarray}
\label{eq:nonlinnoisepol}
\hat{P}_{\mathrm{NL},m}^{(N)}(\vect{r}) &=& \frac{1}{i\hbar}
\sqrt{\frac{\hbar\varepsilon_0}{\pi}} \int
d\vect{0}\,d\vect{2}\,d\vect{3}\,
\frac{\sqrt{\mathrm{Im}\,\varepsilon(\vect{0})}}{\omega_2+\omega_3}
\alpha_{njk}(\vect{0},\vect{2},\vect{3})
\delta(\vect{r}-\vect{s}) \hat{f}_j(\vect{2}) \hat{f}_k(\vect{3})
\nonumber \\ && 
+\frac{1}{i\hbar}
\sqrt{\frac{\hbar\varepsilon_0}{\pi}} \int
d\vect{0}\,d\vect{1}\,d\vect{3}\,
\frac{\sqrt{\mathrm{Im}\,\varepsilon(\vect{0})}}{\omega_1-\omega_3}
\alpha_{imk}^\ast(\vect{1},\vect{0},\vect{3})
\delta(\vect{r}-\vect{s}) \hat{f}_k^\dagger(\vect{3})
\hat{f}_i(\vect{1}) \nonumber \\ && 
+\mbox{h.c.}
\end{eqnarray}
and the notation $\vect{0}=(\vect{s},\omega)$. 

From Eqs.~(\ref{eq:nonlinpol}) and (\ref{eq:nonlinnoisepol}) it is
hard to see how the coupling tensor $\alpha_{ijk}$ has to be related
to the nonlinear susceptibility. Instead, we invoke comparison with
the definition of the nonlinear polarisation from classical nonlinear
response theory \cite{SchubertWilhelmi},
\begin{equation}
P_{\mathrm{NL},m}(\vect{r},t) = \varepsilon_0
\int\limits_{-\infty}^t d\tau'\,d\tau''\,
\chi^{(2)}_{mrs}(\vect{r},t-\tau',t-\tau'')
E_r(\vect{r},\tau')E_s(\vect{r},\tau'')
+P_{\mathrm{NL},m}^{(N)}(\vect{r},t) 
\end{equation}
and introduce slowly varying electric fields whose (non-overlapping)
amplitudes are centred at the mid-frequencies $\Omega_i$ with
$\Omega_0=\Omega_1+\Omega_2$ such that
\begin{equation}
\label{eq:slowPNL}
\tilde{P}_{\mathrm{NL},m}(\vect{r},\Omega,t) = \varepsilon_0
\chi^{(2)}_{mrs}(\vect{r},\Omega_0,\Omega_1,\Omega_2)
\tilde{E}_r(\vect{r},\Omega_1)\tilde{E}_s(\vect{r},\Omega_2)
+\tilde{P}_{\mathrm{NL},m}^{(N)}(\vect{r},\Omega_0,t) \,.
\end{equation}
Expressing the electric fields in Eq.~(\ref{eq:slowPNL}) in terms of
slowly varying dynamical variables and comparing with the slowly
varying version of Eq.~(\ref{eq:nonlinpol}) yields the sought after
relation between $\alpha_{ijk}$ and $\chi^{(2)}_{ijk}$ as
\cite{Scheel06a,Scheel06b}
\begin{eqnarray}
\label{eq:alphachi}
\alpha_{ijk}(\vect{r},\Omega_0,\vect{s}_2,\Omega_2,\vect{s}_3,
\Omega_3)
&=& \frac{\hbar^2}{i\pi c^2} \sqrt{\frac{\pi}{\hbar\varepsilon_0}}
\sqrt{\frac{\mathrm{Im}\,\varepsilon(\vect{s}_2,\Omega_2)
\mathrm{Im}\,\varepsilon(\vect{s}_3,\Omega_3)}
{\mathrm{Im}\,\varepsilon(\vect{r}_2,\Omega_0)}}
\nonumber \\ && \hspace*{-15ex}
\times H_{li}(\vect{r},\Omega_0) \left[
\frac{\chi_{imn}^{(2)}(\vect{r},\Omega_2,\Omega_3)}
{\varepsilon(\vect{r},\Omega_0)} G_{mj}(\vect{r},\vect{s}_2,\Omega_2)
G_{nk}(\vect{r},\vect{s}_3,\Omega_3) \right]
\end{eqnarray}
where $H_{li}(\vect{r},\Omega_0)$ are the cartesian components of the
Helmholtz operator, or equivalently, the integral operator associated
with the inverse dyadic Green function. 

Reinserted into the nonlinear interaction Hamiltonian (\ref{eq:HNL}), 
and combined with the dynamical variables, Eq.~(\ref{eq:alphachi})
will yield a formulation in terms of electromagnetic field operators.
What becomes immediately clear, though, is that $\hat{H}_\mathrm{NL}$
will not be of the form $\hat{H}_\mathrm{NL}$
$\propto\chi_{ijk}^{(2)}\hat{E}_i^\dagger\hat{E}_j\hat{E}_k$ as
in standard nonlinear optics. Mathematically, the reason is that there
is no third Green tensor in Eq.~(\ref{eq:alphachi}). Instead, an
inverse Green function (or Helmholtz operator) has to be dealt with
which, by formally expanding it into a power series, will lead to
additional contributions to the standard nonlinear interaction. The
physical reason for this behaviour has to be sought in the fact that,
from a microscopic point of view, an effective nonlinear interaction
does not take place in free space but rather inside the absorbing
medium where the local electric field is altered by the presence of
the dielectric material. To account for that, local-field corrections
such as those discussed in Sec.~\ref{sec:localfield} have to be
included which automatically yields additional contributions to the
nonlinear interaction Hamiltonian that are not of the standard form. 


\newpage
\section{Atomic relaxation rates}
\label{sec:relaxation}

The first set of applications we consider in this review regards the
theory of atomic transition rates. In this section we discuss the
influence of dielectric bodies towards atomic relaxation and heating
rates, and some of their experimental ramifications.

To begin, it is necessary to study the dynamics of internal atomic
degrees of freedom in the presence of absorbing magnetoelectric
matter. In Sec.~\ref{sec:interaction} we derived the Hamiltonian of
the system comprising the medium-assisted electromagnetic field, the
atomic system and their mutual interaction (here taken in the electric
dipole approximation) as
\begin{equation}
\hat{H} = \hat{H}_F +\hat{H}_A +\hat{H}_{AF}
\end{equation}
with
\begin{equation}
\hat{H}_F = \sum\limits_{\lambda=e,m} \int d^3r \int\limits_0^\infty 
d\omega \, \hbar\omega\,
\hat{\vect{f}}^\dagger_\lambda(\vect{r},\omega)
\cdot \hat{\vect{f}}_\lambda(\vect{r},\omega)
\end{equation}
the Hamiltonian of the medium-assisted electromagnetic field 
[Eq.~(\ref{eq:multipolarHF})] in the presence of magnetoelectric
bodies. The free atomic Hamiltonian $\hat{H}_A$
[Eq.~(\ref{eq:multipolarHA})] is expanded into atomic energy
eigenstates $|n\rangle$, 
\begin{equation}
\hat{H}_A = \sum\limits_n \hbar\omega_n \hat{A}_{nn}\,,
\end{equation}
with the corresponding eigenenergies $\hbar\omega_n$, and the atomic
flip operators $\hat{A}_{mn}=|m\rangle\langle n|$  obeying
the commutation rules 
\begin{equation}
\left[ \hat{A}_{kl}, \hat{A}_{mn} \right] = \delta_{lm}\hat{A}_{kn}
-\delta_{kn} \hat{A}_{ml} \,.
\end{equation}
The electric-dipole interaction Hamiltonian $\hat{H}_{AF}$ which, in
multipolar coupling, is given by 
$\hat{H}_{AF}=-\hat{\vect{d}}\cdot\hat{\vect{E}}(\vect{r}_A)$ [first
term in Eq.~(\ref{eq:multipolarHAF})], is expanded in terms of the
energy eigenstates as
\begin{equation}
\hat{H}_{AF} = -\sum\limits_{m,n} \vect{d}_{mn} \cdot 
\hat{\vect{E}}(\vect{r}_A) \hat{A}_{mn}
\end{equation}
where $\vect{d}_{mn}=\langle m|\hat{\vect{d}}|n\rangle$ are the matrix
elements of the dipole operator. Recall that we have dropped all
primes on the electromagnetic field operators that indicate the
multipolar coupling.

In similar fashion to the free-space theory outlined in
Sec.~\ref{sec:freespaceHeisenberg}, the internal atomic dynamics is
governed by the solution to the coupled set of Heisenberg's equations
of motion ($\omega_{mn}=\omega_m-\omega_n$)
\begin{eqnarray}
\label{eq:Amndot}
\dot{\hat{A}}_{mn}
&=& \frac{i}{\hbar} \left[ \hat{A}_{mn}, \hat{H} \right]
= i\omega_{mn} \hat{A}_{mn} \nonumber \\ &&
+\frac{i}{\hbar} \sum\limits_k \int\limits_0^\infty d\omega\,
\bigg[
\left( \vect{d}_{nk}\hat{A}_{mk}-\vect{d}_{km}\hat{A}_{kn} \right)
\cdot \hat{\vect{E}}(\vect{r}_A,\omega) \nonumber \\ && \hspace*{12ex}
+\hat{\vect{E}}^\dagger(\vect{r}_A,\omega) \cdot
\left( \vect{d}_{nk}\hat{A}_{mk}-\vect{d}_{km}\hat{A}_{kn} \right)
\bigg]
\end{eqnarray}
and
\begin{equation}
\label{eq:fdot}
\dot{\hat{\vect{f}}}_\lambda(\vect{r},\omega) = \frac{i}{\hbar}
\left[ \hat{\vect{f}}_\lambda(\vect{r},\omega) ,\hat{H} \right]
= -i\omega \hat{\vect{f}}_\lambda(\vect{r},\omega) +\frac{i}{\hbar}
\sum\limits_{m,n} \vect{d}_{mn} \cdot
\tens{G}^\ast_\lambda(\vect{r}_A,\vect{r},\omega) \hat{A}_{mn} \,.
\end{equation}
We formally integrate Eq.~(\ref{eq:fdot}) as
\begin{equation}
\label{eq:formalf}
\hat{\vect{f}}_\lambda(\vect{r},\omega,t) = e^{-i\omega t}
\hat{\vect{f}}_\lambda(\vect{r},\omega) +\frac{i}{\hbar}
\sum\limits_{m,n} \int\limits_0^t d\tau\, e^{-i\omega(t-\tau)}
\vect{d}_{mn} \cdot \tens{G}^\ast_\lambda(\vect{r}_A,\vect{r},\omega)
\hat{A}_{mn}(\tau)
\end{equation}
and reinsert this formal solution into Eq.~(\ref{eq:Amndot}) and
obtain
\begin{eqnarray}
\label{eq:Amndot2}
\dot{\hat{A}}_{mn}(t) &=&i\omega_{mn} \hat{A}_{mn}(t) +\frac{i}{\hbar}
\sum\limits_k \int\limits_0^\infty d\omega\, \bigg\{ e^{-i\omega t}
\left[ \vect{d}_{nk}\hat{A}_{mk}(t)-\vect{d}_{km}\hat{A}_{kn}(t)
\right] \cdot \hat{\vect{E}}(\vect{r}_A,\omega)
\nonumber \\ &&
+e^{i\omega t} \hat{\vect{E}}^\dagger(\vect{r}_A,\omega) \cdot \left[
\vect{d}_{nk}\hat{A}_{mk}(t)-\vect{d}_{km}\hat{A}_{kn}(t) \right]
\bigg\} +\hat{Z}_{mn}(t)
\end{eqnarray}
where
\begin{eqnarray}
\label{eq:Zmn}
\hat{Z}_{mn}(t) &=& -\frac{\mu_0}{\hbar\pi} \sum\limits_{k,l,j}
\int\limits_0^\infty d\omega\,\omega^2 \int\limits_0^t d\tau
\nonumber \\ && \hspace*{-10ex}
\times \bigg\{ \left[ e^{-i\omega(t-\tau)} \hat{A}_{mk}(t)
\hat{A}_{lj}(\tau) -e^{i\omega(t-\tau)} \hat{A}_{lj}(\tau)
\hat{A}_{mk}(t) \right] \vect{d}_{nk} \cdot
\mathrm{Im}\,\tens{G}(\vect{r}_A,\vect{r}_A,\omega) \cdot
\vect{d}_{lj}
\nonumber \\ && \hspace*{-9ex}
-\left[ e^{-i\omega(t-\tau)} \hat{A}_{kn}(t)\hat{A}_{lj}(\tau)
-e^{i\omega(t-\tau)} \hat{A}_{lj}(\tau) \hat{A}_{kn}(t)  \right]
\vect{d}_{kn} \cdot
\mathrm{Im}\,\tens{G}(\vect{r}_A,\vect{r}_A,\omega)
\cdot \vect{d}_{lj} \bigg\} \nonumber \\ &&
\end{eqnarray}
is the zero-point contribution to the internal atomic dynamics due to
the second term in Eq.~(\ref{eq:formalf}). A self-consistent solution
is obtained if Eq.~(\ref{eq:Amndot2}) is formally integrated as
\begin{eqnarray}
\hat{A}_{mn}(t) &=& e^{i\tilde{\omega}_{mn}t} \hat{A}_{mn}(0)
+\frac{i}{\hbar} \sum\limits_k \int\limits_0^\infty d\omega
\int\limits_0^t d\tau \, e^{i\tilde{\omega}_{mn}(t-\tau)}
\nonumber \\ && \times
\bigg\{ e^{-i\omega\tau} \left[ \vect{d}_{nk}\hat{A}_{mk}(\tau)
-\vect{d}_{mk}\hat{A}_{kn}(\tau) \right] \cdot
\hat{\vect{E}}(\vect{r}_A,\omega) 
\nonumber \\ &&
+e^{i\omega\tau} \hat{\vect{E}}^\dagger(\vect{r}_A,\omega) \cdot
\left[ \vect{d}_{nk}\hat{A}_{mk}(\tau)
-\vect{d}_{mk}\hat{A}_{kn}(\tau) \right] \bigg\} \,,
\end{eqnarray}
reinserted into Eq.~(\ref{eq:Amndot2}), and (thermal) expectation
values being taken. The $\tilde{\omega}_{mn}$ are the shifted atomic
transition frequencies. Using the thermal expectation values for the
electromagnetic field operators given in Sec.~\ref{sec:langevin},
Eqs.~(\ref{eq:expectE})--(\ref{eq:EEdagger}), what remains is a set of
coupled differential equations for the atomic quantities,
\begin{equation}
\langle\dot{\hat{A}}_{mn}\rangle_T = i\omega_{mn}
\langle\hat{A}_{mn}\rangle_T +\langle\hat{Z}_{mn}\rangle_T
+\langle\hat{T}_{mn}\rangle_T \,, 
\end{equation}
where the thermal contributions $\langle\hat{T}_{mn}\rangle_T$ read
($\bar{n}_\mathrm{th}=[e^{\hbar\omega/(k_BT)}-1]^{-1}$) 
\begin{eqnarray}
\langle\hat{T}_{mn}\rangle_T &=& \frac{\mu_0}{\hbar\pi}
\sum\limits_{k,l}
\int\limits_0^\infty d\omega\,\omega^2 \bar{n}_\mathrm{th}(\omega)
\int\limits_0^t d\tau\, \left[ e^{-i\omega(t-\tau)}
+e^{i\omega(t-\tau)} \right]  
\nonumber \\ && \hspace*{-18ex}
\times \bigg\{
e^{i\tilde{\omega}_{mk}(t-\tau)} \left[ \langle\hat{A}_{ml}\rangle_T
\vect{d}_{nk}\cdot\mathrm{Im}\,\tens{G}(\vect{r}_A,\vect{r}_A,
\omega)\cdot
\vect{d}_{kl} -\langle\hat{A}_{lk}\rangle_T
\vect{d}_{nk}\cdot\mathrm{Im}\,\tens{G}(\vect{r}_A,\vect{r}_A,
\omega)\cdot
\vect{d}_{lm} \right]
\nonumber \\ && \hspace*{-18ex}
-e^{i\tilde{\omega}_{kn}(t-\tau)} \left[ \langle\hat{A}_{kl}\rangle_T
\vect{d}_{km}\cdot\mathrm{Im}\,\tens{G}(\vect{r}_A,\vect{r}_A,
\omega)\cdot
\vect{d}_{nl} -\langle\hat{A}_{ln}\rangle_T
\vect{d}_{km}\cdot\mathrm{Im}\,\tens{G}(\vect{r}_A,\vect{r}_A,
\omega)\cdot
\vect{d}_{lk} \right] \bigg\} \,,
\nonumber \\ &&
\end{eqnarray}
recall Eq.~(\ref{eq:Zmn}). These sets of equations form the basis for
all following investigations in internal atomic dynamics.


\subsection{Modified spontaneous decay and body-induced Lamb shift,
local-field corrections}
\label{sec:modifiedse}

For weak atom-field coupling, these expressions can be further
evaluated using the Markov approximation (see
Sec.~\ref{sec:freespaceHeisenberg}) in which the expectation values at
times $\tau$ can be related to those at the upper limit $t$ of the
time integrals as 
\begin{equation}
\langle\hat{A}_{mn}(\tau)\rangle_T \simeq
e^{-i\tilde{\omega}_{mn}(t-\tau)} 
\langle\hat{A}_{mn}(t)\rangle_T \,.
\end{equation}
The time integrals themselves are approximated as in the free-space
theory by 
\begin{equation}
\int\limits_0^t d\tau\, e^{-i(\omega-\tilde{\omega}_{mn})(t-\tau)}
\simeq \pi\delta(\omega-\tilde{\omega}_{mn}) 
+i\mathcal{P} \frac{1}{\omega-\tilde{\omega}_{mn}} \,.
\end{equation}

Introducing the atomic density matrix elements $\sigma_{mn}=\langle
m|\hat{\sigma}|n\rangle=\langle\hat{A}_{nm}\rangle$,
the set of differential equations reduces to
\begin{eqnarray}
\dot{\sigma}_{nn}(t) &=& -\Gamma_n \sigma_{nn}(t) 
+\sum\limits_k \Gamma_{kn} \sigma_{kk}(t) \,,\\
\dot{\sigma}_{mn}(t) &=& \left[ -i\tilde{\omega}_{mn} -\frac{1}{2}
\left( 
\Gamma_m + \Gamma_n \right) \right] \sigma_{mn}(t) \,,\quad m\ne n\,.
\end{eqnarray}
These equations resemble closely those obtained in free-space theory.
One can identify individual decay rates $\Gamma_{nk}$ from state
$|n\rangle$ to $|k\rangle$, and total loss rates 
$\Gamma_n=\sum_k \Gamma_{nk}$ of a level $|n\rangle$. The individual
rates contain zero-point [superscript $(Z)$] as well as thermal
[superscript $(T)$] contributions,
$\Gamma_{nk}=\Gamma_{nk}^{(Z)}+\Gamma_{nk}^{(T)}$, with 
\begin{eqnarray}
\label{eq:zeropointgamma}
\Gamma_{nk}^{(Z)} &=& \frac{2\mu_0}{\hbar} \tilde{\omega}_{nk}^2 
\Theta(\tilde{\omega}_{nk}) \vect{d}_{nk} \cdot 
\mathrm{Im}\,\tens{G}(\vect{r}_A,\vect{r}_A,\tilde{\omega}_{nk}) \cdot
\vect{d}_{kn} \,,\\ \label{eq:thermalgamma}
\Gamma_{nk}^{(T)} &=& \frac{2\mu_0}{\hbar} \tilde{\omega}_{nk}^2 
\vect{d}_{nk} \cdot 
\mathrm{Im}\,\tens{G}(\vect{r}_A,\vect{r}_A,|\tilde{\omega}_{nk}|) 
\cdot \vect{d}_{kn} 
\left[ \Theta(\tilde{\omega}_{nk})
\bar{n}_\mathrm{th}(\tilde{\omega}_{nk}) 
+ \Theta(\tilde{\omega}_{kn}) \bar{n}_\mathrm{th}(\tilde{\omega}_{kn})
\right] \,. 
\nonumber \\ &&
\end{eqnarray}
Similarly, the shifted atomic transition frequencies 
$\tilde{\omega}_{mn}=\omega_{mn}+\delta\omega_m-\delta\omega_n$ depend
on the frequency shifts $\delta\omega_m=\sum_k\delta\omega_{nk}$ where
the zero-point and thermal contributions to the level shift induced by
a single level $|k\rangle$ read
\begin{eqnarray}
\label{eq:zeropointshift}
\delta\omega_{nk}^{(Z)} &=& \frac{\mu_0}{\pi\hbar} \mathcal{P} 
\int\limits_0^\infty d\omega\,\omega^2 
\frac{\vect{d}_{nk}\cdot
\mathrm{Im}\,\tens{G}^{(S)}(\vect{r}_A,\vect{r}_A,\omega)
\cdot\vect{d}_{kn}}{\tilde{\omega}_{nk}-\omega} \,,\\
\label{eq:thermalshift} 
\delta\omega_{nk}^{(T)} &=& \frac{\mu_0}{\pi\hbar} \mathcal{P} 
\int\limits_0^\infty d\omega\,\omega^2 \vect{d}_{nk}\cdot
\mathrm{Im}\,\tens{G}(\vect{r}_A,\vect{r}_A,\omega)
\cdot\vect{d}_{kn} \left[
\frac{\bar{n}_\mathrm{th}(\omega)}{\tilde{\omega}_{nk}-\omega}
+\frac{\bar{n}_\mathrm{th}(\omega)}{\tilde{\omega}_{nk}+\omega}\right]
\,.
\end{eqnarray}
In $\delta\omega_{nk}^{(Z)}$, we have explicitly used the scattering
part $\tens{G}^{(S)}(\vect{r}_A,\vect{r}_A,\omega)$ of the dyadic
Green function because we assume that the vacuum-induced Lamb shift
has already been included into the bare atomic transition frequencies
$\omega_{mn}$.


\paragraph{Spontaneous decay of a two-level atom:}
As an instructive example, let us consider a two-level atom with
energy levels $|g\rangle$ and $|e\rangle$ separated by an energy
$\hbar\omega_A=\hbar\omega_{eg}$ which we assume to contain the
free-space Lamb shift. At zero temperature, the spontaneous decay rate
$\Gamma\equiv\Gamma^{(Z)}$ of the excited state $|e\rangle$ is then
given by Eq.~(\ref{eq:zeropointgamma}) as \cite{Scheel99a,Scheel99b} 
\begin{equation}
\label{eq:seDGF}
\Gamma = \frac{2\omega_A^2}{\hbar\varepsilon_0 c^2} \vect{d}
\cdot \mathrm{Im}\,\tens{G}(\vect{r}_A,\vect{r}_A,\omega_A) \cdot
\vect{d}^\ast
\end{equation}
where we have neglected the shift in the atomic transition frequency
$\omega_A$. This is the same result one would obtain using
perturbation theory, i.e. Fermi's Golden Rule. We have noted
previously that the rate of spontaneous decay in free space is
proportional to the strength of the vacuum fluctuations of the
electric field [Eq.~(\ref{eq:FGR})] which, by Eq.~(\ref{eq:EEdagger}),
is now seen to be proportional to the imaginary part of the dyadic
Green function,
\begin{equation}
\langle 0| \hat{\vect{E}}(\vect{r}_A,\omega)
\otimes \hat{\vect{E}}^\dagger(\vect{r}_A,\omega_A) |0\rangle
= \frac{\hbar}{\pi\varepsilon_0} \frac{\omega_A^2}{c^2}
\mathrm{Im}\,\tens{G}(\vect{r}_A,\vect{r}_A,\omega_A)
\delta(\omega-\omega_A) \,,
\end{equation}
which underpins our interpretation of
$\omega^2\mathrm{Im}\,\tens{G}(\vect{r}_A,\vect{r}_A,\omega_A)$ as the
local density of states. The rate $\Gamma_0$ of spontaneous decay in
vacuum, Eq.~(\ref{eq:vacuumse}), is recovered by inserting the
free-space Green tensor $\tens{G}^{(0)}(\vect{r},\vect{r}',\omega)$
into Eq.~(\ref{eq:seDGF}).

To give a nontrivial example of how Eq.~(\ref{eq:seDGF}) can be used
for investigating atom-surface interactions, we consider a two-level
atom placed near a planar dielectric half-space with permittivity
$\varepsilon(\omega)$. The Green function for this structure is known
(see Appendix~\ref{sec:dgf}), and the result in the limit
$z_A\omega/c\ll 1$ reads \cite{ScheelActa99,Yeung96}
\begin{equation}
\label{eq:nearfieldplanar}
\Gamma = \Gamma_0 \frac{3}{8} \left( 1+\frac{|d_z|^2}{|\vect{d}|^2}
\right)
\left( \frac{c}{\omega_Az_A} \right)^3
\frac{\mathrm{Im}\,\varepsilon(\omega_A)}{|\varepsilon(\omega_A)+1|^2}
+\mathcal{O}(z_A^{-1}) \,.
\end{equation}
Hence, a dipole oriented perpendicular to a planar surface decays
twice as fast as a dipole parallel to it. Note the difference to the
spontaneous decay rate near a perfect mirror which approaches finite
values ($2\Gamma_0$ for perpendicular dipole orientation, $0$ for
parallel dipole orientation) in the on-surface limit. In reality,
however, nonradiative decay processes cause the near-field spontaneous
decay rate to diverge.

\begin{figure}[!t!]
\centerline{\includegraphics[width=4cm]{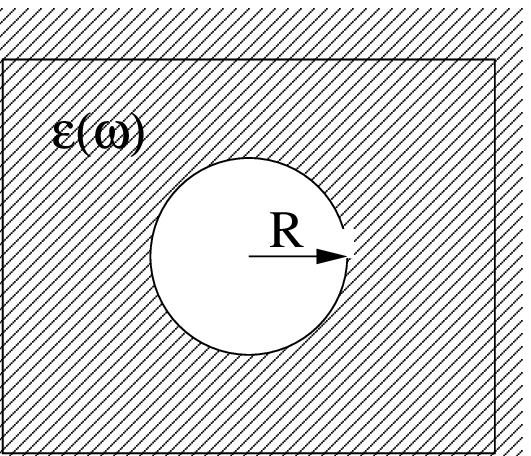}}
\caption{\label{fig:spherical} Atom at the centre of a spherical
microcavity of radius $R_\mathrm{cav}$.} 
\end{figure}

Another important example, which has been used to investigate
modified spontaneous decay inside a dielectric host medium
\cite{Scheel99b,Buch}, is that of an atom in a spherical
microcavity of radius $R_\mathrm{cav}$ (Fig.~\ref{fig:spherical}).
%
%
We use the dyadic Green function presented in Appendix~\ref{sec:dgf}
and note that, if the atom is located at the centre of the
microcavity, only the TM-wave spherical vector wave functions
$\vect{N}_{pm1}(k)$ do not vanish. Inserting the result into
Eq.~(\ref{eq:seDGF}) yields \cite{Scheel99b,Buch}
\begin{equation}
\Gamma = \Gamma_0 \left[ 1+\mathrm{Re}\,r_p^{22}(l=1) \right]
\end{equation}
where the reflection coefficient $r_p^{22}$ for $l=1$ takes the form
\begin{equation}
\label{eq:reflectioncoefficient}
r_p^{22}(l=1) = \frac{[i+\rho(n+1)-i\rho^2n-\rho^3n^2/(n+1)]e^{i\rho}}
{\sin\rho-\rho(\cos\rho+in\sin\rho)+i\rho^2n\cos\rho-
\rho^3(\cos\rho-in\sin\rho)n^2/(n^2-1)}
\end{equation}
[$n=n(\omega)=\sqrt{\varepsilon(\omega)}$ and
$\rho=R_\mathrm{cav}\omega_A/c$]. In the near-field limit, i.e. when
the size of the microcavity is much smaller than the atomic transition
wavelength, we can expand in powers of
$\rho=R_\mathrm{cav}\omega_A/c\ll 1$ and obtain [see also
Eq.~(\ref{eq:G1loc})] 
\begin{equation}
\label{eq:nearfieldspherical}
\Gamma = \Gamma_0 \,\mathrm{Im}\left\{
 \frac{3(\varepsilon-1)}{2\varepsilon+1}
 \left(\frac{c}{\omega_A R_\mathrm{cav}}\right)^3
+\frac{9[4\varepsilon^2-3\varepsilon-1]}{5(2\varepsilon+1)^2}
\left(\frac{c}{\omega R_\mathrm{cav}}\right) 
 +i\frac{9\varepsilon^2 n}{(2\varepsilon+1)^2}
 +\mathcal{O}(R_\mathrm{cav}) \right\} 
\end{equation}
[$\varepsilon\equiv\varepsilon(\omega_A)$,
$n\equiv\sqrt{\varepsilon(\omega_A)}$]. 

The leading terms in both examples, Eqs.~(\ref{eq:nearfieldplanar})
and (\ref{eq:nearfieldspherical}), are proportional to the inverse
cube of the atom-surface distance. This is attributed to resonant
energy transfer to the absorbing dielectric surroundings which is a
nonradiative decay process
\cite{Wylie84,Yeung96,Juzeliunas97,Tomas97,Fleischhauer99}.
The next-to-leading terms, the induction terms, are proportional to
the inverse atom-surface distance and correspond to absorption of real
photons. The nonradiative decay rates make it impossible to hold atoms
or molecules near dielectric or metallic surfaces and use their
internal states for coherent manipulation. The correct
quantum-statistical description of the dielectric bodies also imply
that the suppression of spontaneous decay of a dipole parallel to a
mirror surface (see Sec.~\ref{sec:vacuumse}) cannot be observed.

As the near-field contributions are proportional to the imaginary part
of the permittivity, $\mathrm{Im}\,\varepsilon(\omega_A)$, these terms
will vanish if absorption can be disregarded at the atomic transition
frequency $\omega_A$. In this case, the spontaneous decay rate is
modified to \cite{Glauber91,Scheel99b}
\begin{equation}
\label{eq:glauberlewenstein}
\Gamma = \Gamma_0 \left(
\frac{3\varepsilon(\omega_A)}{2\varepsilon(\omega_A)+1} \right)^2
n(\omega_A) \,.
\end{equation}

The results for a small spherical cavity can also be used to study the
decay of atoms embedded inside a medium. In this real-cavity model,
the cavity implements the local-field correction arising due to the
difference between the macroscopic field and the local field
experienced by the atom. Using the local-field corrected Green tensor
from Appendix~\ref{sec:localfield} one can show that, for weakly
absorbing media, Eq.~(\ref{eq:glauberlewenstein}) remains valid for
arbitrary geometries when written in the form
\begin{equation}
\Gamma_\mathrm{loc} = \Gamma \left(
\frac{3\varepsilon(\omega_A)}{2\varepsilon(\omega_A)+1} \right)^2 \,,
\end{equation}
where $\Gamma$ is the uncorrected decay rate. For absorbing media, the
more general result
\begin{equation}
\Gamma = \Gamma_C +\frac{2\omega_A^2}{c^2\hbar\varepsilon_0} \vect{d}
\cdot \mathrm{Im}\left[ \left(
\frac{3\varepsilon(\omega_A)}{2\varepsilon(\omega_A)+1} \right)^2
\tens{G}^{(S)}(\vect{r}_A,\vect{r}_A,\omega_A) \right] \cdot
\vect{d}^\ast
\end{equation}
holds, where $\Gamma_C$ is the near-field rate
(\ref{eq:nearfieldspherical}) \cite{Dung06}.


\subsection{Sum rules for the local density of states}

One of the principal cornerstones of macroscopic QED
is the validity of Kramers--Kronig relations (\ref{eq:kk}) for the 
response functions such as the dielectric permittivity, the magnetic 
permeability or the generalised conductivity. Equally important are
sum rules that follow from asymptotic limits of these Hilbert
transforms. Examples for sum rules in optics are the optical theorem
that relates the imaginary part of the forward scattering amplitude to
the total scattering cross section \cite{BornWolf}, or the
Thomas--Reiche--Kuhn oscillator strength rule \cite{BetheSalpeter}. A
particularly relevant relation for our purposes is the integral
relation \cite{Altarelli72}
\begin{equation}
\label{eq:nsumrule}
\int\limits_0^\infty d\omega\, \left[ \eta(\omega)-1 \right] =0
\end{equation}
satisfied by the real part $\eta(\omega)$ of the complex index of
refraction $n(\omega)=\sqrt{\varepsilon(\omega)\mu(\omega)}
=\eta(\omega)+i\kappa(\omega)$. This follows from the superconvergence
theorem for Hilbert transform pairs \cite{Frye63}. It means that the
propagation properties of light in a dielectric medium are
redistributed in such a way that, averaged over the whole frequency
axis, the (real part of the) refractive index is the same as in
vacuum.

Because macroscopic QED is inherently based on linear
response theories, it is conceivable that quantities such as
spontaneous decay rates would obey certain sum rules, too. Recall
from  Sec.~\ref{sec:vacuumse} that the rate of spontaneous decay and
the Lamb shift form a Hilbert transform pair. We will now try to
derive a sum rule for the rate of spontaneous decay, and investigate
integrals of the form
\begin{equation}
\int\limits_0^\infty d\omega \frac{\Gamma-\Gamma_0}{\Gamma_0}
= \int\limits_0^\infty d\omega\, \frac{6\pi c}{\omega |\vect{d}|^2} \,
\vect{d}\cdot
\mathrm{Im}\,\tens{G}^{(S)}(\vect{r}_A,\vect{r}_A,\omega) \cdot
\vect{d}^\ast \,,
\end{equation}
where $\Gamma_0$ is the free-space decay rate (\ref{eq:vacuumse}) and
$\tens{G}^{(S)}(\vect{r}_A,\vect{r}_A,\omega)$ the scattering part of
the dyadic Green function. Hence, we are seeking to compute the
integral
\begin{equation}
\label{eq:sumruleintegral}
\int\limits_0^\infty d\omega \, \frac{c}{\omega} 
\mathrm{Im}\,\tens{G}^{(S)}(\vect{r}_A,\vect{r}_A,\omega)
= \mathrm{Im} \int\limits_{-\infty}^\infty d\omega \,
\frac{c}{2\omega}
\tens{G}^{(S)}(\vect{r}_A,\vect{r}_A,\omega)
\end{equation}
where we have used the Schwarz reflection principle
(\ref{eq:SchwarzG}).

For the sake of definiteness, we imagine the radiating atom being
placed inside a dielectric structure such as the spherical microcavity
sketched in Fig.~\ref{fig:spherical}. In fact, any generic situation
will involve an atom in free space surrounded by some dielectric
material, although not necessarily in a spherically symmetric way. The
scattering part of the dyadic Green function for any arrangement of
dielectric bodies can always be expanded into a Born series (see
Appendix~\ref{sec:bornseries}) as
\begin{eqnarray}
\tens{G}^{(S)}(\vect{r}_A,\vect{r}_A,\omega) &=& \frac{\omega^2}{c^2} 
\int_{V_\chi} d^3s'\,\chi(\vect{s}',\omega) 
\tens{G}^{(0)}(\vect{r}_A,\vect{s}',\omega) \cdot
\tens{G}^{(0)}(\vect{s}',\vect{r}_A,\omega) \nonumber \\ &&
\hspace*{-20ex} 
+\left( \frac{\omega^2}{c^2} \right)^2 \iint_{V_\chi} d^3s'd^3s''\,
\chi(\vect{s}',\omega) \chi(\vect{s}'',\omega) 
\tens{G}^{(0)}(\vect{r}_A,\vect{s}',\omega) \cdot
\tens{G}^{(0)}(\vect{s}',\vect{s}'',\omega) \cdot
\tens{G}^{(0)}(\vect{s}'',\vect{r}_A,\omega)
\nonumber \\ && \hspace*{-20ex} + \ldots
\end{eqnarray}
where $\tens{G}^{(0)}(\vect{r},\vect{r}',\omega)$ is the free-space
Green tensor and $\chi(\vect{r},\omega)$ the dielectric susceptibility
of the material surrounding the radiating atom. The integration
extends over the total volume of the body, but excludes the location
of the atom. The free-space Green tensor can be read off from 
Eqs.~(\ref{eq:longitudinalG}) and (\ref{eq:transverseG}), setting 
$q(\omega)=\omega/c$. The first thing to note is that all terms
containing the $\delta$ function involving the location $\vect{r}_A$
of the atom do not contribute to the Born series.

As a function of $\omega$, the free-space Green tensor has single and 
double poles at $\omega=0$ whose contributions to the integral 
(\ref{eq:sumruleintegral}) give rise to contributions that can be
computed as follows. Concentrating on the first term in the Born
series expansion, we need to look at contributions of the form
$f(\omega)=\sum_nc_n\chi(\vect{s}',\omega)e^{2i\omega\rho/c}/\omega^n$
with $n=1,2,3$ whose residues at $\omega=0$ are \cite{Scheel08}
\begin{equation}
\mathrm{Res}\,f(\omega)|_{\omega=0} 
= \sum\limits_{n=1}^3 \frac{c_n}{(n-1)!} \sum\limits_{m=0}^{n-1}
\frac{(2i)^m}{m!} \chi^{(n-1-m)}(\vect{s}',0) \,.
\end{equation}
These terms arise from a short-distance or, equivalently,
low-frequency expansion of the Green tensor. It is seen from
Eq.~(\ref{eq:longitudinalG}) that they can be traced back to the
longitudinal part of the free-space Green tensor. Hence, all these
pole contributions, in any order of the Born series, arise from either
purely longitudinal terms 
$\propto\omega^{2j}\tens{G}^{(0)\|}(\vect{r}_A,\vect{s}',
\omega)\cdots$
$\tens{G}^{(0)\|}(\vect{s}^{(j+1)},\vect{r}_A,\omega)$ or from those
in which one (and only one) of the longitudinal Green tensors has been
replaced by a transverse part 
$\tens{G}^{(0)\perp}(\vect{s}^{(i)},\vect{s}^{(i+1)},\omega)$ 
\cite{Scheel08}. Subtracting these terms from the total scattering
Green tensor,
\begin{eqnarray}
\label{eq:modifiedG}
\tens{G}^{(S)'}(\vect{r}_A,\vect{r}_A,\omega) &=&
\tens{G}^{(S)}(\vect{r}_A,\vect{r}_A,\omega) 
\nonumber \\ && \hspace*{-10ex}
-\frac{\omega^2}{c^2} \int_{V_\chi} d^3s'\,\chi(\vect{s}',\omega) 
\tens{G}^{(0)\|}(\vect{r}_A,\vect{s}',\omega) \cdot
\tens{G}^{(0)\|}(\vect{s}',\vect{r}_A,\omega) 
\nonumber \\ && \hspace*{-10ex}
-\frac{\omega^2}{c^2} \int_{V_\chi} d^3s'\,\chi(\vect{s}',\omega) 
\tens{G}^{(0)\|}(\vect{r}_A,\vect{s}',\omega) \cdot
\tens{G}^{(0)\perp}(\vect{s}',\vect{r}_A,\omega) 
\nonumber \\ && \hspace*{-10ex}
-\frac{\omega^2}{c^2} \int_{V_\chi} d^3s'\,\chi(\vect{s}',\omega) 
\tens{G}^{(0)\perp}(\vect{r}_A,\vect{s}',\omega) \cdot
\tens{G}^{(0)\|}(\vect{s}',\vect{r}_A,\omega) 
\nonumber \\ && \hspace*{-10ex}
-\left( \frac{\omega^2}{c^2} \right)^2 \iint_{V_\chi} d^3s'd^3s'' 
\cdots \,,
\end{eqnarray}
and noting that all other contributions containing non-negative powers
of $\omega$ vanish after contour integration, it becomes obvious that
\begin{equation}
\int\limits_0^\infty d\omega\, \frac{c}{\omega} \,\mathrm{Im}\,
\tens{G}^{(S)'}(\vect{r}_A,\vect{r}_A,\omega) =0 \,.
\end{equation}
This in turn means that we can define a modified spontaneous decay
rate $\Gamma'$ which is constructed from 
$\tens{G}^{(S)'}(\vect{r}_A,\vect{r}_A,\omega)$, that obeys the sum
rule
\begin{equation}
\label{eq:modifiedsumrule}
\int\limits_0^\infty d\omega \frac{\Gamma'-\Gamma_0}{\Gamma_0} =0\,.
\end{equation}
From its construction, it is clear that $\Gamma'$ excludes
dipole-dipole interactions. Because nonradiative contributions to the
decay rate have been subtracted from $\Gamma$, we can also interpret
the sum rule (\ref{eq:modifiedsumrule}) as a conservation of the
integrated local density of states associated with photonic final
states. Recalling our discussion following Eq.~(\ref{eq:nsumrule})
this means that a magnetoelectric medium merely redistributes the
photonic density of states across the frequency axis, but remain at
the free-space level on average.

Finally, we make the connection to local-field corrections. The
local-field corrected single-point Green tensor, obtained after
discarding the terms containing $R_\mathrm{cav}$ in
Eq.~(\ref{eq:G1loc}), is one example of a modified Green tensor
(\ref{eq:modifiedG}) that leads to a valid sum rule for spontaneous
decay rates. In this way, a connection is established to the theory in
Ref.~\cite{Barnett96} which is strictly valid only in nonabsorbing
materials. 
%
%
\subsection{Heating of polar molecules}
\label{sec:heating}
In a previous section we argued that at very small atom-body
distances, nonradiative decay will dominate. Given that this would
require the atom or molecule to be held at distances much smaller than
the (optical) wavelength, this effect will hardly be seen in a
controlled experiment. The situation is changed dramatically when the
transition wavelength in question is very large, say longer than a
centimetre. This is the case for polar molecules since the spacings
between neighbouring rotational and vibrational levels are relatively
large. Molecules whose projection $\Lambda$ of the total orbital
angular momentum $\hat{\vect{L}}$ vanishes ($\Lambda=0$) are best
described by Hund's coupling case (b) \cite{0813}. In this scheme, the
molecular eigenstates $|S,N,J,M\rangle|v\rangle$ are characterised by
the quantum numbers $J$ and $M$ of the total angular momentum and its
projection on the space-fixed $z$-axis, the total spin quantum number
$S$ ($\hat{\vect{S}}$), the rotational quantum number $N$
($\hat{\vect{N}}=\hat{\vect{J}}-\hat{\vect{S}}$) and the
vibrational quantum number $v$. For deeply bound states, the
rotational and vibrational eigenenergies are \cite{0813}
\begin{equation}
\label{eq:Erot}
E_N=h B_\mathrm{e} N(N+1),\, N=0,1,\ldots\qquad
E_v=h\omega_\mathrm{e}
 \bigl(v+{\textstyle\frac{1}{2}}\bigr),\, v=0,1,\ldots
\end{equation}
where $B_\mathrm{e}$ and $\omega_\mathrm{e}$ are the rotational and 
vibrational constants, typical values of which are listed in
Tab.~\ref{tab:heating}. 
\begin{table}[!t!]
\begin{center}
\begin{tabular}{ccccccc}
\hline
Spec.
&$B_\mathrm{e}(\mathrm{GHz}$)
&$\omega_\mathrm{e}(\mathrm{THz})$
&$\mu_\mathrm{e}(10^{-30}\mathrm{Cm})$
&$\mu'_\mathrm{e}(10^{-21}\mathrm{C})$
&$\tau_\mathrm{r}(\mathrm{s})$&$\tau_\mathrm{v}(\mathrm{s})$\\
\hline
LiH&$222$
&$42.1$
&$19.6$
&$60.5$
&$2.1$&$25$\\
CaF&$10.5$
&$18.4$
&$10.2$
&$172$
&$3,\!400$&$4.7$\\
BaF&$6.30$
&$14.1$
&$11.7$
&$285$
&$7,\!200$&$1.8$\\
YbF&$7.20$
&$15.2$
&$13.1$
&$195$
&$4,\!400$&$4.1$\\
LiRb&$6.60$
&$5.55$
&$13.5$
&$21.4$
&$4,\!900$&$128$\\
NaRb&$2.03$
&$3.21$
&$11.7$
&$12.6$
&$70,\!000$&$1,\!400$\\
KRb&$1.15$
&$2.26$
&$0.667$
&$1.89$
&$6.7\!\times\!10^{7}$&$120,\!000$\\
LiCs&$5.80$
&$4.92$
&$21.0$
&$28.4$
&$2,\!600$&$80$\\
NaCs&$17.7$
&$2.94$
&$19.5$
&$21.4$
&$330$&$580$\\
KCs&$92.8$
&$1.98$
&$8.61$
&$6.93$
&$62$&$12,\!000$\\
RbCs&$0.498$
&$1.48$
&$7.97$
&$4.41$
&$2.5\!\times\!10^6$&$63,\!000$\\
\end{tabular}
\end{center}
\caption{
\label{tab:heating}
Properties of various diatomic radicals (electronic ground state,
rotation and vibration constants, dipole moment and its derivative at
equilibrium bond length, reduced mass) and life times
$\tau_\mathrm{r}$ and $\tau_\mathrm{v}$ of their rovibrational ground
states against rotational and vibrational heating at room
temperature ($T=293\mathrm{K}$) in free space \cite{Heating}. 
}
\end{table}

Even at room temperature, the first few rotationally and
vibrationally excited states may be considerably populated due to
heating out of the ground state, placing severe limits on the
coherent manipulation of polar molecules in their ground states.
According to Eqs.~(\ref{eq:thermalgamma}), the respective heating
rate is given by
\begin{equation}
\label{eq:gsheating}
\Gamma=\frac{2\mu_0}{\hbar}\sum_k\omega_{k0}^2
\bar{n}_\mathrm{th}(\omega_{k0})
\vect{d}_{0k}\sprod\operatorname{Im}
 \ten{G}(\vect{r}_A,\vect{r}_A,\omega_{k0})
 \sprod\vect{d}_{k0},
\end{equation}
where frequency shifts [Eqs.~(\ref{eq:zeropointshift}) and
(\ref{eq:thermalshift})] can typically be neglected. For a molecule in
free space, use of the Green tensor~(\ref{eq:freespaceG}) leads to
\cite{Heating,0772}
\begin{equation}
\label{eq:gsheatingfree}
\Gamma_0=\sum_k\Gamma_{0k}
=\sum_k\frac{\omega_{0k}^3|\vect{d}_{0k}|^2}
 {3\pi\hbar\varepsilon_0c^3}\,
 \bar{n}_\mathrm{th}(\omega_{k0}).
\end{equation}
In order to evaluate these rates, the relevant dipole matrix elements
need to be determined. For rotational transitions, this can best be
done using Hund's case (a) basis
$|S,\Lambda,\Sigma,\Omega,J,M\rangle$, where $\Lambda$,
$\Sigma$ and $\Omega$ denote the projections of $\hat{\vect{L}}$,
$\hat{\vect{S}}$ and $\hat{\vect{J}}$ onto the internuclear axis
($\Omega =\Lambda +\Sigma$). In this basis, one has \cite{0813}
\begin{eqnarray}
\label{eq:dipoleMe1}
\vect{d}_{mn}&=&\langle\Omega JM|\hat{\vect{d}}|\Omega'J'M'\rangle
 =\mu_\mathrm{e}\langle \Omega JM|\hat{\vect{u}}|\Omega'J'M'\rangle
 \nonumber\\
&=&\mu_\mathrm{e}\biggl[(u_{mn}^{-1}-u_{mn}^{+1})
 \frac{\vect{e}_{x}}{\sqrt{2}}
 +(u_{mn}^{-1}+u_{mn}^{+1})\frac{\mi\,\vect{e}_{y}}{\sqrt{2}}
 +u_{mn}^{\,0}\vect{e}_{z}\biggr],
\end{eqnarray}
where $\mu_\mathrm{e}$ is the molecular dipole moment at the
equilibrium internuclear separation,
$\hat{\vect{u}}=\hat{\vect{r}}/|\hat{\vect{r}}|$ and
\begin{equation}
\label{eq:dipoleMe2}
u_{mn}^{q}=(-1)^{M-\Omega}\sqrt{(2J+1)(2J'+1)}
 \begin{pmatrix}
  J & 1 & J' \\ -M & q & M'
 \end{pmatrix}
 \begin{pmatrix}
 J & 1 & J' \\ -\Omega  & 0 & \Omega'
 \end{pmatrix}.
\end{equation}
Relating the two bases via
\begin{equation}
\label{eq:expansion}
|S,N,J,M\rangle =\sum_{\Omega =-S}^S(-1)^{J-S}\sqrt{2N+1}
\begin{pmatrix}J&S&N\\ \Omega&-\Omega&0\end{pmatrix}
|\Omega,J,M\rangle
\end{equation}
one finds 
\begin{equation}
\label{eq:drotation}
\sum_{k}\vect{d}_{0k}\tprod\vect{d}_{k0}
=\tfrac{1}{3}\mu_\mathrm{e}^{2}\ten{I}
\end{equation}
for the molecules under consideration. For (ro-)vibrational heating,
the dipole matrix elements follow from \cite{0813}
\begin{gather}
\label{eq:dvibration1}
\langle v\,\Omega JM|\hat{\vect{d}}|v'\,\Omega'J M'\rangle
 =\mu_\mathrm{e}'\langle\Omega JM|\hat{\vect{u}}|\Omega'J'M'\rangle
 \langle v|\hat{q}|v'\rangle,\\
\label{eq:dvibration2}
\langle v=1|\hat{q}|v'=0\rangle=\sqrt{\frac{\hbar}{4\pi
m\omega_\mathrm{e}}}
\end{gather}
($\mu_\mathrm{e}'$: derivative of the dipole moment at equilibrium
bond length, $m$: reduced mass) to be
\begin{equation}
\label{eq:dvibration}
\sum_{k}\vect{d}_{0k}\tprod\vect{d}_{k0}
=\frac{\hbar\mu_\mathrm{e}'^{2}}{12\pi m\omega_\mathrm{e}}\,\ten{I}.
\end{equation}

Combining the above results, one can calculate life times
$\tau_0=\Gamma_0^{-1}$ of polar molecules in free space against
rotational and vibrational heating out of the ground state at room
temperature ($T=293\mathrm{K}$) which are listed in
Tab.~\ref{tab:heating}. Since the rotational transition frequencies
lie typically well below the maximum of the thermal spectrum
($17\mathrm{THz}$ at $T=293\mathrm{K}$), rotational heating mostly
affects light molecules like LiH whose transition frequency is
largest. For this molecule, rotational heating severely limits the
life time of the ground state to about 2 seconds. Vibrational heating,
on the contrary, mostly affects the fluorides whose vibrational
transition frequencies are very close to the peak of the thermal
spectrum; associated lifetimes lie in the range of a few seconds only.
The rotational and vibrational heating rates exhibit strong
temperature-dependences via the thermal photon number, so the impact
of both heating channels can be considerably reduced by lowering the
environment temperature. 

Just like spontaneous decay, heating can be considerably enhanced
when molecules are placed close to surfaces. Using the decomposition
$\ten{G}=\ten{G}^{(0)}+\ten{G}^{(S)}$ of the Green tensor into its
free-space part and the scattering part accounting for the presence of
the surface and recalling Eqs.~(\ref{eq:drotation}) and
(\ref{eq:dvibration}), the ground state heating
rate~(\ref{eq:gsheatingfree}) can be written as \cite{Heating}
\begin{eqnarray}
\label{eq:gsheatingplate}
\Gamma(z_A)=\Gamma_0\left[1+\frac{2\pi
c}{\omega_{k0}}\operatorname{Im}\trace
\ten{G}^{(S)}(\vect{r}_A,\vect{r}_A,\omega_{k0})\right]
\end{eqnarray}
($z_A$: atom-surface separation), where $\ten{G}^{(S)}$ is given in
App.~\ref{sec:layeredmedia}. Note that due to our neglect of the
frequency shifts, Eqs.~(\ref{eq:zeropointshift}) and
(\ref{eq:thermalshift}), the heating rate separates into the
temperature-dependent factor $\Gamma_0$ as given by
Eq.~(\ref{eq:gsheatingfree}) and a purely position-dependent part. 

\begin{figure}[!t!]
\begin{center}
\includegraphics[width=0.7\linewidth]{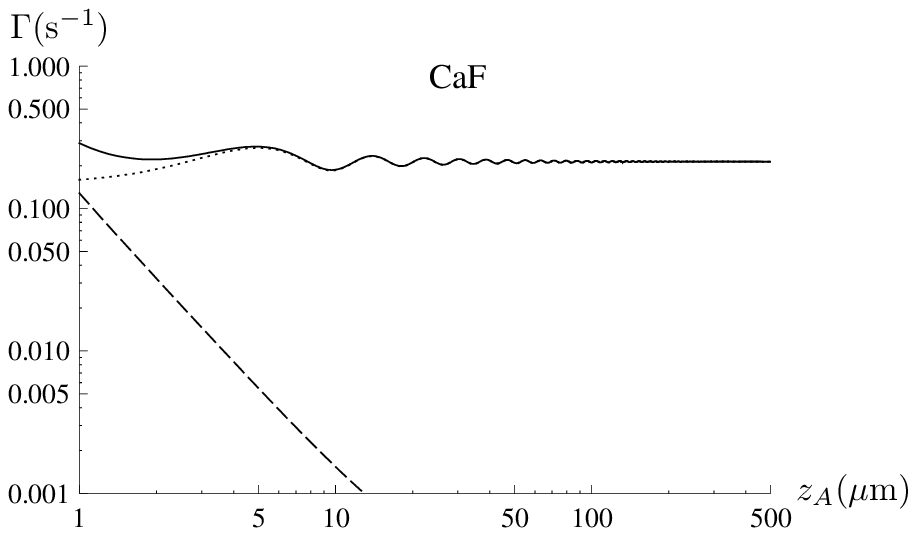}\\[4ex]
\includegraphics[width=0.7\linewidth]{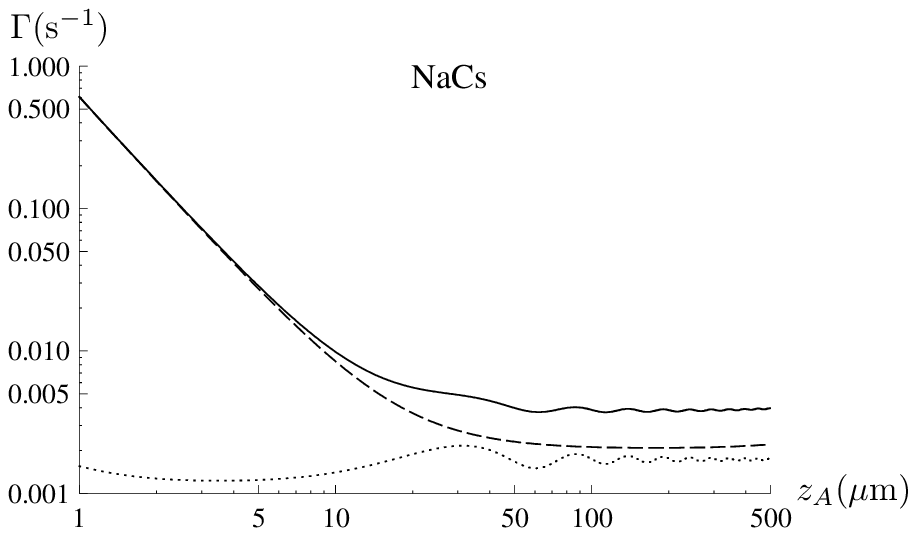}
\end{center}
\caption{
\label{fig:heating}
Heating rates for CaF and NaCs as a function of distance from a gold
surface ($\omega_P=1.37\!\times\!10^{16}\mathrm{rad}/\mathrm{s}$,
$\gamma=4.12\!\times\!10^{13}\mathrm{rad}/\mathrm{s}$). Solid lines:
total heating rate. Dotted lines: vibrational excitation rate. Dashed
lines: rotational excitation rate \cite{Heating}.}
\end{figure}%

In the nonretarded limit,
$z_A|\sqrt{\varepsilon(\omega_{nk})}|\omega_{nk}/c\ll 1$, the
position-dependence is approximately given by \cite{Heating}
\begin{equation}
\label{eq:gsheatingnret}
\Gamma(z_A)=\Gamma_0\biggl(1+\frac{z_\mathrm{nr}^{3}}{z_{\!A}^3}
\biggr),\qquad 
z_\mathrm{nr}=\frac{c}{\omega_{k0}}\,
\sqrt[3]{\frac{\operatorname{Im}\varepsilon(\omega_{k0})}
 {2|\varepsilon(\omega_{k0})+1|^2}}.
\end{equation}
For metals with Drude permittivity
\begin{equation}
\label{eq:epsilonmetal}
\varepsilon(\omega)=1-\frac{\omega_P^2}
 {\omega(\omega+\mi\gamma)}
\end{equation}
and for sufficiently small transition frequencies
$\omega_{k0}\ll\gamma\le\omega_P$, one has
\begin{equation}
\label{eq:znr}
z_\mathrm{nr}=c\,
\sqrt[3]{\frac{\gamma}{2\omega_P^2\omega_{k0}^2}}
\end{equation}
and the condition
$z_A|\sqrt{\varepsilon(\omega_{nk})}|\omega_{nk}/c\ll 1$ is not even
valid for very small atom-surface separations. Instead, the heating
rate is well approximated by the empirical formula \cite{Heating}
\begin{equation}
\label{eq:Mike}
\Gamma(z_A)=\Gamma_0\biggl(1+\frac{z_\mathrm{c}^2}{z_{\!A}^2}
 +\frac{z_\mathrm{nr}^{3}}{z_{\!A}^3}\biggr),\qquad
z_\mathrm{c}\simeq\frac{3c}{4}\,
\sqrt[4]{\frac{\gamma}{2\omega_P^2\omega_{k0}^3}}
\end{equation}
which is valid for distances $z_A\le z_\mathrm{c}$ ($z_\mathrm{c}$
being the critical distance for which surface-induced heating becomes
comparable to free-space heating).

In the opposite retarded limit, $z_A\omega_{nk}/c\gg 1$, the
ground-state heating rate is approximately given by \cite{Heating}
\begin{eqnarray}
\label{eq:gsheatingret}
\Gamma(z_A)&=&\Gamma_0
 \Biggl[1+\frac{c}{2z_A\omega_{k0}}\,
 \operatorname{Im}\Biggl(
 \frac{1-\sqrt{\varepsilon(\omega_{k0})}}
 {1+\sqrt{\varepsilon(\omega_{k0})}}\,
 e^{2\mi z_A\omega_{k0}/c}\Biggr)\Biggr]\nonumber\\
&=&\Gamma_0\biggl[1-\frac{c}{2z_A\omega_{k0}}\,
 \sin\biggl(\frac{2z_A\omega_{nk}}{c}\biggr)\biggr]
\end{eqnarray}
where the second equality holds for good conductors. The
distance-dependence is thus governed by attenuated oscillations away
from the surface where the oscillation period equals twice the
molecular transition wavelength. 

The spatial dependence of the molecular heating rates over the entire
distance rates is shown in Fig.~\ref{fig:heating} for CaF and NaCs. It
is seen that vibrational heating dominates for CaF and results in
rapid oscillations of the heating rate as a function of distance. For
short distances, rotational heating strongly increases and begins to
contribute to the total heating rate. For NaCs, rotational heating
slightly dominates for moderate distances, although the oscillations
associated with vibrational heating are still manifest in the total
heating rate. At distance smaller than about $10\mu\mathrm{m}$,
rotational heating becomes strongly dominant. The results show that
molecular heating can strongly increase in close proximity to
surfaces, thus placing severe limits on the miniaturisation of
molecular traps.


\subsection{Spin-flip rates}

\begin{figure}[!t!]
\includegraphics[height=4cm]{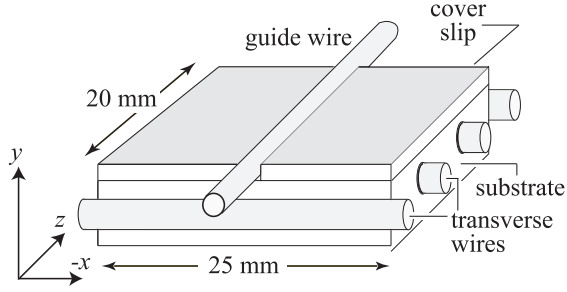}
\hfill
\includegraphics[height=4cm]{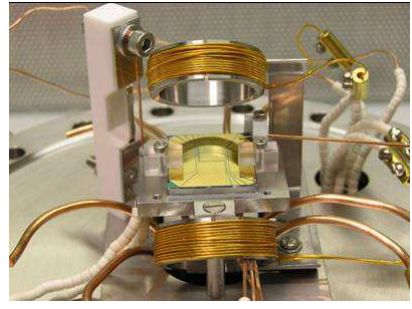}
\caption{\label{fig:wire} Schematic set-up of a wire trap creating a 
confining trapping potential for low-field seeking atoms (left
figure) (picture taken from Ref.~\cite{Jones03}). Typical
experimental set-up using a $Z$-shaped wire and a reflective gold
surface (right figure) [picture courtesy of E.A.~Hinds].}
\end{figure}

In cold-atom physics, where microengineered magnetoelectric or
metallic structures are designed to magnetically trap ultracold atoms
(e.g. ${}^{87}$Rb) in a well-defined Zeeman sublevel of their
respective hyperfine ground states (cf.~Fig.~\ref{fig:wire}), typical
transition frequencies range from $100\,\mbox{kHz}\ldots
10\,\mbox{MHz}$ (for reviews, see
\cite{HindsHughes,Folman,FortaghZimmermann}). 
%
%
The equivalent free-space wavelengths are thus on the order of
several metres. As typical atom-surface distances are in the
sub-millimetre range, the atoms are in the deep near field of the
relevant (magnetic) transitions.

\begin{figure}[!t!]
\begin{minipage}{5cm}
\vspace{0pt}
\includegraphics[height=5cm]{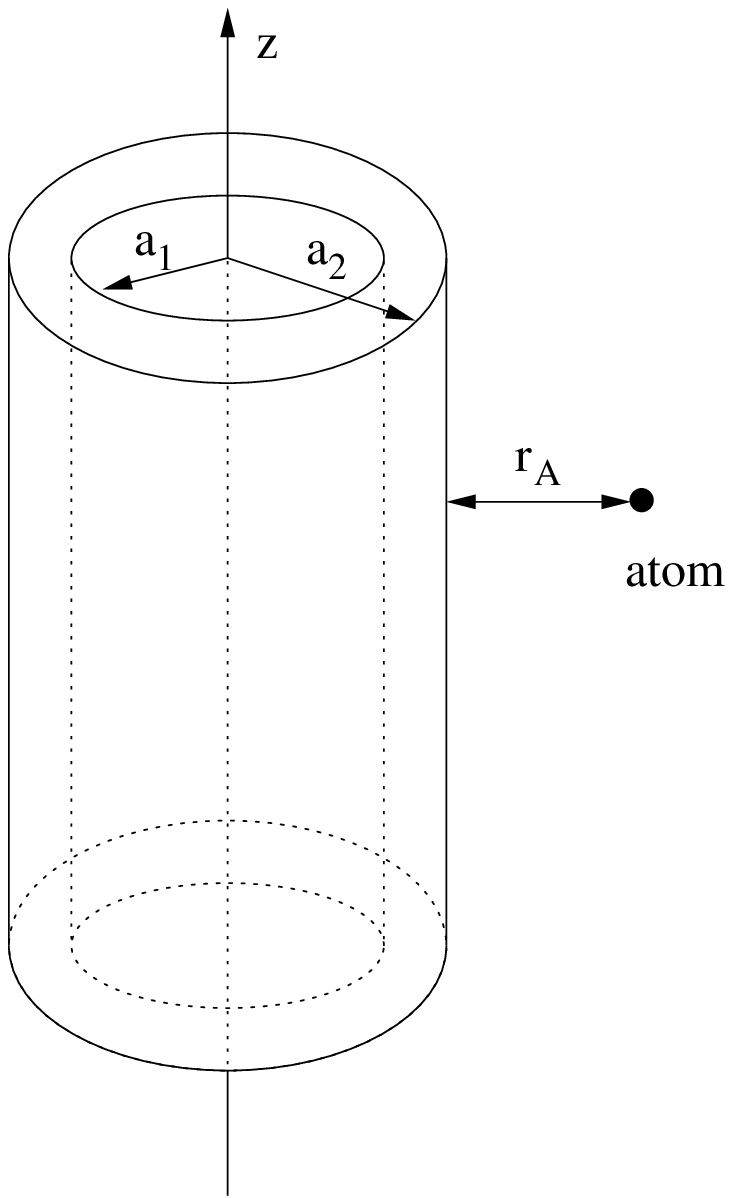}
\end{minipage}
\hspace*{1cm}
\begin{minipage}{5cm}
\includegraphics[width=5cm,angle=-90]{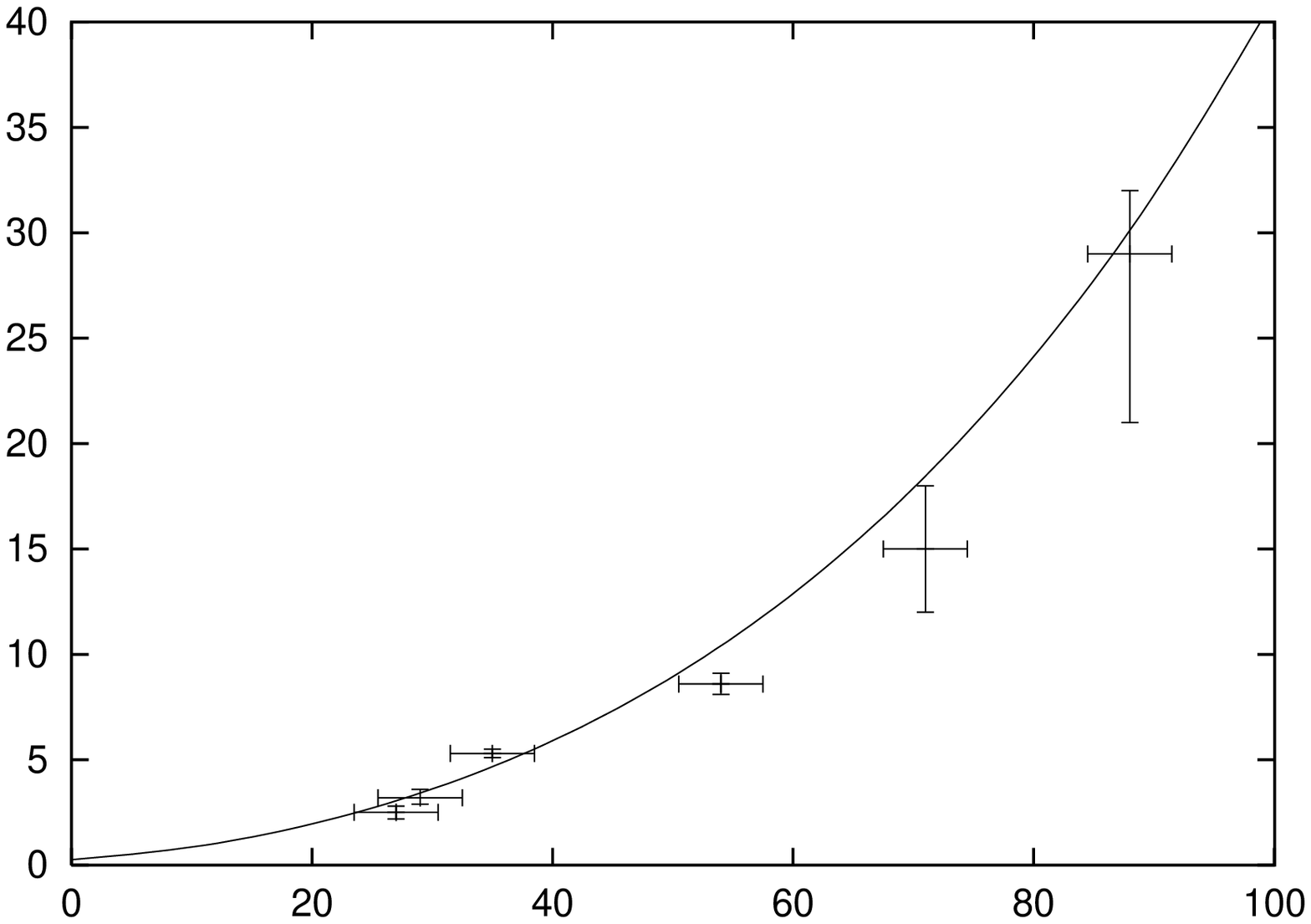}
\begin{picture}(100,10)(0,0)
\put(90,5){$z(\mu\mbox{m})$}
\put(-20,90){$\tau(\mbox{s})$}
\end{picture}
\end{minipage}
\caption{\label{fig:lifetime} Trapping lifetime $\tau=1/\Gamma$ as a
function of the atom-surface distance $z$ (right figure). Experimental
data taken from Ref.~\cite{Jones03}, theoretical curve taken from
Ref.~\cite{Rekdal04}. The geometry is depicted on the left. The wire
consisted of a Cu core (inner radius $a_1=185\,\mu\mbox{m}$) and a
$55\,\mu\mbox{m}$ thick Al cladding.} 
\end{figure}

In complete analogy to electric-dipole transitions, we can
investigate magnetic-dipole transitions using the approximate
interaction Hamiltonian (\ref{eq:multipolarHAF})
\begin{equation}
\hat{H}_{AF} = -\hat{\vect{m}} \cdot \hat{\vect{B}}(\vect{r}_A) \,.
\end{equation}
From the statistical properties of the electromagnetic field and our
previously used arguments regarding Fermi's Golden Rule we already
know that the spin transition rate is proportional to the strength of
the (thermal) magnetic field fluctuations,
$\Gamma\propto\langle\hat{\vect{B}}(\vect{r}_A,\omega)\otimes$
$\!\hat{\vect{B}}^\dagger(\vect{r}_A,\omega_A)\rangle_T$. If we assume
an atom to be in its electronic ground state, the magnetic moment
vector $\hat{\vect{m}}$ is proportional to the electronic spin
operator $\hat{\vect{S}}$ (the nuclear spin operator $\hat{\vect{I}}$
is smaller by a factor of $m_e/m_p$, i.e. the ratio between electron
and proton mass; we also assume that the ground state has $L=0$). It
can then be shown that the transition rate between two magnetic
sublevels $|i\rangle$ and $|f\rangle$ is \cite{Rekdal04}
\begin{eqnarray}
\label{eq:spinflip}
\Gamma &=& \frac{2\mu_0}{\hbar} \vect{m} \cdot \mathrm{Im}\left[
\curl\tens{G}(\vect{r}_A,\vect{r}_A,\omega_A)\times 
\overleftarrow{\bm{\nabla}}'\right]\cdot \vect{m}^\ast
\left[ \bar{n}_\mathrm{th}(\omega_A)+1 \right]
 \\ &=&
\frac{2\mu_0(\mu_Bg_S)^2}{\hbar} \langle f|\hat{\vect{S}}|i\rangle 
\cdot \mathrm{Im}\left[
\curl\tens{G}(\vect{r}_A,\vect{r}_A,\omega_A)\times
\overleftarrow{\bm{\nabla}}'\right]\cdot \langle
i|\hat{\vect{S}}|f\rangle
\left[ \bar{n}_\mathrm{th}(\omega_A)+1 \right] \nonumber
\end{eqnarray}
[$\vect{m}=\langle f|\hat{\vect{m}}|i\rangle$, $\mu_B$: Bohr magneton,
$g_S\approx 2$]. Note that this rate corresponds to the rates
associated with electric transitions as given by
Eqs.~(\ref{eq:zeropointgamma}), (\ref{eq:thermalgamma}) by means of a
duality transformation $\vec{d}\leftrightarrow\vec{m}/c$,
$\varepsilon\leftrightarrow\mu$, cf.\ the transformation properties
(\ref{dgftrans1}) and (\ref{dgftrans2}) of the Green tensor given in
App.~\ref{sec:dualgreen}. An experiment using ${}^{87}$Rb atoms in
their $|F=2,m_F=2\rangle$ hyperfine ground state has revealed spin
flip lifetimes on the order of seconds for distances between
$20\ldots 100\mu\mbox{m}$ \cite{Jones03}. Figure~\ref{fig:lifetime}
shows the experimental data together with the theoretical predictions
according to Eq.~(\ref{eq:spinflip}) where the Green function for a
three-layered cylindrical medium was used \cite{Rekdal04} (see also
App.~\ref{sec:cylindricaldgf}). 
%

It is instructive to investigate certain asymptotic regimes in which
analytical approximations to the expected spin flip lifetime can be
given. For this purpose, we consider an atom at a distance $d$ away
from a planar metallic surface with skin depth $\delta$ and with
thickness $h$. Its Green tensor can be found in
App.~\ref{sec:planardgf}. The skin depth $\delta$ is related to
the dielectric permittivity by
$\varepsilon(\omega)=2ic^2/(\omega^2\delta^2)$. Together with the
transition wavelength $\lambda$, there are four different length
scales involved, of which $\lambda$ is by far the longest and can be
taken to be infinite. From the remaining three length scales one can
find three experimentally relevant extreme cases that can be
summarised in the following expression \cite{Henkel99,Scheel05}:
\begin{equation}
\label{eq:planarscaling}
\frac{\tau}{\tau_0} = \left( \frac{8}{3} \right)^2 
\frac{1}{\bar{n}_\mathrm{th}+1} \left( \frac{\omega}{c} \right)^3
\left\{ \begin{array}{ll}
\displaystyle\frac{d^4}{3\delta} \,, & \delta\ll d,h\,,\\[5pt]
\displaystyle\frac{\delta^2d}{2} \,, & \delta,h\gg d\,,\\[5pt]
\displaystyle\frac{\delta^2d^2}{2h} \,, & \delta\gg d\gg h\,.
\end{array} \right.
\end{equation}
Here, $\tau_0=1/\Gamma_0$ is the trapping lifetime in free space
which, for a transition frequency $\omega_A=2\pi\,400$kHz, amounts to
$3\cdot 10^{25}$s \cite{Purcell46}. For fixed values of ($d,h$)
there are two distinct functional dependencies on the skin depth
$\delta$. These two scaling laws can be supported by rather intuitive
explanations. If the skin depth is larger than the remaining length
scales (i.e. the material is more dielectric), the magnetic field
fluctuations weaken. As their strength is proportional to the
imaginary part of the permittivity and hence the squared inverse skin
depth, the lifetime increases quadratically with the skin depth
$\delta$ [second and third lines of Eq.~(\ref{eq:planarscaling})]. If,
on the other hand, the skin depth is the smallest length scale, the
effective volume diminishes from which fluctuations emanate which
leads to a lifetime increase with decreasing skin depth [first line in
Eq.~(\ref{eq:planarscaling})]. 
%
\begin{figure}[!t!]
\begin{minipage}{6cm}
\includegraphics[height=6cm,angle=-90]{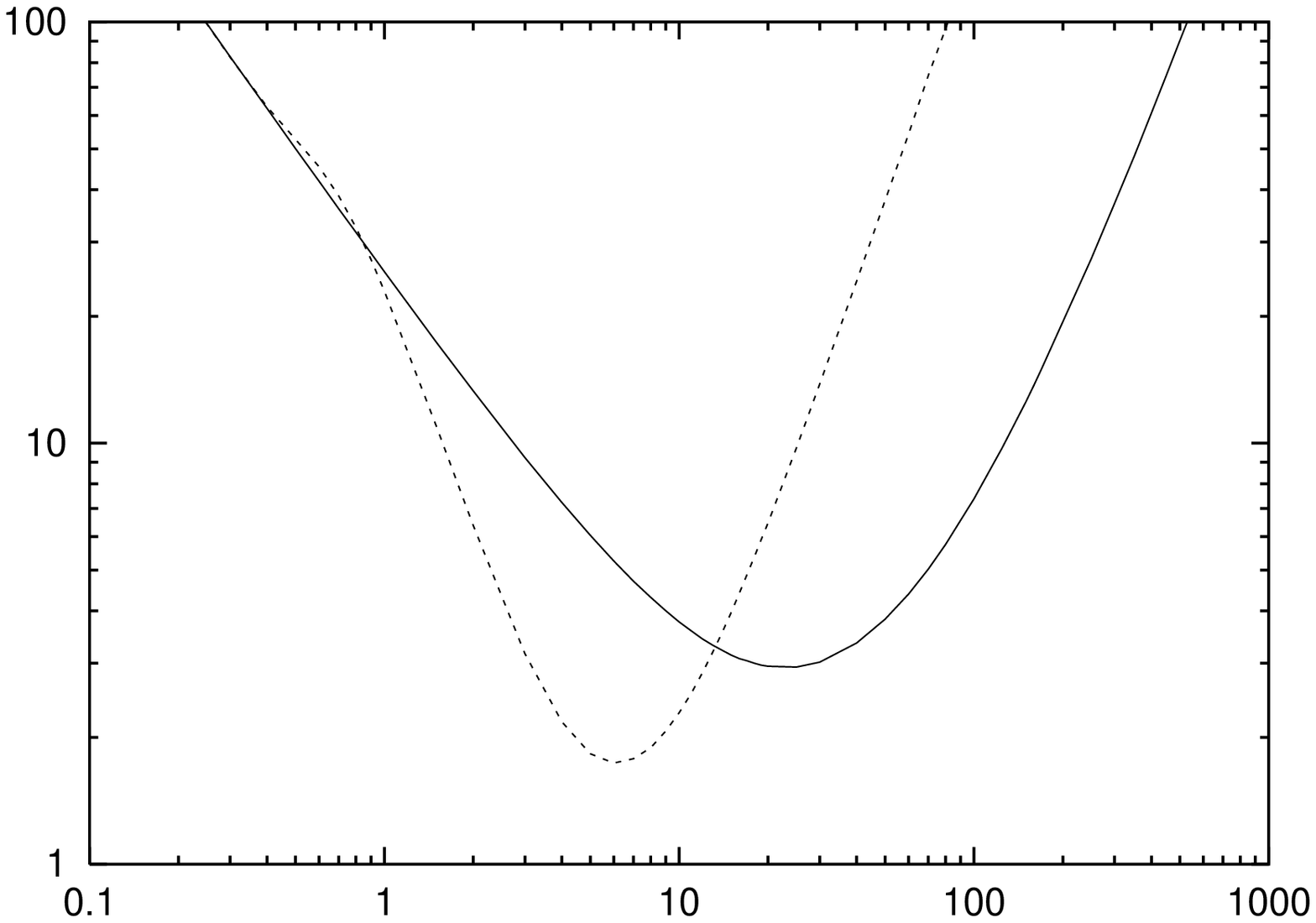}
\begin{picture}(100,10)(0,0)
\put(80,5){$\delta(\mu\mbox{m})$}
\put(-12,70){$\tau(\mbox{s})$}
\end{picture}
\end{minipage}
\hfill
\begin{minipage}{6.5cm}
\includegraphics[width=6.5cm]{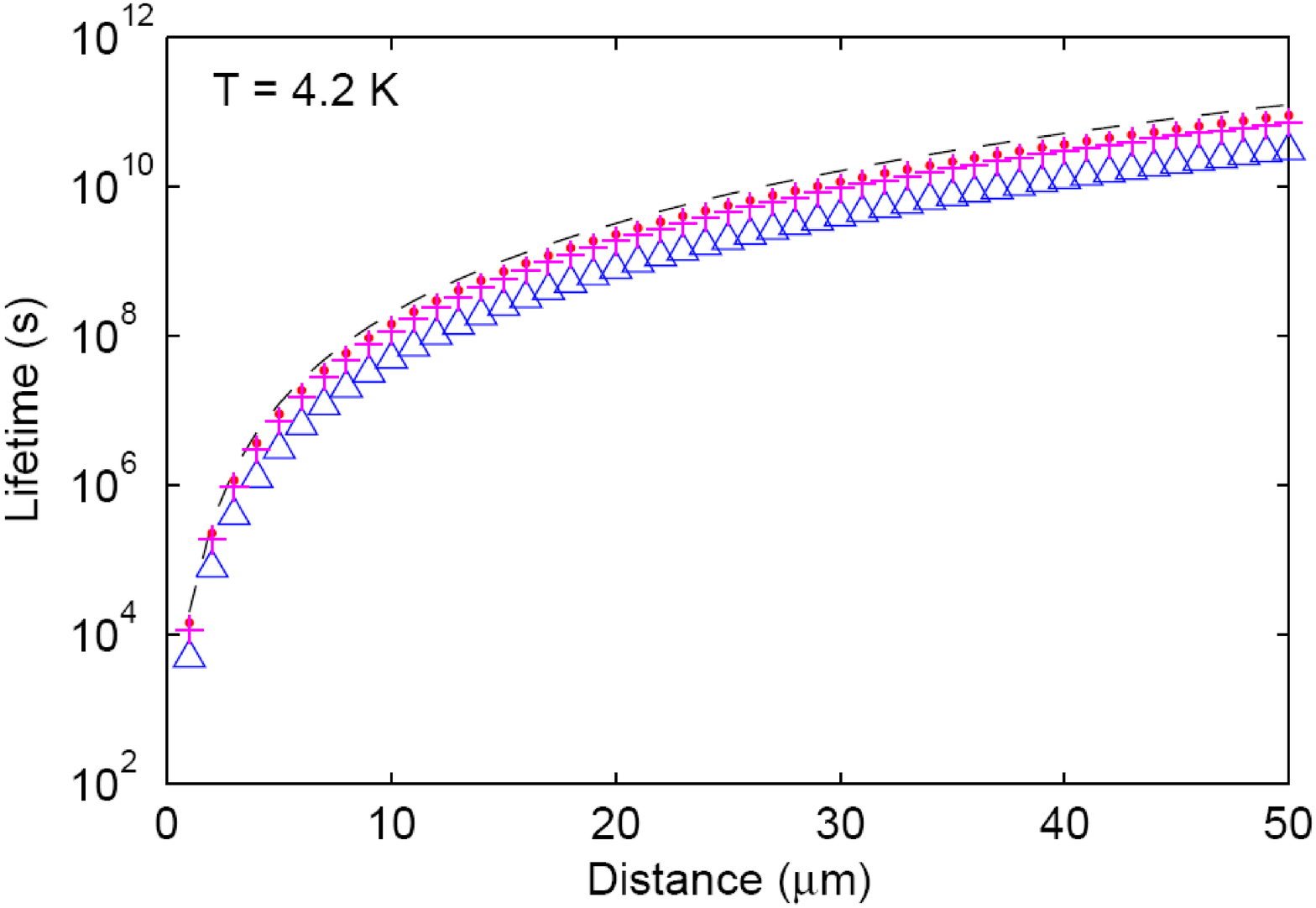}
\end{minipage}
\caption{\label{fig:skindepth} Left panel: Trapping lifetime as a
function of skin depth for a thin substrate layer ($h=1\mu\mbox{m}$,
dashed line) and a half-space (solid line). The atom-surface distance
was chosen as $d=50\mu\mbox{m}$, the transition frequency is
$\omega_A=2\pi\,560\mbox{kHz}$ \cite{Scheel05}. Right panel: Spin flip
lifetime $\tau$ near a superconducting Nb slab ($T_c=9.2$K) as a
function of atom-surface distance. Dashed line: two-fluid model;
symbols: Eliashberg theory with various elastic scattering rates
\cite{Hohenester07}.}
\end{figure}
%
This in turn means that there is a pronounced lifetime minimum when
the skin depth is on the order of the atom-surface distance
(Fig.~\ref{fig:skindepth}). To either side of this minimum, the
expected lifetime increases drastically. However, choosing a surface
material with larger skin depth limits its capability to generate
strong enough magnetic traps. On the other hand, natural materials
with skin depths below $50\,\mu\mbox{m}$ at room temperature and
$1\,\mbox{MHz}$ are impossible to find.

A possible solution is to make use of superconductors on the
assumption that due to the supercurrent magnetic field fluctuations
are shielded away from the vacuum-superconductor interface, and hence
spin transitions are suppressed. For a bulk superconductor with
conductivity
$\sigma(\omega)=\sigma'(\omega)+i\sigma''(\omega)$ and
$\sigma''(\omega)\ll\sigma'(\omega)$, we can rewrite the first line of
Eq.~(\ref{eq:planarscaling}) as \cite{Hohenester07}
\begin{equation}
\frac{\tau}{\tau_0} = \left( \frac{8}{3} \right)^2
\frac{(\omega\mu_0)^{1/2}d^4}{\bar{n}_\mathrm{th}+1}
\left( \frac{\omega}{c} \right)^3
\frac{[\sigma''(\omega)]^{3/2}}{\sigma'(\omega)} \,.
\end{equation}

In the London two-fluid model \cite{London35} one assumes that the
two types of charge carriers, \textit{n}ormal and
\textit{s}uperconducting, react to an external field according to
Ohm's law $\mathbf{j}_n=\sigma_n\mathbf{E}$ and the London relation
$\Lambda\frac{\partial\mathbf{j}_s}{\partial t}=\mathbf{E}$, 
respectively. As a function of temperature, the fraction of normal
conducting electrons follows the Gorter--Casimir expression
$n_n(T)/n_0=(T/T_c)^4$ \cite{Gorter34}. With the plasma frequency
$\omega_P$ and the elastic scattering rate $\gamma$ of the electrons,
the conductivity can be written in the form 
\begin{equation}
\sigma(\omega) = \varepsilon_0\omega_P^2 \left\{ \frac{1}{\gamma}
\left( \frac{T}{T_c}\right)^4 +\frac{i}{\omega} \left[ 1- \left(
\frac{T}{T_c}\right)^4\right] \right\} \,.
\end{equation}
Although the two-fluid model does not fully capture the rich dynamics
of superconductors (as it neglects coherence effects and dissipation),
it gives a relatively accurate and intuitive picture of the strength
of the magnetic field fluctuations. More elaborate models such as the
Eliashberg theory prove that with regards to spin transitions the
two-fluid model is perfectly adequate \cite{Hohenester07}. In
Fig.~\ref{fig:skindepth} we show the distance-dependence of the spin
flip lifetime for the two-fluid model of superconducting Nb (dashed
line) and the corresponding results for an Eliashberg calculation with
varying elastic scattering rates (symbols).


\newpage
\section{Dispersion forces}
\label{sec:dispersion}

Dispersion forces such as Casimir forces between bodies
\cite{Casimir48,0057}, Casimir--Polder (CP) forces between atoms
and bodies \cite{0022,0030} and van der Waals (vdW) forces between
atoms \cite{0030,0374} are effective electromagnetic forces that
arise as immediate consequences of correlated quantum ground-state
fluctuations. The total Lorentz force on an arbitrary macroscopic or
atomic charge distribution characterised by a charge density
$\hat{\rho}$ and a current density $\hat{\vect{j}}$ occupying a
volume $V$ is given by 
\begin{equation}
\label{eq:Lorentz}
\hat{\vect{F}}
 =\int_{V}\dif^3r\,\left[
 \hat{\rho}(\vect{r})\hat{\vect{E}}(\vect{r})
 +\hat{\vect{j}}(\vect{r})\vprod\hat{\vect{B}}(\vect{r})\right].
\end{equation}

To see how this quantum force acquires a nonzero average even in the
absence of external electromagnetic fields due to correlated
zero-point fluctuations, let us consider the example of a neutral,
stationary ground-state atom $A$. Expressing the atomic charge and
current densities in terms of polarisation and magnetisation [recall
Eqs.~(\ref{eq:classicalpolarization}) and
(\ref{eq:classicalmagnetization})], the Lorentz force can equivalently
be represented as \cite{Buhmann04,0012,0696} 
\begin{equation}
\label{eq:Lorentz2}
\hat{\vect{F}}
=\grad_A \int\dif^3r\,\Bigl[
 \hat{\vect{P}}_A(\vect{r})
 \sprod\hat{\vect{E}}(\vect{r})
 +\hat{\vect{M}}_A(\vect{r})
 \sprod\hat{\vect{B}}(\vect{r})\Bigr]
\end{equation}
($\grad_A$: derivative with respect to the atomic centre-of-mass
position). The quantum averages of both the electric and the magnetic
field vanish in the absence of external fields,
$\langle\hat{\vect{E}}\rangle=\langle\hat{\vect{B}}\rangle=\veczero$,
and for an unpolarised and unmagnetised atom so do the atomic
polarisation and magnetisation,
$\langle\hat{\vect{P}}_A\rangle=\langle\hat{\vect{M}}_A\rangle
=\veczero$. In the absence of correlations, this would imply that the
average net force on the atom is vanishing,
$\langle\hat{\vect{F}}\rangle=\vect{0}$. However, both the
electromagnetic and atomic fields are subject to nonvanishing
zero-point fluctuations, $\langle\hat{\vect{E}}^2\rangle$,
$\langle\hat{\vect{B}}^2\rangle$, $\langle\hat{\vect{P}}_A^2\rangle$,
$\langle\hat{\vect{M}}_A^2\rangle\neq 0$. These quantities are
mutually correlated and thus lead to a nonvanishing dispersion force:
For a nonmagnetic atom, one can show that this force is given by
(\cite{0046}, cf.\ also Sec.~\ref{sec:CPgroundPerturbative} below) 
\begin{eqnarray}
\label{eq:CPlinearresponse}
\vect{F}
& =&\frac{\hbar\mu_0}{4\pi\mi}\grad_A \!\!
 \int_0^\infty\!\!\dif\omega\,\omega^2
 \operatorname{Im} \bigl[ \alpha(\omega)\trace
 \ten{G}^{(S)}(\vect{r}_{A},\vect{r}_{A},\omega) \bigr]
\\ &=& 
\frac{\hbar\mu_0}{4\pi\mi}\grad_A \!\!
 \int_0^\infty\!\!\dif\omega\,\omega^2
 \bigl[
 \operatorname{Im}\alpha(\omega)\trace\operatorname{Re}
 \ten{G}^{(S)}(\vect{r}_{A},\vect{r}_{A},\omega)
 +\operatorname{Re}\alpha(\omega)\trace\operatorname{Im}
 \ten{G}^{(S)}(\vect{r}_{A},\vect{r}_{A},\omega)\bigr]. \nonumber
\end{eqnarray}
In accordance with the fluctuation-dissipation theorem, the real and
imaginary parts of the atomic and field response functions (i.e.
atomic polarisability and the Green tensor of the electromagnetic
field) represent the reactive and fluctuating behaviours of these
systems. Hence, the first term in the above equation correspond to
the atomic zero-point fluctuations ($\operatorname{Im}\alpha$) giving
rise to an induced electromagnetic field ($\operatorname{Re}\ten{G}$),
while the second term is due to the fluctuations of the
electromagnetic field ($\operatorname{Im}\ten{G}$) giving rise to an
induced atomic polarisation ($\operatorname{Re}\alpha$). Both
fluctuation sources thus contribute equally to the CP force. This
interpretation is in contrast to the mode summation picture in which
only the electromagnetic field fluctuations are taken into
account, and the only role of the matter is to provide the perfect
boundary conditions.

It is worth noting that for small separations, dispersion forces are
primarily due to the atomic zero-point fluctuations which interact
via the instantaneous Coulomb interaction. Dispersion forces where
first postulated by J.~D. van der Waals \cite{vdW} and theoretically
analysed by F.~London \cite{0374} and J.~E. Lennard-Jones \cite{0022}
in this nonretarded limit. The zero-point fluctuations of the
transverse electromagnetic field become important in the retarded
limit of large separations, as was shown by H.~B.~G. Casimir and
D.~Polder \cite{0030}. To honour these two major steps, one often uses
the notion vdW forces for all nonretarded dispersion forces on atoms
and the term CP force for fully retarded ones --- in contrast
to the naming convention adopted throughout this work. 
%
%
\subsection{Casimir forces}
\label{sec:lorentzcasimir}
The Casimir force (for general literature and reviews, see
Refs.~\cite{0375,Milonni,0378,0136,0135,0696,0753}) on a body of
permittivity $\varepsilon(\vec{r},\omega)$ and permeability
$\mu(\vec{r},\omega)$ occupying a volume $V$ can be found by
calculating the ground-state expectation value of the Lorentz
force~(\ref{eq:Lorentz})
\begin{equation}
\label{eq:CasimirVolume1}
\vect{F}=\int_{V}\dif^3r\,\left\{\langle\{0\}|\left[
 \hat{\rho}(\vect{r})\hat{\vect{E}}(\vect{r}')
 +\hat{\vect{j}}(\vect{r})\vprod\hat{\vect{B}}(\vect{r}')\right]
 |\{0\}\rangle\right\}_{\vect{r'}\to\vect{r}}
\end{equation}
($|\{0\}\rangle$: ground-state of the body-assisted electromagnetic
field) where the coincidence limit $\vect{r'}\to\vect{r}$ must be
performed in such a way that unphysical divergent self-force
contributions are discarded after the vacuum expectation value has
been evaluated. Using the relations
\begin{eqnarray}
\label{eq:rhomat}
\hat{\underline{\rho}}(\vect{r},\omega)
 &=&-\varepsilon_{0}
 \bm{\nabla}\sprod\left\{[\varepsilon(\vect{r},\omega)-1]
 \hat{\underline{\vect{E}}}(\vect{r},\omega)\right\}
 +\frac{1}{\mi\omega}\,\bm{\nabla}\sprod
 \hat{\underline{\vect{j}}}_\mathrm{N}(\vect{r},\omega),\\
\label{eq:jmat}
\hat{\underline{\vect{j}}}(\vect{r},\omega)
&=&-\mi\omega \varepsilon_{0}
 [\varepsilon(\vect{r},\omega)-1]
 \hat{\underline{\vect{E}}}(\vect{r},\omega)
 \nonumber\\[.5ex]
&&+\bm{\nabla}\vprod\left\{\kappa_{0}
 [1-\kappa(\vect{r},\omega)]
 \hat{\underline{\vect{B}}}(\vect{r},\omega)\right\}
 +\hat{\underline{\vect{j}}}_\mathrm{N}(\vect{r},\omega),
\end{eqnarray}
the field expansions~(\ref{eq:Eexpansion}), (\ref{eq:Bexpansion}) and
the expectation values (\ref{eq:expectE})--(\ref{eq:BBdagger}), one
finds for a homogeneous body at zero temperature \cite{0663}
\begin{align}
\label{eq:CasimirVolume}
\vect{F}
=&\;
 \frac{\hbar}{\pi}\int_{V}\dif^3r\int_{0}^\infty\dif\omega\,
 \biggl(\frac{\omega^2}{c^2}\bm{\nabla}\sprod
 \mathrm{Im}\ten{G}^{(S)}(\vect{r},\vect{r},\omega)
 \nonumber\\
&\hspace{19ex}+\trace\biggl\{\ten{I}\vprod
 \biggl[\bm{\nabla}\vprod\bm{\nabla}\vprod\,
 -\frac{\omega^2}{c^2}\biggr]
 \mathrm{Im}\ten{G}^{(S)}(\vect{r},\vect{r},\omega)\vprod
 \overleftarrow{\bm{\nabla}}'\biggr\}
 \biggr)\nonumber\\
=&\;-\frac{\hbar}{\pi}\int_{V}\dif^3r\int_{0}^\infty\dif\xi\,
 \biggl(\frac{\xi^2}{c^2}\bm{\nabla}\sprod
 \ten{G}^{(S)}(\vect{r},\vect{r},\mi\xi)
 \nonumber\\
&\hspace{17ex}
-\trace\biggl\{\ten{I}\vprod
\biggl[\bm{\nabla}\vprod\bm{\nabla}\vprod\,
 +\frac{\xi^2}{c^2}\biggr]
 \ten{G}^{(S)}(\vect{r},\vect{r},\mi\xi)\vprod
 \overleftarrow{\bm{\nabla}}'\biggr\}\biggr),
\end{align}
where the coincidence limit has been performed by replacing the Green
tensor with its scattering part. Here and in the following, the
gradients $\bm{\nabla}$ and $\overleftarrow{\bm{\nabla}}'$ are
understood to act on the first and second arguments of the Green
tensor, respectively.

Alternatively, the Casimir force can be equivalently expressed in
terms of a surface integral rather than a volume integral. To that
end, one makes use of the relation
\begin{equation}
\label{eq:LorentzStress}
\hat{\rho}(\vect{r})\hat{\vect{E}}(\vect{r})
 +\hat{\vect{j}}(\vect{r})\vprod\hat{\vect{B}}(\vect{r})
=\bm{\nabla}\sprod\hat{\ten{T}}(\vect{r})
 -\varepsilon_{0}\frac{\partial}{\partial t}
 \left[\hat{\vect{E}}(\vect{r})\vprod\hat{\vect{B}}(\vect{r})\right]
\end{equation}
with the Maxwell stress tensor
\begin{equation}
\label{eq:stresstensor}
\hat{\ten{T}}(\vect{r})
=\varepsilon_0\hat{\vect{E}}(\vect{r})
 \tprod\hat{\vect{E}}(\vect{r})
 +\mu_0^{-1}\hat{\vect{B}}(\vect{r})
 \tprod\hat{\vect{B}}(\vect{r})
 -\textstyle{\frac{1}{2}}\left[\varepsilon_0
 \hat{\vect{E}}^2(\vect{r})
 +\mu_0^{-1}\hat{\vect{B}}^2(\vect{r})\right]\ten{I} \,.
\end{equation}
This can be used to rewrite Eq.~(\ref{eq:CasimirVolume1}) in the form
\begin{multline}
\label{3.23}
\vect{F}=\int_{\partial V}\dif\vect{a}
 \sprod\langle\{0\}|\bigl\{\varepsilon_0
 \hat{\vect{E}}(\vect{r})\tprod\hat{\vect{E}}(\vect{r'})
 +\mu_0^{-1}\hat{\vect{B}}(\vect{r})\tprod\hat{\vect{B}}(\vect{r'})
 \\[.5ex]
-\textstyle{\frac{1}{2}}\bigl[\varepsilon_0
 \hat{\vect{E}}(\vect{r})\sprod\hat{\vect{E}}(\vect{r'})
 +\mu_0^{-1}\hat{\vect{B}}(\vect{r})\sprod\hat{\vect{B}}(\vect{r'})
 \bigr]\ten{I}\bigr\}|\{0\}\rangle
\end{multline}
[note that the last term in Eq.~(\ref{eq:LorentzStress}) does not
contribute in the stationary case]. Evaluating the field expectation
values according to Eqs.~(\ref{eq:expectE})--(\ref{eq:BBdagger}), one
finds that ($T=0$) \cite{0198} 
\begin{eqnarray}
\label{eq:CasimirSurface}
\vect{F}&=&\frac{\hbar}{\pi}\int_{0}^{\infty} \dif\omega
 \int_{\partial V}\dif\vect{a}\sprod
 \biggl\{\biggl[\frac{\omega^2}{c^2}\,
 \mathrm{Im}\ten{G}(\vect{r},\vect{r},\omega)
 -\bm{\nabla}\vprod
 \mathrm{Im}\ten{G}(\vect{r},\vect{r},\omega)
 \vprod\overleftarrow{\bm{\nabla}}'\biggr]\nonumber\\
&&\qquad-\frac{1}{2}\trace\biggl[\frac{\omega^2}{c^2}\,
 \mathrm{Im}\ten{G}(\vect{r},\vect{r},\omega)
 -\bm{\nabla}\vprod
 \mathrm{Im}\ten{G}(\vect{r},\vect{r},\omega)
 \vprod\overleftarrow{\bm{\nabla}}'\biggr]\ten{I}\biggr\}
 \nonumber\\
&=&-\frac{\hbar}{\pi}\int_{0}^{\infty} \dif\xi
 \int_{\partial V}\dif\vect{a}\sprod\biggl\{
 \biggl[\frac{\xi^2}{c^2}\,\ten{G}(\vect{r},\vect{r},\mi\xi)
 +\bm{\nabla}\vprod\ten{G}(\vect{r},\vect{r},\mi\xi)
 \vprod\overleftarrow{\bm{\nabla}}'\biggr]\nonumber\\
&&\qquad-\frac{1}{2}\trace
 \biggl[\frac{\xi^2}{c^2}\,\ten{G}(\vect{r},\vect{r},\mi\xi)
 +\bm{\nabla}\vprod\ten{G}(\vect{r},\vect{r},\mi\xi)
 \vprod\overleftarrow{\bm{\nabla}}'\biggr]\ten{I}\biggr\}.
\end{eqnarray}

The general expressions (\ref{eq:CasimirVolume}) and
(\ref{eq:CasimirSurface}) can be used to calculate Casimir forces
between bodies of arbitrary shape. In particular,
Eq.~(\ref{eq:CasimirSurface}) in connection with the Green tensor for
planar multilayered dielectrics (App.~\ref{sec:planardgf})
immediately leads to the famous Lifshitz formula \cite{0057}
for the Casimir force between two dielectric half-spaces
\cite{Raabe03}. In the limit of perfectly conducting plates, one
recovers the mode summation formula (\ref{eq:modecasimir}).

%
\subsection{Casimir--Polder forces}
\label{sec:CPground}
Similarly to the Casimir force, the Casimir--Polder force on a single
atom (for a general overview, see
Refs.~\cite{0375,0310,0431,0697,Milonni}) can be calculated as an
effective Lorentz force starting from expression~(\ref{eq:Lorentz}),
where the electromagnetic field acts on the atomic charge and current
distributions. In general, this force is a time-dependent quantity
which can only be found by solving the coupled atom-field dynamics
\cite{0696,0012,Buhmann04}. In this 
section, we restrict our attention to the force on a ground-state atom
with the body-assisted field being in its vacuum state, and we ignore
the effect of motion on the CP force. For this stationary problem, the
CP force can alternatively be derived from the atom-field coupling
energy, following the approach originally used by Casimir and Polder
\cite{0030}. 

Such an approach can be justified by means of a Born--Oppenheimer
approximation by assuming that the fast internal (electronic) motion
effectively decouples from the slow centre-of-mass motion. To see
this, we express the total multipolar
Hamiltonian~(\ref{eq:multipolarHamiltonian}) in terms of the
centre-of-mass momentum 
$\hat{\vec{p}}_A=\sum_{\alpha\in A}\hat{\vect{p}}_\alpha$ and
the internal momenta
$\hat{\overline{\vect{p}}}_\alpha=\hat{\vect{p}}_\alpha
-(m_\alpha/m_A)\hat{\vect{p}}_A$ to obtain 
\begin{equation}
\label{eq:Hsingle}
\hat{H}=\frac{\hat{\vect{p}}_A^2}{2m_A}
 +\hat{H}_A^\mathrm{int}
 +\hat{H}_F+\hat{H}_{AF},
\end{equation}
where 
\begin{equation}
\label{eq:Hint}
\hat{H}_A^\mathrm{int}
 =\frac{\hat{\overline{\vect{p}}}{}_{\alpha}^2}{2m_{\alpha}}
 +\frac{1}{2\varepsilon_0}\int\dif^3r\,
 \hat{\vect{P}}^2_A(\vect{r})
=\sum_k\hbar\omega_k\hat{A}_{kk}
\end{equation}
[$\hat{A}_{kk}=|k\rangle\langle k|$]
is the internal atomic Hamiltonian associated with the electronic
motion and the field Hamiltonian $\hat{H}_\mathrm{F}$ is given by
Eq.~(\ref{eq:multipolarHF}). The influence of atomic motion on the
atom-field interaction can be discarded by formally letting
$m_A\to\infty$, leading to
\begin{eqnarray}
\label{eq:HAFnomotion}
\hat{H}_{AF}
&=&-\int\dif^3r\,\hat{\vect{P}}_A(\vect{r})
 \sprod\hat{\vect{E}}(\vect{r})
 -\int\dif^3r\,\hat{\vect{M}}_A(\vect{r})
 \sprod\hat{\vect{B}}(\vect{r}) \nonumber \\
&&+\sum_{\alpha\in A}\frac{1}{2 m_\alpha}
 \biggl[\int\dif^3r\,\hat{\vect{\Theta}}_\alpha(\vect{r})
 \vprod\hat{\vect{B}}(\vect{r})\biggr]^2,
\end{eqnarray}
where the three terms describe the electric, paramagnetic and
diamagnetic interactions with the electromagnetic field,
respectively. Having thus separated the internal and centre-of-mass
motion as far as possible, we can apply the Born--Oppenheimer
approximation by integrating out the internal motion for given values
of $\hat{\vect{r}}_A$ and $\hat{\vect{p}}_A$, leading to an effective
Hamiltonian for the centre-of-mass motion, 
\begin{equation}
\label{eq:Heff}
\hat{H}_\mathrm{eff} 
 =\frac{\hat{\vect{p}}_A^2}{2m_A}
 +E+\Delta E.
\end{equation}
Here, $E$ is the energy of the uncoupled atom-field system and
$\Delta E$ is the energy shift due to the atom-field coupling
$\hat{H}_{A\mathrm{F}}$. Since we have neglected the influence of
motion on the atom-field interaction, the energy shift does not
depend on $\hat{\vect{p}}_A$ so that we may write
\begin{equation}
\label{eq:DeltaE}
\Delta E=\Delta E_0+\Delta E(\hat{\vect{r}}_A),
\end{equation}
where $\Delta E_0$ is the well-known (free-space) Lamb
shift~(\ref{eq:vacuumLamb}) \cite{Milonni}. By means of the
commutation relations
$\bigl[\hat{\vect{r}}_{\!A},\hat{\vect{p}}_A\bigr]=\mi\hbar\tens{I}$,
the effective Hamiltonian thus generates the following equations of
motion for the centre-of-mass coordinate:
\begin{gather}
\label{eq:HeffEOM1}
m_A\dot{\hat{\vect{r}}}_A
 =\frac{1}{\mi\hbar}\Bigl[m_A\hat{\vect{r}}_A,
 \hat{H}_\mathrm{eff}\Bigr]
 =\hat{\vect{p}}_A,\\
\label{eq:HeffEOM2}
\hat{\vect{F}}=m_A\ddot{\hat{\vect{r}}}_A
 =\frac{1}{\mi\hbar}\Bigl[m_A\dot{\hat{\vect{r}}}_A, 
 \hat{H}_\mathrm{eff}\Bigr]
 =-\vect{\nabla}_AU(\hat{\vect{r}}_A)
\end{gather}
where the CP potential
\begin{equation}
\label{eq:CPpotentialdef}
U(\hat{\vect{r}}_A)
 =\Delta E(\hat{\vect{r}}_A)
\end{equation}
is the position-dependent part of the energy shift, as suggested by
Casimir and Polder. In many cases, the centre-of-mass motion is
effectively classical and position and velocity in
Eqs.~(\ref{eq:HeffEOM1}) and (\ref{eq:HeffEOM2}) can be treated as
$c$-number parameters. In the following, it is not necessary to
distinguish the classical from the quantum case and we drop the
operator hats for convenience.
%
%
\subsubsection{Perturbation theory}
\label{sec:CPgroundPerturbative}
For weak atom-field coupling, the ground-state energy shift
(\ref{eq:DeltaE}) can be obtained from a perturbative calculation.
While the coupling Hamiltonian~(\ref{eq:HAFnomotion}) presents very
general basis for this purpose, the calculation can often be
simplified by applying the long-wavelength approximation (which is
valid provided that the atom-body separations are large with respect
to the atomic radius) and neglecting the weak diamagnetic
interaction, so that [cf. Eq.~(\ref{eq:multipolarHAF})]
\begin{equation}
\label{eq:HAFpert}
\hat{H}_{AF}=-\hat{\vect{d}}\sprod\hat{\vect{E}}(\vect{r}_A)
 -\hat{\vect{m}}\sprod\hat{\vect{B}}(\vect{r}_A)
\end{equation}
This Hamiltonian being linear in both atomic and field variables,
$\Delta E$ is to leading order given by the second-order energy shift
\begin{equation}
\label{eq:Shiftpert}
\Delta E
 =\sum_{I\neq G} \frac{\langle 0| \hat{H}_{AF} | I \rangle \langle
 I | \hat{H}_{AF} | 0 \rangle }{E_G-E_I}\,,
\end{equation}
where $|G\rangle$ $\!=$ $\!|0\rangle|\{0\}\rangle$ denotes the
(uncoupled) ground state of
$\hat{H}_A^\mathrm{int}+\hat{H}_\mathrm{F}$ and the relevant
intermediate states
$|I\rangle=|k\rangle|\vec{1}_\lambda(\vect{r},\omega)\rangle$
are those where the atom is excited, with a single-quantum excitation
of the field being present, $|\vec{1}_\lambda(\vect{r},\omega)\rangle
=\hat{\vect{f}}_\lambda^\dagger(\vect{r},\omega)|\{0\}\rangle$. The
formal sum in Eq.~(\ref{eq:Shiftpert}) thus represents discrete
summations over $\lambda$ and the vector index as well as integrations
over $\vec{r}$ and $\omega$.
\begin{figure}[!t!]
\begin{center}
\includegraphics[width=0.8\linewidth]{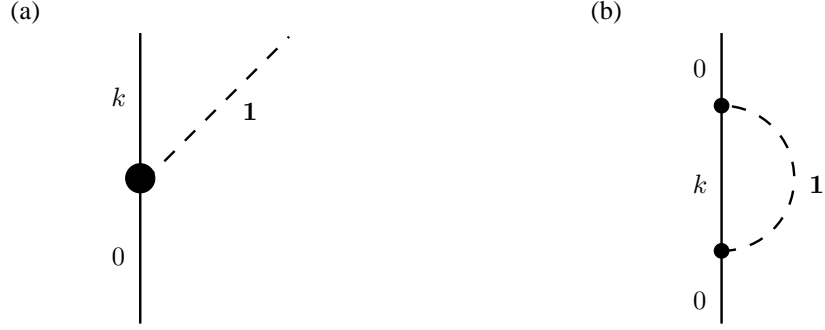}
\end{center}
\caption{
\label{fig:firstorder}
Schematic representation of single-photon interactions (a) and the
second-order energy shift (b). Solid lines represent atomic states
and dashed lines stand for photons. We do not distinguish electric
and magnetic interactions.
}
\end{figure}%
Recalling the field expansions~(\ref{eq:Eexpansion}) and
(\ref{eq:Bexpansion}), the matrix elements of the electric and
magnetic dipole interactions are found to be 
\begin{eqnarray}
\label{eq:dEElement}
\langle 0|\langle\{0\}|
 \hat{\vect{d}}\sprod\hat{\vect{E}}(\vect{r}_A)
 |\vec{1}_\lambda(\vect{r},\omega)\rangle
 |k\rangle
&=&\vect{d}_{0k}\sprod
 \ten{G}_\lambda(\vect{r}_A,\vect{r},\omega),\\
\label{eq:mBElement}
\langle 0|\langle\{0\}|
 \hat{\vect{m}}\sprod\hat{\vect{B}}(\vect{r}_A)
 |\vec{1}_\lambda(\vect{r},\omega)\rangle
 |k\rangle
&=&
 \frac{\vect{m}_{0k}\sprod\grad_A\vprod
 \ten{G}_\lambda(\vect{r}_A,\vect{r},\omega)}{\mi\omega}
\end{eqnarray}
The matrix elements and the second-order energy shift are
schematically depicted in Fig.~\ref{fig:firstorder}. As the energy
shift is quadratic in $\hat{H}_{AF}$, it contains purely electric,
purely magnetic and mixed electric--magnetic contributions. The latter
are proportional to $\vect{d}_{0k}\tprod\vect{m}_{k0}$, so for
nonchiral atoms, they can be excluded by means of a parity argument
($\hat{\vect{d}}$ is odd and $\hat{\vect{m}}$ is even
under parity). After substitution of Eqs.~(\ref{eq:dEElement}) and
(\ref{eq:mBElement}), we thus only retain the purely electric and
magnetic terms, which upon using the integral
relation~(\ref{Glambdaintegral}) leads to
\begin{multline}
\label{eq:Shiftresult}
\Delta E
 =-\frac{\mu_0}{\pi}\sum_k\int_0^\infty
 \frac{\dif\omega}{\omega_{k0}+\omega}\,\Bigl[\omega^2
 \vect{d}_{0k}\sprod\mathrm{Im}\,
 \ten{G}(\vect{r}_A,\vect{r}_A,\omega)\sprod\vect{d}_{k0}\\
\vect{m}_{0k}\sprod\curl
 \operatorname{Im}\,\ten{G}(\vect{r}_A,\vect{r}_A,\omega)
 \vprod\overleftarrow{\grad}'
 \sprod\vect{m}_{k0}\Bigr] \,.
\end{multline}

The Casimir--Polder potential~(\ref{eq:CPpotentialdef}) can be
extracted from this energy shift by discarding the
position-independent contribution associated with the bulk part of
the Green tensor $\ten{G}^{(0)}$ and only retaining its scattering
part $\ten{G}^{(S)}$. The result can be further simplified by writing
$\operatorname{Im}\,\ten{G}$ $\!=$ $(\ten{G}-\ten{G}^\ast)/(2\mi)$,
making use of the Schwarz reflection principle,
Eq.~(\ref{eq:SchwarzG}), and transforming the integrals along the real
axis into ones along the purely imaginary axis
(cf. Ref.~\cite{Buhmann04}). The resulting ground-state CP potential
is given by
\cite{0375,Buhmann04,0012,0046,0035,0039,0050,Wylie84,0439}
\begin{equation}
\label{eq:Usingle}
U(\vect{r}_A)=U_e(\vect{r}_A)+U_m(\vect{r}_A),
\end{equation}
with
\begin{eqnarray}
\label{eq:Ue}
U_e(\vect{r}_A)
 &=&\frac{\hbar\mu_0}{2\pi}
 \int_0^\infty\dif\xi\,\xi^2
 \trace\bigl[\bm{\alpha}(\mi\xi)\sprod
 \ten{G}^{(S)}(\vect{r}_A,\vect{r}_A,\mi\xi)\bigr]\nonumber\\
&=&\frac{\hbar\mu_0}{2\pi}\int_0^\infty\dif\xi\,\xi^2\alpha(\mi\xi)
 \trace\ten{G}^{(S)}(\vect{r}_A,\vect{r}_A,\mi\xi),\\
\label{eq:Um}
U_m(\vect{r}_A)
 &=&\frac{\hbar\mu_0}{2\pi}
 \int_0^\infty\dif\xi\,\trace\left[\bm{\beta}(\mi\xi)
 \sprod\curl\ten{G}^{(S)}(\vect{r}_A,\vect{r}_A,\mi\xi)
 \vprod\overleftarrow{\grad}'\right]\nonumber\\
&=&\frac{\hbar\mu_0}{2\pi}
 \int_0^\infty\dif\xi\,\beta(\mi\xi)\trace\left[
 \curl\ten{G}^{(S)}(\vect{r}_A,\vect{r}_A,\mi\xi)
 \vprod\overleftarrow{\grad}'\right] ,
\end{eqnarray}
denoting the electric and magnetic parts of the potential and
\begin{eqnarray}
\label{eq:alpha}
\bm{\alpha}(\omega)
&=&\lim_{\epsilon\to 0}\frac{2}{\hbar}\sum_k
 \frac{\omega_{k0}
\vect{d}_{0k}\tprod\vect{d}_{k0}}
 {\omega_{k0}^2-\omega^2-\mi\omega\epsilon}\nonumber\\
&=&\lim_{\epsilon\to 0}\frac{2}{3\hbar}\sum_k
 \frac{\omega_{k0}
|\vect{d}_{0k}|^2}
 {\omega_{k0}^2-\omega^2-\mi\omega\epsilon}\,\ten{I}
 =\alpha(\omega)\ten{I}\\
\label{eq:beta}
\bm{\beta}(\omega)
&=&\lim_{\epsilon\to 0}\frac{2}{\hbar}\sum_k
 \frac{
\omega_{k0}
\vect{m}_{0k}\tprod\vect{m}_{k0}}
 {\omega_{k0}^2-\omega^2-\mi\omega\epsilon}\nonumber\\
&=&\lim_{\epsilon\to 0}\frac{2}{3\hbar}\sum_k
 \frac{
\omega_{k0}
|\vect{m}_{0k}|^2}
 {\omega_{k0}^2-\omega^2-\mi\omega\epsilon}\,\ten{I}
 =\beta(\omega)\ten{I}
\end{eqnarray}
being the atomic polarisability and magnetisability. The second lines
of equalities in Eqs.~(\ref{eq:Ue})--(\ref{eq:beta}) above hold for
isotropic atoms.

When considering CP forces on atoms that are embedded in a body or a
medium, one has to account for the fact that the local
electromagnetic field interacting with the atom differs from the
macroscopic one employed in our derivation. This difference gives
rise to local--field corrections which can be implemented via the
real-cavity model by assuming the atom to be surrounded by a small
free-space cavity (recall the discussion in
Sec.~\ref{sec:modifiedse}). In order to apply this procedure to our
results~(\ref{eq:Ue}) and (\ref{eq:Um}), one has to replace the Green 
tensor by its local-field corrected counterpart as given in
App.~\ref{sec:localfield} which, after discarding position-independent
terms, results in \cite{0739,Hassan08}
\begin{eqnarray}
\label{eq:Ueloc}
U_e(\vect{r}_A)
&\!=&\!\frac{\hbar\mu_0}{2\pi}\int_0^\infty\dif\xi\,
 \xi^2\alpha(\mi\xi)
 \biggl[\frac{3\varepsilon(\mi\xi)}{2\varepsilon(\mi\xi)+1}
 \biggr]^2
 \trace\ten{G}^{(S)}(\vect{r}_A,\vect{r}_A,\mi\xi),\\
\label{eq:Umloc}
U_m(\vect{r}_A)
&\!=&\!\frac{\hbar\mu_0}{2\pi}
 \int_0^\infty\dif\xi\,\beta(\mi\xi)
 \biggl[\frac{3}{2\mu(\mi\xi)+1}\biggr]^2\trace\left[
 \curl\ten{G}^{(S)}(\vect{r}_A,\vect{r}_A,\mi\xi)
 \vprod\overleftarrow{\grad}'\right],
\end{eqnarray}
where $\varepsilon(\omega)=\varepsilon(\vect{r}_A,\omega)$ and
$\mu(\omega)=\mu(\vect{r}_A,\omega)$ denote the permittivity and
permeability of the host body at the position of the atom. For an
atom situated in free space, the local-field correction factors are
equal to unity and one recovers Eqs.~(\ref{eq:Ue}) and (\ref{eq:Um}).

It is instructive to study the behaviour of the total CP potential
under a duality transformation, which in this case amounts to a
simultaneous global exchange $\alpha\leftrightarrow\beta/c^2$ and
$\varepsilon\leftrightarrow\mu$ (see Tab.~\ref{tab:duality}). Using
the associated transformation laws~(\ref{dgftrans1}) and
(\ref{dgftrans2}) of the Green tensor as given in
App.~\ref{sec:dualgreen}, one sees that the duality transformation
results in an exchange $U_e\leftrightarrow U_m$ of the free-space
potentials~(\ref{eq:Ue}) and (\ref{eq:Um}), so that the total
potential is invariant with respect to a duality transformation
\cite{BuhmannScheel08}. The transformation laws~(\ref{dgfloctrans1})
and (\ref{dgfloctrans2}) for the local-field corrected Green tensor
imply that the same is true for an embedded atom, provided that
local-field corrections are taken into account. The duality invariance
of the CP potential is very useful when considering specific
geometries: Once the electric CP potential of an atom in a particular
magnetoelectric environment has been calculated, that of a magnetic
one can be obtained by simply replacing $\alpha\rightarrow\beta/c^2$
and exchanging $\varepsilon\leftrightarrow\mu$.
%
%
\subsubsection{Atom in front of a plate}
\label{sec:CPPlanar}
\begin{figure}[!t!]
\noindent\vspace*{-2ex}
\begin{center}
\includegraphics[width=0.6\linewidth]{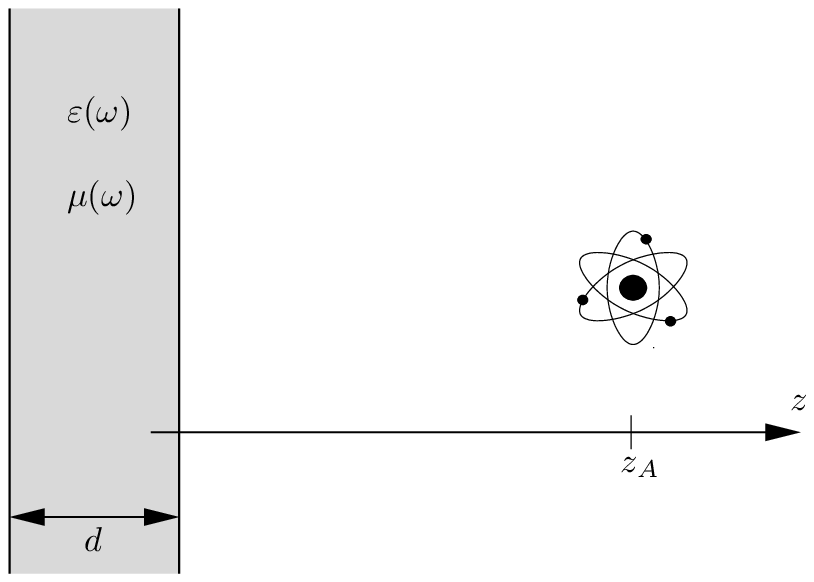}
\end{center}
\caption{
\label{fig:plateschematic}
An atom interacting with a magnetoelectric plate.}
\end{figure}

Let us apply the general results to an isotropic atom which is placed
above ($z_A>0$) a magnetoelectric plate of thickness $d$,
permittivity $\varepsilon(\omega)$ and permeability $\mu(\omega)$,
see Fig.~\ref{fig:plateschematic}.
Using the Green tensor of this very simple geometry as given
App.~\ref{sec:planardgf}, the electric CP potential~(\ref{eq:Ue})
reads \cite{0012,0019}
\begin{equation}
\label{eq:UeHS}
U_e(z_A)=\frac{\hbar\mu_0}{8\pi^2}
 \int_0^\infty\dif\xi\,\xi^2\alpha(\mi\xi)
 \int_{\xi/c}^\infty\dif\kappa_z\,
 \,\me^{-2\kappa_zz_A}
 \biggl[r_s
 -\biggl(2\,\frac{\kappa_z^2c^2}{\xi^2}-1\biggr)r_p\biggr]
\end{equation}
where $r_s$ and $r_p$ are the reflection coefficients of the half
space for $s$- and $p$-polarised waves and
$\kappa_z=\operatorname{Im}k_z$, with $\vec{k}$ being the wave vector
of these waves in free space. By virtue of the duality invariance, we
can obtain the magnetic potential from the above expression by
replacing $\alpha\rightarrow\beta/c^2$ and exchanging
$\varepsilon\leftrightarrow\mu$, which is equivalent to an exchange of
$r_s$ and $r_p$ [cf. Eq.~(\ref{eq:fresnel}) in
App.~\ref{sec:planardgf}]
\cite{0275,Hassan08}:
\begin{equation}
\label{eq:UmHS}
U_m(z_A)=\frac{\hbar\mu_0}{8\pi^2}
 \int_0^\infty\dif\xi\,\xi^2\,\frac{\beta(\mi\xi)}{c^2}
 \int_{\xi/c}^\infty\dif\kappa_z\,
 \,\me^{-2\kappa_zz_A}
 \biggl[r_p
 -\biggl(2\,\frac{\kappa_z^2c^2}{\xi^2}-1\biggr)r_s\biggr].
\end{equation}
%
%
\paragraph{Perfect mirror:}
\label{sec:CPMirror} 
A perfect mirror is realized for a perfectly conducting plate
($\varepsilon\to\infty$) or an infinitely permeable one
($\mu\to\infty$), in which cases the reflection coefficients
are given by $r_s=-r_p=\mp 1$, respectively. In this
particularly simple case, the $\kappa_z$-integrals in
Eqs.~(\ref{eq:UeHS}) and (\ref{eq:UmHS}) can be performed, so that
the total CP potential~(\ref{eq:Usingle}) reads 
 \begin{equation}
\label{eq:UHSperfect}
U(z_A)
=\mp\frac{\hbar}{16\pi^2\varepsilon_0z_A^3}
 \int_0^\infty\dif\xi
 \biggl[\alpha(\mi\xi)-\frac{\beta(\mi\xi)}{c^2}\biggr]
 \me^{-2\xi z_A/c}
 \biggl[1+2\,\frac{\xi z_A}{c}
 +2\,\frac{\xi^2z_A^2}{c^2}\biggr],
\end{equation}
where the upper (lower) is valid for the perfectly conducting
(infinitely permeable) mirror. This result includes the famous
Casimir--Polder potential of a polarisable atom in front of a
perfectly conducting wall \cite{0030}. In the retarded limit
$z_A\gg c/\omega_\mathrm{min}$ ($\omega_\mathrm{min}$: minimum of all
relevant atomic transition frequencies), the $\xi$-integral is
effectively limited to a region where the approximations
$\alpha(\mi\xi)$ $\!\simeq$ $\alpha_A(0)$ and $\beta_A(\mi\xi)$
$\!\simeq$ $\beta(0)$ are valid and after integration one obtains
\cite{0095}
\begin{equation}
\label{eq:UHSperfectret}
U(z_A)=
\mp\frac{3\hbar c\alpha(0)}{32\pi^2\varepsilon_0z_A^4}
\pm\frac{3\hbar \beta(0)}{32\pi^2\varepsilon_0 cz_A^4}\,.
\end{equation}
In the non-retarded limit $z_A\ll c/\omega_\mathrm{max}$
($\omega_\mathrm{max}$: maximum of all relevant atomic transition
frequencies), the factors $\alpha_A(\mi\xi)$ and $\beta_A(\mi\xi)$
limits the $\xi$-integral in Eq.~(\ref{eq:UHSperfect}) to a range
where we may approximately set $e^{-2\xi z_A/c}$ $\!\simeq$ $\!1$ and
neglect the second and third terms in the square brackets. The
integral can then be performed with the aid of the
definitions~(\ref{eq:alpha}) and (\ref{eq:beta})
\begin{equation}
\label{eq:UHSperfectnret}
U(z_A)
=\mp\frac{\langle\hat{\vect{d}}^2\rangle}
 {48\pi\varepsilon_0z_A^3}
 \pm\frac{\langle\hat{\vect{m}}^2\rangle}
 {48\pi\varepsilon_0c^2z_A^3}\,,
\end{equation}
with the first term being the Lennard-Jones potential \cite{0022}.

\begin{figure}[!t!]
\noindent\vspace*{-2ex}
\begin{center}
\includegraphics[width=\linewidth]{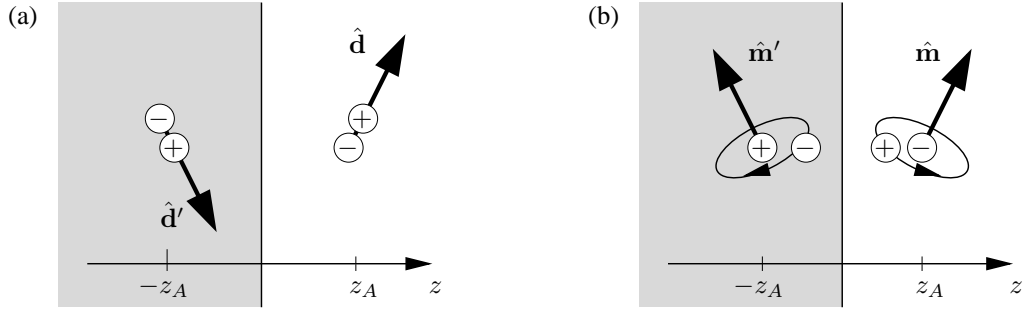}
\end{center}
\caption{
\label{fig:imagedipole}
Image dipole construction for an (a) electric (b) magnetic dipole
in front of a perfectly conducting plate.}
\end{figure}

The results~(\ref{eq:UHSperfect})--(\ref{eq:UHSperfectnret}) show that
a perfectly conducting plate attracts polarisable atoms and repels
magnetisable ones. For the nonretarded limit, this behaviour can be
made plausible by noting that the CP potential is due to the
interaction of the atomic dipole moment with its image in the plate
\cite{0022}. The image of an electric dipole moment 
$\hat{\vect{d}}$, located at a distance $z_A$ away from a perfectly
conducting plate, is constructed by a reflection at the $xy$-plane,
together with an interchange of positive and negative charges and is
hence given by $\hat{\vect{d}}'=(-\hat{d}_x,-\hat{d}_y,\hat{d}_z)$,
cf.\ Fig~\ref{fig:imagedipole}(a). 
The average interaction energy of the dipole and its image is hence
given by \cite{0001}
\begin{equation}
\label{eq:elimage}
U_e(z_A)
 =\frac{1}{2}\,\frac{\langle\hat{\vect{d}}\sprod
 \hat{\vect{d}}'
 -3\hat{d}_z\hat{d}_z'\rangle}
 {4\pi\varepsilon_0(2z_A)^3}
 =-\frac{\langle\hat{\vect{d}}^2\rangle}
 {48\pi\varepsilon_0z_A^3}\,,
\end{equation}
in agreement with Eq.~(\ref{eq:UHSperfectnret}), where the
factor $1/2$ accounts for the fact that the second dipole is induced
by the first one. 

On the contrary, a magnetic dipole $\hat{\vect{m}}$ behaves like a
pseudo-vector under reflection, so its image is given by
$\hat{\vect{m}}'=(\hat{m}_x,\hat{m}_y,-\hat{m}_z)$, cf.\
Fig~\ref{fig:imagedipole}(b). The interaction energy of the magnetic
dipole and its image reads
\begin{equation}
\label{eq:magimage}
U_m(z_A)
 =\frac{1}{2}\,\frac{\langle\hat{\vect{m}}
 \sprod\hat{\vect{m}}'
 -3\hat{m}_z\hat{m}_z'\rangle}
 {4\pi\varepsilon_0(2z_A)^3}
 =\frac{\langle\hat{\vect{m}}^2\rangle}
 {48\pi\varepsilon_0z_A^3}\,,
\end{equation}
again in agreement with Eq.~(\ref{eq:UHSperfectnret}). The different
signs of the CP potential associated with polarisable/magnetisable
atoms can thus be understood from the different reflection behaviour
of electric and magnetic dipoles.  
%
%
\paragraph{Half space:}
\label{sec:CPHS} 
Let us next consider a semi-infinite magnetoelectric half space of
finite permittivity and permeability, which is a good model for
plates whose thickness is large with respect to the atom-surface
separation. Using the reflection coefficients~(\ref{eq:fresnel}) in
App.~\ref{sec:planardgf}, the electric and magnetic
potentials~(\ref{eq:UeHS}) and (\ref{eq:UmHS}) take the forms
\cite{0012,0019,0043,0392,0275,Hassan08}
\begin{eqnarray}
\label{eq:UeHS1}
U_e(z_A)&=&\frac{\hbar\mu_0}{8\pi^2}
 \int_0^\infty\dif\xi\,\xi^2\alpha(\mi\xi)
 \int_{\xi/c}^\infty\dif\kappa_z\,
 \,\me^{-2\kappa_zz_A}
 \biggl[\frac{\mu(\mi\xi)\kappa_z-\kappa_{1z}}
 {\mu(\mi\xi)\kappa_z+\kappa_{1z}}\nonumber\\
&&-\biggl(2\,\frac{\kappa_z^2c^2}{\xi^2}-1\biggr)
 \frac{\varepsilon(\mi\xi)\kappa_z-\kappa_{1z}}
 {\varepsilon(\mi\xi)\kappa_z+\kappa_{1z}}\biggr]\\
\label{eq:UmHS1}
U_m(z_A)&=&\frac{\hbar\mu_0}{8\pi^2}
 \int_0^\infty\dif\xi\,\xi^2\,\frac{\beta(\mi\xi)}{c^2}
 \int_{\xi/c}^\infty\dif\kappa_z\,
 \,\me^{-2\kappa_zz_A}\biggl[
 \frac{\varepsilon(\mi\xi)\kappa_z-\kappa_{1z}}
 {\varepsilon(\mi\xi)\kappa_z+\kappa_{1z}}\nonumber\\
&&-\biggl(2\,\frac{\kappa_z^2c^2}{\xi^2}-1\biggr)
 \frac{\mu(\mi\xi)\kappa_z-\kappa_{1z}}
 {\mu(\mi\xi)\kappa_z+\kappa_{1z}}\biggr]
\end{eqnarray}
with
\begin{equation}
\label{eq:kappa1}
\kappa_{1z}=\operatorname{Im}k_{1z}
=\sqrt{\frac{\xi^2}{c^2}\bigl[\varepsilon(\mi\xi)
 \mu(\mi\xi)-1\bigr]+\kappa^2}\,,
\end{equation} 
where $\vec{k}_1$ is the wave vector inside the half space. In the
retarded limit $z_A\gg c/\omega_\mathrm{min}$ (with
$\omega_\mathrm{min}$ being the minimum of all relevant atom and
medium resonance frequencies) the potentials take the asymptotic forms
\cite{0330}
\begin{eqnarray}
\label{eq:UeHSret}
U_e(z_{A})&=&-\frac{3\hbar c\alpha(0)}{64\pi^2\varepsilon_0z_{A}^4}
 \int_{1}^\infty\dif v\,
 \biggl[\Bigl(\frac{2}{v^2}-\frac{1}{v^4}\Bigr)
 \frac{\varepsilon(0)v-\sqrt{n^2(0)-1+v^2}}
 {\varepsilon(0)v+\sqrt{n^2(0)-1+v^2}}\nonumber\\
&&-\,\frac{1}{v^4}\,
 \frac{\mu(0)v-\sqrt{n^2(0)-1+v^2}}
 {\mu(0)v+\sqrt{n^2(0)-1+v^2}}\,\biggr]\,,\\
\label{eq:UmHSret}
U_m(z_{A})&=&-\frac{3\hbar\beta(0)}{64\pi^2\varepsilon_0c z_{A}^4}
 \int_{1}^\infty\dif v\,
 \biggl[\Bigl(\frac{2}{v^2}-\frac{1}{v^4}\Bigr)
 \frac{\mu(0)v-\sqrt{n^2(0)-1+v^2}}
 {\mu(0)v+\sqrt{n^2(0)-1+v^2 }}\nonumber\\
&&-\,\frac{1}{v^4}\,
 \frac{\varepsilon(0)v-\sqrt{n^2(0)-1+v^2}}
 {\varepsilon(0)v+\sqrt{n^2(0)-1+v^2}}\,\biggr]
\end{eqnarray}
[$n(0)=\sqrt{\varepsilon(0)\mu(0)}$], while in the nonretarded limit
$n(0)z_A\ll c/\omega_\mathrm{max}$ ($\omega_\mathrm{max}$: maximum of
all relevant atom and medium resonance frequencies), they are well
approximated by
\begin{eqnarray}
\label{eq:UeHSnret1}
U_e(z_{A})&=&-\frac{\hbar}{16\pi^2\varepsilon_0z_{A}^3}
 \int_0^\infty\dif\xi\,\alpha(\mi\xi)\,
 \frac{\varepsilon(\mi\xi)-1}{\varepsilon(\mi\xi)+1}\,,\\
\label{eq:UmHSnret1}
U_m(z_{A})&=&\frac{\hbar\mu_0}{32\pi^2z_{A}}
 \int_0^\infty\dif\xi\,\xi^2\,\frac{\beta(\mi\xi)}{c^2}\,
 \frac{[\varepsilon(\mi\xi)-1][\varepsilon(\mi\xi)+3]}
 {\varepsilon(\mi\xi)+1}
\end{eqnarray}
for a dielectric half space and by
\begin{eqnarray}
\label{eq:UeHSnret2}
U_e(z_{A})&=&\frac{\hbar\mu_0}{32\pi^2\varepsilon_0z_{A}}
 \int_0^\infty\dif\xi\,\xi^2\alpha(\mi\xi)\,
  \frac{[\mu(\mi\xi)-1][\mu(\mi\xi)+3]}{\mu(\mi\xi)+1}\,,\\
\label{eq:UmHSnret2}
U_e(z_{A})&=&-\frac{\hbar}{16\pi^2\varepsilon_0z_{A}^3}
 \int_0^\infty\dif\xi\,\frac{\beta(\mi\xi)}{c^2}\,
 \frac{\mu(\mi\xi)-1}{\mu(\mi\xi)+1}
\end{eqnarray}
for a purely magnetic one. One can summarise these results by stating
that the CP potential is attractive for two objects of the same
(electric/magnetic) nature, e.g. a polarisable atom interacting with a
dielectric half space, while being repulsive for two objects of
different nature.

\begin{figure}[!t!]
\begin{center}
\includegraphics[width=0.7\linewidth]{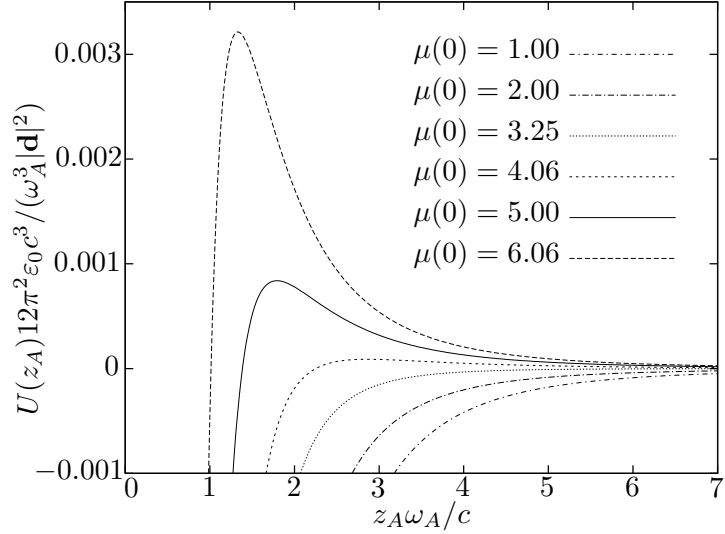}
\end{center}
\caption{
\label{fig:halfspace}
The potential of a polarisable ground-state two-level atom
(transition frequency $\omega_A$, dipole matrix element $\vect{d}$)
in front of a magnetoelectric half space is shown as a function of
the distance between the atom and the half space for different values
of $\mu(0)$ ($\omega_{Pe}/\omega_{10}=0.75$,
$\omega_{Te}/\omega_{10}=1.03$, $\omega_{Tm}/\omega_{10}=1$,
$\gamma_{e}/\omega_{10}=\gamma_{m}/\omega_{10}=0.001$) \cite{0019}.
}
\end{figure}

In contrast to what is suggested by the findings for a perfect
mirror, the different cases furthermore lead to different power laws
in the nonretarded limit, with attractive potentials being
proportional to $1/z_A^3$ and repulsive ones following a much weaker
$1/z_A$ behaviour. When either the atom or the half space
simultaneously exhibit electric and magnetic properties, both
attractive and repulsive force components are present, which may be
combined to form a potential barrier. This is illustrated in
Fig.~\ref{fig:halfspace}, where we show the potential~(\ref{eq:UeHS1})
of a polarisable ground-state two-level atom near a magnetoelectric
half space, whose permittivity and permeability have been modelled by
the single-resonance forms 
\begin{equation}
\label{eq:epsilonmusingle}
\varepsilon(\omega)=1+\frac{\omega_{Pe}^2}{\omega^2_{Te}
-\omega^2-\mi\omega\gamma_e}\,,\qquad
\mu(\omega)=1+\frac{\omega_{Pm}^2}{\omega^2_{Tm}
-\omega^2-\mi\omega\gamma_m}\,.
\end{equation}
%
It is seen that repulsive force components may lead to a potential
barrier at intermediate distances while attractive forces always
dominate close to the surface due to their stronger power law. For a
polarisable atom, repulsive force components are associated with the
magnetic properties of the half space, so the barrier increases in
height as $\mu(0)$ increases. It follows from the retarded
limit~(\ref{eq:UeHSret}) that the threshold for barrier formation is
$\mu(0)-1\ge3.29[\varepsilon(0)-1]$ for a weakly magnetoelectric
half space and $\mu(0)\ge 5.11\varepsilon(0)$ for a strongly
magnetoelectric one.
%
%
\paragraph{Plate of finite thickness:}
\label{sec:CPplate} 
Finally, we consider a plate of arbitrary thickness $d$. Use
of the appropriate reflection coefficients~(\ref{eq:multifresnel}) in
Eqs.~(\ref{eq:UeHS}) and (\ref{eq:UmHS}) leads to the potentials
\cite{0019}
\begin{eqnarray}
\label{eq:Ueplate}
U_e(z_{A})
&=&\frac{\hbar\mu_0}{8\pi^2}
 \int_0^\infty\dif\xi\,\xi^2\alpha(\mi\xi)
 \int_{\xi/c}^\infty\dif \kappa_z\,\me^{-2\kappa_zz_{A}}
 \nonumber\\
&&\times\biggl\{\frac{[\mu^2(\mi\xi)\kappa_z^2-\kappa_{1z}^2]
 \tanh(\kappa_{1z}d)}
 {2\mu(\mi\xi)\kappa_z\kappa_{1z}
 +[\mu^2(\mi\xi)\kappa_z^2+\kappa_{1z}^2]\tanh(\kappa_{1z}d)}
 \nonumber\\
&&-\,\biggl(2\,\frac{\kappa_z^2c^2}{\xi^2}-1\biggr)
 \frac{[\varepsilon^2(\mi\xi)\kappa_z^2-\kappa_{1z}^2]
 \tanh(\kappa_{1z}d)}
 {2\varepsilon(\mi\xi)\kappa_z \kappa_{1z}+
 [\varepsilon^2(\mi\xi)\kappa_z^2+\kappa_{1z}^2]
 \tanh(\kappa_{1z}d)}\biggr\},
\end{eqnarray}
\begin{eqnarray}
\label{eq:Umplate}
U_m(z_{A})
&=&\frac{\hbar\mu_0}{8\pi^2}
 \int_0^\infty\dif\xi\,\xi^2\,\frac{\beta(\mi\xi)}{c^2}\,
 \int_{\xi/c}^\infty\dif \kappa_z\,\me^{-2\kappa_zz_{A}}
 \nonumber\\
&&\times\biggl\{\frac{[\varepsilon^2(\mi\xi)\kappa_z^2-\kappa_{1z}^2]
 \tanh(\kappa_{1z}d)}
 {2\varepsilon(\mi\xi)\kappa_z\kappa_{1z}
 +[\varepsilon^2(\mi\xi)\kappa_z^2+\kappa_{1z}^2]\tanh(\kappa_{1z}d)}
 \nonumber\\
&&-\,\biggl(2\,\frac{\kappa_z^2c^2}{\xi^2}-1\biggr)
 \frac{[\mu^2(\mi\xi)\kappa_z^2-\kappa_{1z}^2]
 \tanh(\kappa_{1z}d)}
 {2\mu(\mi\xi)\kappa_z \kappa_{1z}+
 [\mu^2(\mi\xi)\kappa_z^2+\kappa_{1z}^2]
 \tanh(\kappa_{1z}d)}\biggr\}.
\end{eqnarray}
%
\begin{figure}[!t!]
\begin{center}
\includegraphics[width=0.7\linewidth]{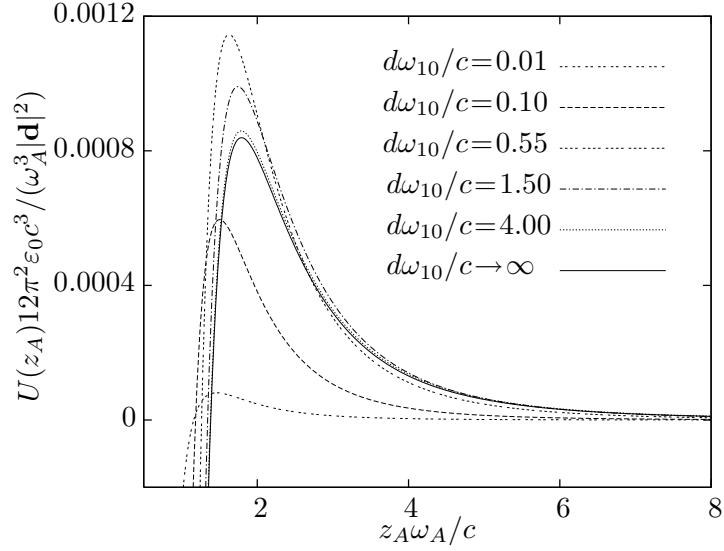}
\end{center}
\caption{
\label{fig:plate}
The potential of a polarisable ground-state two-level atom in front of
a magnetoelectric plate is shown as a function of the atom-plate
separation for different values of the plate thickness $d$ [$\mu(0)$
$\!=$ $\!5$; whereas all other parameters are the same as in
Fig.~\ref{fig:halfspace}] \cite{0019}.
}
\end{figure}%
For a sufficiently thick plate, $d\gg z_A$, one may approximate
$\tanh(\kappa_{1z}d)\simeq 1$ to recover the half-space
results~(\ref{eq:UeHS1}) and (\ref{eq:UmHS1}). In the opposite limit
of a thin plate, $n(0)d\ll z_A$, the approximation $\kappa_{1z}d\ll 1$
results in
\begin{eqnarray}
\label{eq:Uethin}
U_e(z_{A})
&=&\frac{\hbar\mu_0d}{8\pi^2}
 \int_0^\infty\dif\xi\,\xi^2\alpha(\mi\xi)
 \int_{\xi/c}^\infty\dif \kappa_z\,\me^{-2\kappa_zz_{A}}
 \biggl[\frac{\mu^2(\mi\xi)\kappa^2-\kappa_{1z}^2}
 {2\mu(\mi\xi)\kappa}\nonumber\\
&&-\,\biggl(2\,\frac{\kappa_z^2c^2}{\xi^2}-1\biggr)
 \frac{\varepsilon^2(\mi\xi)\kappa_z^2-\kappa_{1z}^2}
 {2\varepsilon(\mi\xi)\kappa_z}\biggr]\,\\
\label{eq:Umthin}
U_m(z_{A})
&=&\frac{\hbar\mu_0d}{8\pi^2}
 \int_0^\infty\dif\xi\,\xi^2\,\frac{\beta(\mi\xi)}{c^2}\,
 \int_{\xi/c}^\infty\dif \kappa_z\,\me^{-2\kappa_zz_{A}}
 \biggl[\frac{\varepsilon^2(\mi\xi)\kappa^2-\kappa_{1z}^2}
 {2\varepsilon(\mi\xi)\kappa}\nonumber\\
&&-\,\biggl(2\,\frac{\kappa_z^2c^2}{\xi^2}-1\biggr)
 \frac{\mu^2(\mi\xi)\kappa_z^2-\kappa_{1z}^2}
 {2\mu(\mi\xi)\kappa_z}\biggr].
\end{eqnarray} 
As for the half space, these potentials reduce to simple power laws
for large and small atom-surface separations. In the retarded limit,
the thin-plate potentials read
\begin{eqnarray}
\label{eq:Uethinret}
U_e(z_{A})=-\frac{\hbar c\alpha(0)d}
 {160\pi^2\varepsilon_0z_{A}^5}\,
 \biggl[\frac{14\varepsilon^2(0)-9}{\varepsilon(0)}
 -\frac{6\mu^2(0)-1}{\mu(0)}\biggr]\,,\\
\label{eq:Umthinret}
U_m(z_{A})=-\frac{\hbar\beta(0)d}
 {160\pi^2\varepsilon_0cz_{A}^5}\,
 \biggl[\frac{14\mu^2(0)-9}{\mu(0)}
 -\frac{6\varepsilon^2(0)-1}{\varepsilon(0)}\biggr]\,,
\end{eqnarray}
while for nonretarded distances, they become
\begin{eqnarray}
\label{eq:Uethinnret1}
U_e(z_{A})&=&-\frac{3\hbar d}{64\pi^2\varepsilon_0z_{A}^4}
 \int_0^\infty\dif\xi\,\alpha(\mi\xi)\,
 \frac{\varepsilon^2(\mi\xi)-1}{\varepsilon(\mi\xi)}\,,\\
\label{eq:Umthinnret1}
U_m(z_{A})&=&\frac{\hbar\mu_0 d}{64\pi^2z_{A}^2}
 \int_0^\infty\dif\xi\,\xi^2\,\frac{\beta(\mi\xi)}{c^2}\,
 \frac{[\varepsilon(\mi\xi)-1][3\varepsilon(\mi\xi)+1]}
 {\varepsilon(\mi\xi)}
\end{eqnarray}
for a purely electric plate and
\begin{eqnarray}
\label{eq:Uethinnret2}
U_e(z_{A})&=&\frac{\hbar\mu_0 d}{64\pi^2z_{A}^2}
 \int_0^\infty\dif\xi\,\xi^2\alpha(\mi\xi)\,
 \frac{[\mu(\mi\xi)-1][3\mu(\mi\xi)+1]}{\mu(\mi\xi)}\,,\\
\label{eq:Umthinnret2}
U_m(z_{A})&=&-\frac{3\hbar d}{64\pi^2\varepsilon_0z_{A}^4}
 \int_0^\infty\dif\xi\,\frac{\beta(\mi\xi)}{c^2}\,
 \frac{\mu^2(\mi\xi)-1}{\mu(\mi\xi)}\\
\end{eqnarray}
for a purely magnetic one. A comparison with the findings of
Sec.~\ref{sec:CPHS} shows that when moving from a thick to a thin
plate, all power laws are increased by one inverse power as a result
of the linear dependence on $d/z_A$. This similarity is due to the
common microscopic origins of both forces (Sec.~\ref{sec:relation}).

The general behaviour of the CP potential for a thin plate being very
similar to that for thick plates, one may expect that repulsive force
components may lead to potential barriers regardless of the plate
thickness. This is confirmed in Fig.~\ref{fig:plate}, where we
display the potential~(\ref{eq:Ueplate}) of a polarisable atom near
plates of various thicknesses. We have chosen sufficiently strong
magnetic properties to ensure the existence of a potential barrier.
It is very low for a thin plate, reaches a maximal height for some
intermediate thickness and then lowers slowly towards the asymptotic
half space value as the thickness is further increased.
%
%
\subsubsection{Atom in front of a sphere}
\label{sec:CPSphere}
\begin{figure}[!t!]
\noindent\vspace*{-2ex}
\begin{center}
\includegraphics[width=0.6\linewidth]{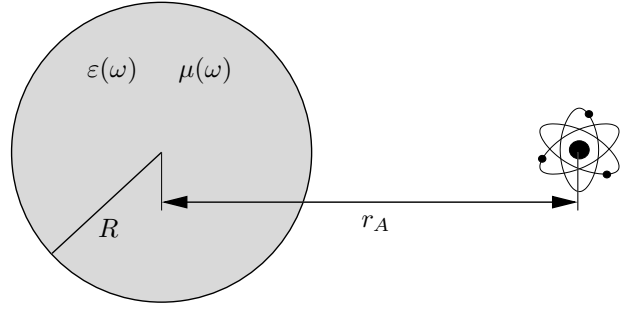}
\end{center}
\caption{
\label{fig:sphereschematic}
An atom interacting with a magneto-electric sphere.}
\end{figure}

As a second application of the general theory, let us consider an
atom placed above at distance $r_A$ from the centre of a
magneto-electric sphere of radius $R$, permittivity
$\varepsilon(\omega)$ and permeability $\mu(\omega)$,
see Fig.~\ref{fig:sphereschematic}.
After substitution of the Green tensor~(\ref{eq:g11}) from
App.~\ref{sec:sphericaldgf}, the electric potential~(\ref{eq:Ue}) in
the presence of the sphere is given by \cite{0017}
\begin{multline}
\label{eq:UeSphere}
U_e(r_A)=-\frac{\hbar\mu_0}{8\pi^2c}
 \int_0^\infty\dif\xi\,
 \xi^3\alpha(\mi\xi)\sum_{l=1}^\infty(2l+1)
 \Biggl\{\bigl[h^{(1)}_l(kr_\mathrm{A})\bigr]^2r_s^l
 +l(l+1)\\
\times \biggl[\frac{h^{(1)}_l(kr_\mathrm{A})}
 {kr_\mathrm{A}}\biggr]^2r_p^l
 +\,\biggl[\frac{1}{kr_\mathrm{A}}\,
 \frac{\dif[r_\mathrm{A}h^{(1)}_l(kr_\mathrm{A})]}
 {\dif r_\mathrm{A}}\biggr]^2r_p^l\Biggr\},
\end{multline}
where the sphere's reflection coefficients for $s$- and $p$-polarised
spherical waves read
\begin{eqnarray}
\label{eq:rsSphere}
r_s^l
 &=&-\frac{\mu(\mi\xi)\bigl[zj_l(z)\bigr]'j_l(z_1)
 -j_l(z)\bigl[z_1j_l(z_1)\bigr]'}
 {\mu(\mi\xi)\bigl[z h_l^{(1)}(z)\bigr]'j_l(z_1)
 -h_l^{(1)}(z)\bigl[z_1j_l(z_1)\bigr]'}\,,\\
\label{eq:rpSphere}
r_p^l
 &=&-\frac{\varepsilon(\mi\xi)\bigl[zj_l(z)\bigr]'j_l(z_1)
 -j_l(z)\bigl[z_1j_l(z_1)\bigr]'}
 {\varepsilon(\mi\xi)\bigl[z h_l^{(1)}(z)\bigr]'j_l(z_1)
 -h_l^{(1)}(z)\bigl[z_1j_l(z_1)\bigr]'}
\end{eqnarray}
[$z=kR$, $k=\mi\xi/c$, $z_1=k_1 R$, $k_1=n(\mi\xi)k$,
$n(\omega)=\sqrt{\varepsilon(\omega)\mu(\omega)}$,
$h_l^{(1)}(z)$: spherical Hankel functions of the first kind,
$j_l(z)$: spherical Bessel functions]. Using the duality invariance
discussed in Sec.~\ref{sec:CPgroundPerturbative}, the
respective magnetic potential~(\ref{eq:Um}) can be obtained by making
the replacements $\alpha\rightarrow\beta/c^2$ and
$\varepsilon\leftrightarrow\mu$, which as in the case of the half
space amounts to an exchange $r_s\leftrightarrow r_p$:
\begin{multline}
\label{eq:UmSphere}
U_m(r_A)=-\frac{\hbar\mu_0}{8\pi^2c}
 \int_0^\infty\dif\xi\,
 \xi^3\,\frac{\beta(\mi\xi)}{c^2}\sum_{l=1}^\infty(2l+1)
 \Biggl\{\bigl[h^{(1)}_l(kr_\mathrm{A})\bigr]^2r_p^l
 +l(l+1)\\
\times \biggl[\frac{h^{(1)}_l(kr_\mathrm{A})}
 {kr_\mathrm{A}}\biggr]^2r_s^l
 +\,\biggl[\frac{1}{kr_\mathrm{A}}\,
 \frac{\dif[r_\mathrm{A}h^{(1)}_l(kr_\mathrm{A})]}
 {\dif r_\mathrm{A}}\biggr]^2r_s^l\Biggr\}.
\end{multline}

In the limit of a large sphere, $R\gg z_A=r_A-R$, the main
contribution to the sums in Eqs.~(\ref{eq:UeSphere}) and
(\ref{eq:UmSphere}) is due to terms with large $n$ and one recovers
the half-space results~(\ref{eq:UeHSnret1})--(\ref{eq:UmHSnret2}), as
expected. For a sufficiently small sphere, $n(0)R\ll r_\mathrm{A}$,
the sums effectively reduce to terms with $l=1$, and one finds
\begin{eqnarray}
\label{eq:UeSmallSphere}
U_e(r_A)&=&-\frac{\hbar}{32\pi^3\varepsilon_0^2r_A^6}
 \int_0^\infty\dif\xi\,\alpha(\mi\xi)\biggl[\alpha_\odot(\mi\xi)
 g(\xi r_A/c)-\frac{\beta_\odot(\mi\xi)}{c^2}\,
 h(\xi r_A/c)\biggr],\\
\label{eq:UmSmallSphere}
U_m(r_A)&=&-\frac{\hbar}{32\pi^3\varepsilon_0^2r_A^6}
 \int_0^\infty\dif\xi\,\frac{\beta(\mi\xi)}{c^2}
 \biggl[\frac{\beta_\odot(\mi\xi)}{c^2}\,
 g(\xi r_A/c)-\alpha_\odot(\mi\xi)
 h(\xi r_A/c)\biggr]
\end{eqnarray}
with $g(x)=2e^{-2x}(3+6x+5x^2+2x^3+x^4)$ and
$h(x)=2x^2e^{-2x}(1+2x+x^2)$, where we have introduced the
small sphere's polarisability \cite{0001}
\begin{equation}
\label{eq:aplhasphere}
\alpha_\odot(\omega)
 =4\pi\varepsilon_0R^3
 \,\frac{\varepsilon(\omega)-1}{\varepsilon(\omega)+2}
\end{equation}
as well as its magnetisability
\begin{equation}
\label{eq:betasphere}
\beta_\odot(\omega)
 =\frac{4\pi R^3}{\mu_0}
 \,\frac{\mu(\omega)-1}{\mu(\omega)+2}\,.
\end{equation}
It is worth noting that in the small-sphere limit the electric and
magnetic properties of the sphere completely separate. Furthermore,
as will be seen in Sec.~\ref{sec:vdWbulk} below, the
potentials~(\ref{eq:UeSmallSphere}) and (\ref{eq:UmSmallSphere}) of an
atom interacting with a small sphere have exactly the same form as
that of two atoms, with the sphere's polarisability and
magnetisability replacing those of the second atom. In the retarded
limit $r_A\gg c/\omega_\mathrm{min}$, the sphere potentials further
reduce to
\begin{eqnarray}
\label{eq:UeSmallSphereret}
U_e(r_A)&=&-\frac{23\hbar c\alpha(0)\alpha_\odot(0)}
 {64\pi^3\varepsilon_0^2r_A^7}
 +\frac{7\hbar \alpha(0)\beta_\odot(0)}
 {64\pi^3\varepsilon_0^2c^2r_A^7}\,,\\
\label{eq:UmSmallSphereret}
U_m(r_A)&=&-\frac{23\hbar c\beta(0)\beta_\odot(0)}
 {64\pi^3\varepsilon_0^2c^4r_A^7}
 +\frac{7\hbar \beta(0)\alpha_\odot(0)}
 {64\pi^3\varepsilon_0^2c^2r_A^7}\,,
\end{eqnarray}
while in the retarded limit $r_A\ll c/\omega_\mathrm{max}$ one has
\begin{eqnarray}
\label{eq:UeSmallSpherenret}
U_e(r_A)&=&-\frac{3\hbar}{16\pi^3\varepsilon_0^2r_A^6}
 \int_0^\infty\dif\xi\,\alpha(\mi\xi)\alpha_\odot(\mi\xi)
+\frac{\hbar}{16\pi^3c^3\varepsilon_0^2r_A^4}
 \int_0^\infty\dif\xi\,\xi^2
 \alpha(\mi\xi)\,\frac{\beta_\odot(\mi\xi)}{c^2}\,,\nonumber \\ \\
\label{eq:UmSmallSpherenret}
U_m(r_A)&=&-\frac{3\hbar}{16\pi^3\varepsilon_0^2r_A^6}
 \int_0^\infty\dif\xi\,\frac{\beta(\mi\xi)}{c^2}\,
 \frac{\beta_\odot(\mi\xi)}{c^2}
 +\frac{\hbar}{16\pi^3c^3\varepsilon_0^2r_A^4}
 \int_0^\infty\dif\xi\,\xi^2\,
 \frac{\beta(\mi\xi)}{c^2}\,\alpha_\odot(\mi\xi). \nonumber \\
\end{eqnarray}
The hierarchy of signs and power laws of these potentials is closely
analogous to that found for a half space or a thin plate: A
polarisable atom is attracted to an electric sphere and repelled from
a magnetic one, while the findings for a magnetisable atom are exactly
opposite. Again, attractive and repulsive potentials follow the same
($1/r_A^7$) power laws in the retarded limit, while for short
distances attractive potentials (with their $1/r_A^6$ power law)
dominate over the repulsive ($1/r_A^4$) ones. 

%
\subsection{Van der Waals forces}
\label{sec:vdWground}
The simultaneous interaction of two atoms with the electromagnetic
field leads to the vdW force between them (for general literature
and reviews, see
\cite{CraigThiru,0696,0375,0261,0608,0620,0323,0832}), which may
sensitively depend on their magneto-electric environment. In close
analogy to the single-atom case, velocity-independent vdW forces
between ground-state atoms can be derived from an atom-field coupling
energy. Starting point is the two-atom generalisation
\begin{equation}
\label{eq:Hdouble}
\hat{H}=\frac{\hat{\vect{p}}_A^2}{2m_A}
 +\frac{\hat{\vect{p}}_B^2}{2m_B}
 +\hat{H}_A^\mathrm{int}+\hat{H}_B^\mathrm{int}
 +\hat{H}_F+\hat{H}_{AF}+\hat{H}_{BF}
\end{equation}
of the Hamiltonian~(\ref{eq:Hsingle}), where the internal dynamics of
the atoms is given by Hamiltonians of the type~(\ref{eq:Hint}) and
each atom individually interacts with the electromagnetic field via
an interaction Hamiltonian of the form~(\ref{eq:HAFnomotion}).
Applying the Born--Oppenheimer approximation by integrating the
internal atomic dynamics for given centre-of-mass positions
$\hat{\vect{r}}_A$, $\hat{\vect{r}}_B$ and momenta
$\hat{\vect{p}}_A$, $\hat{\vect{p}}_B$, one obtains the effective
Hamiltonian
\begin{equation}
\label{eq:Heffdouble}
\hat{H}_\mathrm{eff} 
 =\frac{\hat{\vect{p}}_A^2}{2m_A}+\frac{\hat{\vect{p}}_A^2}{2m_A}
 +E+\Delta E.
\end{equation}
Here, $E$ denotes the energy of the two uncoupled atoms and the field,
and the energy shift
\begin{equation}
\label{eq:DeltaEdouble}
\Delta E=\Delta E_0
 +\Delta E(\hat{\vect{r}}_A)+\Delta E(\hat{\vect{r}}_B)
 +\Delta E(\hat{\vect{r}}_A,\hat{\vect{r}}_B),
\end{equation}
can be separated into a position-independent part (which contains the
Lamb shifts of both atoms), two parts depending only on the positions
of one of the atoms and a genuine two-atom part. The
Hamiltonian~(\ref{eq:Heffdouble}) generates the following equations
of motion for atom $A$:
\begin{gather}
\label{eq:HeffdoubleEOM1}
m_A\dot{\hat{\vect{r}}}_A
 =\frac{1}{\mi\hbar}\Bigl[m_A\hat{\vect{r}}_A,
 \hat{H}_\mathrm{eff}\Bigr]
 =\hat{\vect{p}}_A,\\
\label{eq:HeffdoubleEOM2}
\hat{\vect{F}}_A
 =m_A\ddot{\hat{\vect{r}}}_A
 =\frac{1}{\mi\hbar}\Bigl[m_A\dot{\hat{\vect{r}}}_A, 
 \hat{H}_\mathrm{eff}\Bigr]
 =-\vect{\nabla}_AU(\hat{\vect{r}}_A)
  -\vect{\nabla}_AU(\hat{\vect{r}}_A,\hat{\vect{r}}_B)
\end{gather}
(similarly for atom $B$). The atom is thus subject to two forces, the
CP force, which as discussed in the previous Sec.~\ref{sec:CPground}
can be derived from the CP potential~(\ref{eq:CPpotentialdef}), and
the vdW force which is due to the additional atom. The associated vdW
potential is given by the two-atom part of the energy shift
\begin{equation}
\label{eq:vdWpotentialdef}
U(\hat{\vect{r}}_A,\hat{\vect{r}}_B)
 =\Delta E(\hat{\vect{r}}_A,\hat{\vect{r}}_B)
\end{equation}
and it describes not only the direct, free-space interaction of the
two atoms but also accounts for modifications of this interactions
due to the presence of magnetoelectrics. As a consequence, Newton's
third law $\hat{\vect{F}}_{AB}=-\hat{\vect{F}}_{BA}$ does not
necessarily hold for the vdW force
$\hat{\vect{F}}_{AB}=-\vect{\nabla}_AU(\hat{\vect{r}}_A,\hat{\vect{r}}
_B)$,
due to the contribution of the bodies to the momentum balance.
%
%
\subsubsection{Perturbation theory}
\label{sec:vdWgroundPerturbative}
\begin{figure}[!t!]
\begin{center}
\includegraphics[width=0.8\linewidth]{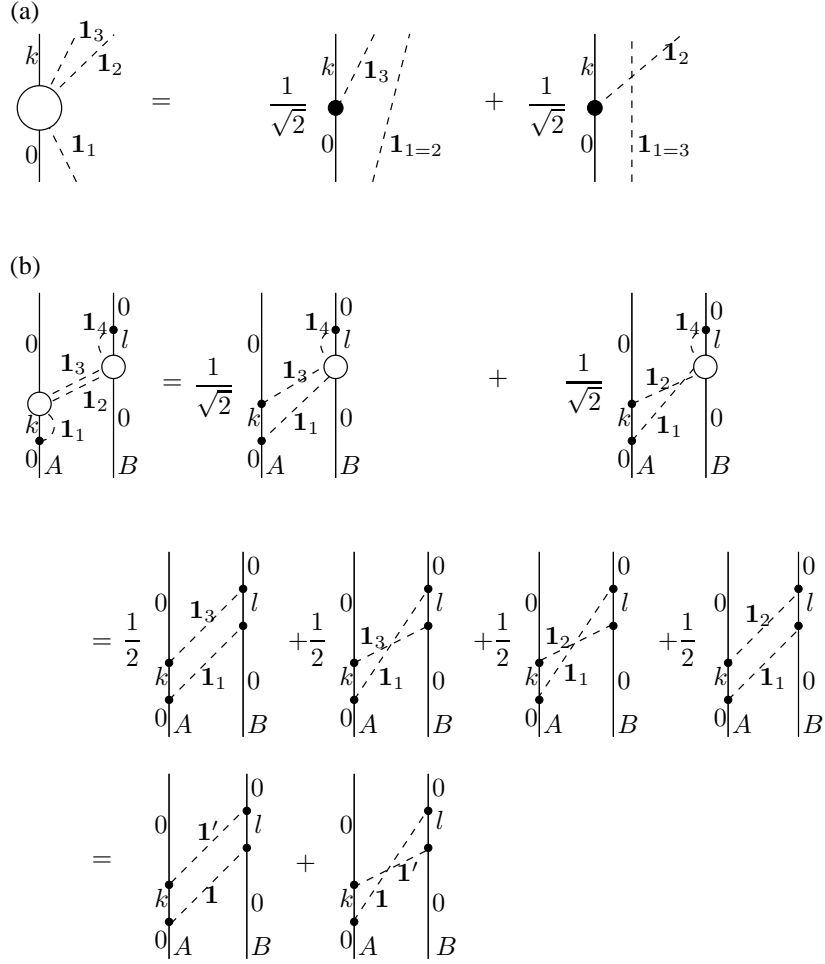}
\end{center}
\caption{
\label{fig:secondorder}
Schematic representation of three-photon interactions (a) and the
contribution (1) to the fourth-order energy shift (b). Solid lines
represent atomic states and dashed lines stand for photons. We do not
distinguish electric and magnetic interactions.
}
\end{figure}

In close analogy to the single-atom case, the vdW potential can for
sufficiently weak atom-field coupling be obtained from a
perturbative calculation of the energy shift. With each atom being
linearly coupled to the electromagnetic field via an
interaction~(\ref{eq:HAFpert}), two-atom contributions start to
appear in the fourth-order energy shift
\begin{multline}
\label{eq:fourthorder}
\Delta E
=\sum_{I,II,III\neq 0}
 \frac{\langle G|\hat{H}_{AF}\!+\!\hat{H}_{BF}|III\rangle
 \langle III|\hat{H}_{AF}\!+\!\hat{H}_{BF}|II\rangle}
 {(E_G-E_{III})}\\
\times\frac{\langle II|\hat{H}_{AF}
 \!+\!\hat{H}_{BF}|I\rangle
 \langle I|\hat{H}_{AF}
 \!+\!\hat{H}_{BF}|0\rangle}
 {(E_G-E_{II})(E_G-E_{I})}\,,
\end{multline}
where $|G\rangle$ $\!=$ $\!|0_A\rangle|0_B\rangle|\{0\}\rangle$
is now the (uncoupled) ground state of 
$\hat{H}_A^\mathrm{int}+\hat{H}_B^\mathrm{int}+\hat{H}_\mathrm{F}$.
In order to give a nonvanishing contribution to the energy shift, the
intermediate states $|I\rangle$ and $|III\rangle$ must be such that
one of the atoms and a single quantum of the fundamental fields are
excited, while three possibilities exist for the intermediate states
$|II\rangle$: Either both atoms are in the ground state and two field
quanta are excited
[$|\vec{1}_\lambda(\vect{r},\omega)
\vec{1}_{\lambda'}(\vect{r}',\omega')\rangle
=\frac{1}{\sqrt{2}}\hat{\vect{f}}_\lambda^\dagger(\vect{r},\omega)
\hat{\vect{f}}_{\lambda'}^\dagger(\vect{r}',\omega')|\{0\}\rangle$],
or
both atoms are excited and the field is in its ground state or both
atoms and two field quanta are excited. When invoking the additional
requirement that each atom must undergo exactly two transitions,
there is a total of ten possible combinations of intermediate states,
which are listed in Tab.~\ref{tab:intermediatestates}.
\begin{table}[!t!]
\begin{center}
\begin{tabular}{clll}
\hline
 Case  & $|I\rangle$    &
 \hspace{1ex}
 $|II\rangle$   &
 \hspace{1ex}
 $\hspace{-1ex}
 |III\rangle$ \\
\hline
($1$)
& $|k_A,0_B\rangle |\vect{1}_1\rangle$
      &\hspace{1ex} $|0_A,0_B\rangle
        |\vect{1}_2,\vect{1}_3\rangle$
      & \hspace{1ex}$|0_A,l_B\rangle |\vect{1}_4\rangle$ \\
($2$)
      & $|k_A,0_B\rangle |\vect{1}_1\rangle$
      & \hspace{2ex}$|k_A,l_B\rangle |\{0\}\rangle$
      &\hspace{1ex}$|k_A,0_B\rangle |\vect{1}_2\rangle$\\
($3$)
      &$|k_A,0_B\rangle |\vect{1}_1\rangle$
      & \hspace{2ex}$|k_A,l_B\rangle |\{0\}\rangle$
      & \hspace{1ex}$|0_A,l_B\rangle |\vect{1}_2\rangle$\\
($4$)
      &$|k_A,0_B\rangle |\vect{1}_1\rangle$
      & \hspace{2ex}$|k_A,l_B\rangle
        |\vect{1}_2,\vect{1}_3\rangle$
      & \hspace{1ex}$|k_A,0_B\rangle |\vect{1}_4\rangle$\\
($5$)
      & $|k_A,0_B\rangle |\vect{1}_1\rangle$
      & \hspace{2ex}$|k_A,l_B\rangle
        |\vect{1}_2,\vect{1}_3\rangle$
      & \hspace{1ex}$|0_A,l_B\rangle |\vect{1}_4\rangle$\\
($6$)
     & $|0_A,l_B\rangle |\vect{1}_1\rangle$
      & \hspace{2ex}$|0_A,0_B\rangle
        |\vect{1}_2,\vect{1}_3\rangle$
      & \hspace{1ex}$|k_A,0_B\rangle |\vect{1}_4\rangle$\\
($7$)
      & $|0_A,l_B\rangle |\vect{1}_1\rangle$
      &\hspace{1ex} $|k_A,l_B\rangle |\{0\}\rangle$
      & \hspace{1ex}$|k_A,0_B\rangle |\vect{1}_2\rangle$\\
($8$)
      & $|0_A,l_B\rangle |\vect{1}_1\rangle$
      & \hspace{1ex} $|k_A,l_B\rangle |\{0\}\rangle$
      & \hspace{1ex}$|0_A,l_B\rangle |\vect{1}_2\rangle$\\
($9$)
      & $|0_A,l_B\rangle |\vect{1}_1\rangle$
      & \hspace{2ex}$|k_A,l_B\rangle
        |\vect{1}_2,\vect{1}_3\rangle$
      & \hspace{1ex}$|k_A,0_B\rangle |\vect{1}_4\rangle$\\
($10$)
      &$|0_A,l_B\rangle |\vect{1}_1\rangle$
      & \hspace{2ex}$|k_A,l_B\rangle
        |\vect{1}_2,\vect{1}_3\rangle$
      & \hspace{1ex}$|0_A,l_B\rangle
        |\vect{1}_4\rangle$\\
\hline
\end{tabular}
\end{center}
\caption{
\label{tab:intermediatestates}
Intermediate states contributing to the two-atom vdW interaction,
where we have used the short-hand notations $|\vect{1}_\mu\rangle=
|\mathit{1}_{\lambda_\mu i_\mu}(\vect{r}_\mu,\omega_\mu)\rangle$,
$|\vect{1}_\mu\vect{1}_\nu\rangle=
|\mathit{1}_{\lambda_\mu i_\mu}(\vect{r}_\mu,\omega_\mu)
\mathit{1}_{\lambda_\nu i_\nu}(\vect{r}_\nu,\omega_\nu)\rangle$.
}
\end{table}
The fourth-order energy shift~(\ref{eq:fourthorder}) thus involves
both transitions from zero- to single-quantum excitations and those
between single- and two-quantum excitations of the electromagnetic
field. The former are given by Eqs.~(\ref{eq:dEElement}) and
(\ref{eq:mBElement}), the latter can, upon recalling the
field expansions~(\ref{eq:Eexpansion}) and (\ref{eq:Bexpansion}), be
found to be  
\begin{align}
\label{eq:dEElement2}
&\langle k_A|
 \langle\mathit{1}_{\lambda_1i_1}(\vect{r}_1,\omega_1)|
 \hat{\vect{d}}_A\sprod\hat{\vect{E}}(\vect{r}_A)|
 \textit{1}_{\lambda_2i_2}(\vect{r}_2,\omega_2)
 \textit{1}_{\lambda_3i_3}(\vect{r}_3,\omega_3)\rangle|0_A\rangle
 \nonumber\\
&\quad=\frac{\delta_{(13)}}{\sqrt{2}}
 \bigl[\vect{d}_{A}^{k0}\sprod
 \ten{G}_{\lambda_2}(\vect{r}_A,\vect{r}_2,\omega_2)\bigr]_{i_2}
 +\frac{\delta_{(12)}}{\sqrt{2}}
 \bigl[\vect{d}_{A}^{k0}\sprod
 \ten{G}_{\lambda_3}(\vect{r}_A,\vect{r}_3,\omega_3)\bigr]_{i_3}
 \,,
\end{align}
\begin{align}
\label{eq:mBElement2}
&\langle k_A|
 \langle\mathit{1}_{\lambda_1i_1}(\vect{r}_1,\omega_1)|
 \hat{\vect{m}}_A\sprod\hat{\vect{B}}(\vect{r}_A)|
 \textit{1}_{\lambda_2i_2}(\vect{r}_2,\omega_2)
 \textit{1}_{\lambda_3i_3}(\vect{r}_3,\omega_3)\rangle|0_A\rangle
 \nonumber\\
&\quad=\frac{\delta_{(13)}}{\sqrt{2}}\,
 \frac{\bigl\{\vect{m}_{A}^{k0}\!\sprod\!\bm{\nabla}_A\!\vprod\!
 \ten{G}_{\lambda_2}(\vect{r}_A,\vect{r}_2,\omega_2)
 \bigr\}_{i_2}}{\mi\omega_2}
 +\frac{\delta_{(12)}}{\sqrt{2}}\,
 \frac{\bigl\{\vect{m}_{A}^{k0}\!\sprod\!\bm{\nabla}_A\!\vprod\!
 \ten{G}_{\lambda_3}(\vect{r}_A,\vect{r}_3,\omega_3)
 \bigr\}_{i_3}}{\mi\omega_3}
\end{align}
with 
\begin{equation}
\label{eq:multidelta}
\delta_{(\mu\nu)}
 =\delta_{\lambda_\mu\lambda_\nu}
 \delta_{i_\mu i_\nu}(\vect{r}_\mu-\vect{r}_\nu)
 \delta(\omega_\mu-\omega_\nu).
\end{equation}
Comparison with Eqs.~(\ref{eq:dEElement}) and (\ref{eq:mBElement})
reveals that the matrix element for the one- to two-photon transition
is equivalent to the two possible combinations of one photon being
created and one propagating freely. This is schematically depicted in
Fig.~\ref{fig:secondorder}(a).

The various contributions to the vdW potential can be calculated by
substituting the intermediate states from
Tab.~\ref{tab:intermediatestates} and the matrix
elements~(\ref{eq:dEElement}), (\ref{eq:mBElement}),
(\ref{eq:dEElement2}) and (\ref{eq:mBElement2}) into
Eq.~(\ref{eq:fourthorder}). Let us begin with the intermediate-state
combination (1) from Tab.~\ref{tab:intermediatestates}, the
calculation of which is schematically represented in
Fig.~\ref{fig:secondorder}(b). After expanding the product in
Eq.~(\ref{eq:fourthorder}), evaluating the delta
functions~(\ref{eq:multidelta}) and making use of the integral
relation~(\ref{Glambdaintegral}), one obtains
\begin{align}
\label{eq:deltaE(1)}
&\Delta E_{(1)}=-\frac{\mu_0^2}{\hbar\pi^2}
     \sum_{k,l}\int_0^\infty\dif\omega\int_0^\infty
     \dif\omega'\left(\frac{1}{D_{(1a)}}+
     \frac{1}{D_{(1b)}}\right) 
\nonumber\\
&
\times\biggl\{\omega^2\omega^{\prime 2}\vect{d}_A^{0k}\sprod
 \operatorname{Im}\ten{G}(\vect{r}_A,\vect{r}_B,\omega)
 \sprod\vect{d}_B^{0l}
 \vect{d}_A^{0k}\sprod
 \operatorname{Im}\ten{G}(\vect{r}_A,\vect{r}_B,\omega')
 \sprod\vect{d}_B^{0l}\nonumber\\
&+\omega\omega'\left[
 \vect{d}_A^{0k}\sprod\mathrm{Im}\,
 \ten{G}(\vect{r}_A,\vect{r}_B,\omega)\vprod
 \overleftarrow{\grad}'\sprod\vect{m}_B^{0l}\right]
 \left[\vect{m}_B^{0l}\sprod
 \curl\mathrm{Im}\,\ten{G}(\vect{r}_B,\vect{r}_A,\omega')
 \sprod\vect{d}_A^{0k}\right]\nonumber\\
&+\omega\omega'\left[
 \vect{m}_A^{0k}\sprod\curl\mathrm{Im}\,
 \ten{G}(\vect{r}_A,\vect{r}_B,\omega)
 \sprod\vect{d}_B^{0l}\right]
 \left[\vect{d}_B^{0l}\sprod
 \mathrm{Im}\,\ten{G}(\vect{r}_B,\vect{r}_A,\omega')
 \vprod\overleftarrow{\grad}'\sprod\vect{m}_A^{0k}\right]
 \nonumber\\
&+\!\left[
 \vect{m}_A^{0k}\sprod\curl\mathrm{Im}\,
 \ten{G}(\vect{r}_A,\vect{r}_B,\omega)\vprod\overleftarrow{\grad}'
 \sprod\vect{m}_B^{0l}\right]
 \left[\vect{m}_B^{0l}\sprod\curl
 \mathrm{Im}\,\ten{G}(\vect{r}_B,\vect{r}_A,\omega')
 \overleftarrow{\grad}'\sprod\vect{m}_A^{0k}\right].
\end{align}
Note that the single- to two-photon transition~(\ref{eq:dEElement2})
[or (\ref{eq:mBElement2})] with its two possible processes
[Fig.~\ref{fig:secondorder}(a)] enters the energy shift
quadratically, so after expanding the product, four different terms
arise, see Fig.~\ref{fig:secondorder}(b). They can be grouped into
pairs of equal terms, so one ends up with only two distinct
contributions. After assuming real dipole matrix elements, these two
only differ in their frequency denominators $D_{(1a)}$ and $D_{(1b)}$
which are given in Tab.~\ref{tab:denominators}.
\begin{table*}[!t!]
\begin{center}
\begin{tabular}{cll}
\hline
 Case  & Sign & Denominator\\
\hline
($1$)&$+$  
      & $D_{\mathrm {(1a)}}=(\omega_A^{k0}+\omega)
      (\omega+\omega')(\omega_B^{l0}+\omega')$,  \\
      {}
&$+$
      & $D_{(1b)}=(\omega_A^{k0}+\omega)
      (\omega+\omega')(\omega_B^{l0}+\omega)$  \\
($2$)&$\pm$ 
      & $D_{(2)}=(\omega_A^{k0}+\omega)
         (\omega_A^{k0}+\omega_B^{l0})
            (\omega_A^{k0}+\omega')$\\
($3$)&$+$
      & $D_{(3)}=(\omega_A^{k0}+\omega)
        (\omega_A^{k0}+\omega_B^{l0})
        (\omega_B^{l0}+\omega')$\\
($4$)&$\pm$
      & $D_{(4a)}=(\omega_A^{k0}+\omega)
         (\omega_A^{k0}+\omega_B^{l0}+
         \omega+\omega')
     (\omega_A^{k0}+\omega')$\\
($5$)&$\pm$
      & $D_{(5a)}=(\omega_A^{k0}+\omega)
         (\omega_A^{k0}+\omega_B^{l0}+\omega'+
         \omega')
     (\omega_B^{k0}+\omega)$\\
($6$)&$+$
            & $D_{(6a)}=(\omega_B^{l0}+\omega)
      (\omega+\omega')(\omega_A^{k0}+\omega')$,  \\
&$+$
           & $D_{(6b)}=(\omega_B^{l0}+\omega)
      (\omega+\omega')(\omega_A^{k0}+\omega)$  \\
($7$)&$+$
      & $D_{(7)}=(\omega_B^{l0}+\omega)
        (\omega_A^{k0}+\omega_B^{l0})
        (\omega_A^{k0}+\omega')$\\
($8$)&$\pm$
        & $D_{(8)}=(\omega_B^{l0}+\omega)
         (\omega_A^{k0}+\omega_B^{l0})
            (\omega_B^{l0}+\omega')$\\
($9$)&$\pm$
      & $D_{(9a)}=(\omega_B^{l0}+\omega)
         (\omega_A^{k0}+\omega_B^{l0}+\omega+\omega')
     (\omega_A^{k0}+\omega)$\\
($10$)&$\pm$
      & $D_{(10a)}=(\omega_B^{l0}+\omega)
         (\omega_A^{k0}+\omega_B^{l0}+\omega+\omega')
     (\omega_B^{l0}+\omega')$\\
\hline
\end{tabular}
\end{center}
\caption{
\label{tab:denominators}
Signs and frequency denominators associated with the
intermediate-state combinations given in
Tab.~\ref{tab:intermediatestates}.
}
\end{table*}
%
The energy shift~(\ref{eq:deltaE(1)}) contains electric-electric
contributions (where both atoms undergo purely electric transitions),
magnetic-magnetic contributions (where both atoms undergo magnetic
transitions) and mixed electric-magnetic and magnetic-electric
ones. As in the single-atom case, we have assumed nonchiral atoms and
discarded all those contributions where at least one atom undergoes an
electric and a magnetic transition.

\begin{figure}[!t!]
\begin{center}
\includegraphics[width=\linewidth]{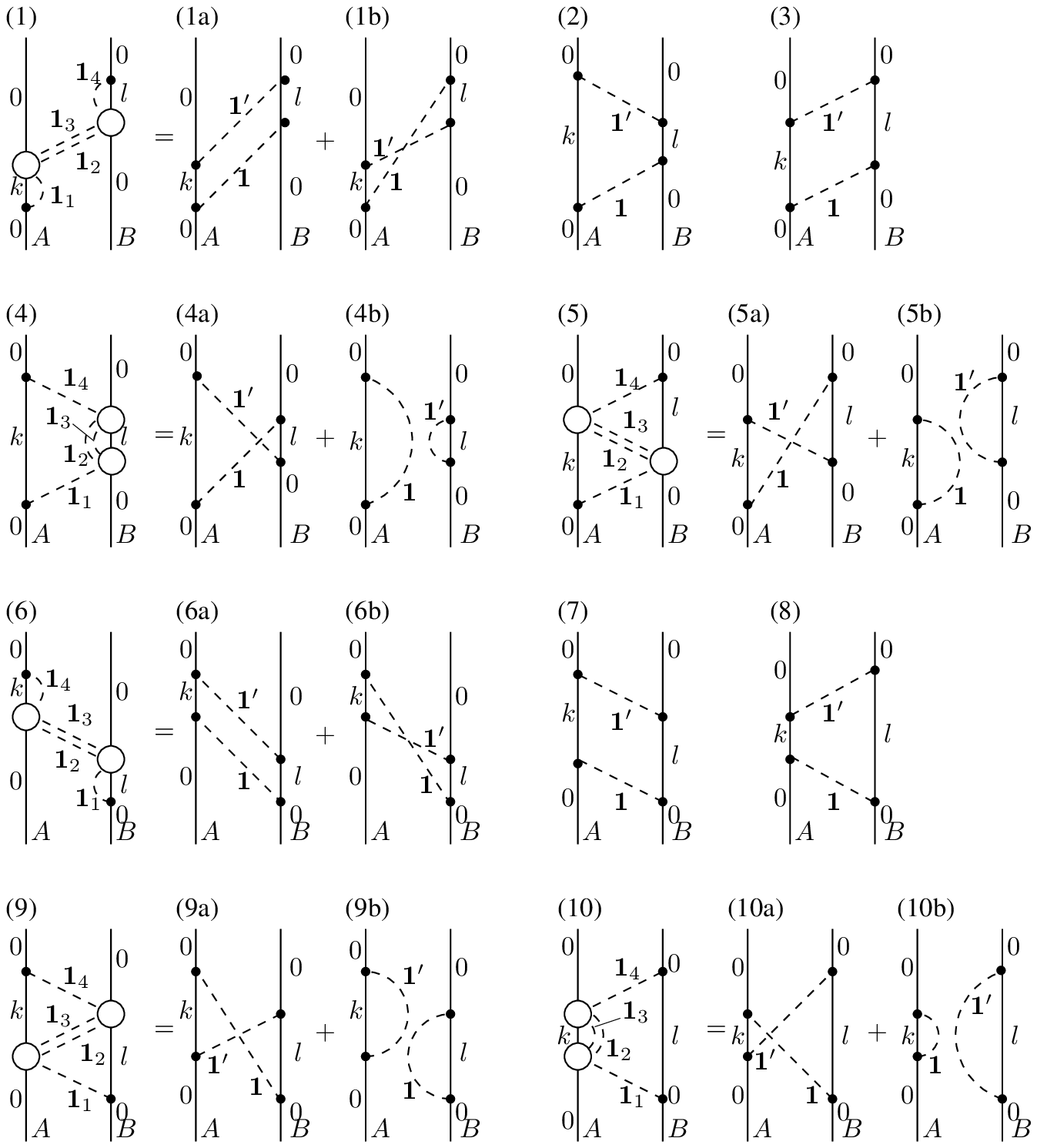}
\end{center}
\caption{
\label{fig:fourthorder}
Schematic representation of all two-atom contributions to the
fourth-order energy shift.
}
\end{figure}

The contributions to the energy shift which are associated with the
remaining intermediate-state combinations listed in
Tab.~\ref{tab:intermediatestates} can be calculated in a similar way;
all contributions are depicted in Fig.~\ref{fig:fourthorder}
(cf.~also Ref.~\cite{CraigThiru}). 
The contributions (4)--(6), (9) and (10) contain two-photon states, so
they correspond to two distinct terms, similar to the
contribution~(1) studied above. However, for the contributions (4),
(5), (9) and (10), one of those two terms separates into two
single-atom processes [diagrams (4b), (5b), (9b) and (10b) of
Fig.~\ref{fig:fourthorder}] and does hence not contribute to the vdW
potential. For real dipole matrix elements, all genuine two-atom
contributions only differ from Eq.~(\ref{eq:deltaE(1)}) by the
frequency denominators $D_{(i)}$ and possibly also by the sign of the
terms in the third and fourth lines of the equation; signs and
denominators are listed in Tab.~\ref{tab:denominators}. The total vdW
potential $U(\vect{r}_A,\vect{r}_B)
 =\Delta E(\vect{r}_A,\vect{r}_B)=\sum_i\Delta E_{(i)}$
can be obtained by summing all contributions with the aid of the
identity
\begin{align}
\label{eq:denominatorsum}
&\int_0^\infty\dif\omega\int_0^\infty
     \dif\omega'\biggl[
 \frac{1}{D_{(1a)}}+\frac{1}{D_{(1b)}}\pm\frac{1}{D_{(2)}}
 +\frac{1}{D_{(3)}}\pm\frac{1}{D_{(4)}}\pm\frac{1}{D_{(5)}}
  \nonumber\\
&\qquad+\frac{1}{D_{(6a)}}+\frac{1}{D_{(6b)}}
 +\frac{1}{D_{(7)}}\pm\frac{1}{D_{(8)}}\pm\frac{1}{D_{(9)}}
 \pm\frac{1}{D_{(10)}}\biggr] f(\omega,\omega') \nonumber\\
&=\int_0^\infty\dif\omega\int_0^\infty\dif\omega'
\frac{4(\omega_A^k+\omega_B^l+\omega)}
{(\omega_A^k+\omega_B^l)(\omega_A^k+\omega)(\omega_{B}^l+\omega)}
\left( \frac{1}{\omega+\omega'}\mp\frac{1}{\omega-\omega'}\right)
f(\omega,\omega')
\end{align}
which holds because the remaining parts $f(\omega,\omega')$ of the
integrands in Eq.~(\ref{eq:deltaE(1)}) are symmetric with respect to
an interchange of $\omega$ and $\omega'$.
The $\omega'$-integral can be performed with the help of
\begin{multline}
\label{eq:omega'integral}
\int_0^\infty \dif\omega'\omega'
 \left(\frac{1}{\omega+\omega'} + \frac{1}{\omega-\omega'}\right)
 \operatorname{Im}\,\ten{G}(\vect{r}_B,\vect{r}_A,\omega')\\
 =-\frac{\pi}{2}\,\omega [\ten{G}(\vect{r}_B,\vect{r}_A,\omega)+
 \ten{G}^\ast(\vect{r}_B,\vect{r}_A,\omega)] \,.
\end{multline}
Transforming the $\omega$-integral to run along the positive imaginary
axis, one finds \cite{0375,0036,0092,0009,Hassan08} 
\begin{equation}
\label{eq:vdWpotential}
U(\vect{r}_A,\vect{r}_B) =
U_{ee}(\vect{r}_A,\vect{r}_B) +
U_{em}(\vect{r}_A,\vect{r}_B)
+U_{me}(\vect{r}_A,\vect{r}_B) +
U_{mm}(\vect{r}_A,\vect{r}_B),
\end{equation}
\begin{align}
\label{eq:Uee}
U_{ee}(\vect{r}_A,\vect{r}_B)
 =&-\frac{\hbar\mu_0^2}{2\pi}\int_0^\infty
 \dif\xi\,\xi^4
 \trace\bigl[\bm{\alpha}_A(\mi\xi)\sprod
 \ten{G}(\vect{r}_A,\vect{r}_B,\mi\xi)
 \sprod\bm{\alpha}_B(\mi\xi)\sprod
 \ten{G}(\vect{r}_B,\vect{r}_A,\mi\xi)\bigr]\nonumber\\
=&-\frac{\hbar\mu_0^2}{2\pi}\int_0^\infty
 \dif\xi\,\xi^4\alpha_A(\mi\xi)\alpha_B(\mi\xi)
 \trace\bigl[
 \ten{G}(\vect{r}_A,\vect{r}_B,\mi\xi)\sprod
 \ten{G}(\vect{r}_B,\vect{r}_A,\mi\xi)\bigr],
\end{align}
\begin{align}
\label{eq:Uem}
&U_{em}(\vect{r}_A,\vect{r}_B)\nonumber\\
&=-\frac{\hbar\mu_0^2}{2\pi}\int_0^\infty
 \dif\xi\,\xi^2
 \trace\Bigl[\bm{\alpha}_A(\mi\xi)\sprod
 \ten{G}(\vect{r}_A,\vect{r}_B,\mi\xi)\vprod\overleftarrow{\grad}'
 \sprod\bm{\beta}_B(\mi\xi)\sprod\curl
 \ten{G}(\vect{r}_B,\vect{r}_A,\mi\xi)\Bigr]\nonumber\\
&=-\frac{\hbar\mu_0^2}{2\pi}\int_0^\infty
 \dif\xi\,\xi^2\alpha_A(\mi\xi)\beta_B(\mi\xi)
 \trace\Bigl\{ \bigl[
 \ten{G}(\vect{r}_A,\vect{r}_B,\mi\xi)\vprod
 \overleftarrow{\grad}'\bigr] \sprod \bigl[
 \curl\ten{G}(\vect{r}_B,\vect{r}_A,\mi\xi) \bigr]\Bigr\},
\end{align}
\begin{align}
\label{eq:Ume}
&U_{me}(\vect{r}_A,\vect{r}_B)\nonumber\\
&=-\frac{\hbar\mu_0^2}{2\pi}\int_0^\infty
 \dif\xi\,\xi^2
 \trace\Bigl[\bm{\beta}_A(\mi\xi)\sprod\curl
 \ten{G}(\vect{r}_A,\vect{r}_B,\mi\xi)
 \sprod\bm{\alpha}_B(\mi\xi)\sprod
 \ten{G}(\vect{r}_B,\vect{r}_A,\mi\xi)
 \vprod\overleftarrow{\grad}'\Bigr]\nonumber\\
&=-\frac{\hbar\mu_0^2}{2\pi}\int_0^\infty
 \dif\xi\,\xi^2\beta_A(\mi\xi)\alpha_B(\mi\xi)
 \trace\Bigl[
 \curl\ten{G}(\vect{r}_A,\vect{r}_B,\mi\xi)
 \sprod
 \ten{G}(\vect{r}_B,\vect{r}_A,\mi\xi)
 \vprod\overleftarrow{\grad}'\Bigr],
\end{align}
\begin{align}
\label{eq:Umm}
&U_{mm}(\vect{r}_A,\vect{r}_B)\nonumber\\
&=-\frac{\hbar\mu_0^2}{2\pi}\int_0^\infty
 \dif\xi
 \trace\Bigl[\bm{\beta}_A(\mi\xi)\!\sprod\!
 \grad\!\vprod\!
 \ten{G}(\vect{r}_A,\vect{r}_B,\mi\xi)
 \!\vprod\!\overleftarrow{\grad}'
 \!\sprod\!\bm{\beta}_B(\mi\xi)\!\sprod\!
 \grad\!\vprod\!\ten{G}(\vect{r}_B,\vect{r}_A,\mi\xi)
 \!\vprod\!\overleftarrow{\grad}'\Bigr]\nonumber\\
&=-\frac{\hbar\mu_0^2}{2\pi}\int_0^\infty
 \dif\xi\beta_A(\mi\xi)\beta_B(\mi\xi) \nonumber \\ & \qquad\times
 \trace\Bigl\{ \bigl[
 \curl
 \ten{G}(\vect{r}_A,\vect{r}_B,\mi\xi)
 \vprod\overleftarrow{\grad}' \bigr]
 \sprod \bigl[ \curl
 \ten{G}(\vect{r}_B,\vect{r}_A,\mi\xi)
 \vprod\overleftarrow{\grad}' \bigr]\Bigr\},
\end{align}
where the atomic polarisabilities and magnetisabilities are given by
Eqs.~(\ref{eq:alpha}) and (\ref{eq:beta}) and the respective second
lines of these equalities hold for isotropic atoms. In close
similarity to the single-atom CP
potential~(\ref{eq:Usingle})--(\ref{eq:Um}), the two-atom vdW 
potential can hence be expressed in terms of the atomic response
functions and the Green tensor of the electromagnetic field, where the
latter connects the positions of the two atoms.

In order to treat the interaction of atoms which are embedded in a
magnetoelectric, we have to generalise the above results by taking
into account local-field effects. In close analogy to the single-atom
case, this can be achieved via the real-cavity model by substituting
the respective local-field corrected Green tensors (\ref{eq:G2loc}),
(\ref{eq:G2magelloc})--(\ref{eq:G2magmagloc}) from
App.~\ref{sec:localfield} into Eqs.~(\ref{eq:Uee})--(\ref{eq:Umm})
resulting in \cite{0739,Hassan08}
\begin{align}
\label{eq:Ueeloc}
U_{ee}(\vect{r}_A,\vect{r}_B)
 =&-\frac{\hbar\mu_0^2}{2\pi}\int_0^\infty
 \dif\xi\,\xi^4\alpha_A(\mi\xi)\alpha_B(\mi\xi)
 \biggl[\frac{3\varepsilon_A(\mi\xi)}{2\varepsilon_A(\mi\xi)+1}
 \biggr]^2
 \biggl[\frac{3\varepsilon_B(\mi\xi)}{2\varepsilon_B(\mi\xi)+1}
 \biggr]^2 \nonumber\\
&\times\trace\bigl[
 \ten{G}(\vect{r}_A,\vect{r}_B,\mi\xi)\sprod
 \ten{G}(\vect{r}_B,\vect{r}_A,\mi\xi)\bigr],
\end{align}
\begin{align}
\label{eq:Uemloc}
U_{em}(\vect{r}_A,\vect{r}_B)
=&-\frac{\hbar\mu_0^2}{2\pi}\int_0^\infty
 \dif\xi\,\xi^2\alpha_A(\mi\xi)\beta_B(\mi\xi)
 \biggl[\frac{3\varepsilon_A(\mi\xi)}{2\varepsilon_A(\mi\xi)+1}
 \biggr]^2
 \biggl[\frac{3}{2\mu_B(\mi\xi)+1}\biggr]^2\nonumber\\
&\times \trace\Bigl\{ \bigl[
 \ten{G}(\vect{r}_A,\vect{r}_B,\mi\xi)\vprod
 \overleftarrow{\grad}'\bigr] \sprod \bigl[
 \curl\ten{G}(\vect{r}_B,\vect{r}_A,\mi\xi) \bigr] \Bigr\},
\end{align}
\begin{align}
\label{eq:Umeloc}
U_{me}(\vect{r}_A,\vect{r}_B)
=&-\frac{\hbar\mu_0^2}{2\pi}\int_0^\infty
 \dif\xi\,\xi^2\beta_A(\mi\xi)\alpha_B(\mi\xi)
 \biggl[\frac{3}{2\mu_A(\mi\xi)+1}\biggr]^2
 \biggl[\frac{3\varepsilon_B(\mi\xi)}{2\varepsilon_B(\mi\xi)+1}
 \biggr]^2
 \nonumber\\
&\times \trace\Bigl[
 \curl\ten{G}(\vect{r}_A,\vect{r}_B,\mi\xi)
 \sprod 
 \ten{G}(\vect{r}_B,\vect{r}_A,\mi\xi)
 \vprod\overleftarrow{\grad}' \Bigr],
\end{align}
\begin{align}
\label{eq:Ummloc}
U_{mm}(\vect{r}_A,\vect{r}_B)
=&-\frac{\hbar\mu_0^2}{2\pi}\int_0^\infty
 \dif\xi\beta_A(\mi\xi)\beta_B(\mi\xi)
 \biggl[\frac{3}{2\mu_A(\mi\xi)+1}\biggr]^2
 \biggl[\frac{3}{2\mu_B(\mi\xi)+1}\biggr]^2\nonumber\\
&\times \trace\Bigl\{ \bigl[ \curl
 \ten{G}(\vect{r}_A,\vect{r}_B,\mi\xi)
 \vprod\overleftarrow{\grad}' \bigr] 
 \sprod \bigl[\curl
 \ten{G}(\vect{r}_B,\vect{r}_A,\mi\xi)
 \vprod\overleftarrow{\grad}' \bigr] \Bigr\},
\end{align}
where $\varepsilon_A(\omega)=\varepsilon(\vect{r}_A,\omega)$ and
$\mu_A(\omega)=\mu(\vect{r}_A,\omega)$ (and similarly for atom $B$). 

The behaviour of the vdW potential under a duality transformation 
$\alpha\leftrightarrow\beta/c^2$, $\varepsilon\leftrightarrow\mu$
follows immediately from the respective transformation laws of the
Green tensor derived in Apps.~\ref{sec:dualgreen} and
\ref{sec:localfield}. Using
Eqs.~(\ref{dgftrans1})--(\ref{dgftrans4}), one sees that the
free-space potentials~(\ref{eq:Uee})--(\ref{eq:Umm}) transform into
one another according to $U_{ee}\leftrightarrow U_{mm}$,
$U_{em}\leftrightarrow U_{me}$, so that the total vdW
potential~(\ref{eq:vdWpotential}) is duality invariant
\cite{BuhmannScheel08}. The same is true for embedded atoms when
including local-field corrections, as
Eqs.~(\ref{dgfloctrans3})--(\ref{dgfloctrans6}) show. The duality
invariance can be exploited when applying the general potentials to
specific geometries; e.g., after calculation of $U_{ee}$ for a
certain magnetoelectric body, $U_{mm}$ can be obtained by replacing
$\alpha\rightarrow\beta/c^2$ and exchanging
$\varepsilon\leftrightarrow\mu$.
%
%
\subsubsection{Two atoms inside a bulk medium}
\label{sec:vdWbulk}
\begin{figure}[!t!]
\noindent\vspace*{-2ex}
\begin{center}
\includegraphics[width=0.6\linewidth]{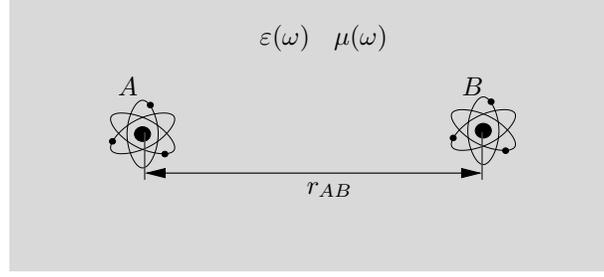}
\end{center}
\caption{
\label{fig:bulkschematic}
Interaction of two atoms embedded in a bulk magneto-dielectric
medium.}
\end{figure}

Let us first consider the vdW potential of two isotropic atoms that
are embedded in an infinite homogeneous bulk medium of permittivity
$\varepsilon(\omega)$ and permeability $\mu(\omega)$
(Fig.~\ref{fig:bulkschematic}). 
Using the bulk Green tensor~(\ref{eq:bulkG}), the local-field
corrected potentials~(\ref{eq:Ueeloc}) and (\ref{eq:Uemloc}) take the
forms \cite{0739,Hassan08,0669,0829}
\begin{align}
\label{eq:Ueebulk}
U_{ee}(\vect{r}_A,\vect{r}_B)
 =&-\frac{\hbar}{16\pi^3\varepsilon_0^2r_{AB}^6}
 \int_0^\infty\dif\xi\,\alpha_A(\mi\xi)\alpha_B(\mi\xi)
 \,\frac{81\varepsilon^2(\mi\xi)}
 {[2\varepsilon(\mi\xi)+1]^4}\,
 g[n(\mi\xi)\xi r_{AB}/c],\\
\label{eq:Uembulk}
U_{em}(\vect{r}_A,\vect{r}_B)
 =&\,\frac{\hbar\mu_0^2}{16\pi^3r_{AB}^4}
 \int_0^\infty\dif\xi\,\xi^2
 \alpha_A(\mi\xi)\beta_B(\mi\xi)
 \,\frac{81\varepsilon^2(\mi\xi)\mu^2(\mi\xi)}
 {[2\varepsilon(\mi\xi)+1]^2[2\mu(\mi\xi)+1]^2}\,\nonumber\\
&\times h[n(\mi\xi)\xi r_{AB}/c],
\end{align}
[$r_{AB}=|\vec{r}_A-\vec{r}_B|$;
$n(\omega)=\sqrt{\varepsilon(\omega)\mu(\omega)}$, refractive index
of the medium] with $g(x)=\me^{-2x}(3+6x+5x^2+2x^3+x^4)$ and
$h(x)=\me^{-2x}(1+2x+x^2)$.
Making use of duality invariance, we can apply the replacements
$\alpha\leftrightarrow\beta/c^2$, $\varepsilon\leftrightarrow\mu$ to
directly infer the remaining two potentials
\cite{0739,Hassan08,0669,0829}
\begin{align}
\label{eq:Umebulk}
U_{me}(\vect{r}_A,\vect{r}_B)
 =&\frac{\hbar\mu_0^2}{16\pi^3r_{AB}^4}
 \int_0^\infty\dif\xi\,\xi^2
 \beta_A(\mi\xi)\alpha_B(\mi\xi)
 \,\frac{81\mu^2(\mi\xi)\varepsilon^2(\mi\xi)}
 {[2\mu(\mi\xi)+1]^2[2\varepsilon(\mi\xi)+1]^2}\,\nonumber\\
&\times h[n(\mi\xi)\xi r_{AB}/c],\\
\label{eq:Ummbulk}
U_{mm}(\vect{r}_A,\vect{r}_B)
 =&-\frac{\hbar\mu_0^2}{16\pi^3r_{AB}^6}
 \int_0^\infty\dif\xi\,\beta_A(\mi\xi)\beta_B(\mi\xi)
 \,\frac{81\mu^2(\mi\xi)}
 {[2\mu(\mi\xi)+1]^4}\,
 g[n(\mi\xi)\xi r_{AB}/c].
\end{align}
These results and their retarded and nonretarded limits
given below generalise the well-known free-space potentials
\cite{0095,0089,0094,0121,0096}; in particular, $U_{ee}$ reduces to
the famous Casimir--Polder potential of two polarisable ground state
atoms in free space for $\varepsilon=\mu=1$ \cite{0030}. In free
space, the potential between two electric or two magnetic atoms is
attractive and that between an electric and a magnetic one is
repulsive, in agreement with the general heuristic rule that
dispersion forces between objects of the same electric/magnetic nature
are attractive and those between objects of opposite nature are
repulsive. 

An inspection of Eqs.~(\ref{eq:Ueebulk})--(\ref{eq:Ummbulk}) reveals
that a bulk magneto-electric medium can influence the strengths of the
various two-atom interactions, but cannot change their signs. In order
to discuss the effect of the medium in more detail, it is useful to
consider the limits of large and small interatomic separations. In the
retarded limit $r_{AB}\gg c/\omega_\mathrm{min}$, the potentials are
well approximated by
\begin{eqnarray}
\label{eq:Ueebulkret}
U_{ee}(\vect{r}_A,\vect{r}_B)
&=&-\frac{23\hbar c\alpha_A(0)\alpha_B(0)}
 {64\pi^3\varepsilon_0^2r_{AB}^7}\,
 \frac{81\varepsilon^2(0)}{n(0)[2\varepsilon(0)+1]^4}\,,\\
\label{eq:Uembulkret}
U_{em}(\vect{r}_A,\vect{r}_B)
&=&\frac{7\hbar c\mu_0\alpha_A(0)\beta_B(0)}
 {64\pi^3\varepsilon_0 r_{AB}^7}\,
 \frac{81\varepsilon(0)\mu(0)}
 {n(0)[2\varepsilon(0)+1]^2[2\mu(0)+1]^2}\,,\\
\label{eq:Umebulkret}
U_{me}(\vect{r}_A,\vect{r}_B)
&=&\frac{7\hbar c\mu_0\beta_A(0)\alpha_B(0)}
 {64\pi^3\varepsilon_0 r_{AB}^7}\,
 \frac{81\mu(0)\varepsilon(0)}
 {n(0)[2\mu(0)+1]^2[2\varepsilon(0)+1]^2}\,,\\
\label{eq:Ummbulkret}
U_{mm}(\vect{r}_A,\vect{r}_B)
&=&-\frac{23\hbar c\mu_0^2\beta_A(0)\beta_B(0)}{64\pi^3r_{AB}^7}
 \,\frac{81\mu^2(0)}{n(0)[2\mu(0)+1]^4}\,.
\end{eqnarray}
In this case, the influence of the medium on all four types of
potentials is very similar: The coupling of each atom to the field is
screened by a factor $3\varepsilon(0)/[2\varepsilon(0)+1]^2$ for
polarisable atoms, and a factor $3\mu(0)/[2\mu(0)+1]^2$ for
magnetisable atoms; in addition, the reduced speed of light in the
medium leads to a further reduction of the potential by a factor
$n(0)$. In the nonretarded limit $r_{AB}\ll c/\omega_\mathrm{max}$,
the medium-assisted potentials simplify to
\begin{align}
\label{eq:Ueebulknret}
U_{ee}(\vect{r}_A,\vect{r}_B)
=&-\frac{3\hbar}{16\pi^3\varepsilon_0^2r_{AB}^6}
 \int_0^\infty\dif\xi\,\alpha_A(\mi\xi)\alpha_B(\mi\xi)
 \, \frac{81\varepsilon^2(\mi\xi)}{[2\varepsilon(\mi\xi)+1]^4}\,,\\
\label{eq:Uembulknret}
U_{em}(\vect{r}_A,\vect{r}_B)
=&\;\frac{\hbar\mu_0^2}{16\pi^3r_{AB}^4}
 \int_0^\infty\dif\xi\,\xi^2
 \alpha_A(\mi\xi)\beta_B(\mi\xi)
 \,\frac{81\varepsilon^2(\mi\xi)\mu^2(\mi\xi)}
 {[2\varepsilon(\mi\xi)+1]^2[2\mu(\mi\xi)+1]^2}\,,\\
\label{eq:Umebulknret}
U_{me}(\vect{r}_A,\vect{r}_B)
=&\;\frac{\hbar\mu_0^2}{16\pi^3r_{AB}^4}
 \int_0^\infty\dif\xi\,\xi^2
 \beta_A(\mi\xi)\alpha_B(\mi\xi)
 \,\frac{81\varepsilon^2(\mi\xi)\mu^2(\mi\xi)}
 {[2\varepsilon(\mi\xi)+1]^2[2\mu(\mi\xi)+1]^2}\,,\\
\label{eq:Ummbulknret}
U_{mm}(\vect{r}_A,\vect{r}_B)
=&-\frac{3\hbar\mu_0^2}{16\pi^3l^6}
 \int_0^\infty\dif\xi\,\beta_A(\mi\xi)\beta_B(\mi\xi)\,
 \frac{81\mu^2(\mi\xi)}{[2\mu(\mi\xi)+1]^4}\,.
\end{align}
The potentials $U_{ee}$ and $U_{mm}$ are thus again reduced, where in
the nonretarded limit, $U_{ee}$ is only influenced by the electric
properties of the medium and $U_{mm}$ only by the magnetic ones. On
the contrary, the mixed potentials $U_{em}$ and $U_{me}$ are enhanced
by a factor of up to $81/16$ in the nonretarded limit.
%
%
\subsubsection{Two atoms near a half space}
\label{sec:vdWHS}
\begin{figure}[!t!]
\noindent\vspace*{-2ex}
\begin{center}
\includegraphics[width=0.8\linewidth]{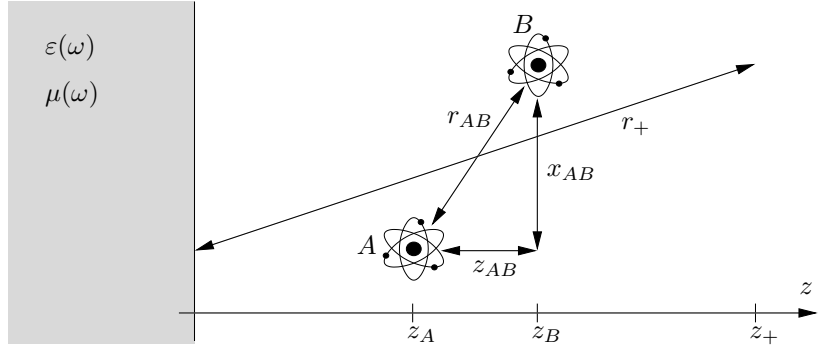}
\end{center}
\caption{
\label{fig:plateschematictwo}
Interaction of two atoms in the presence of a magnetoelectric half
space.}
\end{figure}

A modification of the vdW interaction does not only occur for atoms
embedded in a medium, but can also be induced by a
magnetoelectric body placed near to atoms in free space. To see
this, let us consider the body-assisted vdW interaction of two atoms
placed above ($z_A,z_B>0$) a semi-infinite half space of permittivity
$\varepsilon(\omega)$ and permeability $\mu(\omega)$, see
Fig.~\ref{fig:plateschematictwo}.
For simplicity, we assume both atoms to be isotropic and nonmagnetic,
so that $U=U_{ee}$ as given by Eq.~(\ref{eq:Uee}). Separating the
Green tensor into its bulk (free-space) and and scattering parts
according to $\ten{G}=\ten{G}^{(0)}+\ten{G}^{(S)}$, the potential
reads
\cite{0009}
\begin{align}
\label{eq:Ueeexpand}
U(\vect{r}_A,\vect{r}_B)
 =&-\frac{\hbar\mu_0^2}{2\pi}\int_0^\infty
 \dif\xi\,\xi^4\alpha_A(\mi\xi)\alpha_B(\mi\xi)
 \trace\bigl[
 \ten{G}^{(0)}(\vect{r}_A,\vect{r}_B,\mi\xi)\sprod
 \ten{G}^{(0)}(\vect{r}_B,\vect{r}_A,\mi\xi)\bigr]\nonumber\\
&-\frac{\hbar\mu_0^2}{2\pi}\int_0^\infty
 \dif\xi\,\xi^4\alpha_A(\mi\xi)\alpha_B(\mi\xi)
 \trace\bigl[
 \ten{G}^{(0)}(\vect{r}_A,\vect{r}_B,\mi\xi)\sprod
 \ten{G}^{(S)}(\vect{r}_B,\vect{r}_A,\mi\xi)\nonumber\\
&\hspace{28ex}+\ten{G}^{(S)}(\vect{r}_A,\vect{r}_B,\mi\xi)\sprod
 \ten{G}^{(0)}(\vect{r}_B,\vect{r}_A,\mi\xi)\bigr]\nonumber\\
&-\frac{\hbar\mu_0^2}{2\pi}\int_0^\infty
 \dif\xi\,\xi^4\alpha_A(\mi\xi)\alpha_B(\mi\xi)
 \trace\bigl[
 \ten{G}^{(S)}(\vect{r}_A,\vect{r}_B,\mi\xi)\sprod
 \ten{G}^{(S)}(\vect{r}_B,\vect{r}_A,\mi\xi)\bigr].
\end{align}
The first term is the free-space potential
$U_0(\vect{r}_A,\vect{r}_B)$ which is due to a direct 
exchange of two photons between the two atoms according to one of the
various processes depicted in Fig.~\ref{fig:fourthorder}. The second
and third terms represent the body-induced modification of the
potential; they are due to processes where one or both of the two
exchanged photons are scattered of the surface of the half space
before being absorbed. The free-space part of the potential is just a
special case of the bulk-medium potential~(\ref{eq:Ueebulk})
calculated in Sec.~\ref{sec:vdWbulk}, while the body-induced change
can be found by employing the free-space Green
tensor~(\ref{eq:freespaceG}) from App.~\ref{sec:bulkdgf} together with
the scattering tensor of the half space from
App.~\ref{sec:planardgf}. As the resulting expressions are rather
involved due to the large number of geometric parameters, we only give
analytical formulae for some simple special cases.
%
%
\paragraph{Perfect mirror:}
\label{sec:vdWMirror} 
We first consider a perfect mirror, with the reflection coefficients
being given by $r_s=-r_p=-1$ for a perfectly conducting plate and by 
$r_s=-r_p=1$ for an infinitely permeable one. In the retarded limit
$z_A,z_B,r_{AB}\gg c/\omega_\mathrm{min}$, the vdW potential in the
special case $x_{AB}\ll z_A+z_B=z_+$ can be given in closed form
\cite{0009,0367,0679}
\begin{equation}
\label{eq:vdWmirrorret}
U(\vect{r}_A,\vect{r}_B)
=-\frac{23\hbar c\alpha_A(0)\alpha_B(0)}{64\pi^3\varepsilon_0^2}
\biggl[\frac{1}{r_{AB}^7}
 \mp\frac{32}{23}\frac{x_{AB}^2+6r_{AB}^2}{r_{AB}^3z_+(r_{AB}+z_+)^5}
+\frac{1}{z_+^7}\biggr]
\end{equation}
($r_{AB}=|\vec{r}_A-\vec{r}_B|$, $x_{AB}=|x_A-x_B|$)
where the first term in square brackets is the free-space potential
due to the direct exchange of two photons and the second and third
terms represent the mirror-induced modification of the potential due
to scattering of one or two photons of the mirror surface (where the
different signs refer to the two cases of a conducting or permeable
mirror). An infinitely permeable plate thus always leads to an
enhancement of the retarded interaction of two polarisable atoms
whereas two cases need to be distinguished for a perfectly conducting
one: When the atoms are aligned parallel to the plate
($z_{AB}=|z_A-z_B|=0$), Eq.~(\ref{eq:vdWmirrorret}) shows that the
potential is always reduced due to the presence of the plate, while
for a perpendicular alignment ($x_{AB}=0$), the potential of reduced
if  $z_B/z_A\lesssim 4.90$ (atom $A$ being closer to the plate than
atom $B$).

In the nonretarded limit $z_A,z_B,r_{AB}\ll c/\omega_\mathrm{max}$,
the vdW potential in the presence of a perfect mirror is found to be
\cite{0009,0367,0679}
\begin{multline}
\label{eq:vdWmirrornret}
U(\vect{r}_A,\vect{r}_B) 
=-\frac{3\hbar}{16\pi^3\varepsilon_0^2}
 \int_0^\infty\dif\xi\,\alpha_A(\mi\xi)\alpha_B(\mi\xi)\\
\times\bigg[\frac{1}{r_{AB}^6}
\mp\frac{4x_{AB}^4-2z_{AB}^2z_+^2+x_{AB}^2(z_+^2+z_{AB}^2)}
 {3r_{AB}^5r_+^5}
+\frac{1}{r_+^6}\bigg]
\end{multline}
($r_+=\sqrt{x_{AB}^2+z_+^2}\,$). For a parallel alignment of the two
atoms, a perfectly conducting plate thus reduces the vdW potential of
two polarisable atoms while a permeable plate leads to an enhancement.
In particular in the on-surface limit $z_+\to 0$, the reduction and
enhancement factors with respect to the free-space potential are given
by $2/3$ and $10/3$. On the contrary, for a vertical alignment the
potential is reduced by a conducting plate while for a permeable
plate it is reduced if $z_B/z_A\lesssim 14.82$ (atom $A$ being
closer to the plate than atom $B$).

The enhancement and reduction of the nonretarded vdW interaction due
to a perfect mirror can be understood from the interaction of the
fluctuating atomic dipole moments $A$ and $B$ and their images ${A}'$
and ${B}'$ in the plate, with
\begin{equation}
\label{eq:Himage}
\hat H_\mathrm{dipole}=\hat V_{AB}+\hat V_{AB'}+\hat V_{BA'}
\end{equation}
being the corresponding interaction Hamiltonian. Here, $\hat{V}_{AB}$
denotes  the direct interaction between dipole $A$ and dipole $B$,
while $\hat V_{AB'}$ and $\hat V_{BA'}$ denote the indirect
interaction between each dipole and the image induced by the other one
in the plate. The leading contribution to the two-atom energy shift is
of second order in $\hat{H}_\mathrm{dipole}$,
\begin{multline}
\label{eq:energyshiftimage}
\Delta E(\vec{r}_A,\vec{r}_B)=-\sum_{(k,l)\neq(0,0)}
\frac{\langle 0_A|\langle 0_B|\hat H_\mathrm{int}|l_B\rangle
 |k_A\rangle\langle k_A|\langle l_B|\hat H_\mathrm{int}
 |0_B\rangle|0_A\rangle}
 {\hbar(\omega_A^{k0}+\omega_B^{l0})}\,.
\end{multline}
The three terms in Eq.~(\ref{eq:vdWmirrornret}) can be identified as
different contributions to this energy shift: The first term is due
to contributions that are quadratic in the direct interaction $\hat
V_{AB}$ and is always attractive due to the global minus sign in
Eq.~(\ref{eq:energyshiftimage}). The third term which is associated
with quadratic contributions of indirect dipole-image interactions
$\hat V_{AB'}$ or $\hat V_{BA'}$ is attractive for the same reason and
thus always acts as an enhancement of the free-space potential. The
second term in Eq.~(\ref{eq:vdWmirrornret}) is due to mixed
direct-indirect interactions, its sign is negative for attractive
dipole-image interactions (thus tending to enhance the free-space
potential) and positive for repulsive dipole-image interactions
(leading to a reduction). The overall effect of the perfect mirror
thus depends on the attraction or repulsion of the dipole-image
interaction and, in the case of repulsion, on the relative
strength of the mixed contributions compared to the purely indirect
one.

\begin{figure}[!t!]
\noindent\vspace*{-2ex}
\begin{center}
\includegraphics[width=\linewidth]{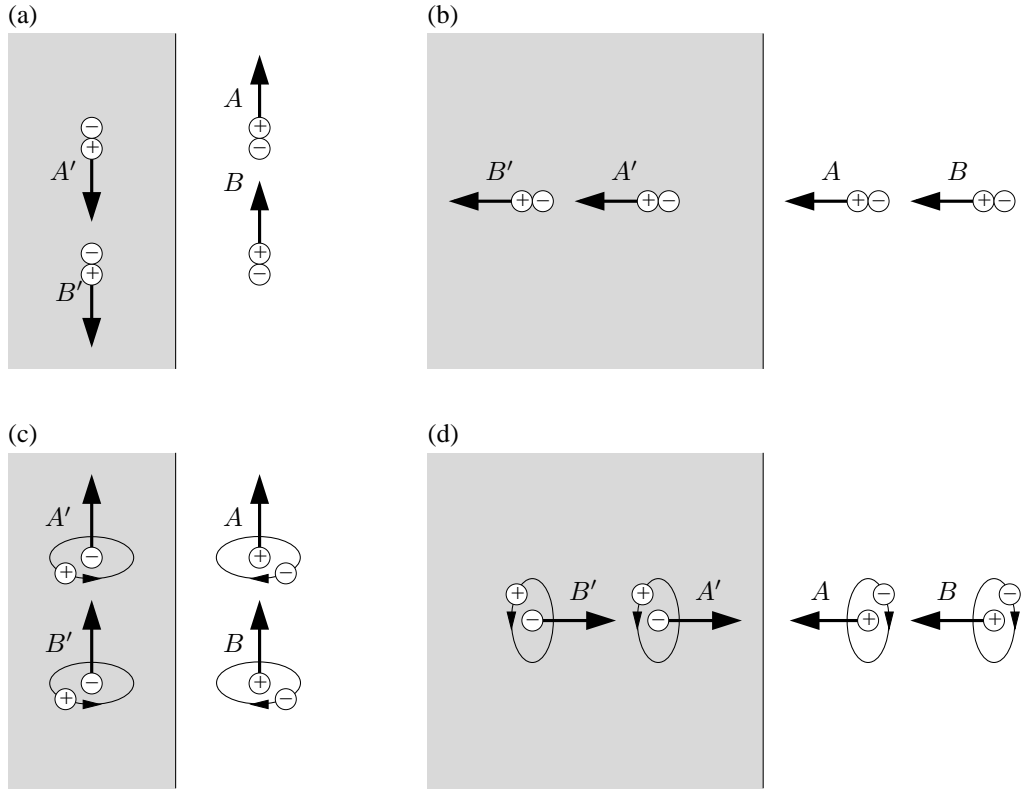}
\end{center}
\caption{
\label{fig:imagetwo}
Image dipole construction for two (a,b) electric and (c,d) magnetic
dipoles in front of a perfectly conducting plate.}
\end{figure}

We begin our investigations with the case of two (polarisable)
atoms placed aligned parallel to a perfectly conducting plane. The
image-dipole construction [Fig.~\ref{fig:imagetwo}(a)] reveals that
the indirect interactions $\hat V_{AB'}$ and $\hat V_{BA'}$ are
repulsive, explaining why the second term in
Eq.~(\ref{eq:vdWmirrornret}) acts to reduce the interaction. 
Equation~(\ref{eq:vdWmirrornret}) shows that it is always dominant
over the enhancing purely indirect interaction, leading to an overall
reduction of the free-space potential. The image-dipole construction
further reveals that the quantity $r_+$ entering the mirror-induced
modification of the potential is simply the distance between an atom
and the image of the other atom. Let us consider next the case of
perpendicular alignment of the two atoms. As seen from the
image-dipole construction [Fig.~\ref{fig:imagetwo}(b)], the indirect
interactions are attractive in this case, so that the mirror enhances
the potential, in agreement with Eq.~(\ref{eq:vdWmirrornret}).

\begin{figure}[!t!]
\noindent
\begin{center}
\includegraphics[width=0.6\linewidth]{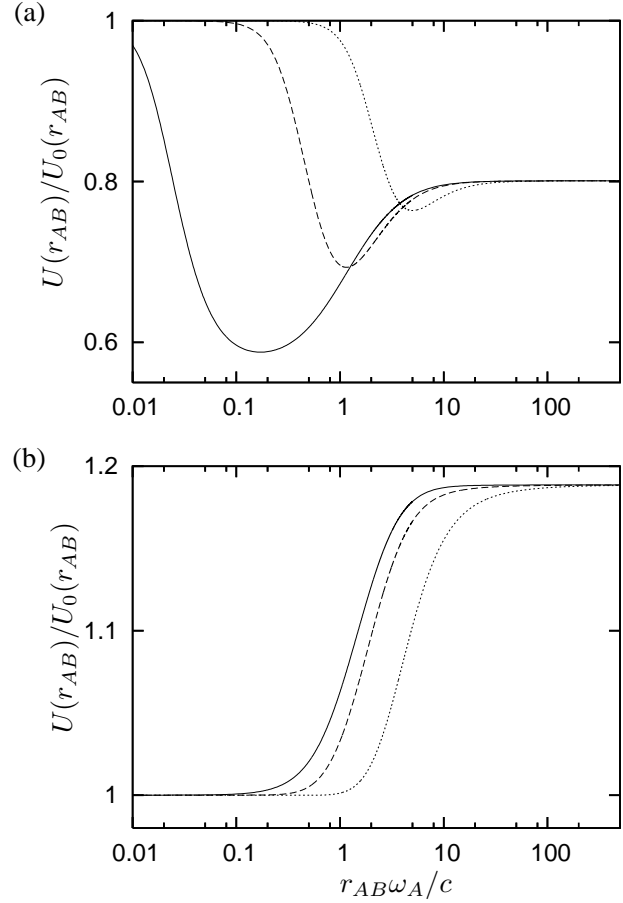}
\end{center}
\caption{
\label{fig:parallel}
The vdW potential for two identical two-level atoms (transition
frequency $\omega_A$) aligned parallel ($z_{AB}=0$) to the surface of
an (a) purely dielectric half space ($\omega_{Pe}/\omega_A=3$,
$\omega_{Te}/\omega_A=1$, $\gamma_e/\omega_A=0.001$)
(b) purely magnetic half space ($\omega_{Pm}/\omega_A=3$,
$\omega_{Tm}/\omega_A=1$, $\gamma_m/\omega_A=0.001$) is
shown as a function of the interatomic separation $r_{AB}$. The
potential is normalised with respect to the free-space potential
$U_0(r_{AB})$. The atom-half-space separations are
$z_A=z_B=0.01c/\omega_A$ (solid line), $0.2c/\omega_A$ (dashed line),
and $c/\omega_A$ (dotted line) \cite{0009}.
}
\end{figure}
\begin{figure}[!t!]
\noindent
\begin{center}
\includegraphics[width=0.6\linewidth]{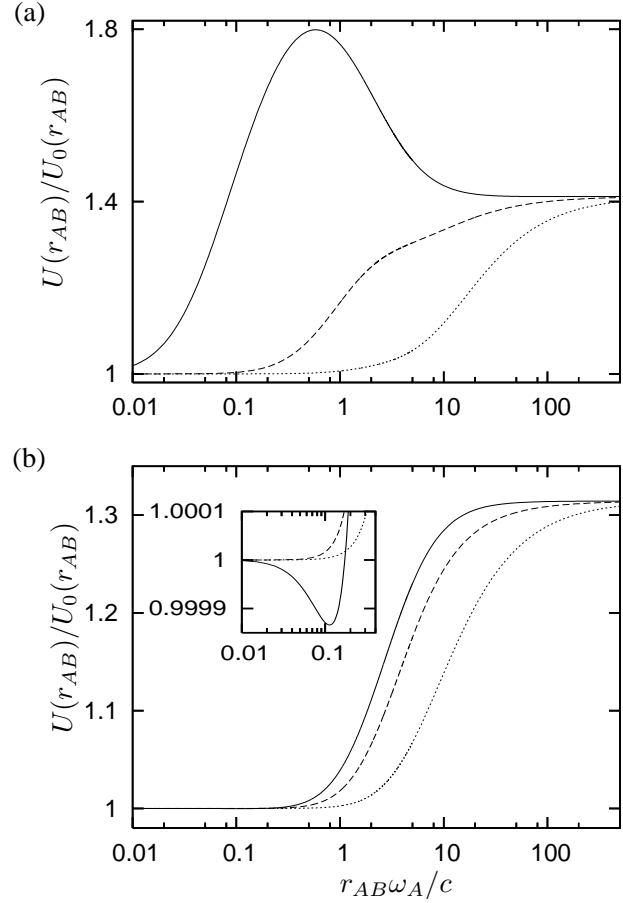}
\end{center}
\caption{
\label{fig:perpendicular}
The vdW potential for two two-level atoms aligned perpendicular
($x_{AB}=0$) to an (a) purely dielectric half space and (b) purely
magnetic half space is shown as a function of the interatomic
separation $r_{AB}$. The distance between atom $A$ (which is closer to
the surface of the half space than atom $B$) and the surface is
$z_A=0.01c/\omega_{10}$ (solid line), $0.2c/\omega_{10}$ (dashed
line), and $c/\omega_{10}$ (dotted line). All other parameters are the
same as in Fig.~\ref{fig:parallel} \cite{0009}.
}
\end{figure}

%
The case of two polarisable atoms near an infinitely permeable plate
can be addressed by studying the interaction of two magnetisable atoms
near a perfectly conducting plate instead, since the two problems are
equivalent by virtue of duality. Noting that magnetic dipoles behave
as pseudovectors under spatial reflection, one finds that for parallel
alignment of the two atoms the dipole-image interaction is attractive
[Fig.~\ref{fig:imagetwo}(c)] like the dipole-dipole interaction, so
all mirror-induced terms enhance the free-space potential as predicted
by Eq.~(\ref{eq:vdWmirrornret}). For perpendicular alignment the
indirect dipole-image interaction is seen to be repulsive
[Fig.~\ref{fig:imagetwo}(d)], so the mixed direct--indirect
interaction tends to reduce the interaction [cf. the second term in
Eq.~(\ref{eq:vdWmirrornret})] while the purely indirect one acts
towards an enhancement. For small atom-atom separations, the former
clearly dominates due to the enhanced strength of the direct
interaction, leading to a reduction of the overall potential while for
large atom-atom separations, an enhancement due to the influence of
the purely indirect interaction may be expected. The two cases are
obviously separated by the inequality given below
Eq.~(\ref{eq:vdWmirrornret}).
%
%
\paragraph{Magnetodielectric half space:}
\label{sec:vdWmeHS} 
For a semi-infinite half space of finite permittivity
$\varepsilon(\omega)$ and permeability $\mu(\omega)$, the vdW
potential~(\ref{eq:Ueeexpand}) is found by substitution of the
free-space Green tensor~(\ref{eq:freespaceG}) and the scattering
Green tensor from App.~\ref{sec:planardgf} together with the
reflection coefficients~(\ref{eq:fresnel}) of the half space. The
arising integrals over frequency and wave-vector components have to be
solved numerically in order to obtain the half-space assisted
potential for arbitrary interatomic and atom-half space
distances. The results are shown in Figs.~\ref{fig:parallel} and
\ref{fig:perpendicular} for two identical two-level atoms with the
permittivity and permeability of the half space being given by the
single-resonance models~(\ref{eq:epsilonmusingle}), where we display
the relative modification of the vdW potential with respect to its
free-space value $U_0$ as found from Eq.~(\ref{eq:Ueebulk}).

Figure~\ref{fig:parallel} shows the case of the two atoms being
aligned parallel to the surface of the half space ($z_{AB}=0$). In
agreement with the findings for a perfect mirror,
Eqs.~(\ref{eq:vdWmirrorret}) and (\ref{eq:vdWmirrornret}), the
potential is always reduced by a purely dielectric half space
[Fig.~\ref{fig:parallel}(a)] and enhanced by a purely magnetic one
[Fig.~\ref{fig:parallel}(b)]. The relative reduction/enhancement
becomes noticeable as soon as the interatomic separation becomes
comparable to the atom-surface distance, it saturates for large
interatomic separations. The figure further shows that the relative
reduction for a dielectric half space has a pronounced minimum, the
enhancement due to a magnetic one increases monotonically with
interatomic distance.

In Fig.~\ref{fig:perpendicular}, we study the case of perpendicular
alignment ($x_{AB}=0$). In agreement with the perfect-mirror
results~(\ref{eq:vdWmirrorret}) and (\ref{eq:vdWmirrornret}), a
purely dielectric half space is found to always enhance the
potential. The enhancement sets in when interatomic and atom-surface
distances become equal, reaches a maximum in some cases (i.e.,
whenever the atom-surface separation of the closer atom $A$ is
sufficiently small) and saturates for large interatomic distances. At
first glance, the findings for a purely magnetic half-space seem to
indicate a global enhancement which monotonically increases with the
interatomic separation, thus being in contradiction with the
perfect-mirror result~(\ref{eq:vdWmirrornret}). However, a closer look
[cf.~inset in Fig.~\ref{fig:perpendicular}(b)] reveals that for very
small interatomic and atom-surface separations a reduction can indeed
be found, as predicted by Eq.~(\ref{eq:vdWmirrornret}). 
%
%
\subsubsection{Two atoms near a sphere}
\label{sec:vdWSphere}
\begin{figure}[!t!]
\noindent\vspace*{-2ex}
\begin{center}
\includegraphics[width=0.6\linewidth]{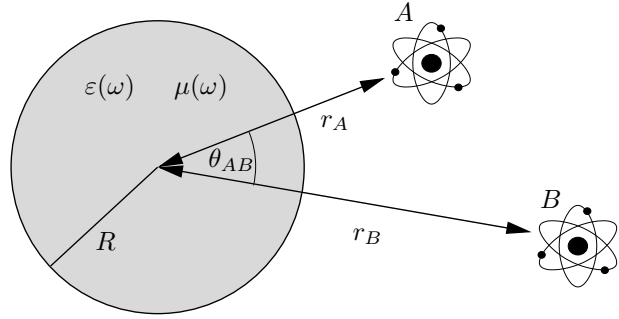}
\end{center}
\caption{
\label{fig:sphereschematictwo}
Two atoms interacting with a magneto-electric sphere.}
\end{figure}

In order to demonstrate the effect of different geometries on the vdW
potential, consider next to polarisable atoms placed at distances
$r_A$ and $r_B$ from the centre of a magneto-electric sphere of radius
$R$, permittivity $\varepsilon(\omega)$ and permeability
$\mu(\omega)$, with the separation angle of the two atoms being
denoted by $\theta_{AB}$, see Fig.~\ref{fig:sphereschematictwo}.
The sphere-assisted vdW potential is given by
Eq.~(\ref{eq:Ueeexpand}) where the scattering Green tensor
$\ten{G}^{(S)}$ for the sphere can be found in
App.~\ref{sec:sphericaldgf}. For a small sphere ($R\ll r_A,r_B$) in
the nonretarded limit ($r_A,r_B,r_{AB}\ll c/\omega_{max}$), one can
show that the sphere-induced part $U_\odot=U-U_0$ of the potential
reduces to \cite{0830}
\begin{equation}
\label{eq:Axilrod}
U_\odot(\vec{r}_A,\vec{r}_B)=\frac{3\hbar\bigl[1-3
 (\vec{e}_A\sprod\vec{e}_B)
 (\vec{e}_B\sprod\vec{e}_{AB})
 (\vec{e}_{AB}\sprod\vec{e}_A)\bigr]}
 {64\pi^4\varepsilon_0^3r_A^3r_A^3r_{AB}^3}
 \int_0^{\infty} \dif\xi\, 
  \alpha_A(\mi\xi)\alpha_B(\mi\xi)\alpha_\odot(\mi\xi)
\end{equation}
[$\vec{e}_A=\vec{r}_A/r_A$, $\vec{e}_B=\vec{r}_B/r_B$,
$\vec{e}_{AB}=(\vec{r}_A-\vec{r}_B)/r_{AB}$] where
$\alpha_\odot(\omega)$ is the polarisability of the sphere as given by
Eq.~(\ref{eq:aplhasphere}). When replacing the sphere by a third atom
$C$ ($\alpha_\odot\to\alpha_C$), our results coincides with the
Axilrod--Teller potential of three atoms \cite{0084}. The sign of the
three-body potential~(\ref{eq:Axilrod}) depends on the geometric
arrangement of the three objects. For instance, an attractive
potential is found when they are placed in a straight line while the
potential is repulsive when they form an equilateral triangle.

\begin{figure}[!t!]
\noindent
\begin{center}
\includegraphics[width=0.6\linewidth]{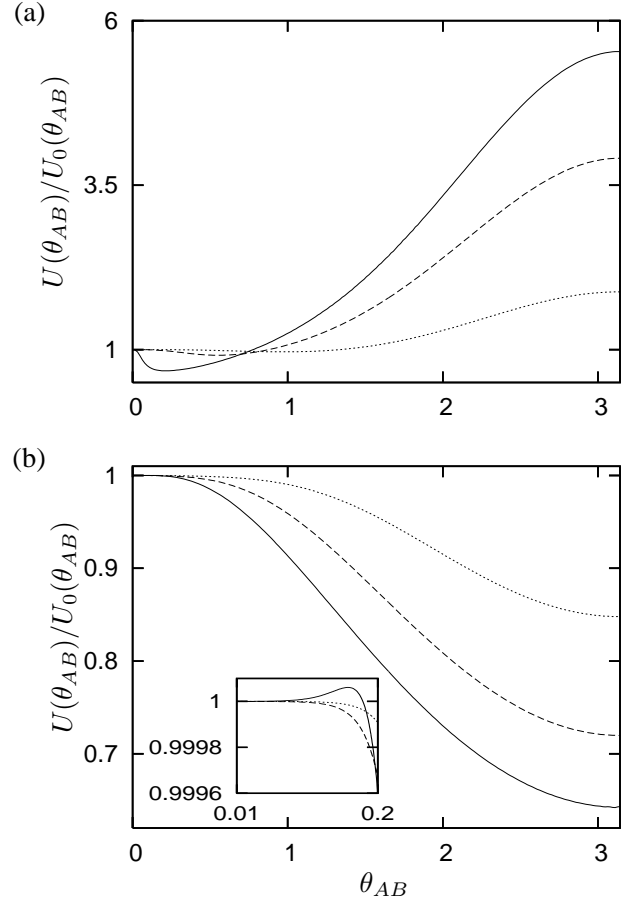}
\end{center}
\caption{
\label{fig:angular}
The vdW potential for two identical two-level atoms placed at equal
distance ($r_A=r_B$) from the centre of an (a) purely dielectric
sphere (b) purely magnetic sphere is shown as a function of the
interatomic angle $\theta_{AB}$. The potential is normalised with
respect to the free-space potential
$U_0(\theta_{AB})$. The sphere radius is $R=c/\omega_A$ and the
distances of the atoms from the centre of the sphere are
$r_A=r_B=1.03c/\omega_A$ (solid line), $1.3c/\omega_A$ (dashed line),
and $2c/\omega_A$ (dotted line). The medium parameters are the same
as in Fig.~\ref{fig:parallel} \cite{0830}.
}
\end{figure}

In Fig.~\ref{fig:angular}, we show the total vdW potential of two
identical two-level atoms placed at equal distance from a purely
dielectric or purely magnetic sphere whose magnetoelectric response is
given by Eqs.~(\ref{eq:epsilonmusingle}).
For small separation angles and atom-sphere separations, the
sphere-induced modification of the vdW potential is very similar to
our findings for two atoms placed in parallel alignment with respect
to a half space: A dielectric sphere reduces the potential, while a
magnetic one leads to an enhancement. However, the figure also shows
that the behaviour is reversed as the angular separation grows. For
an electric sphere, this enhancement of the potential for large
angular separations can be understood from the fact that the
Axilrod--Teller potential is attractive when the atoms and the sphere
are situated on a straight line. 
%
%
\subsection{Relation between dispersion forces}
\label{sec:relation}
The three types of dispersion forces considered in the previous three
sections have a common origin and are thus closely related. To see
this, let us start from the Casimir force~(\ref{eq:CasimirVolume}) on
a dielectric body of volume $V_1$ which is situated in free space and
interacts with a second dielectric body of volume $V_2$. Assuming the
first body to consist of a dilute gas of atoms [number density
$\eta_1(\vec{r})$, polarisability $\alpha(\omega)$] such that the
linearised Clausius--Mosotti law \cite{0001}
\begin{equation}
\label{eq:Clausiuslinear}
\varepsilon(\vec{r},\omega)-1=\frac{\eta(\vec{r})\alpha(\omega)}
{\varepsilon_0}
\end{equation}
holds, one can make use of the linear Born expansion given in
App.~\ref{sec:bornseries} to show that \cite{0663}
\begin{equation}
\label{eq:Casimirexpand}
\vect{F}
 =-\int_{V_1}\dif^3r\,\eta(\vect{r})
 \bm{\nabla}U(\vect{r})
\end{equation}
where $U(\vect{r})$ is the Casimir--Polder potential~(\ref{eq:Ue}).
To leading order in the atomic polarisability, the Casimir force on
the body is simply the sum of the CP forces on the atoms contained
inside it. One can repeat the exercise for the second body to find
\begin{equation}
\label{eq:CPexpand}
U(\vect{r}_A)=\int_{V_2}\dif^3r\,\eta(\vect{r})
 U(\vect{r}_A,\vect{r}),
\end{equation}
so the CP potential $U(\vect{r})$ of each atom in body 1 with body 2
is due to its vdW interactions~(\ref{eq:Uee}) with the atoms in body
2. Combining these results, one has \cite{0663}
\begin{equation}
\label{eq:Casimirexpand2}
\vect{F}=-\int_{V_1}\dif^3r\,\eta(\vect{r})
 \int_{V_2}\dif^3r'\,\eta(\vect{r}')
 \bm{\nabla}U(\vect{r},\vect{r}') \,.
\end{equation}
Hence, to leading order the Casimir force between the two bodies is a
sum over all possible vdW forces between atoms in body 1 and atoms in
body 2. For weakly dielectric bodies, both Casimir and CP forces may
thus be regarded as a consequence of two-atom vdW forces. These
results naturally generalise to weakly magnetic bodies and magnetic
atoms.

For bodies with a stronger magnetoelectric response, this simple
additivity breaks down due to the influence of many-atom
interactions \cite{0087,Golestanian05,0113,0020}. For instance, using
the exact Clausius--Mosotti law \cite{0001}
\begin{equation}
\label{eq:Clausius}
\varepsilon(\vect{r},\omega)-1
=\frac{\eta(\vect{r})\alpha(\omega)/\varepsilon_{0}}
 {1-\eta(\vect{r})\alpha(\omega)/(3\varepsilon_{0})} \,,
\end{equation}
together with the full Born expansion, one can show that the CP
potential~(\ref{eq:CPexpand}) generalises to \cite{0113,0020}
\begin{equation}
\label{eq:CPfullexpand}
U(\vect{r}_A)
 =\sum_{K=1}^\infty\frac{1}{K!}
 \idotsint\dif^3r_1\,\eta(\vect{r}_1) \cdots
 \dif^3r_K\,\eta(\vect{r}_K) \,
 U(\vect{r}_A,\vect{r}_1,\ldots,\vect{r}_K)
\end{equation}
where $U(\vect{r}_1,\ldots,\vect{r}_N)$ denotes $N$-atom vdW
potentials \cite{0090,0091}. The full CP potential of a single atom in
the presence of a body is just due to the whole hierarchy of its
$2,3,\ldots N$-atom interactions with the body atoms.

To illustrate the close relation between dispersion forces, let us
look at a few simple examples, most of which have already been studied
throughout this Sec.~\ref{sec:dispersion}. In
Tab.~\ref{tab:dispersion}, we list signs and leading power laws of
the dispersion forces between two atoms (Sec.~\ref{sec:vdWbulk}), an
atom interacting with a small sphere (Sec.~\ref{sec:CPSphere}), a thin
ring \cite{0020}, a thin plate (Sec.~\ref{sec:CPPlanar}) and a
semi-infinite half space (Sec.~\ref{sec:CPPlanar}), and that between
two half spaces \cite{0134,0133}.
\begin{table}[!t!]
\begin{center}
 \begin{tabular}{|c||c|c|c|c|}
\hline
 Distance $\rightarrow$
 &\multicolumn{2}{c|}{Retarded}&\multicolumn{2}{c|}{Nonretarded} \\
\hline
 Object combination $\rightarrow$
 &${e}\leftrightarrow {e}$
 &${e}\leftrightarrow {m}$
 &${e}\leftrightarrow {e}$
 &${e}\leftrightarrow {m}$\\
\hline
 Dual object combination $\rightarrow$
 &${m}\leftrightarrow {m}$
 &${m}\leftrightarrow {e}$
 &${m}\leftrightarrow {m}$
 &${m}\leftrightarrow {e}$\\
\hline\hline
 \hspace*{1ex}(a)\hspace*{4.4cm}\begin{picture}(1,1)
 \put(-117,-15){\includegraphics[width=4cm]{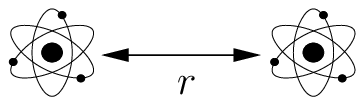}}
 \end{picture}
 &\parbox{5ex}{$$-\frac{1}{r^8}$$}
 &\parbox{5ex}{$$+\frac{1}{r^8}$$}
 &\parbox{5ex}{$$-\frac{1}{r^7}$$}
 &\parbox{5ex}{$$+\frac{1}{r^5}$$}\\
\hline
 \hspace*{1ex}(b)\hspace*{4.4cm}\begin{picture}(1,1)
 \put(-117,-15){\includegraphics[width=4cm]{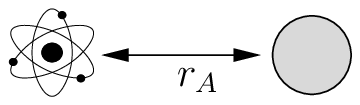}}
 \end{picture}
 &\parbox{5ex}{$$-\frac{1}{r_{A}^8}$$}
 &\parbox{5ex}{$$+\frac{1}{r_{A}^8}$$}
 &\parbox{5ex}{$$-\frac{1}{r_{A}^7}$$}
 &\parbox{5ex}{$$+\frac{1}{r_{A}^5}$$}\\
\hline
 \hspace*{1ex}(c)\hspace*{4.4cm}\begin{picture}(1,1)
 \put(-117,-15){\includegraphics[width=4cm]{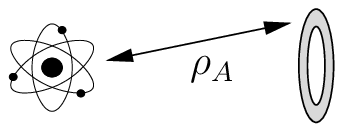}}
 \end{picture}
 &\parbox{5ex}{$$-\frac{1}{\rho_{A}^8}$$}
 &\parbox{5ex}{$$+\frac{1}{\rho_{A}^8}$$}
 &\parbox{5ex}{$$-\frac{1}{\rho_{A}^7}$$}
 &\parbox{5ex}{$$+\frac{1}{\rho_{A}^5}$$}\\
\hline
 \hspace*{1ex}(d)\hspace*{4.4cm}\begin{picture}(1,1)
 \put(-117,-15){\includegraphics[width=4cm]{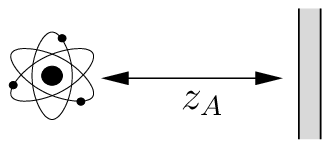}}
 \end{picture}
 &\parbox{5ex}{$$-\frac{1}{z_{A}^6}$$}
 &\parbox{5ex}{$$+\frac{1}{z_{A}^6}$$}
 &\parbox{5ex}{$$-\frac{1}{z_{A}^5}$$}
 &\parbox{5ex}{$$+\frac{1}{z_{A}^3}$$}\\
\hline
 \hspace*{1ex}(e)\hspace*{4.4cm}\begin{picture}(1,1)
 \put(-117,-15){\includegraphics[width=4cm]{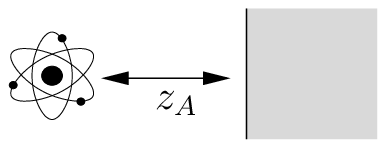}}
 \end{picture}
 &\parbox{5ex}{$$-\frac{1}{z_{A}^5}$$}
 &\parbox{5ex}{$$+\frac{1}{z_{A}^5}$$}
 &\parbox{5ex}{$$-\frac{1}{z_{A}^4}$$}
 &\parbox{5ex}{$$+\frac{1}{z_{A}^2}$$}\\
\hline
 \hspace*{1ex}(f)\hspace*{4.4cm}\begin{picture}(1,1)
 \put(-117,-15){\includegraphics[width=4cm]{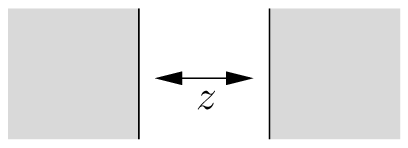}}
 \end{picture}
 &\parbox{5ex}{$$-\frac{1}{z^4}$$}
 &\parbox{5ex}{$$+\frac{1}{z^4}$$}
 &\parbox{5ex}{$$-\frac{1}{z^3}$$}
 &\parbox{5ex}{$$+\frac{1}{z}$$}\\
\hline
\end{tabular}
\end{center}
\vspace*{2ex}
\caption{%
Asymptotic power laws for the forces between (a) two atoms, (b) an
atom and a small sphere, (c) an atom and a thin ring, (d) an atom an a
thin plate, (e) an atom and a half space and (f) for the force per
unit area between two half spaces. In the table heading, ${e}$
stands for an electric object and ${m}$ for a magnetic one. The
signs $+$ and $-$ denote repulsive and attractive forces,
respectively.}
\label{tab:dispersion}
\end{table}%
For weakly magnetoelectric objects, the rows (b)--(f) of the table
follow by pairwise summation of the vdW forces displayed in row (a):
Summation over the compact volumes of a small sphere (b) or
a thin ring (c) does not change the respective power laws, while
summation over an infinite volume lowers the leading inverse power
according to the number of infinite dimensions. So, the leading
inverse powers are lowered by two and three for the interaction of an
atom with a thin plate of infinite lateral extension (d) and a half
space (e), respectively. The power laws for the force between two half
spaces (f) can then be obtained from the atom-half-space force
(e) by integrating over the three infinite dimensions where
integration over $z$ lowers the leading inverse powers by one while
the trivial integrations over $x$ and $y$ yield an infinite force,
i.e., a finite force per unit area. Note that all of these power laws
remain valid for objects with stronger magnetoelectric bodies, so we
can conclude that many-atom interactions do not change the asymptotic
power laws.

The common features of all dispersion forces listed in the table are
as follows: Forces between objects of the same (electric/magnetic)
nature are attractive while those between objects of opposite nature
are repulsive. The combinations ${e}\leftrightarrow {e}$ and
${m}\leftrightarrow{m}$ obviously lead to the same behaviour of the
force as an immediate consequence of duality invariance, the same hold
for the combinations ${e}\leftrightarrow{m}$ and
${m}\leftrightarrow{e}$. Attractive and repulsive forces generally
follow the same power law in the retarded limit, while in the
nonretarded limit attractive forces are stronger by two inverse powers
in the object separation.
%
%
\subsection{Thermal effects}
\label{sec:thermalCP}

The Casimir--Polder force acting on a stationary atom at finite
temperatures can be derived from the Lorentz force (\ref{eq:Lorentz2})
which for a stationary nonmagnetic atom in electric dipole
approximation takes the simple form
\begin{equation}
\vect{F}(\vect{r}_A,t) = \left\langle \left[ \grad\hat{\vect{d}} \cdot
\hat{\vect{E}}(\vect{r}) \right]_{\vect{r}=\vect{r}_A} \right\rangle
\,.
\end{equation}
This expression can be evaluated using the statistical averages of the
relevant electromagnetic field operators given in
Eqs.~(\ref{eq:expectE})--(\ref{eq:EEdagger}). We use the
multipolar-coupling Hamiltonian (\ref{eq:Hmultipolar}) in
electric-dipole approximation to solve the coupled dynamics of the
electromagnetic field and the atom which, upon using the solutions to
the equations of motion for the internal atomic dynamics
(Sec.~\ref{sec:relaxation}), reads as \cite{Buhmann08}
\begin{align}
&\vect{F}(\vect{r}_A,t) = \frac{i\mu_0}{\pi} \sum\limits_{n,k}
\int\limits_0^\infty d\omega\,\omega^2 \left .\grad \vect{d}_{nk}
\cdot \mathrm{Im}\,\tens{G}^{(S)}(\vect{r},\vect{r}_A,\omega) \cdot
\vect{d}_{kn} \right|_{\vect{r}=\vect{r}_A}
\int\limits_0^t d\tau\, \langle \hat{A}_{nn}(\tau) \rangle
\nonumber \\ &\quad \times
\left\{ \bar{n}_\mathrm{th}(\omega)
e^{[i(\omega+\omega_{nk})-(\Gamma_n+\Gamma_k)/2](t-\tau)} +\left[
\bar{n}_\mathrm{th}(\omega)+1  \right]
e^{[-i(\omega-\omega_{nk})-(\Gamma_n+\Gamma_k)/2](t-\tau)} \right\}
\nonumber \\ &\quad+\mbox{c.c.} 
\end{align}
After evaluating the $\tau$ integral in Markov approximation and the
$\omega$ integral using countour-integral techniques, the thermal
Casimir--Polder force on an atom in an incoherent superposition of
energy eigenstates is given by
$\vect{F}(\vect{r}_A,t)=\sum_n\sigma_{nn}(t)\vect{F}_n(\vect{r}_A)$
with force components
\begin{align}
\label{eq:thermalCP}
&\vect{F}_n(\vect{r}_A) =-\mu_0k_\mathrm{B}T\sum_{N=0}^\infty
\bigl(1-{\textstyle\frac{1}{2}}\delta_{N0}\bigr)\xi_N^2
\nonumber \\ &\quad \times
\grad\trace \bigl\{
[\bm{\alpha}_n(i\xi_N) +\bm{\alpha}_n(-i\xi_N)]
\cdot \tens{G}^{(S)}(\vect{r}_A,\vect{r},i\xi_N) \bigr\}
\bigr|_{\vect{r}=\vect{r}_A} \nonumber \\ &\quad
+\mu_0\sum_k \bigl\{\Theta(\tilde{\omega}_{nk}) \Omega^2_{nk}
[\bar{n}_\mathrm{th}(\Omega_{nk})+1] \grad \vect{d}_{nk} \cdot
\tens{G}^{(S)}(\vect{r},\vect{r}_A,\Omega_{nk})
\cdot \vect{d}_{kn}\bigr|_{\vect{r}=\vect{r}_A}
\nonumber \\ &\quad
-\Theta(\tilde{\omega}_{kn})\Omega^{\ast
2}_{kn}\bar{n}_\mathrm{th}(\Omega_{kn}^\ast)
\grad\vect{d}_{nk}\cdot
\tens{G}^{(S)}(\vect{r},\vect{r}_A,\Omega_{kn}^\ast)
\cdot \vect{d}_{kn}\bigr|_{\vect{r}=\vect{r}_A}
 +\mathrm{c.c.}\bigr\}
\end{align}
[$\Omega_{nk}=\tilde{\omega}_{nk}+i(\Gamma_n+\Gamma_k)/2$] and atomic
polarisability
\begin{equation}
\bm{\alpha}_n(\omega) = \frac{1}{\hbar}\sum_k\biggl[
\frac{\vect{d}_{nk} \otimes \vect{d}_{kn}}{-\Omega_{nk}-\omega}
+\frac{\vect{d}_{kn} \otimes \vect{d}_{nk}}{-\Omega_{nk}^\ast+\omega}
\biggr] \,.
\end{equation}

One important feature of the result (\ref{eq:thermalCP}) is the
appearance of resonant force contributions proportional to
$\bar{n}_\mathrm{th}(\omega_{nk})$ and
$\bar{n}_\mathrm{th}(\omega_{nk})+1$ which are due to absorption and
emission of photons, respectively. Even for ground-state atoms there
exists a resonant force contribution 
\begin{equation}
\vect{F}_0^\mathrm{res}(\vect{r}_A) =
-\frac{1}{3}\mu_0\omega_{k0}^2\bar{n}_\mathrm{th}(\omega_{k0})
|\vect{d}_{0k}|^2 \grad_A 
\trace \mathrm{Re}\,\tens{G}^{(S)}(\vect{r}_A,\vect{r}_A,\omega_{k0})
\end{equation}
associated with absorption of thermal photons at the atomic transition
frequency $\omega_{k0}$. This result is in contrast to the frequently
used Lifshitz result \cite{Lifshitz61}
\begin{equation}
\label{eq:LifshitzCP}
\vect{F}_0^\mathrm{Lifshitz}(\vect{r}_A) = -\mu_0k_\mathrm{B}T
\sum_{N=0}^\infty \bigl(1-{\textstyle\frac{1}{2}}\delta_{N0}\bigr)
\xi_N^2 \alpha(i\xi_N) \grad_A \trace
\tens{G}^{(S)}(\vect{r}_A,\vect{r}_A,i\xi_N)
\end{equation}
which only contains the non-resonant force contributions. Clearly,
because of the thermalisation of the atom, these resonant forces can
only be observed on time scales that are short compared to the inverse
ground-state heating rates $\Gamma_{0k}^{-1}$. As their magnitude
scales with the mean thermal number
$\bar{n}_\mathrm{th}(\omega_{k0})$, the atomic transition frequency
must not be too large in comparison with the ambient temperature,
$\hbar\omega_{k0}\lesssim k_BT$. Ideal candidates to observe these
resonant force contributions would therefore be polar molecules whose
vibrational and rotational frequencies are small enough to yield large
thermal photon numbers at room temperature, and whose heating times
can reach several seconds (see also Sec.~\ref{sec:heating})
\cite{Heating}.

The atom thermalises after times much longer than $\Gamma_{0k}^{-1}$
and reaches its stationary (thermal) state
\begin{equation}
\hat{\sigma}_T
=\frac{\exp\left[-\sum_n\hbar\tilde{\omega}_n\hat{A}_{nn}/
(k_\mathrm{B}T)\right]}{\trace\,\exp\left[-\sum_n\hbar\tilde{\omega}_n
\hat{A}_{nn}/(k_\mathrm{B}T)\right]}
\,, \quad \tilde{\omega}_n=\omega_n+\delta\omega_n
\end{equation}
in the long-time limit. In this limit all resonant force contributions
cancel, and the Casimir--Polder force can be written in the form
(\ref{eq:LifshitzCP}) only if the atomic polarisability has been
identified with the thermal polarisability
\begin{equation}
\alpha_T(\omega)=\sum_n\sigma_{T,nn}\alpha_n(\omega) \,.
\end{equation}
Without this identification, the equilibrium force
(\ref{eq:LifshitzCP}) is generally larger than the one obtained from
Eq.~(\ref{eq:thermalCP}). The microscopic degrees of freedom of an
atomic system thus give rise to a rich additional structure of CP
forces, which is not explicitly present in the macroscopic Casimir
force.


\newpage
\section{Cavity QED effects}
\label{sec:cavityQED}

In this final section, we turn our attention to the application of
macroscopic QED to strong atom-field coupling scenarios. This is of
particular interest in cases when the atom or molecule is confined in
resonator-like structures in which the photon round-trip time is short
compared to the atomic relaxation time. 


\subsection{Quantum-state extraction from a high-$Q$ cavity}
\label{sec:extraction}

The interaction of a two-level atom with light in a resonant cavity is
of fundamental importance to our understanding of quantum optics, and 
provides a playground to test ideas of quantum information processing 
such as the generation of entangled quantum states of two or more
atoms \cite{Hagley97} or the realization of single-photon sources. In
a previous section (Sec.~\ref{sec:jc}) we described the idealised
situation of perfectly coherent atom-light interaction that led to the
Jaynes--Cummings model. Within this model it is possible to completely
transfer the excitation of a two-model atom to the cavity field and
back coherently. The idea to use a high-$Q$ cavity to
deterministically generate single photons in well-defined quantum
states is at the heart of cavity QED \cite{Kuhn02,McKeever04}. Here we
will examine the closely related problem of how to extract a photon
from a cavity within the framework of macroscopic QED which provides
us with the means to treat leaky cavities in a quantum-theoretically
consistent way \cite{Khanbekyan07}. 

\begin{figure}[!b!]
\centerline{\includegraphics[width=8cm]{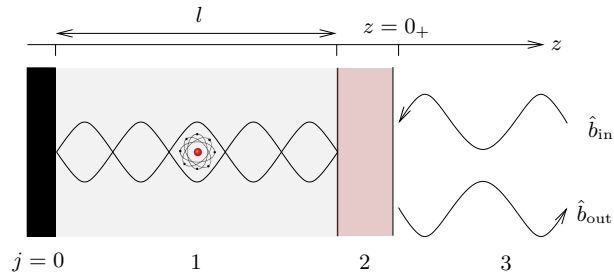}}
\caption{\label{fig:lossycavity} The semi-transparent
mirror of the cavity (region 2) is modeled by a dielectric plate,
and the atom inside the cavity (region 1) can be embedded
in some dielectric medium. Figure taken from
Ref.~\cite{Khanbekyan07}.}
\end{figure}
%
For simplicity, we will focus on one-dimensional cavities such as in
Fig.~\ref{fig:lossycavity} where region 0 denotes a perfect mirror,
region 2 the semi-transparent mirror, region 1 the cavity interspace,
and region 3 the free space surrounding the system. In order to
calculate the response of the atom-cavity system, we assume
that the atom is initially being prepared in its excited state
$|e\rangle$ and the cavity field is in its ground state
$|\{0\}\rangle$. Then, the state vector at a later time can be
expanded into
\begin{equation}
|\psi(t)\rangle = C_e(t)|e\rangle|\{0\}\rangle +\int dz 
\int\limits_0^\infty d\omega\, C_g(z,\omega,t) e^{-i\omega t}
|g\rangle|1(z,\omega)\rangle \,.
\end{equation}
The Schr\"odinger equation with the electric-dipole Hamiltonian
(\ref{eq:multipolarHAF}) yields coupled equations of motion for the
unknown coefficients $C_e(t)$ and $C_g(z,\omega,t)$ as
\begin{eqnarray}
\label{eq:ce}
\dot{C}_e(t) &=& - \frac{d}{\sqrt{\pi\hbar\varepsilon_0\mathcal{A}}}
\int\limits_0^\infty d\omega \,\frac{\omega^2}{c^2} \int dz\,
\sqrt{\mathrm{Im}\,\varepsilon(z,\omega)} G(z_A,z,\omega)
C_g(z,\omega,t)
e^{-i(\omega-\omega_A)t} \,,\nonumber \\ && \\
\label{eq:cg}
\dot{C}_g(z,\omega,t) &=& 
\frac{d^\ast}{\sqrt{\pi\hbar\varepsilon_0\mathcal{A}}}
\frac{\omega^2}{c^2} \sqrt{\mathrm{Im}\,\varepsilon(z,\omega)} 
G^\ast(z_A,z,\omega) C_e(t) e^{i(\omega-\omega_A)t}
\end{eqnarray}
[$\mathcal{A}$: cross-sectional area of the cavity]. Inserting the
formal solution of Eq.~(\ref{eq:cg}) into Eq.~(\ref{eq:ce}) yields an
integro-differential equation 
\begin{equation}
\label{eq:integrodiff}
\dot{C}_e(t) = \int\limits_0^t dt'\, K(t-t') C_e(t')
\end{equation}
with the integral kernel
\begin{equation}
\label{eq:kernel}
K(t) = -\frac{|d|^2}{\pi\hbar\varepsilon_0\mathcal{A}} 
\int\limits_0^\infty d\omega\,\frac{\omega^2}{c^2} \,
e^{-i(\omega-\omega_A)t} \mathrm{Im}\,G(z_A,z_A,\omega) \,.
\end{equation}

The spectral response is carried by the Green function
$G(z,z',\omega)$ which has to be determined by adjusting the boundary
conditions at the interfaces between the regions of piecewise constant
permittivity. Because of multiple scattering inside the cavity region
1, the Green  function has poles at the complex frequencies
$\Omega_k=\omega_k-i\Gamma_k/2$ where the condition
\begin{equation}
D_1(\Omega_k) = 1+r_{13}(\Omega_k)
e^{2i\varepsilon_1(\Omega_k)\Omega_kl/c} =0 
\end{equation}
[$r_{13}$: generalised Fresnel reflection coefficient
(\ref{eq:multifresnel})] is met. We assume that the line widths
$\Gamma_k$ are much smaller than the line separations
$\Gamma_k\ll(\omega_{k+1}-\omega_{k-1})/2$ so that the integration
over frequency in the kernel can be restricted to subintervals
$\Delta_k=[(\omega_{k-1}+\omega_k)/2,(\omega_k+\omega_{k+1})/2]$. Near
a cavity resonance, the time integral in Eq.~(\ref{eq:integrodiff})
can then be performed in the Markov approximation to provide the
cavity-induced shift of the atomic transition frequency as
\cite{Khanbekyan07}
\begin{equation}
\delta\omega = \sum\limits_{k'}
\frac{\alpha_{k'}}{4|\tilde{\omega}_A-\Omega_{k'}|^2} \left[
\tilde{\omega}_A \omega_{k'} -|\Omega_{k'}|^2
-\frac{\tilde{\omega}_A\Gamma_{k'}}{4\pi}
\ln\left(\frac{\omega_{k'}}{\omega_A} \right) \right] \,,
\end{equation}
with
\begin{equation}
\alpha_k =
\frac{4|d|^2}{\hbar\varepsilon_0\mathcal{A} l}
\sin^2\left[ \frac{\omega_k z_A}{c}\right]
\end{equation}
[$\tilde{\omega}_A=\omega_A-\delta\omega$]. Within these
approximations, the kernel function $\tilde{K}(t)$ for the
excited-state probability amplitude
$\tilde{C}_e(t)=C_e(t)e^{-i\delta\omega t}$ is then
\begin{equation}
\tilde{K}(t) = -\frac{1}{4} \alpha_k \Omega_k
e^{-i(\Omega_k-\tilde{\omega}_A)t} \,.
\end{equation}

\begin{figure}[!t!]
\centerline{\includegraphics[width=8cm]{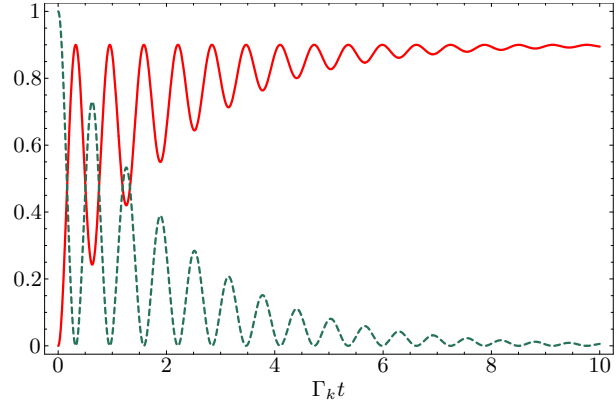}}
\caption{\label{fig:efficiency} The efficiency of single-photon Fock
state preparation $\eta(t)$ (solid curve) and the excited-state
probability $|\tilde{C}_e(t)|^2$ (dashed curve) for
$\omega_k=2\times 10^8\Gamma_k$, $\alpha_k=5\times 10^{-7}\Gamma_k$,
$\omega_k-\tilde{\omega}_A=0.1\Gamma_k$, and the radiative damping
rate $\gamma_k=0.9\Gamma_k$. Taken from Ref.~\cite{Khanbekyan07}.}
\end{figure}

The solution for $\tilde{C}_e(t)$ is the prerequisite for the
calculation of the photon extraction efficiency $\eta(t)$ from the
cavity \cite{Khanbekyan07}, which is defined as the probability of
having the outgoing field prepared in a single-photon Fock state. The
result for a typical situation is shown in Fig.~\ref{fig:efficiency}.
%
%
For very long times, a sufficiently high-$Q$ cavity and almost exact
resonance, the extraction efficiency approaches 
$\eta(t)\to\gamma_{k\mathrm{rad}}/\Gamma_k$. Here, the
$\gamma_{k\mathrm{rad}}$ is the radiative damping rate
$\gamma_{k\mathrm{rad}}=c/(2l)|T_k|^2$ and $T_k$ the transmission
coefficient $T_k=t_{13}(\Omega_k)e^{i\omega_kl/c}$. The decay rate
$\Gamma_k$ of a cavity resonance is the sum of the radiative damping
rate $\gamma_{k\mathrm{rad}}$ and the absorptive damping rate
$\gamma_{k\mathrm{abs}}=c/(2l)|A_k|^2$ where $A_k$ is the absorption
coefficient at the cavity resonance frequency $\omega_k$. Hence, the
extraction efficiency in the stationary limit is
\cite{Khanbekyan04,Cui05} 
\begin{equation}
\eta(t) \to
\frac{\gamma_{k\mathrm{rad}}}{\gamma_{k\mathrm{rad}}+\gamma_{k\mathrm{
abs}}}
\,.
\end{equation}
For intermediate times, the efficiency shows an oscillating (tidal)
behaviour. Thus, during each emission/re-absorption cycle within the
cavity, the single-photon wave function builds up outside the cavity
until the steady-state solution (empty cavity) is reached.


\subsection{Spherical microcavities}

Spherical microresonators possess a rich structure of field resonances
to which an atom can be coupled when placed inside (whispering gallery
modes) or outside (surface-guided modes) the resonator.


\subsubsection{Atom inside a spherical microcavity}

\begin{figure}[!t!]
\centerline{\includegraphics[width=3cm]{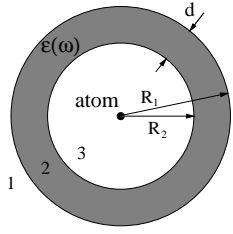}}
\caption{\label{fig:hollowsphere} Scheme of the spherical cavity with
outer radius $R_1$ and inner radius $R_2$.}
\end{figure}

\begin{figure}[!t!]
\includegraphics[width=0.49\textwidth]{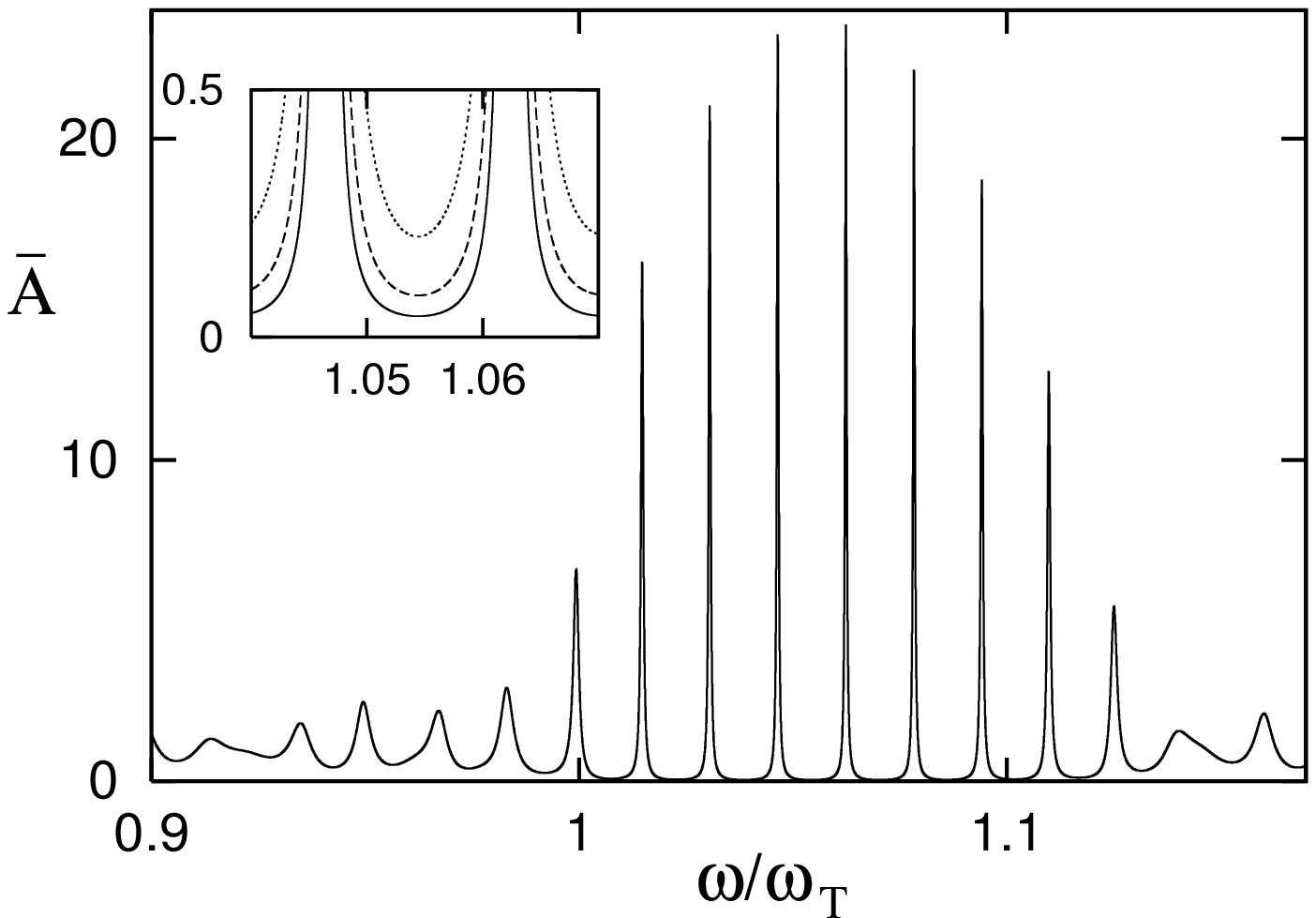}
\hfill
\includegraphics[width=0.49\textwidth]{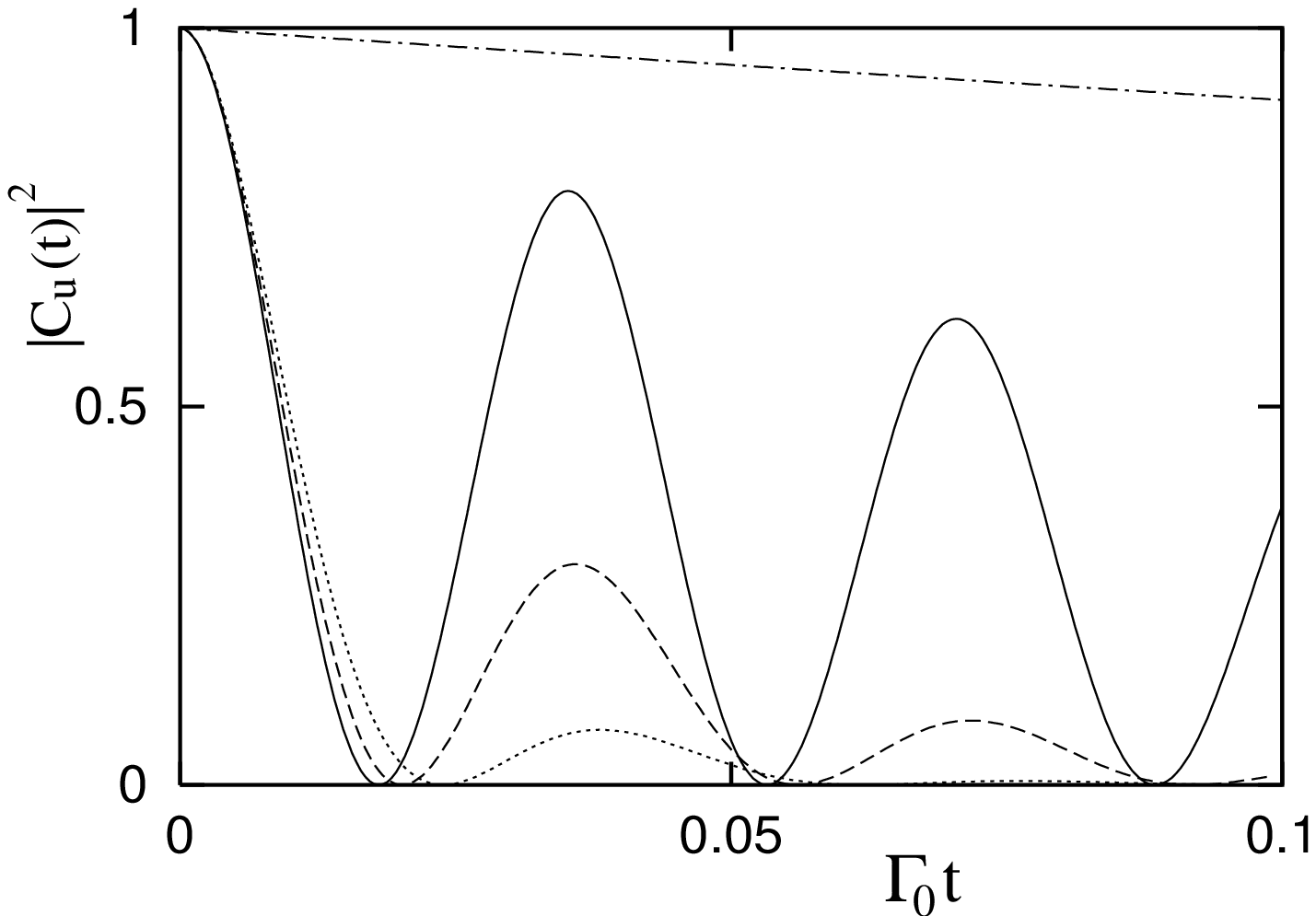}
\caption{\label{fig:cavityresonance} Spontaneous decay rate
$\Gamma/\Gamma_0$ as a function of frequency (left
panel). Excited-state probability $|C_e(t)|^2$ as a function of decay
time (right panel). For parameters, see text. Figures taken from
Ref.~\cite{Dung00}.}
\end{figure}

In cases in which the atom is embedded in a spherical microcavity,
we can distinguish between large and small cavities, weak and strong
coupling. In Sec.~\ref{sec:modifiedse} we have discussed the case in
which an atom was placed inside a very small cavity. This led us to
the notion of local-field corrections. We now have a brief look at a
cavity whose radius is much larger than the relevant atomic transition
frequency. In view of the three-layered structure depicted in
Fig.~\ref{fig:hollowsphere}, we thus require that $R_2\omega_A/c\gg
1$.
%
%
Then, the rate of spontaneous decay can be calculated from
Eq.~(\ref{eq:reflectioncoefficient}) in the limit of thick cavity
walls as \cite{Dung00}
\begin{equation}
\Gamma \simeq \Gamma_0 \mathrm{Re} \left[ 
\frac{n(\omega_A)-i\tan(R_2\omega_A/c)}{
1-in(\omega)\tan(R_2\omega_A/c)} 
\right] \,.
\end{equation}
The left panel in Fig.~\ref{fig:cavityresonance} depicts the
normalised decay rate $\Gamma/\Gamma_0$ as a function of frequency. We
assume a single-resonance Drude--Lorentz model
(\ref{eq:epsilonmusingle}) with $\omega_P=0.5\omega_T$ and
$\gamma=0.01\omega_T$. The microresonator has parameters
$R_2=30\lambda_T$ and $R_1-R_2=\lambda_T$. One observes 
narrow-band enhancement ($\Gamma/\Gamma_0\gg 1$) alternating with
broadband suppression ($\Gamma/\Gamma_0\ll 1$) of spontaneous
decay. The narrow peaks are located at the cavity resonances. 
%

When the atomic transition frequency approaches one of the cavity
resonances, the atom-field coupling becomes stronger. At resonance,
when $\omega_A=\omega_C$ where $\omega_C$ is one of the resonance
frequencies of the microcavity, the time evolution for the
excited-state occupation probability is the one of a damped oscillator
with damping constant $\delta\omega_C$, the width of the cavity
resonance. For small material absorption,
$\gamma\ll\omega_T,\omega_P,\omega_P^2/\omega_T$, the cavity resonance
line can be approximated by a Lorentzian with width
\begin{equation}
\delta\omega_C = \frac{c\Gamma_0}{R_2\Gamma(\omega_C)} \,.
\end{equation}
In the strong-coupling regime when the condition
$\Gamma(\omega_C)\gg\delta\omega_C$ is fulfilled, the excited-state
probability
\begin{equation}
|C_e(t)|^2 = e^{-\delta\omega_Ct}
\cos^2 \left(\sqrt{\frac{\Gamma(\omega_C)\delta\omega_C}{2}} t\right)
\end{equation}
shows damped Rabi oscillations with Rabi frequency
$\Omega=\sqrt{2\Gamma(\omega_C)\delta\omega_C}$. The right panel in
Fig.~\ref{fig:cavityresonance} shows the temporal evolution of the
excited-state occupation probability for $\omega_A=1.046448\omega_T$,
$\Gamma_0\lambda_T/(2c)=10^{-6}$, and $\gamma=10^{-4}\omega_T$ (solid
line), $\gamma=5\times 10^{-4}\omega_T$ (dashed line), and
$\gamma=10^{-3}\omega_T$ (dotted line). The other parameters are the
same as those used for the curves in the left panel. For comparison,
the exponential decay in free space is also shown (dashed-dotted
line).


\subsubsection{Atom outside a spherical microcavity}
\label{sec:guidedmodes}

Atoms need not be located inside a closed dielectric or metallic
structure in order to be coupled strongly to the electromagnetic
field. Instead of whispering gallery modes that are due to total
internal reflection inside a sphere, surface-guided modes can
be excited on the outside surface of a spherical microcavity
\cite{Goetzinger06}. Assuming again a Drude--Lorentz model
for the permittivity of the dielectric sphere that features a band
gap between the transverse resonance frequency $\omega_T$ and the
longitudinal resonance frequency
$\omega_L=\sqrt{\omega_T^2+\omega_P^2}$. It is clear that for
frequencies below $\omega_T$ ($\omega_A<\omega_T$) whispering gallery
modes are excited, whereas for frequencies within the band gap
($\omega_T<\omega_A<\omega_L$) surface-guided modes are excited. 

\begin{figure}[!t!]
\includegraphics[width=0.49\textwidth]{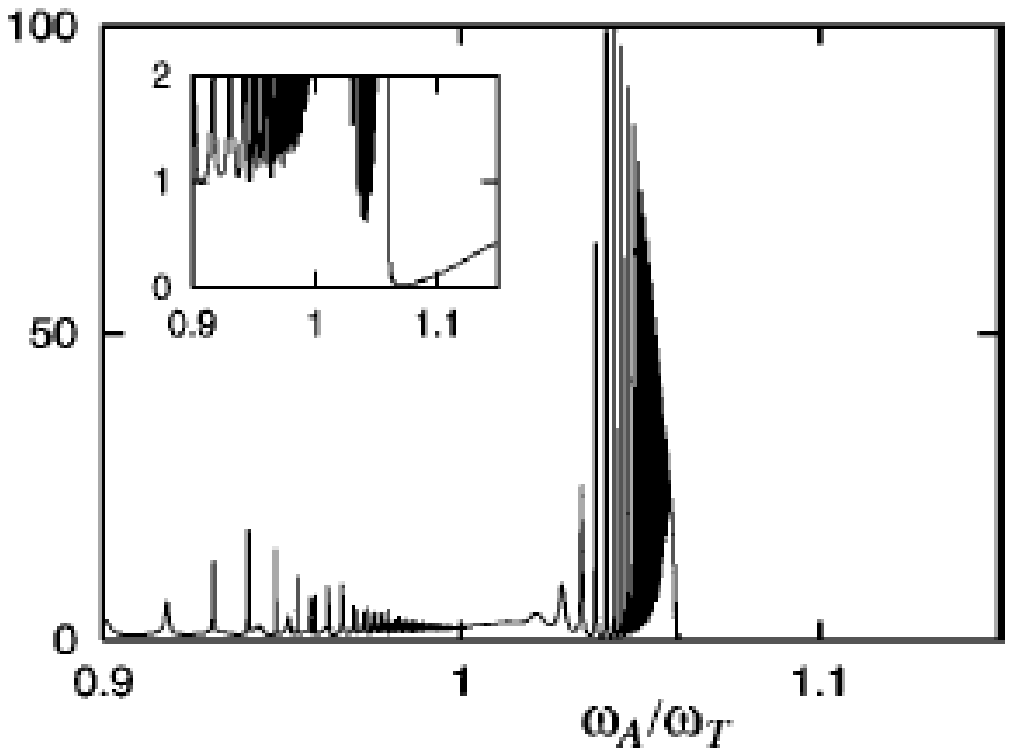}
\hfill
\includegraphics[width=0.49\textwidth]{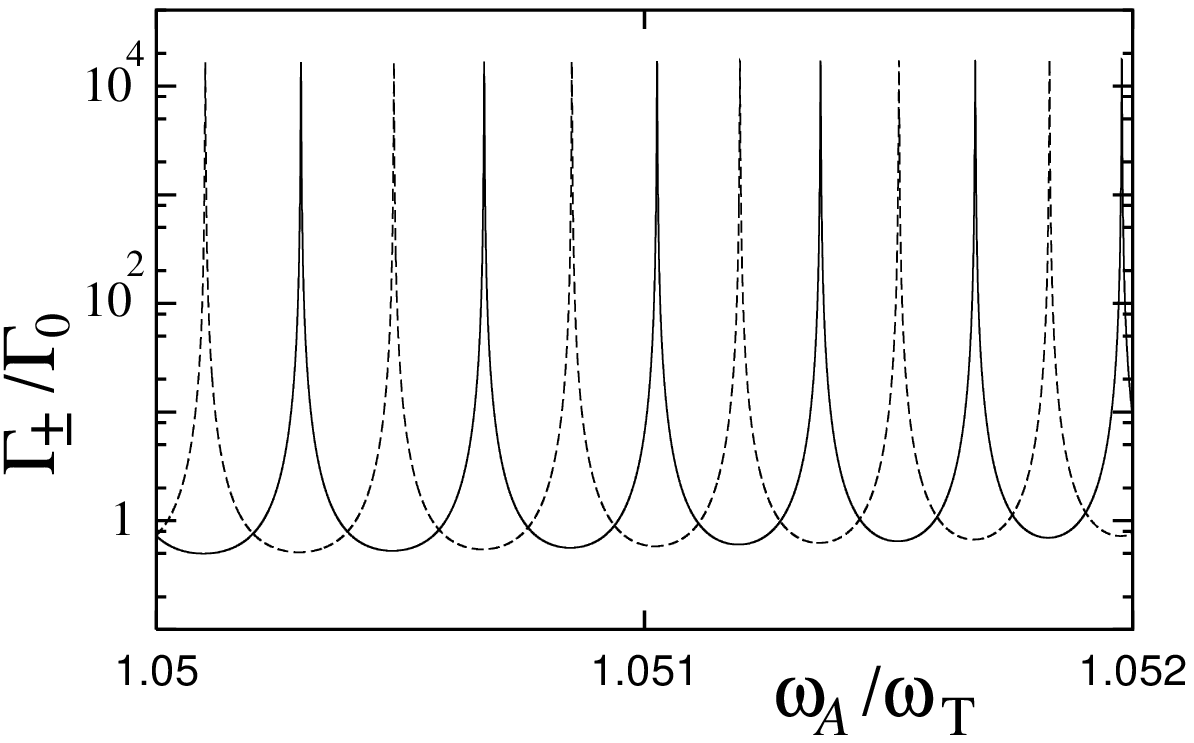}
\caption{\label{fig:strongdecay} Left panel: Decay rate as a function
of the atomic transition frequency for a radially oscillating dipole
near a microsphere of radius $R=2\lambda_A$. The chosen parameters are
$\omega_P=0.5\omega_T$, $\gamma=10^{-4}\omega_T$,
$\Delta r=0.1\lambda_A$. Figure taken from Ref.~\cite{Dung01}. Right
panel: Frequency dependence of the decay rates $\Gamma_\pm$ for two
radially oriented dipoles at
$\Delta r_A=\Delta r_B=0.02\lambda_A$. The other parameters are
$\omega_P=0.5\omega_T$, $\gamma=10^{-6}\omega_T$,
$R=10\lambda_T$ \cite{Dung02}.}
\end{figure}

Figure~\ref{fig:strongdecay} shows the rate of spontaneous decay of a 
radially oriented dipole very close ($z_A=r_A-R=0.1\lambda_A$)
to a microsphere of radius $R=2\lambda_A$.
%
%
In this near-field limit, the decay rate is approximately given by 
Eq.~(\ref{eq:nearfieldplanar}) for a planar interface,
\begin{equation}
\frac{\Gamma^\perp}{\Gamma_0} =
\frac{3\mathrm{Im}\,\varepsilon(\omega_A)}
{4|\varepsilon(\omega_A)+1|^2} \left( \frac{c}{\omega_Az_A}
\right)^3
+\mathcal{O}(1) \,.
\end{equation}
The strong enhancement in the band-gap region
($\omega_T<\omega_A<\omega_L$) is due to surface-guided modes.

These surface excitations can be used to induce entanglement between
two atoms located diametrically opposite around the microsphere
\cite{Goetzinger06,Dung02}. If the two atoms $A$ and $B$ are initially
prepared in their respective excited and ground states, their combined
initial state can be written in terms of the (entangled) coherent
superpositions
$|\pm\rangle=(|e_A,g_B\rangle\pm|g_A,e_B\rangle)/\sqrt{2}$ which are 
initially equally excited. Associated with these superpositions are
decay rates
\begin{eqnarray}
\frac{\Gamma_\pm}{\Gamma_0} &=& \frac{3}{2} \sum\limits_{l=1}^\infty
\mathrm{Re} \bigg\{ \frac{l(l+1)(2l+1)}{(k_Ar_A)^2} h_l^{(1)}(k_Ar_A)
\nonumber \\ && \times
\left[ j_l(k_Ar_A) +r_{p,l}^{(11)}(\omega_A) h_l^{(1)}(k_Ar_A)
\right] \left[ 1\mp(-1)^l \right] \bigg\}
\end{eqnarray}
[$k_A=\omega_A/c$; $j_l(z)$ and $h_l^{(1)}(z)$, spherical Bessel and
Hankel functions; $r_{p,l}^{(11)}(\omega_A)$, generalised reflection
coefficient given in Sec.~\ref{sec:sphericaldgf}]. If the atomic
transition frequency $\omega_A$ coincides with one of the microsphere
resonances $l$, the single-atom decay rate can be approximated by
\cite{Dung01}
\begin{equation}
\frac{\Gamma}{\Gamma_0} \simeq \frac{3}{2} l(l+1)(2l+1) \mathrm{Re}
\left\{ \left[ \frac{h_l^{(1)}(k_Ar_A)}{k_Ar_A} \right]^2
r_{p,l}^{(11)}(\omega_A)  \right\}
\end{equation}
and therefore the decay rates of the superposition states reduce to
$\Gamma_\pm\simeq\Gamma[1\mp(-1)^l]$. This means that if $l$ is even
(odd), the state $|-\rangle$ ($|+\rangle$) decays much faster than the
state $|+\rangle$ ($|-\rangle$). The frequency dependence of
$\Gamma_\pm$ is shown on the right panel in
Fig.~\ref{fig:strongdecay}.
Hence, depending on $l$ one of the superposition states $|\pm\rangle$
is superradiant while the other is subradiant. After the superradiant
combination has decayed, the two atoms reside in the subradiant
entangled state $|+\rangle$ or $|-\rangle$. 


\newpage
\section{Outlook}

The theory of macroscopic quantum electrodynamics is a powerful tool
that provides the link between isolated atomic ensembles and absorbing
solid-state surfaces. It extends the well-established field of
free-space quantum optics to new horizons. It is remarkable that, at
least within the framework of causal linear response, one finds a
Hamiltonian description of the medium-assisted electromagnetic field
that does not have to fall back onto a master equation description.
Instead, unitary evolution equations can be found that take into
account absorption processes as well. Examples of such unitary
evolutions include the absorbing beam splitter
(Sec.~\ref{sec:lossybeamsplitter}) and internal atomic dynamics
(Sec.~\ref{sec:relaxation}). In many cases, macroscopic quantum 
electrodynamics can provide more information than perturbative 
calculations, as exemplified by the theory of thermal Casimir--Polder 
forces (Sec.~\ref{sec:thermalCP}). 

Obvious extensions towards nonlinear interaction Hamiltonians 
(Sec.~\ref{sec:nonlinear}) will be further studied to bridge the gap
between classical descriptions of nonlinear processes to include
both linear and nonlinear absorption, and to establish the connection
with microscopic models of effective nonlinear interactions. Such a
theory is likely to provide detailed information on the quantum state
of light that emerges from, e.g. spontaneous parametric
down-conversion crystals, and on the effect of higher-order
nonlinearities.

As it is evident from our collection of examples, for the time being
we have always assumed that all macroscopic bodies as well as all
atomic systems are at rest. Such a restriction is clearly not
necessary. Centre-of-mass motion of atomic systems can be accounted
for by either imposing a classical trajectory or more consistently by
solving Newton's equations of motion in addition to Maxwell's
equations. To assume a classical trajectory for an atom or a molecule
means to neglect the backaction onto the motion which in many cases
can be justified. 

For example, from our discussion in Sec.~\ref{sec:dispersion} it is
clear that an atom moving parallel to a planar dielectric surface will
not only experience a Casimir--Polder force in the direction
perpendicular to its motion, but also in the direction of its motion.
This effect is sometimes called quantum friction and can be
understood in two ways. In our intuitive picture of Casimir--Polder
forces being caused by forces between the atomic dipole and image
dipoles in the magnetoelectric material, one can view quantum friction
as a drag effect acting on the image dipoles due to finite
conductivity or even resonances. Another way to view quantum friction
is as an interaction with the surface plasmon at the magnetoelectric
interface.

Taking the quantum-mechanical character of atomic position and
momentum into account will lead one to a theory of motional heating
that is important in particular in ion traps. The opposite effect of
cooling to the motional ground state, enhanced by coupling to a
resonant microcavity, can also be understood within the framework of
macroscopic quantum electrodynamics.

Eventually, understanding how motion affects dispersion forces acting
on isolated atomic systems will lead to a theory of a quantum theory
of moving dielectric materials which thus far only exists for
strictly non-absorbing materials. This in of fundamental interest as
it makes contact with the low-energy limit of effective action in
quantum field theory.


\begin{ack}
This work was financially supported partially by the UK Engineering
and Physical Sciences Research Council and the Alexander~von~Humboldt
foundation. The authors are grateful to all our colleagues who
contributed to the theory of macroscopic quantum electrodynamics, in
particular D.-G.~Welsch who was involved in developing this theory
from the very beginning, J.A.~Crosse, H.T.~Dung, R.~Fermani,
M.~Khanbekyan, L.~Kn\"oll, C.~Raabe, H.~Safari, A.~Sambale,
M.R.~Tarbutt, A.~Tip, and M.S.~Toma\v{s}. We thank E.A.~Hinds and
P.L.~Knight at Imperial College London for their continuous support,
and the staff at the Gloucester Arms for hospitality during our
Doppelkopf sessions. 
\end{ack}


\appendix


\newpage
\section{Dyadic Green functions}
\label{sec:dgf}
        
The notorious problem one perpetually faces in macroscopic QED is to
find the Green tensor or dyadic Green function (DGF) associated with
the classical electromagnetic scattering problem. This is not unusual
as the DGF generalizes the usual mode expansion in free space to the
solution of a more general boundary value problem. In what follows we
will restrict ourselves to the special case of spatially local
magnetoelectric materials. Extensions to materials with spatial
dispersion can be found in Ref.~\cite{Raabe07}.

We will need to look for the fundamental solution to the Helmholtz
equation
\begin{equation}
\label{eq:a1}
\biggl[\curl\kappa(\vect{r},\omega)\curl
-\frac{\omega^2}{c^2} \varepsilon(\vect{r},\omega)\biggr]
\vect{E}(\vect{r},\omega) = \mi\mu_0\omega \vect{j}(\vect{r},\omega) 
\end{equation}
with the dielectric permittivity
$\varepsilon(\vect{r},\omega)$ and inverse magnetic permeability
$\kappa(\vect{r},\omega)$ $=\mu^{-1}(\vect{r},\omega)$ being
arbitrary. The solution to Eq.~(\ref{eq:a1}) can be written in terms
of the DGF as
\begin{equation}
\label{eq:fieldsource}
\vect{E}(\vect{r},\omega) = \mi\mu_0\omega \int
d^3r'\,\tens{G}(\vect{r},\vect{r}',\omega) \cdot
\vect{j}(\vect{r}',\omega) 
\end{equation}
where the dyadic Green function satisfies
\begin{equation}
\label{eq:a2}
\biggl[\curl\kappa(\vect{r},\omega)\curl
-\frac{\omega^2}{c^2} \varepsilon(\vect{r},\omega)\biggr]
\tens{G}(\vect{r},\vect{r}',\omega) =
\ten{\delta}(\vect{r}-\vect{r}').
\end{equation}
Together with the boundary condition that
$\tens{G}(\vect{r},\vect{r}',\omega)$ vanishes as
$|\vect{r}-\vect{r}'|\to\infty$, Eq.~(\ref{eq:a2}) has a unique
solution provided that the strict inequalities
$\mathrm{Im}\,\varepsilon(\vect{r},\omega)>0$ and
$\mathrm{Im}\,\mu(\vect{r},\omega)>0$ hold. Physically, these
requirements mean that all dielectric materials have to be passive,
i.e. absorbing, media.


\subsection{General properties}
\label{sec:dgfproperties}

The dyadic Green function inherits the analytical properties of the
permittivity and permeability. That is, it is meromorphic in the upper
complex frequency half-plane, and the Schwarz reflection principle
\begin{equation}
\label{eq:analyticity}
\tens{G}^\ast(\vect{r},\vect{r}',\omega) =
\tens{G}(\vect{r},\vect{r}',-\omega^\ast)
\end{equation}
holds. The Onsager reciprocity theorem requires that the DGF
satisfies the additional relation
\begin{equation}
\label{eq:reciprocity}
\tens{G}(\vect{r}',\vect{r},\omega) =
\tens{G}^\trans(\vect{r},\vect{r}',\omega) \,.
\end{equation}
Particularly useful is the following integral relation involving
products of Green functions,
\begin{eqnarray}
\label{eq:magicformula}
\lefteqn{
\int d^3s \bigg\{
\frac{\omega^2}{c^2}\,\mathrm{Im}\,\varepsilon(\vect{s},\omega)
\tens{G}(\vect{r},\vect{s},\omega) \cdot
\tens{G}^\ast(\vect{s},\vect{r}',\omega)
} \nonumber \\ && 
+\mathrm{Im}\,\kappa(\vect{s},\omega) \left[
\tens{G}(\vect{r},\vect{s},\omega) \times \overleftarrow \grad \right]
\cdot \left[ \curl \tens{G}^\ast(\vect{s},\vect{r}',\omega) \right] 
\bigg\} =  
\mathrm{Im}\,\tens{G}(\vect{r},\vect{r}',\omega) \,.
\end{eqnarray}
This relation is essentially the linear fluctuation-dissipation
theorem, written in terms of the dyadic Green function.


\subsection{Duality relations}
\label{sec:dualgreen}

Another useful property of the DGF is its behaviour under a
duality transformation. The dual DGF is defined by the equation
\begin{equation}
\label{eq:defdualdgf}
\biggl[\curl\kappa^\star(\vect{r},\omega)\curl
-\frac{\omega^2}{c^2}\varepsilon^\star(\vect{r},\omega)\biggr]
\tens{G}^\star(\vect{r},\vect{r}',\omega) =
\ten{\delta}(\vect{r}-\vect{r}')
\end{equation}
with $\varepsilon^\star=\mu=1/\kappa$,
$\kappa^\star=1/\mu^\star=1/\varepsilon$, i.e., it is the solution to
the Helmholtz equation with $\varepsilon$ and $\mu$ exchanged.
By applying the duality transformation to the field
expansions~(\ref{eq:meE})--(\ref{eq:meH}), using the transformation
rules listed in Tab.~\ref{tab:duality} and comparing coefficients, one
can easily verify the following relations between the original DGF and
its dual \cite{BuhmannScheel08}:
\begin{eqnarray}
\label{dgftrans1}
\frac{\omega^2}{c^2}\,\ten{G}^\star(\vect{r},\vect{r}',\omega)
 &=&-\kappa(\vect{r},\omega)
 \curl\ten{G}(\vect{r},\vect{r}',\omega)\vprod
 \overleftarrow{\grad}'
 \kappa(\vect{r}',\omega)
\nonumber\\ &&
 -\kappa(\vect{r},\omega)
 \bm{\delta}(\vect{r}-\vect{r}'),\\
\label{dgftrans2}
\curl\ten{G}^\star(\vect{r},\vect{r}',\omega)\vprod
 \overleftarrow{\grad}'
&=&-\varepsilon(\vect{r},\omega)\,\frac{\omega^2}{c^2}\,
 \ten{G}(\vect{r},\vect{r}',\omega)
 \varepsilon(\vect{r}',\omega)
+\varepsilon(\vect{r},\omega)\bm{\delta}(\vect{r}-\vect{r}'), \\
\label{dgftrans3}
\curl\ten{G}^\star(\vect{r},\vect{r}',\omega)
&=&-\varepsilon(\vect{r},\omega)
 \ten{G}(\vect{r},\vect{r}',\omega)\vprod\overleftarrow{\grad}'
 \kappa(\vect{r}',\omega),\\
\label{dgftrans4}
\ten{G}^\star(\vect{r},\vect{r}',\omega)\vprod\overleftarrow{\grad}'
&=&-\kappa(\vect{r},\omega)
 \curl\ten{G}(\vect{r},\vect{r}',\omega)
 \varepsilon(\vect{r}',\omega)\,,
\end{eqnarray}

In the next subsections, we will list the explicit forms of dyadic
Green functions for important highly symmetric geometric arrangements
of magnetoelectric bodies.

\subsection{Bulk material}
\label{sec:bulkdgf}

The simplest situation one can envisage is one of a homogeneous,
isotropic dielectric medium. In this case, the dielectric permittivity
$\varepsilon(\omega)$ and magnetic permeability $\mu(\omega)$ do not
depend on the spatial position. This means that the magnetoelectric
material and hence the dyadic Green function that describes it must be
translationally invariant. Thus, the DGF can only depend on the
difference between the coordinates $\bm{\varrho}=\vect{r}-\vect{r}'$, 
$\tens{G}(\vect{r},\vect{r}',\omega)\equiv\tens{G}(\bm{\varrho},
\omega)$.
We can therefore use Fourier transform techniques to solve
Eq.~(\ref{eq:a2}). Defining
\begin{equation}
\tens{G}(\vect{k},\omega) = \int \frac{d^3k}{(2\pi)^{3/2}} \,
\tens{G}(\bm{\varrho},\omega) e^{-i\vect{k}\cdot\bm{\varrho}} \,,
\end{equation}
the Fourier transformed Helmholtz equation is
\begin{equation}
-\vect{k}\times\vect{k}\times \tens{G}(\vect{k},\omega)
-\frac{\omega^2}{c^2} \varepsilon(\omega)\mu(\omega)
\tens{G}(\vect{k},\omega) = \mu(\omega) \tens{I} \,.
\end{equation}
This is now an algebraic equation that can be easily solved. To do so,
we decompose the unit tensor into its transverse and longitudinal
parts with respect to the wave vector $\vect{k}$,
\begin{equation}
\tens{I} = \left( \tens{I}-\frac{\vect{k}\otimes\vect{k}}{k^2} \right)
+\frac{\vect{k}\otimes\vect{k}}{k^2} \,.
\end{equation}
Note that
\begin{equation}
-\vect{k}\times\vect{k}\times \equiv k^2 \left(
\tens{I}-\frac{\vect{k}\otimes\vect{k}}{k^2} \right)
\end{equation}
is transverse. Hence, the Green tensor is, after inserting into the
inverse Fourier transform,
\begin{equation}
\tens{G}(\bm{\varrho},\omega) =  \int \frac{d^3k}{(2\pi)^{3/2}} \,
e^{i\vect{k}\cdot\bm{\varrho}} \mu(\omega) \left[
\frac{c^2}{k^2c^2-\omega^2\varepsilon(\omega)\mu(\omega)} \left(
\tens{I}-\frac{\vect{k}\otimes\vect{k}}{k^2} \right)
-\frac{c^2}{\omega^2\varepsilon(\omega)\mu(\omega)}
\frac{\vect{k}\otimes\vect{k}}{k^2} 
\right]
\end{equation}
which can be transformed back into configuration space using contour
integral methods. Noting  that the dielectric permittivity
$\varepsilon(\omega)$ and the magnetic permeability $\mu(\omega)$ have
no poles or zeros in the upper complex frequency half-plane, the
result is
\begin{equation}
\label{eq:bulkG}
\tens{G}(\bm{\varrho},\omega) = \left[ \grad\otimes\grad
+\frac{\omega^2}{c^2} \varepsilon(\omega)\mu(\omega) \tens{I} \right]
\frac{e^{iq(\omega)\varrho}\mu(\omega)}{4\pi q^2(\omega)\varrho}
\end{equation}
[$q^2(\omega)=\frac{\omega^2}{c^2}\varepsilon(\omega)\mu(\omega)$].
After evaluating the derivatives one arrives at a decomposition into
transverse and longitudinal parts as
\begin{eqnarray}
\label{eq:longitudinalG}
\tens{G}^\|(\bm{\varrho},\omega) &=& -\frac{\mu(\omega)}{4\pi q^2}
\left[
\frac{4\pi}{3} \delta(\bm{\varrho})\tens{I} +\left( \tens{I}
-\frac{3\bm{\varrho}\otimes\bm{\varrho}}{\varrho^2} \right)
\frac{1}{\varrho^2} \right] \,,\\
\label{eq:transverseG}
\tens{G}^\perp(\bm{\varrho},\omega) &=& \frac{\mu(\omega)}{4\pi q^2}
\left\{ \left( \tens{I}
-\frac{3\bm{\varrho}\otimes\bm{\varrho}}{\varrho^2}
\right) +q^3 \left[ \left( \frac{1}{q\varrho} +\frac{i}{(q\varrho)^2}
-\frac{1}{(q\varrho)^3} \right) \tens{I}
\right.\right. \nonumber \\ && \hspace*{4ex} \left.\left.
- \left( \frac{1}{q\varrho}
+\frac{3i}{(q\varrho)^2} -\frac{3}{(q\varrho)^3} \right)
\frac{\bm{\varrho}\otimes\bm{\varrho}}{\varrho^2} \right]
e^{iq\varrho} \right\} \,.
\end{eqnarray}
In particular, in free space where $\varepsilon(\omega)=1$, we find
that
\begin{equation}
\label{eq:freespaceG}
\tens{G}^{(0)}(\bm{\varrho},\omega) = -\frac{c^2}{3\omega^2}
\tens{\delta}(\bm{\varrho}) +\frac{\omega}{4\pi c} \left[
f\left(\frac{c}{\omega\varrho}\right)\tens{I} 
+g\left(\frac{c}{\omega\varrho}\right)
\frac{\bm{\varrho}\otimes\bm{\varrho}}{\varrho^2} \right] 
e^{i\varrho\omega/c}
\end{equation}
with $f(x)=x+ix^2-x^3$ and $g(x)=x+3ix^2-3x^3$.


\subsection{Layered media: planar, cylindrical, spherical}
\label{sec:layeredmedia}

A more challenging situation arises if one considers magnetoelectric
materials that are structured to form boundaries of 
certain symmetry. What we envisage here are examples of
stratified materials in planar, cylindrical and spherical
geometries. The idea is to expand the dyadic Green function in terms
of vector wave functions associated with the symmetry of the
problem. The success of this procedure is directly related to the
problem of separability of the Helmholtz operator
$\curl\mu^{-1}(\vect{r},\omega)\curl$
$-\frac{\omega^2}{c^2}\varepsilon(\vect{r},\omega)$. There
are only a few coordinate systems in which this operator is separable
\cite{MoonSpencer}, namely the cartesian, cylindrical, spherical and
spheroidal coordinate systems. Hence, only there can the solution
to the three-dimensional Helmholtz operator be reduced to seeking
the scalar solution of a one-dimensional wave equation.

The scalar Helmholtz equation takes the form
\begin{equation}
\left( \Delta+k^2 \right) \psi(\vect{k},\vect{r}) =0
\end{equation}
[$k^2=\frac{\omega^2}{c^2}\varepsilon(\vect{r},\omega)$].
Table~\ref{tab:pilot} lists the scalar eigenfunctions that are regular
at the origin for cartesian, cylindrical and spherical coordinate
systems.

\begin{table}[!t!]
\begin{center}
\begin{tabular}{|l|c|l|}
\hline
coordinate system & pilot vector & scalar eigenfunctions \\
\hline\hline
cartesian ($x,y,z$) & $\vect{e}_z$ &
$\psi(\vect{k},\vect{r}) = e^{ik_xx}e^{ik_yy}e^{ik_zz}$ \\ \hline
cylindrical ($r,\varphi,z$) & $\vect{e}_z$ &
$\psi_n(k_r,k_z,\vect{r}) = J_n(k_rr)e^{in\varphi}e^{ik_zz}$ \\ \hline
spherical ($r,\theta,\varphi$) & $\vect{e}_r$ &
$\psi_{l,m}(k,\vect{r}) = j_l(kr)P_l^m(\cos\theta)e^{im\varphi}$ \\
\hline
\end{tabular}
\end{center}
\caption{\label{tab:pilot} Pilot vectors and scalar eigenfunctions
that are regular at the origin for cartesian, cylindrical and
spherical coordinate systems. The functions $J_n(kr)$ and $j_l(kr)$
denote cylindrical and spherical Bessel functions, respectively. The
$P_l^m(\cos\theta)$ are associated Legendre polynomials.}
\end{table}

The scalar wave function $\psi(\vect{k},\vect{r})$ is then used to
construct the irrotational ($\vect{L}$) and solenoidal ($\vect{M}$ and
$\vect{N}$) eigenfunctions with respect to a pilot vector $\vect{c}$
as \cite{Chew}
\begin{eqnarray}
\vect{L}(\vect{k},\vect{r}) &=& \grad \psi(\vect{k},\vect{r}) \,,\\
\vect{M}(\vect{k},\vect{r}) &=& \curl
\left[\psi(\vect{k},\vect{r})\vect{c}\right] 
\,,\\
\vect{N}(\vect{k},\vect{r}) &=& \frac{1}{k}\curl\curl
\left[\psi(\vect{k},\vect{r})\vect{c}\right] \,.
\end{eqnarray}
Because of the orthogonality of the scalar eigenfunctions,
\begin{equation}
\int d^3r\, \psi(\vect{k},\vect{r})\psi(-\vect{k}',\vect{r})
=(2\pi)^3 \delta(\vect{k}-\vect{k}')\,,
\end{equation}
the vector wave function $\vect{M}$, $\vect{N}$ and $\vect{L}$ are
mutually orthogonal.

Due to the Helmholtz theorem, these three types of eigenfunctions form
a complete set of basis functions for the Helmholtz operator. The
dyadic Green tensor, written in terms of these eigenfunctions, takes
the form of a Fourier integral 
\begin{equation}
\tens{G}(\vect{r},\vect{r}',\omega)  = \int d^3k\, \left[
\vect{M}(\vect{k},\vect{r})\otimes\vect{a}(\vect{k})
+\vect{N}(\vect{k},\vect{r})\otimes\vect{b}(\vect{k})
+\vect{L}(\vect{k},\vect{r})\otimes\vect{c}(\vect{k}) \right]
\end{equation}
with yet unknown expansion coefficients $\vect{a}(\vect{k})$,
$\vect{b}(\vect{k})$ and $\vect{c}(\vect{k})$ which can be obtained
using the mutual orthogonality of the vector wave functions. The
interested reader is referred to the excellent textbook \cite{Chew}
for further details.

\begin{figure}[!t!]
\centerline{\includegraphics[width=8cm]{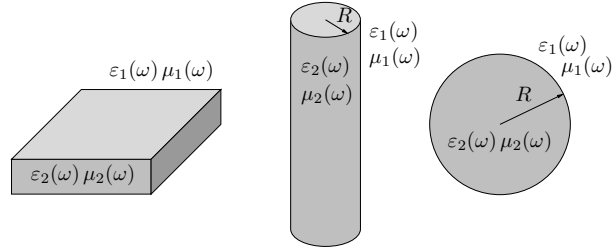}}
\caption{\label{fig:shapes} Inhomogeneous dielectrics for which the
dyadic Green function is analytically known.}
\end{figure}

For us, it suffices to know that these vector wave functions can be
used to determine the (unbounded) dyadic Green function in various
coordinate systems. Our interest is now focussed onto some physically
relevant situations in which two (or more) dielectric materials form a
planarly, cylindrically or spherically layered structure. More
specifically, if one were to consider only two adjoining materials,
one would be looking at two half-spaces joined by a planar interface,
an infinitely long wire or a dielectric sphere, respectively (see
Fig.~\ref{fig:shapes}).

The unbounded DGFs for each individual dielectric material have to be
joined at the common interface(s) across which the transverse field
components of electric and magnetic fields have to be continuous. Let
$\vect{n}$ be a unit vector normal to the interface
$\vect{r}=\vect{r}_S$ between two dielectrics. Then Maxwell's
equations imply that at the interface
\begin{eqnarray}
\vect{n}\times\tens{G}^{(fs)}
&=&\vect{n}\times\tens{G}^{[(f+1)s]}\,,\\ 
\frac{1}{\mu_f}\vect{n}\times\curl\tens{G}^{(fs)}
&=&\frac{1}{\mu_{f+1}}\vect{n}\times\curl\tens{G}^{[(f+1)s]}\,,
\end{eqnarray}
where the superscripts denote that the $s$ource point $\vect{r}'$
connects to the $f$ield point $\vect{r}$ via a Green function
$\tens{G}^{(fs)}$. The dyadic Green function can then be decomposed
into two parts, an unbounded DGF
$\tens{G}^{(0)}(\vect{r},\vect{r}',\omega)$ that represents direct
propagation from $\vect{r}'$ to $\vect{r}$ in an unbounded medium, and
a scattering part $\tens{G}^{(S)}(\vect{r},\vect{r}',\omega)$ that
describes the contributions from multiply reflected and transmitted
waves,
\begin{equation}
\tens{G}^{(fs)}(\vect{r},\vect{r}',\omega)
=\tens{G}^{(0,fs)}(\vect{r},\vect{r}',\omega) \delta_{fs}
+\tens{G}^{(S,fs)}(\vect{r},\vect{r}',\omega) \,.
\end{equation}
Mathematically speaking, the scattering part of the DGF has to be
included in order to fix the boundary conditions at the medium
interfaces, whereas the DGF of the unbounded medium is responsible for
the correct boundary conditions at infinity. 

\subsubsection{Planarly layered media}
\label{sec:planardgf}

There are two distinct methods for finding the dyadic Green function
for planarly layered media. Here we will describe the Weyl expansion
method \cite{Tomas95,Dung98} that gives very compact expressions for
the DGF. The second method using vector wave functions \cite{LiLW94a}
is somewhat less transparent and will be reserved for cylindrically
and spherically layered media.

The Weyl expansion is based on the translational invariance of the
dyadic Green function with respect to the directions parallel to the
planar interface,
\begin{equation}
\tens{G}(\vect{r},\vect{r}',\omega) = \int \frac{d^2k_\|}{(2\pi)^2}
e^{i\vect{k}_\|\cdot(\bm{\rho}-\bm{\rho}')} \mu(\omega)
\tens{G}(\vect{k}_\|,z,z',\omega)
\end{equation}
[$\vect{r}=(\bm{\rho},z)$]. The matrix components for the scattering
part of the DGF (we drop the superscript $S$ for notational
convenience) associated with reflection from the interface are given
by \cite{Tomas95,Dung98} 
\begin{eqnarray}
G^{(11)}_{xx}(\vect{k}_\|,z,z',\omega) &=&
\frac{i}{2k_{1z}}e^{ik_{1z}(|z|+|z'|)}
\left[ -r_p(\omega) \frac{k_{1z}^2k_x^2}{k_1^2k_\|^2} +r_s(\omega)
\frac{k_y^2}{k_\|^2} \right] \,,
\\
G^{(11)}_{xy}(\vect{k}_\|,z,z',\omega) &=&
\frac{i}{2k_{1z}}e^{ik_{1z}(|z|+|z'|)}
\left[ -r_p(\omega) \frac{k_{1z}^2k_xk_y}{k_1^2k_\|^2} -r_s(\omega)
\frac{k_xk_y}{k_\|^2} \right] \,,
\\
G^{(11)}_{xz}(\vect{k}_\|,z,z',\omega) &=&
\frac{i}{2k_{1z}}e^{ik_{1z}(|z|+|z'|)}
\left[ r_p(\omega) \frac{k_{1z}k_x}{k_1^2} \right] \,,
\\
G^{(11)}_{xz}(\vect{k}_\|,z,z',\omega) &=&
\frac{i}{2k_{1z}}e^{ik_{1z}(|z|+|z'|)}
\left[ -\mathrm{sgn}(z') r_p(\omega) \frac{k_\|^2}{k_1^2} \right] \,,
\end{eqnarray}
where $k_i^2=(\omega^2/c^2)\varepsilon_i(\omega)\mu_i(\omega)$ and
$k_{iz}^2=k_i^2-k_\|^2$. The functions $r_s(\omega)$ and $r_p(\omega)$
are the Fresnel coefficients of $s$- and $p$-polarised waves,
\begin{equation}
\label{eq:fresnel}
r_s(\omega) = \frac{\mu_2(\omega)k_{1z}-\mu_1(\omega)k_{2z}}
{\mu_2(\omega)k_{1z}-\mu_1(\omega)k_{2z}}
\,,\qquad
r_p(\omega) =
\frac{\varepsilon_2(\omega)k_{1z}-\varepsilon_1(\omega)k_{2z}} 
{\varepsilon_2(\omega)k_{1z}-\varepsilon_1(\omega)k_{2z}} \,.
\end{equation}
The remaining vector components can be deduced by employing the
replacement rule
$G^{(11)}_{yy}=G^{(11)}_{xx}(k_x\leftrightarrow k_y)$ and the
reciprocity theorem $\tens{G}(\vect{r},\vect{r}',\omega)$
$=\tens{G}^\trans(\vect{r}',\vect{r},\omega)$ which translates into
$\tens{G}^{(11)}(\vect{k}_\|,z,z',\omega)$
$=\tens{G}^{(11)\trans}(-\vect{k}_\|,z',z,\omega)$.

The Fresnel reflection coefficients (\ref{eq:fresnel}) describe a
single planar interface. For more than one interface, say a planar
layer of thickness $d$, the reflection coefficients can be combined
(for both polarisations) as
\begin{equation}
\label{eq:multifresnel}
\tilde{r}_{12} =
\frac{r_{12}+r_{23}e^{2ik_{2z}d}}{1-r_{21}r_{23}e^{2ik_{2z}d}}
\end{equation}
where $r_{ij}$ is the Fresnel coefficient for the interface
between layers $i$ and $j$.

The matrix components of the scattering DGF associated with
transmission through the planar interface are
\begin{eqnarray}
G^{(12)}_{xx}(\vect{k}_\|,z,z',\omega) &=& \frac{i}{2k_{1z}}
e^{ik_{1z}|z|+ik_{2z}|z'|} \left[ t_p(\omega)
\frac{k_{1z}k_{2z}k_x^2}{k_1k_2k_\|^2} +t_s(\omega)
\frac{k_y^2}{k_\|^2} \right] \,, 
\\
G^{(12)}_{xy}(\vect{k}_\|,z,z',\omega) &=& \frac{i}{2k_{1z}}
e^{ik_{1z}|z|+ik_{2z}|z'|} \left[ t_p(\omega)
\frac{k_{1z}k_{2z}k_xk_y}{k_1k_2k_\|^2} +t_s(\omega) 
\frac{k_xk_y}{k_\|^2}
\right] \,,
\\
G^{(12)}_{xz}(\vect{k}_\|,z,z',\omega) &=& \frac{i}{2k_{1z}}
e^{ik_{1z}|z|+ik_{2z}|z'|} \left[ \mathrm{sgn}(z') t_p(\omega)
\frac{k_{1z}k_x}{k_1k_2} \right] \,,
\\
G^{(12)}_{zx}(\vect{k}_\|,z,z',\omega) &=& \frac{i}{2k_{1z}}
e^{ik_{1z}|z|+ik_{2z}|z'|} \left[ \mathrm{sgn}(z') t_p(\omega)
\frac{k_{2z}k_x}{k_1k_2} \right] \,,
\\
G^{(12)}_{zz}(\vect{k}_\|,z,z',\omega) &=& \frac{i}{2k_{1z}}
e^{ik_{1z}|z|+ik_{2z}|z'|} \left[ t_p(\omega)
\frac{k_\|^2}{k_1k_2} \right] \,.
\end{eqnarray}
The transmission coefficients for $s$- and $p$-polarised waves
are
\begin{equation}
t_s(\omega) = \sqrt{\frac{\mu_1(\omega)}{\mu_2(\omega)}} \left[
1+r_s(\omega) \right] \,,\qquad
t_p(\omega) =
\sqrt{\frac{\varepsilon_1(\omega)}{\varepsilon_2(\omega)}} \left[
1+r_p(\omega) \right] \,, 
\end{equation}
The remaining vector components can again by symmetry arguments such
as $G^{(12)}_{yx}=G^{(12)}_{xy}$ and
$G^{(12)}_{yz}=G^{(12)}_{xz}(k_x\leftrightarrow k_y)$,
$G^{(12)}_{zy}=G^{(12)}_{zx}(k_x\leftrightarrow k_y)$ and
$G^{(12)}_{yy}=G^{(12)}_{xx}(k_x\leftrightarrow k_y)$.


\subsubsection{Cylindrically layered media}
\label{sec:cylindricaldgf}

For the two examples that follow here, we use the expansion of the DGF
in terms of vector wave functions. We first define the cylindrical
vector wave functions \cite{Tai,LiLW00} that depend on the radial
($\eta$) and axial ($h$) components of the wave number
[$k^2=\varepsilon(\omega)\mu(\omega)\omega^2/c^2=\eta^2+h^2$] as 
\begin{eqnarray}
\vect{M}_{\ueber{e}{o}n\eta}(h) &\!\!\!=&\!\!\! \left[ \mp
\frac{n}{\varrho}
Z_n(\eta\varrho) \ueber{\sin}{\cos} n\varphi\,
\vect{e}_\varrho
-\frac{dZ_n(\eta\varrho)}{d\varrho} \ueber{\cos}{\sin}n\varphi\,
\vect{e}_\varphi \right] e^{ihz} \,, \\
\vect{N}_{\ueber{e}{o}n\eta}(h) &\!\!\!=&\!\!\! \frac{1}{k} \left[
ih\frac{dZ_n(\eta\varrho)}{d\varrho} \ueber{\cos}{\sin}n\varphi\,
\vect{e}_\varrho
\mp ih\frac{n}{\varrho} Z_n(\eta\varrho) \ueber{\sin}{\cos}n\varphi\,
\vect{e}_\varphi
+\eta^2 Z_n(\eta\varrho) \ueber{\cos}{\sin}n\varphi\,
\vect{e}_z \right] e^{ihz} \,. \nonumber \\
\end{eqnarray}
The trigonometric functions have to be chosen appropriately for
the $e$ven and $o$dd types of functions. In terms of those, the dyadic
Green function for a cylindrical wire can be written in the form
\cite{Tai,LiLW00}
\begin{eqnarray}
\tens{G}^{(11)}(\vect{r},\vect{r}',\omega) &=&
\frac{i\mu_1}{8\pi} \int\limits_{-\infty}^\infty dh
\sum\limits_{n=0}^\infty \frac{(2-\delta_{n0})}{\eta_1^2}
\nonumber \\ && \hspace*{-10ex} \times
\left[ r_{MM}^{11} \vect{M}_{\ueber{e}{o}n\eta_1}^{(1)}(h) \otimes
\vect{M}_{\ueber{e}{o}n\eta_1}^{'(1)}(-h)
+r_{NN}^{11} \vect{N}_{\ueber{e}{o}n\eta_1}^{(1)}(h) \otimes
\vect{N}_{\ueber{e}{o}n\eta_1}^{'(1)}(-h) \right.
\nonumber \\ && \hspace*{-9ex}
+\left. r_{NM}^{11} \vect{N}_{\ueber{o}{e}n\eta_1}^{(1)}(h) \otimes
\vect{M}_{\ueber{e}{o}n\eta_1}^{'(1)}(-h)
+r_{MN}^{11} \vect{M}_{\ueber{o}{e}n\eta_1}^{(1)}(h) \otimes
\vect{N}_{\ueber{e}{o}n\eta_1}^{'(1)}(-h) \right],
\end{eqnarray}
\begin{eqnarray}
\tens{G}^{(21)}(\vect{r},\vect{r}',\omega) &=&
\frac{i\mu_1}{8\pi} \int\limits_{-\infty}^\infty dh
\sum\limits_{n=0}^\infty \frac{(2-\delta_{n0})}{\eta_1^2}
\nonumber \\ && \hspace*{-10ex} \times
\left[ t_{MM}^{21} \vect{M}_{\ueber{e}{o}n\eta_2}(h) \otimes
\vect{M}_{\ueber{e}{o}n\eta_1}^{'(1)}(-h)
+t_{NN}^{21} \vect{N}_{\ueber{e}{o}n\eta_2}(h) \otimes
\vect{N}_{\ueber{e}{o}n\eta_1}^{'(1)}(-h) \right.
\nonumber \\ && \hspace*{-9ex}
+\left. t_{NM}^{21} \vect{N}_{\ueber{o}{e}n\eta_2}(h) \otimes
\vect{M}_{\ueber{e}{o}n\eta_1}^{'(1)}(-h)
+t_{MN}^{21} \vect{M}_{\ueber{o}{e}n\eta_2}(h) \otimes
\vect{N}_{\ueber{e}{o}n\eta_1}^{'(1)}(-h) \right],
\end{eqnarray}
\begin{eqnarray}
\tens{G}^{(12)}(\vect{r},\vect{r}',\omega) &=&
\frac{i\mu_2}{8\pi} \int\limits_{-\infty}^\infty dh
\sum\limits_{n=0}^\infty \frac{(2-\delta_{n0})}{\eta_2^2}
\nonumber \\ && \hspace*{-10ex} \times
\left[ t_{MM}^{12} \vect{M}_{\ueber{e}{o}n\eta_1}^{(1)}(h) \otimes
\vect{M}_{\ueber{e}{o}n\eta_2}'(-h)
+t_{NN}^{12} \vect{N}_{\ueber{e}{o}n\eta_1}^{(1)}(h) \otimes
\vect{N}_{\ueber{e}{o}n\eta_2}'(-h) \right.
\nonumber \\ && \hspace*{-9ex}
+\left. t_{NM}^{12} \vect{N}_{\ueber{o}{e}n\eta_1}^{(1)}(h) \otimes
\vect{M}_{\ueber{e}{o}n\eta_2}'(-h)
+t_{MN}^{12} \vect{M}_{\ueber{o}{e}n\eta_1}^{(1)}(h) \otimes
\vect{N}_{\ueber{e}{o}n\eta_2}'(-h) \right],
\end{eqnarray}
\begin{eqnarray}
\tens{G}^{(22)}(\vect{r},\vect{r}',\omega) &=&
\frac{i\mu_2}{8\pi} \int\limits_{-\infty}^\infty dh
\sum\limits_{n=0}^\infty \frac{(2-\delta_{n0})}{\eta_2^2}
\nonumber \\ && \hspace*{-10ex} \times
\left[ r_{MM}^{22} \vect{M}_{\ueber{e}{o}n\eta_2}(h) \otimes
\vect{M}_{\ueber{e}{o}n\eta_2}'(-h)
+r_{NN}^{22} \vect{N}_{\ueber{e}{o}n\eta_2}(h) \otimes
\vect{N}_{\ueber{e}{o}n\eta_2}'(-h) \right.
\nonumber \\ && \hspace*{-9ex}
+\left. r_{NM}^{22} \vect{N}_{\ueber{o}{e}n\eta_2}(h) \otimes
\vect{M}_{\ueber{e}{o}n\eta_2}'(-h)
+r_{MN}^{22} \vect{M}_{\ueber{o}{e}n\eta_2}(h) \otimes
\vect{N}_{\ueber{e}{o}n\eta_2}'(-h) \right].
\end{eqnarray}
In this compact notation, a superscript ${}^{(1)}$ means that the
function $Z_n$ in the respective vector wave function is a Hankel
function of the first kind, $H_n^{(1)}$. Without that superscript, it
is understood that the Bessel function $J_n$ is used.

The reflection and transmission coefficients are determined by a
$4\times 4$-matrix equation
\begin{equation}
\tens{T}^{(H,V)} = \left[ \tens{F}^{(H,V)}_2 \right]^{-1}
\cdot \tens{F}^{(H,V)}_1
\end{equation}
with
\begin{eqnarray}
\tens{F}^H_j &=& \left( \begin{array}{cccc}
\frac{\partial H_n^{(1)}(\eta_jR)}{\partial R} &
\mp\frac{\zeta_jH_n^{(1)}(\eta_jR)}{R} &
\frac{\partial J_n(\eta_jR)}{\partial R} &
\mp\frac{\zeta_jJ_n(\eta_jR)}{R} \\
0 & \rho_jH_n^{(1)}(\eta_jR) & 0 & \rho_jJ_n(\eta_jR) \\
\pm\frac{\zeta_j\tau_jH_n^{(1)}(\eta_jR)}{R} &
\frac{\tau_j\partial H_n^{(1)}(\eta_jR)}{\partial R} &
\pm\frac{\zeta_j\tau_jJ_n(\eta_jR)}{R} &
\frac{\tau_j\partial J_n(\eta_jR)}{\partial R} \\
\tau_j\rho_jH_n^{(1)}(\eta_jR) & 0 & \rho_jJ_n(\eta_jR) & 0
\end{array} \right) \,,\\
\tens{F}^V_j &=& \left( \begin{array}{cccc}
\pm\frac{\zeta_jH_n^{(1)}(\eta_jR)}{R} &
\frac{\partial H_n^{(1)}(\eta_jR)}{\partial R} &
\pm\frac{\zeta_jJ_n(\eta_jR)}{R} &
\frac{\partial J_n(\eta_jR)}{\partial R} \\
\rho_jH_n^{(1)}(\eta_jR) & 0 & \rho_jJ_n(\eta_jR) & 0 \\
\frac{\tau_j\partial H_n^{(1)}(\eta_jR)}{\partial R} &
\mp\frac{\zeta_j\tau_jH_n^{(1)}(\eta_jR)}{R} &
\frac{\tau_j\partial J_n(\eta_jR)}{\partial R} &
\mp\frac{\zeta_j\tau_jJ_n(\eta_jR)}{R} \\
0 & \tau_j\rho_jH_n^{(1)}(\eta_jR) & 0 & \tau_j\rho_jJ_n(\eta_jR)
\end{array}\right) \,,
\end{eqnarray}
and the abbreviations
\begin{equation}
\tau_j=\sqrt{\frac{\varepsilon_j}{\mu_j}}
\,,\qquad
\zeta_j=\frac{ihn}{k_j}
\,,\qquad
\rho_j=\frac{\eta_j^2}{k_j} \,.
\end{equation}
With these preparations, the reflection and transmission coefficients
can be derived as
\begin{eqnarray}
\left( \begin{array}{c}
r^{11}_{MM,NN} \\ r^{11}_{NM,MN}
\end{array}\right)
&=& -\left( \begin{array}{cc}
T_{11}^{(H,V)} & T_{12}^{(H,V)} \\
T_{21}^{(H,V)} & T_{22}^{(H,V)}
\end{array}\right)^{-1}
\left( \begin{array}{c}
T_{13}^{(H,V)} \\ T_{23}^{(H,V)}
\end{array}\right)
\,, \\
\left( \begin{array}{c}
t^{21}_{MM,NN} \\ t^{21}_{NM,MN}
\end{array}\right)
&=& \left( \begin{array}{cc}
T_{31}^{(H,V)} & T_{32}^{(H,V)} \\
T_{41}^{(H,V)} & T_{42}^{(H,V)}
\end{array}\right)
\left( \begin{array}{c}
r^{11}_{MM,NN} \\ r^{11}_{NM,MN}
\end{array}\right)
+\left( \begin{array}{c}
T_{33}^{(H,V)} \\ T_{43}^{(H,V)}
\end{array}\right)
\,, \\
\left( \begin{array}{c}
t^{12}_{MM,NN} \\ t^{12}_{NM,MN}
\end{array}\right)
&=& \left( \begin{array}{cc}
T_{11}^{(H,V)} & T_{12}^{(H,V)} \\
T_{21}^{(H,V)} & T_{22}^{(H,V)}
\end{array}\right)^{-1}
\left( \begin{array}{c}
1 \\ 0
\end{array}\right)
\,, \\
\left( \begin{array}{c}
r^{22}_{MM,NN} \\ r^{22}_{NM,MN}
\end{array}\right)
&=& \left( \begin{array}{cc}
T_{31}^{(H,V)} & T_{32}^{(H,V)} \\
T_{41}^{(H,V)} & T_{42}^{(H,V)}
\end{array}\right)
\left( \begin{array}{c}
t^{12}_{MM,NN} \\ t^{12}_{NM,MN}
\end{array}\right)
\,.
\end{eqnarray}
Note that, in contrast to planarly or spherically layered media, $s$-
and $p$-polarised waves, represented by the $\vect{M}$ and $\vect{N}$
vector wave functions, becomes mixed for all angular momenta $n>0$.


\subsubsection{Spherically layered media}
\label{sec:sphericaldgf}

Similarly to the cylindrically layered case, we use the expansion of
the dyadic Green function in terms of vector wave functions.
The spherical vector wave functions are given by \cite{LiLW94b}
\begin{eqnarray}
\vect{M}_{\ueber{e}{o} ml}(k) &\!\!\!=&\!\!\! \mp \frac{m}{\sin\Theta}
z_l(kr) P_l^m(\cos\Theta)
\ueber{\sin}{\cos}m\varphi\,
\vect{e}_{\Theta}
-z_l(kr) \frac{dP_l^m(\cos\Theta)}{d\Theta}
\ueber{\cos}{\sin}m\varphi\,
\vect{e}_{\varphi} , \nonumber \\ \\
\vect{N}_{\ueber{e}{o} ml}(k) &\!\!\!=&\!\!\! \frac{l(l+1)}{kr}
z_l(kr)
P_l^m(\cos\Theta)
\ueber{\cos}{\sin}m\varphi\,
\vect{e}_r \nonumber \\ &\!\!\!&\!\!\! \hspace*{-2ex}
+\frac{z_l'(kr)}{kr} \left[ \frac{dP_l^m(\cos\Theta)}{d\Theta}
\ueber{\cos}{\sin}m\varphi\,
\vect{e}_{\Theta}
\mp\frac{m}{\sin\Theta} P_l^m(\cos\Theta)
\ueber{\sin}{\cos}m\varphi\,
\vect{e}_{\varphi} \right] .
\end{eqnarray}
With the help of those, the dyadic Green function for a dielectric
sphere is \cite{LiLW94b}
\begin{eqnarray}
\label{eq:g11}
\tens{G}^{(11)}(\vect{r},\vect{r}',\omega) &=&
\frac{i\mu_1k_1}{4\pi}
\sum\limits_{e,o} \sum\limits_{l=1}^\infty \sum\limits_{m=0}^l
(2-\delta_{l0}) \frac{2l+1}{l(l+1)} \frac{(l-m)!}{(l+m)!}
\nonumber \\ && \hspace*{-6ex} \times
\left[ r_s^{11} \vect{M}^{(1)}_{\ueber{e}{o}ml}(k_1) \otimes
\vect{M}^{'(1)}_{\ueber{e}{o}ml}(k_1)
+r_p^{11} \vect{N}^{(1)}_{\ueber{e}{o}ml}(k_1) \otimes
\vect{N}^{'(1)}_{\ueber{e}{o}ml}(k_1) \right] \,,\\
\tens{G}^{(21)}(\vect{r},\vect{r}',\omega) &=&
\frac{i\mu_1k_1}{4\pi}
\sum\limits_{e,o} \sum\limits_{l=1}^\infty \sum\limits_{m=0}^l
(2-\delta_{l0}) \frac{2l+1}{l(l+1)} \frac{(l-m)!}{(l+m)!}
\nonumber \\ && \hspace*{-6ex} \times
\left[ t_s^{21} \vect{M}_{\ueber{e}{o}ml}(k_2) \otimes
\vect{M}^{'(1)}_{\ueber{e}{o}ml}(k_1)
+t_p^{21} \vect{N}_{\ueber{e}{o}ml}(k_2) \otimes
\vect{N}^{'(1)}_{\ueber{e}{o}ml}(k_1) \right] \,,\\
\tens{G}^{(12)}(\vect{r},\vect{r}',\omega) &=&
\frac{i\mu_2k_2}{4\pi}
\sum\limits_{e,o} \sum\limits_{l=1}^\infty \sum\limits_{m=0}^l
(2-\delta_{l0}) \frac{2l+1}{l(l+1)} \frac{(l-m)!}{(l+m)!}
\nonumber \\ && \hspace*{-6ex} \times
\left[ t_s^{12} \vect{M}^{(1)}_{\ueber{e}{o}ml}(k_1) \otimes
\vect{M}'_{\ueber{e}{o}ml}(k_2)
+t_p^{12} \vect{N}^{(1)}_{\ueber{e}{o}ml}(k_1) \otimes
\vect{N}'_{\ueber{e}{o}ml}(k_2) \right] \,,\\
\tens{G}^{(22)}(\vect{r},\vect{r}',\omega) &=&
\frac{i\mu_2k_2}{4\pi}
\sum\limits_{e,o} \sum\limits_{l=1}^\infty \sum\limits_{m=0}^l
(2-\delta_{l0}) \frac{2l+1}{l(l+1)} \frac{(l-m)!}{(l+m)!}
\nonumber \\ && \hspace*{-6ex} \times
\left[ r_s^{22} \vect{M}_{\ueber{e}{o}ml}(k_2) \otimes
\vect{M}'_{\ueber{e}{o}ml}(k_2)
+r_p^{22} \vect{N}_{\ueber{e}{o}ml}(k_2) \otimes
\vect{N}'_{\ueber{e}{o}ml}(k_2) \right] \,.
\end{eqnarray}
For the following, it is convenient to define the Ricatti functions
\begin{equation}
\eta_l(k_iR):=\frac{1}{x} \left.\frac{d[xj_l(x)]}{dx}\right|_{x=k_iR}
\\,\qquad
\xi_l(k_iR):=\frac{1}{x}
\left.\frac{d[xh^{(1)}_l(x)]}{dx}\right|_{x=k_iR} \,.
\end{equation}
In terms of those, the Mie scattering coefficients read
\begin{eqnarray}
r_s^{(11)} &=&
-\frac{\mu_1k_2\eta_l(k_2R)j_l(k_1R)-\mu_2k_1\eta_l(k_1R)j_l(k_2R)}%
{\mu_1k_2\eta_l(k_2R)h^{(1)}_l(k_1R)-\mu_2k_1\xi_l(k_1R)j_l(k_2R)}
\,,\\
r_p^{(11)} &=&
-\frac{\mu_1k_2\eta_l(k_1R)j_l(k_2R)-\mu_2k_1\eta_l(k_2R)j_l(k_1R)}%
{\mu_1k_2\xi_l(k_1R)j_l(k_2R)-\mu_2k_1\eta_l(k_2R)h^{(1)}_l(k_1R)} \,,
\end{eqnarray}
\begin{eqnarray}
r_s^{(22)} &=&
-\frac{\mu_1k_2\xi_l(k_2R)h^{(1)}_l(k_1R)-\mu_2k_1\xi_l(k_1R)h^{(1)}
_l(k_2R)}%
{\mu_1k_2\eta_l(k_2R)h^{(1)}_l(k_1R)-\mu_2k_1\xi_l(k_1R)j_l(k_2R)}
\,,\\
r_p^{(22)} &=&
-\frac{\mu_1k_2\xi_l(k_1R)h^{(1)}_l(k_2R)-\mu_2k_1\xi_l(k_2R)h^{(1)}
_l(k_1R)}%
{\mu_1k_2\xi_l(k_1R)j_l(k_2R)-\mu_2k_1\eta_l(k_2R)h^{(1)}_l(k_1R)} \,,
\end{eqnarray}
\begin{eqnarray}
t_s^{(12)} &=&
\frac{\mu_1k_2[\eta_l(k_2R)h_l^{(1)}(k_2R)-\xi_l(k_2R)j_l(k_2R)]}%
{\mu_1k_2\eta_l(k_2R)h^{(1)}_l(k_1R)-\mu_2k_1\xi_l(k_1R)j_l(k_2R)}
\,,\\
t_p^{(12)} &=&
\frac{\mu_1k_2[\xi_l(k_2R)j_l(k_2R)-\eta_l(k_2R)h_l^{(1)}(k_2R)]}%
{\mu_1k_2\xi_l(k_1R)j_l(k_2R)-\mu_2k_1\eta_l(k_2R)h^{(1)}_l(k_1R)} \,,
\end{eqnarray}
\begin{eqnarray}
t_s^{(21)} &=&
\frac{\mu_2k_1[\eta_l(k_1R)h_l^{(1)}(k_1R)-\xi_l(k_1R)j_l(k_1R)]}%
{\mu_1k_2\eta_l(k_2R)h^{(1)}_l(k_1R)-\mu_2k_1\xi_l(k_1R)j_l(k_2R)}
\,,\\
t_p^{(21)} &=&
\frac{\mu_2k_1[\xi_l(k_1R)j_l(k_1R)-\eta_l(k_1R)h_l^{(1)}(k_1R)]}%
{\mu_1k_2\xi_l(k_1R)j_l(k_2R)-\mu_2k_1\eta_l(k_2R)h^{(1)}_l(k_1R)} \,.
\end{eqnarray}
Note that the transmission coefficients $t_{s,p}$ can be further
simplified by using the Wronski determinant between spherical Bessel
and Hankel functions.

\subsection{Born series expansion}
\label{sec:bornseries}

In most situations, the geometric arrangement of dielectric bodies is
not symmetrical enough to yield a separable Helmholtz operator in
which case an expansion into vector wave functions would be feasible
(Sec.~\ref{sec:layeredmedia}). Instead, we will describe an iterative
method known from quantum mechanics and quantum field theory as Born
(or Dyson) series expansion of the dyadic Green function. This method
applies to an arbitrary arrangement of dielectric bodies but, in
general, converges quickly only if the dielectric contrast between
bodies and the surrounding material is sufficiently small.

\begin{figure}[!t!]
\centerline{\includegraphics[width=5cm]{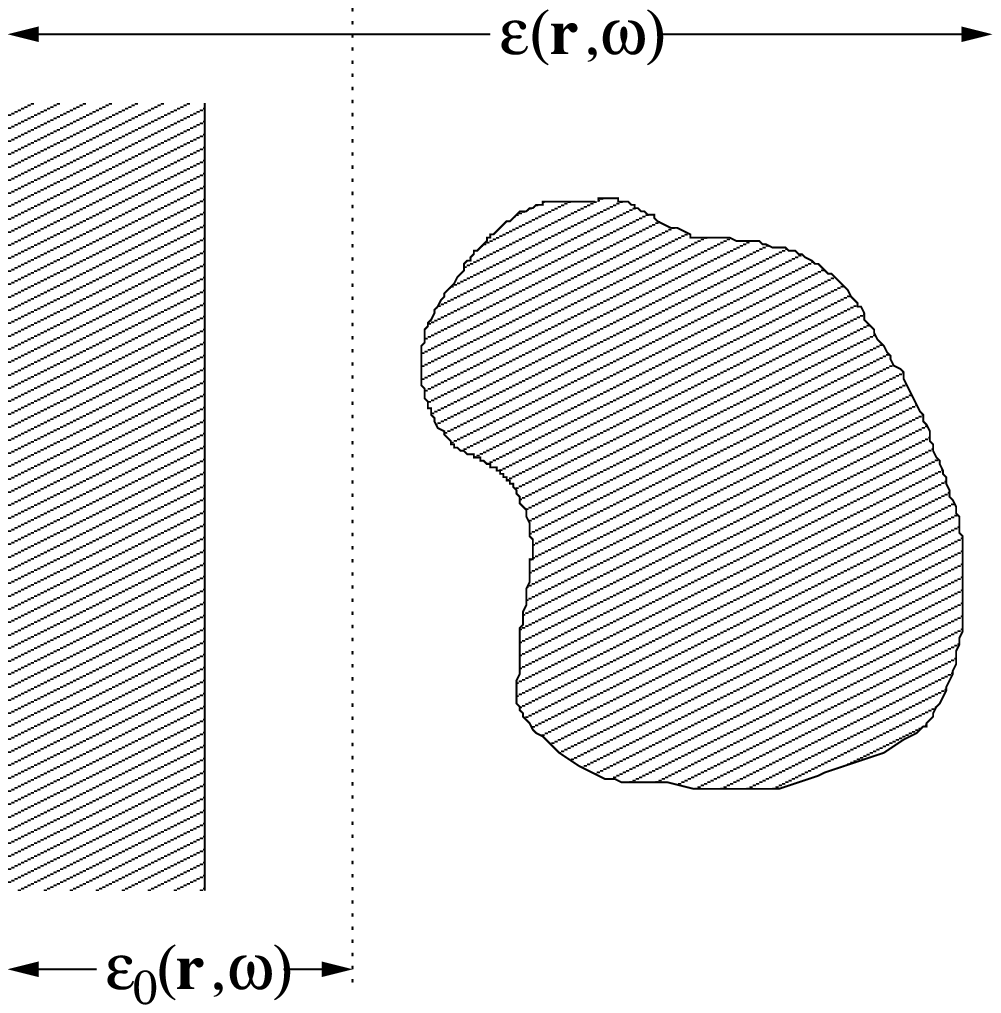}
}
\caption{\label{fig:borndecomposition} Dielectric body of arbitrary
shape in front of a dielectric wall. The whole arrangement of bodies
is described by a permittivity $\varepsilon(\vect{r},\omega)$. The DGF
for the half-space alone [with permittivity
$\varepsilon_0(\vect{r},\omega)$] is known and can be used as starting
point for the Born series.}
\end{figure}

Let us assume that from an arrangement of dielectric bodies, described
by a dielectric permittivity $\varepsilon(\vect{r},\omega)$, one can
separate a part whose DGF $\tens{G}^{(0)}(\vect{r},\vect{r}',\omega)$
is analytically known (e.g. vacuum, bulk material or layered media)
and is described by a permittivity $\varepsilon_0(\vect{r},\omega)$
(see, for example, Fig.~\ref{fig:borndecomposition}).
%
%
The dyadic Green functions are solutions to the respective Helmholtz
equations
\begin{eqnarray}
\curl\curl\tens{G}(\vect{r},\vect{r}',\omega) -\frac{\omega^2}{c^2}
\varepsilon(\vect{r},\omega) \tens{G}(\vect{r},\vect{r}',\omega)
&=& \tens{\delta}(\vect{r}-\vect{r}') \,,\\
\curl\curl\tens{G}^{(0)}(\vect{r},\vect{r}',\omega)
-\frac{\omega^2}{c^2}
\varepsilon_0(\vect{r},\omega)
\tens{G}^{(0)}(\vect{r},\vect{r}',\omega) 
&=& \tens{\delta}(\vect{r}-\vect{r}') \,.
\end{eqnarray}
Subtracting both equations from one another, we find that the
difference between both DGFs [scattering Green function
$\tens{G}^{(S)}(\vect{r},\vect{r}',\omega)$] solves the
inhomogeneous Helmholtz equation
\begin{eqnarray}
\label{eq:born1}
&&
\curl\curl\tens{G}^{(S)}(\vect{r},\vect{r}',\omega)
-\frac{\omega^2}{c^2} \varepsilon_0(\vect{r},\omega)
\tens{G}^{(S)}(\vect{r},\vect{r}',\omega)  \nonumber \\
&&= \frac{\omega^2}{c^2} \delta\varepsilon(\vect{r},\omega) \left[
\tens{G}^{(0)}(\vect{r},\vect{r}',\omega) +
\tens{G}^{(S)}(\vect{r},\vect{r}',\omega) \right]
\end{eqnarray}
where $\delta\varepsilon(\vect{r},\omega)\equiv$
$\varepsilon(\vect{r},\omega)-\varepsilon_0(\vect{r},\omega)$ is the
perturbation from the permittivity $\varepsilon_0(\vect{r},\omega)$.
Equation~(\ref{eq:born1}) bears some resemblance to the Helmholtz
equation (\ref{eq:a1}) for the electric field, with the current
density $i\mu_0\omega\vect{j}(\vect{r},\omega)$ being replaced by the
rhs of Eq.~(\ref{eq:born1}). Its formal solution is therefore
\begin{equation}
\label{eq:born2}
\tens{G}^{(S)}(\vect{r},\vect{r}',\omega) = \int d^3s \,
\tens{G}^{(0)}(\vect{r},\vect{s},\omega) \cdot \frac{\omega^2}{c^2}
\delta\varepsilon(\vect{s},\omega) \left[
\tens{G}^{(0)}(\vect{s},\vect{r}',\omega) +
\tens{G}^{(S)}(\vect{s},\vect{r}',\omega) \right] \,.
\end{equation}
Because the unknown scattering DGF
$\tens{G}^{(S)}(\vect{r},\vect{r}',\omega)$ appears on both sides of
Eq.~(\ref{eq:born2}), making it a Fredholm integral equation of the
second kind, one can solve it iteratively as \cite{0020}
\begin{eqnarray}
\label{eq:bornseries}
\tens{G}^{(S)}(\vect{r},\vect{r}',\omega) &=& \frac{\omega^2}{c^2}
\int
d^3s'\, \tens{G}^{(0)}(\vect{r},\vect{s}',\omega) \cdot
\delta\varepsilon(\vect{s}',\omega)
\tens{G}^{(0)}(\vect{s}',\vect{r}',\omega)
\nonumber \\ && \hspace*{-15ex} +
\left( \frac{\omega^2}{c^2} \right)^2 \iint d^3s'd^3s''\,
\tens{G}^{(0)}(\vect{r},\vect{s}',\omega) \cdot
\delta\varepsilon(\vect{s}',\omega)
\tens{G}^{(0)}(\vect{s}',\vect{s}'',\omega) \cdot
\delta\varepsilon(\vect{s}'',\omega)
\tens{G}^{(0)}(\vect{s}'',\vect{r}',\omega)
\nonumber \\ && \hspace*{-15ex} + \ldots
\end{eqnarray}

\begin{figure}[!t!]
\centerline{\includegraphics[width=0.6\columnwidth]{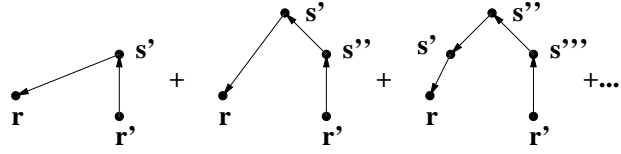}}
\caption{\label{fig:born} Diagrammatic representation of the Born
series, Eq.~(\ref{eq:bornseries}). Each arrow corresponds to the
propagation according to the Green tensor
$\tens{G}^{(0)}$, each intermediate vertex at position
$\vect{s}^{(i)}$ contributes with a weight factor
$\frac{\omega^2}{c^2}\delta\varepsilon(\vect{s}^{(i)},\omega)$. The
integration is over all intermediate positions $\vect{s}^{(i)}$.}
\end{figure}

Equation~(\ref{eq:bornseries}) is known as the Born (or Dyson) series
expansion of the scattering dyadic Green function
$\tens{G}^{(S)}(\vect{r},\vect{r}',\omega)$. It has a simple
diagrammatic representation as shown in Fig.~\ref{fig:born}.
%
%
The field propagates from the source point $\vect{r}'$ to the
observation point $\vect{r}$ via the intermediate positions
$\vect{s}^{(i)}$. Each of the arrows represents the dyadic Green
function $\tens{G}^{(0)}$ (which we assumed to be analytically known),
and each intermediate position (or vertex) $\vect{s}^{(i)}$ carries a
weight factor
$\frac{\omega^2}{c^2}\delta\varepsilon(\vect{s}^{(i)},\omega)$. The
intermediate positions are integrated over, and the number of those
points increases with the order of the iteration.

As the Born series (\ref{eq:bornseries}) is a perturbation series in
$\delta\varepsilon(\vect{r},\omega)$, it is clear that fast
convergence is only guaranteed for small enough permittivity (or
equivalently, refractive index) contrast. It should be noted, however,
that the series does always eventually converge to its unique solution
$\tens{G}^{(S)}(\vect{r},\vect{r}',\omega)$.


\subsection{Local-field corrected Green tensors}
\label{sec:localfield}

\begin{figure}[t]
\centerline{\includegraphics[width=0.8\textwidth]{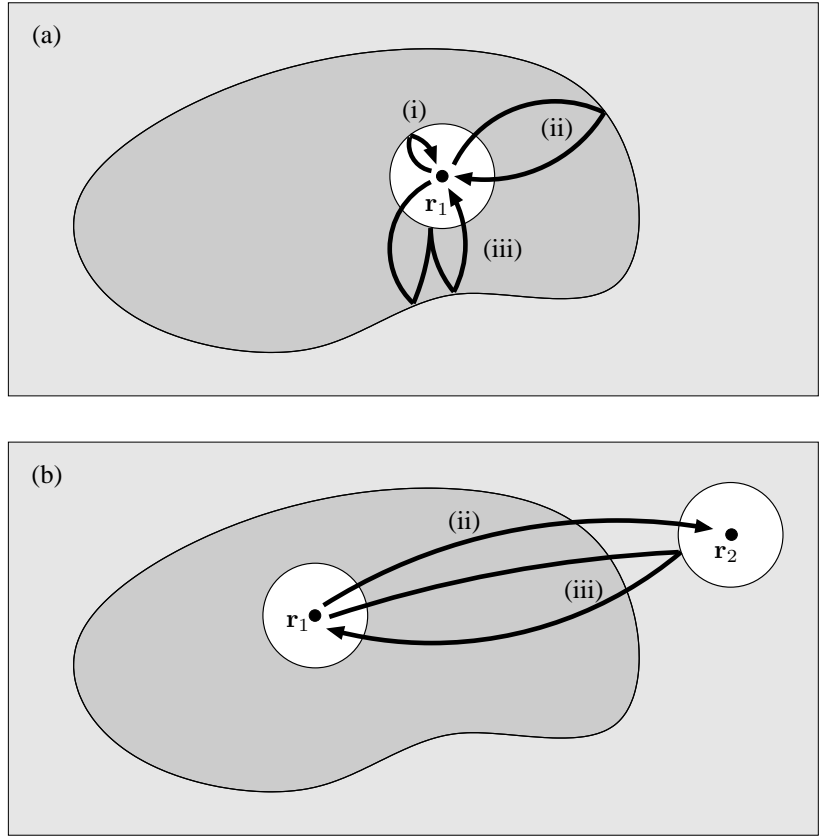}}
\caption{
\label{fig:localfield} 
Real-cavity model for applying local-field corrections to (a) single-
and (b) two-point DGFs. Three typical contributions (i)--(iii) to the 
corrected DGFs are schematically indicated.
}
\end{figure}

While the DGF $\tens{G}(\vect{r},\vect{r}',\omega)$ connects the
macroscopic electric field $\vect{E}(\vect{r})$ with a macroscopic
source $\vect{j}(\vect{r}')$ [recall Eq.~(\ref{eq:fieldsource})], we
are often interested in the coupling of the local electromagnetic
field to microscopic sources such as atoms. For atoms which
are embedded in a magnetoelectric medium, this difference between
microscopic and macroscopic quantities leads to local-field
corrections which can be implemented via the real-cavity
model: One assumes that both field point $\vect{r}_1$ and source point
$\vect{r}_2$ are not situated directly in the medium but surrounded by
small free-space cavities of radius $R_\mathrm{cav}$ (cf.
Fig.~\ref{fig:localfield}). Modifying the original permittivity
$\varepsilon(\vect{r},\omega)$ and (inverse) permeability
$\kappa(\vect{r},\omega)$ describing the present media to
\begin{equation}
\label{eq:mediumlocalfield}
\varepsilon_\mathrm{loc}(\vect{r},\omega),
\kappa_\mathrm{loc}(\vect{r},\omega)
=\begin{cases}
1&\mbox{if }|\vect{r}-\vect{r}_1|<R_\mathrm{cav}
\mbox{ or }|\vect{r}-\vect{r}_2|<R_\mathrm{cav},\\
\varepsilon(\vect{r},\omega),\kappa(\vect{r},\omega)&
\mbox{else},
\end{cases}
\end{equation}
the required local-field corrected DGF $\tens{G}_\mathrm{loc}$ can be
found as the solution to
\begin{equation}
\label{eq:defGloc}
\biggl[\grad\times\kappa_\mathrm{loc}(\vect{r},\omega)\grad\times
-\frac{\omega^2}{c^2}\,\varepsilon_\mathrm{loc}(\vect{r},\omega)
\biggr]
\tens{G}(\vect{r},\vect{r}',\omega) =
\ten{\delta}(\vect{r}-\vect{r}').
\end{equation}
The real-cavity model thus accounts for the fact that an atom
occupies some space within the host medium, where the cavity radius is
of the order of one half the lattice constant of the latter. 

For magnetoelectrics (but not for metals), the condition
$|\sqrt{\epsilon\mu}|R_\mathrm{cav}\omega/c$ $\!\ll$ $\!1$ is
typically valid for the relevant frequencies of the electromagnetic
field. In this case, the effect of the introduced cavities can be
studied in a perturbative way by means of the spherical DGFs given in
Sec.~\ref{sec:sphericaldgf}. One finds that the local-field corrected
single-point DGF is given by \cite{Tomas01,Dung06}
\begin{multline}
\label{eq:G1loc}
\ten{G}^{(S)}_\mathrm{loc}
 (\vect{r}_1,\vect{r}_1,\omega)
=\frac{\omega}{6\pi c}\,\biggl\{
 \frac{3(\varepsilon_1-1)}{2\varepsilon_1+1}\,
 \frac{c^3}{\omega^3R_\mathrm{cav}^3}
+\frac{9[\varepsilon_1^2(5\mu_1-1)-3\varepsilon_1-1]}
 {5(2\varepsilon_1+1)^2}\,
 \frac{c}{\omega R_\mathrm{cav}}\\
 +\mi\biggl[\frac{9\varepsilon_1n_1^3}
 {(2\varepsilon_1+1)^2}
 -1\biggr]\biggr\}\,\ten{1}
 +\biggl(\frac{3\varepsilon_1}
 {2\varepsilon_1+1}\biggr)^2
 \ten{G}^{(S)}(\vect{r}_1,\vect{r}_1,\omega)
\end{multline}
where $\ten{G}$ is the uncorrected DGF, $\varepsilon_i$ $\!=$
$\!\varepsilon(\vec{r}_i,\omega)$ denotes the permittvity of
the unperturbed host medium at the source and field points (similarly
for $\mu$) and $n_i$ $\!=$ $\!\sqrt{\varepsilon_i\mu_i}$ is the
respective refractive index. The first term in Eq.~(\ref{eq:G1loc})
represents contributions to the electric field which are reflected at
the interior of the cavity and never reach the surrounding medium
[type (i) in Fig.~\ref{fig:localfield}(a)]. The second term
represents contributions which are transmitted into the host medium
and, after possible transmissions and reflections, transmitted back
into the cavity [type (ii)]; each of the transmissions through the
cavity surface gives rise to one factor in round brackets. Processes
where the field is backreflected from the outside of the cavity [type
(iii)] have been neglected since they are of higher order in the small
parameter $|\sqrt{\epsilon\mu}|R_\mathrm{cav}\omega/c$.
Similarly, the local-field corrected two-point DGF is found to be
\cite{Dung06}
\begin{equation}
\label{eq:G2loc}
\ten{G}_\mathrm{loc}
 (\vect{r}_1,\vect{r}_2,\omega)
=\frac{3\varepsilon_1}{2\varepsilon_1+1}\,
 \frac{3\varepsilon_2}{2\varepsilon_2+1}\,
 \ten{G}(\vect{r}_1,\vect{r}_2,\omega) 
 \quad\mbox{for }\vect{r}_1\neq\vect{r}_2.
\end{equation}
Equation~(\ref{eq:G2loc}) represents the transmission of the field out
of the first cavity, transmissions and reflections within the host
medium, followed by a transmission into the second cavity [type (ii)
of Fig.~\ref{fig:localfield}(b)]; each of the transmissions at the
cavity surfaces gives rise to one of the two factors. Again,
reflections at the outside of either cavity have been neglected [type
(iii)].

When describing magnetic interactions, the curl of the DGF typically
arises. Local-field corrections in such situations cannot simply be
obtained by taking the curl of Eqs.~(\ref{eq:G1loc}) and
(\ref{eq:G2loc}) above, since the implementation of the local-field
correction via the real-cavity model does not commute with this vector
operation. Instead, the local-field correction has to applied after
taking the curl. One finds the single-point DGF \cite{Hassan08}
\begin{multline}
\label{eq:G1magloc}
\bm{\nabla}\vprod
 \ten{G}^{(S)}(\vect{r}_1,\vect{r}_1,\omega)
 \vprod\overleftarrow{\bm{\nabla}}'
 \big|_\mathrm{loc}\\
=-\frac{\omega^3}{2\pi c^2}\,\biggl\{
 \frac{\mu_1\!-\!1}{2\mu_1\!+\!1}\,
 \frac{c^3}{\omega^3R_\mathrm{cav}^3}
 +\frac{3}{5}\,\frac{\mu_1^2(5\varepsilon_1\!-\!1)
 \!-\!3\mu_1\!-\!1}{(2\mu_1\!+\!1)^2}
 \frac{c}{\omega R_\mathrm{cav}}
 +\mi\biggl[\frac{3\mu_1 n_1^3}
 {(2\mu_1\!+\!1)^2}-\frac{1}{3}\biggr]\biggr\}\ten{I}\\
+\biggl(\frac{3}{2\mu_1\!+\!1}\biggr)^2
 \bm{\nabla}\vprod
 \ten{G}^{(S)}(\vect{r}_1,\vect{r}_1,\omega)
 \vprod\overleftarrow{\bm{\nabla}}'
\end{multline}
and the two-point DGFs \cite{Hassan08}
\begin{gather}
\label{eq:G2magelloc}
\bm{\nabla}\vprod\ten{G}(\vect{r}_1,\vect{r}_2,\omega)
 \big|_\mathrm{loc}
=\frac{3}{2\mu_1+1}\,
 \frac{3\varepsilon_2}{2\varepsilon_2+1}\,
 \ten{G}(\vect{r}_1,\vect{r}_2,\omega)
 \quad\mbox{for }\vect{r}_1\neq\vect{r}_2,\\
\label{eq:G2elmagloc}
\ten{G}(\vect{r}_1,\vect{r}_2,\omega)
 \vprod\overleftarrow{\bm{\nabla}}'\big|_\mathrm{loc}
=\frac{3\varepsilon_1}{2\varepsilon_1+1}\,
 \frac{3}{2\mu_2+1}\,\ten{G}(\vect{r}_1,\vect{r}_2,\omega)
 \quad\mbox{for }\vect{r}_1\neq\vect{r}_2,\\
\label{eq:G2magmagloc}
\bm{\nabla}\vprod\ten{G}(\vect{r}_1,\vect{r}_2,\omega)
 \vprod\overleftarrow{\bm{\nabla}}'\big|_\mathrm{loc}
=\frac{3}{2\mu_1+1}\,
 \frac{3\varepsilon_2}{2\varepsilon_2+1}\,
 \ten{G}(\vect{r}_1,\vect{r}_2,\omega)
 \quad\mbox{for }\vect{r}_1\neq\vect{r}_2.
\end{gather}

The behaviour of the local-field corrected Green tensors under a
duality transformation (cf. Sec.~\ref{sec:dualgreen}) can be easily
derived by combining Eqs.~(\ref{eq:G1loc})--(\ref{eq:G2magelloc})
with Eqs.~(\ref{dgftrans1})--(\ref{dgftrans4}):
\begin{eqnarray}
\label{dgfloctrans1}
\frac{\omega^2}{c^2}\,
 \ten{G}^{(S)\star}_\mathrm{loc}(\vect{r}_1,\vect{r}_1,\omega)
 &=&-\curl\ten{G}^{(S)}(\vect{r}_1,\vect{r}_1,\omega)\vprod
 \overleftarrow{\grad}'\bigr|_\mathrm{loc},\\
\label{dgfloctrans2}
\curl\ten{G}^{(S)\star}(\vect{r}_1,\vect{r}_1,\omega)\vprod
 \overleftarrow{\grad}'\bigr|_\mathrm{loc}
&=&-\frac{\omega^2}{c^2}\,
 \ten{G}^{(S)}_\mathrm{loc}(\vect{r}_1,\vect{r}_1,\omega),\\
\label{dgfloctrans3}
\frac{\omega^2}{c^2}\,
 \ten{G}^{\star}_\mathrm{loc}(\vect{r}_1,\vect{r}_2,\omega)
 &=&-\curl\ten{G}(\vect{r}_1,\vect{r}_2,\omega)\vprod
 \overleftarrow{\grad}'\bigr|_\mathrm{loc}
 \quad\mbox{for }\vect{r}_1\neq\vect{r}_2,\\
\label{dgfloctrans4}
\curl\ten{G}^\star(\vect{r}_1,\vect{r}_2,\omega)\vprod
 \overleftarrow{\grad}'\bigr|_\mathrm{loc}
&=&-\frac{\omega^2}{c^2}\,
 \ten{G}_\mathrm{loc}(\vect{r}_1,\vect{r}_2,\omega)
 \quad\mbox{for }\vect{r}_1\neq\vect{r}_2,\\
\label{dgfloctrans5}
\curl\ten{G}^\star(\vect{r},\vect{r}',\omega)\bigr|_\mathrm{loc}
&=&-\ten{G}(\vect{r}_1,\vect{r}_2,\omega)\vprod\overleftarrow{\grad}'
 \bigr|_\mathrm{loc}\quad\mbox{for }\vect{r}_1\neq\vect{r}_2,\\
\label{dgfloctrans6}
\ten{G}^\star(\vect{r},\vect{r}',\omega)\vprod\overleftarrow{\grad}'
 |_\mathrm{loc}
&=&-\curl\ten{G}(\vect{r}_1,\vect{r}_2,\omega)\bigr|_\mathrm{loc}
 \quad\mbox{for }\vect{r}_1\neq\vect{r}_2.
\end{eqnarray}


\end{document}